\documentclass[a4paper,12pt,twoside]{ThesisStyle}

\usepackage{setspace}
\usepackage[english]{babel}
\usepackage{cite}

\usepackage{amstext}
\usepackage{amsmath}
\usepackage{amssymb}
\usepackage{multirow}
\usepackage[section]{placeins}

\usepackage{epsfig}
\usepackage{epstopdf}
\usepackage{graphicx,color}  %
\usepackage{bm}  %

\usepackage{subfigure}
\usepackage{multirow}
\usepackage{color}
\usepackage{cases}

\usepackage[shortcuts]{extdash} 

\usepackage{dsfont}

\usepackage{MnSymbol} 

\allowdisplaybreaks[1] 

\usepackage{fancyhdr}
\usepackage[left=30mm,right=45mm,top=15mm,bottom=62mm,includehead,headheight=10pt,headsep=15pt]{geometry}

\usepackage{emptypage} 

\usepackage[hyphens]{url}

\definecolor{ColorToUse}{rgb}{0.54 , 0.27 , 0.08 }

\usepackage[colorlinks = true,
            linkcolor = ColorToUse,
            urlcolor  = ColorToUse,
            citecolor = ColorToUse,
            anchorcolor = ColorToUse
,            ,breaklinks]{hyperref}

\begin{document}

\frontmatter

\pagestyle{plain}

\begin{titlepage}

\thispagestyle{empty}

\begin{center}

\begin{tabular}{c}
\hline \\
\LARGE \textbf{Two- and three-dimensional few-body } \\ 
\LARGE \textbf{systems in the universal regime} \\ 
\\
\hline \\
\end{tabular}

\vspace{1.0cm}

\LARGE\textbf{Filipe Furlan Bellotti}\\
\vspace{1cm}
\Large{ Department of Physics and Astronomy \\
 Aarhus University, Denmark }\\

\vspace{1.0cm}

    
    \begin{figure}[h]
    	\centering
		\includegraphics[width=0.6 \textwidth]{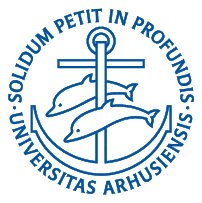};
    \end{figure}


\LARGE\textbf{Dissertation for the degree of \\
Doctor of Philosophy} \\
\vspace{0.5cm}
\LARGE October 2014\\

\end{center}
\end{titlepage} 


\newpage 
\thispagestyle{empty}

\vspace*{12cm}

\noindent Filipe Furlan Bellotti\\
Department of Physics and Astronomy\\
Aarhus University\\
Ny Munkegade, Bldg.~1520\\
8000 {\AA}rhus C\\
Denmark\\
E-mail: filipe@phys.au.dk\\

\newpage


\fbox{
\parbox{\linewidth}{
\vspace{.3cm}
\centering
\begin{minipage}{1.0\linewidth}
\noindent This dissertation has been submitted to the Faculty of Science and Technology at Aarhus University, Denmark, in partial fulfillment of the requirements for the PhD degree in physics. The work presented has been performed in the period from August 2012 to July 2014 under the supervision of Aksel S.\ Jensen from Aarhus University/Denmark and Tobias Frederico from Instituto Tecnol\' ogico de Aeron\' autica/Brazil. 
The work is the result of an agreement on joint supervision of doctoral studies and was carried out at the Department of Physics and Astronomy in Aarhus and at the Department for Natural Sciences in S\~ao Jos\' e dos Campos. 
 
\end{minipage}
\vspace{0.3cm}
}
}

\begin{flushright}
\begin{spacing}{1}
\mbox{}\vfill
{\sffamily\itshape
"...
A boy walks and walking he reaches the wall. \\
And  there, right ahead, the future is waiting for us. \\
And the future is a spacecraft which we try to fly \\
It has neither moment, nor compassion, nor time to arrive. \\
Without asking permission it changes our lives and next it invites us to either laugh or cry. \\

In this road, we are not supposed to either know or see what is coming. \\
Nobody does quite know for sure where it ends. ..."} \\
--- \textsc{Toquinho / Vinicius de Morais / Maurizio Fabrizio / Guido Morra \\
Free translation of part of {\it Aquarela}}
\end{spacing}
\end{flushright}

\begin{spacing}{1.3}

\chapter{Acknowledgment}
The project, which is ended by this thesis, has started approximately three years ago. During the Master studies, my advisor Prof. Tobias Frederico (ITA/Brazil) trusted and gave me the opportunity to visit Aarhus University for three months. I have been very well received and the visit was very fruitful. After getting my M.Sc. degree, Prof. Aksel S. Jensen (AU/Denmark) proposed the joint supervision of the PhD studies. Since then, I have two official advisors and I would like to thank them: Profs. Tobias and Aksel. They have taught me about Physics, Mathematics, numerical calculation, writing papers and reports, oral presentation, politics in the academic area and other uncountable subjects. In practice, they have given me guidelines to become a good, independent and honest Scientist. Thank you so much for all the time spent and patience with me. I am very proud in saying that you were my supervisors.

In practice, I had also three more advisors, who I should properly acknowledge: Dmitri Fedorov and Nikolaj Zinner from from Aarhus University (AU/Denmark) and Marcelo Yamashita from Instituto de F\'isica Te\'orica (IFT-UNESP/Brazil). They were always available to discuss about any topic and their suggestions and comments have improved a lot my formation. In particular, Dmitri has always been exceptionally supportive during my visits to Aarhus, as well as the secretaries Brigitte Henderson, Trine Binderup and Karin Vittrup. They have always helped me with the administrative and logistic issues. A special thank goes for Tobias and Nikolaj for having proofread the thesis.

The enjoyable time I had in Aarhus was in great extension due to the company of Jeremy Armstrong, Artem Volosniev, Oleksandr Marchukov and Jakob Knorborg, who were constantly present apart of the working time. 

At my work place in Brazil, Instituto de Fomento e Coordena\c c\~ao Industrial, I would like to firstly thank Cel. Eng. Augusto Luiz de Castro Otero, who gave me conditions to study without quitting my job. I also appreciate the support from my boss, Dr. Cesar Augusto Botura, and the advices from Dr. Pedro Jos\' e Pomp\' eia.

My parents are the basis of my journey and they know how grateful I am. The time spent in Aarhus has contributed a lot for my personal and professional development, but it was one of the toughest periods I have ever experienced, since I had to spent almost half of the first year of my daughter away from my family. Therefore, the most special acknowledgement goes to my beloved wife M\^ onica, who has also passed through very hard times in order to support me in this project. Without her support, this project would not have been started, constructed and concluded. I am very thankful to my young Cecilia, who has showed me that the mathematical and abstract concept of infinity has a meaning: is the love I feel by her.

\chapter{Abstract}
Macro properties of cold atomic gases are driven by few-body correlations, 
even if the gas has thousands of particles.
Quantum systems composed of two and three particles with attractive 
zero\=/range pairwise interactions are considered for general masses 
and interaction strengths in two and three dimensions (2D and 3D). 
The Faddeev decomposition is used to derive the equations for the 
bound state, 
which is the starting point for the investigation of universal properties 
of few\=/body systems, i.e. those that all potentials with the same 
physics at low energy are able to describe in a model\=/independent form.
In 2D, the number of bound states in a three\=/body system increases 
without bound as the mass of one particle becomes much lighter than the 
other two. 
The analytic form of an effective potential between the heavy particles 
explains the mass\=/dependence on the number of bound energy levels. An exact 
analytic expression for the large\=/momentum asymptotic behavior of the 
spectator function in the Faddeev equation is presented. The spectator function and its 
asymptotic form define the two- and three\=/body contact parameters. 
The two\=/body parameter is found to be independent of the quantum 
state in some specific 2D systems. 
The 2D and 3D momentum distributions have a distinct sub\=/leading form 
whereas the 3D term depends on the mass of the particles.
A model that interpolates between 2D and 3D is proposed and a sharp 
transition in the energy spectrum of three-body systems is found.

\chapter{Resum\'e}
Makroskopiske egenskaber af ultrakolde atomare gasser styres af 
f\aa-legeme korrelationer, det p\aa~trods af at gassen har 
tusindvis af partikler.
Vi studerer kvantesystemer best\aa ende af to og tre partikler med attraktive
to-partikel vekselvirkninger med nul r\ae kkevidde for generelle 
partikelmasser og vekselvirkningsstyrker i to og tre dimensioner (2D og 3D).
Vi benytter Faddeev dekomposition til at udlede ligningerne for 
bundne tilstande, hvilket er udgangspunktet for studier af universielle
egenskaber ved f\aa-legeme systemer. Dvs. de egenskaber som er uafh\ae ngig 
af den specifikke model for potentialer der giver samme fysik ved lav energi.
I 2D vokser antallet af bundne tilstande i et tre-partikel system 
uden gr\ae nse n\aa r en af de tre masser er meget lettere end de
andre to. Et analytisk udtryk for et effektivt potential mellem
de to tunge partikler kan forklare hvorledes antal 
bundne energiniveauer afh\ae nger af massen.
Vi udleder et analytisk udtryk for den s\aa kaldte tilskuer-funktion fra
Faddeev dekompositionen i den gr\ae nse hvor impulsen bliver stor.
Denne tilskuer-funktionen og den asymptotiske opf\o rsel benyttes til 
at bestemme to- og tre-partikel kontakt-parameteren.
To-partikel kontakt parameteren viser sig at v\ae re uafh\ae ngig af
kvantetilstand i nogle bestemte 2D systemer.
Impulsfordelingen i 2D og 3D har en karakteristisk opf\o rsel 
n\aa r man betragter den f\o rste korrektion til den ledende orden 
og i 3D f\aa r man her en korrektion der afh\ae nger af massen af 
partiklerne.
Vi foresl\aa r en model der interpolerer mellem 2D og 3D gr\ae nserne 
og finder en veldefineret og skarp overgang i energispektret for
bundne tilstande af tre partikler.

\end{spacing}

\tableofcontents
\addcontentsline{toc}{chapter}{Table of Contents}

\begin{spacing}{1.3}

\chapter{List of Publications} \label{sec1.1}
\vspace{-0.5cm}
\begin{spacing}{1.0}
\noindent
Bellotti, F. F.; Frederico, T.; Yamashita, M. T.; Fedorov, D. V.; Jensen, A. S.; Zinner,
N. T. Scaling and universality in two dimensions: three-body bound states with
short-ranged interactions. {\bf Journal of Physics B} , v. 44, n. 20, p. 205302, 2011. 
Labtalk: \\*  \url{ http://iopscience.iop.org/0953-4075/labtalk-article/47307}. \\

\noindent
Bellotti, F. F.; Frederico, T.; Yamashita, M. T.; Fedorov, D. V.; Jensen, A. S.; Zinner, N. T.
Supercircle description of universal three-body states in two dimensions. {\bf Physical Review A}, v. 85, p. 025601, 2012. \\

\noindent
Bellotti, F. F.; Frederico, T.; Yamashita, M. T.; Fedorov, D. V.; Jensen, A. S.; Zinner,
N. T. Dimensional effects on the momentum distribution of bosonic trimer states.
{\bf Physical Review A}, v. 87, n. 1, p. 013610, Jan. 2013. \\

\noindent
Bellotti, F. F.; Frederico, T.; Yamashita, M. T.; Fedorov, D. V.; Jensen, A. S.; Zinner,
N. T. Mass-imbalanced three-body systems in two dimensions. {\bf Journal of Physics B},
 v. 46, n. 5, p. 055301, May 2013. 
Labtalk: \\* \url{ http://iopscience.iop.org/0953-4075/labtalk-article/52615}. \\

\noindent
Yamashita, M. T.; Bellotti, F. F.; Frederico, T.; Fedorov, D. V.; Jensen, A. S.; Zinner,
N. T. Single-particle momentum distributions of efimov states in mixed-species systems.
{\bf Physical Review A}, v. 87, p. 062702, Jun 2013. \\

\noindent 
Bellotti, F. F.; Frederico, T.; Yamashita, M. T.; Fedorov, D. V.; Jensen, A. S.; Zinner,
N. T. Contact parameters in two dimensions for general three-body systems. {\bf New
Journal of Physics}, v. 16, n. 1, p. 013048, 2014. \\

\noindent
Bellotti, F. F.; Frederico, T.; Yamashita, M. T.; Fedorov, D. V.; Jensen, A. S.; Zinner,
N. T. Mass-imbalanced three-body systems in 2d: Bound states and the analytical
approach to the adiabatic potential. {\bf Few-Body Systems}, v. 55,
p. 847, 2014. \\

\noindent
Bellotti, F. F.; Frederico, T.; Yamashita, M. T.; Fedorov, D. V.; Jensen, A. S.; Zinner,
N. T. Universality of three-body systems in 2d: Parametrization of the bound states
energies. {\bf Few-Body Systems}, v. 55, p. 1025, 2014. \\

\noindent
Yamashita, M. T.; Bellotti, F. F.; Frederico, T.; Fedorov, D. V.; Jensen, A. S.; Zinner,
N. T. Dimensional crossover transitions of strongly interacting two- and three-boson
systems. {\bf ArXiv e-prints}, Apr. 2014.

\end{spacing}

\listoffigures
\addcontentsline{toc}{chapter}{List of Figures}

\listoftables
\addcontentsline{toc}{chapter}{List of Tables}

\end{spacing}

\mainmatter

\pagestyle{fancy}

\begin{spacing}{1.3}  
\chapter{Introduction} \label{ch1}

In the last decade scientists around the world found experimental
evidences \cite{kraemerNP2006,ferlainoPOJ2010} of a remarkable phenomenon in
few-body systems that was predicted long time ago \cite{efimovYF1970} and  today is
known as the Efimov effect. It corresponds to an accumulation 
of three-boson energy levels when the two-body scattering length tends to infinity.
In the exact limit - when the dimer energy is zero - the energies of successive states are
geometrically spaced obeying a universal ratio.
Experiments \cite{kraemerNP2006} were able to identify few of these Efimov states,
bringing the attention of the physics community to few-body problems
of short-ranged interactions with large scattering lengths.

The experiments were realized using Feshbach resonances in cold
atomic gases (see, e.g., Ref.~\cite{chinRMP2010}). Using this technique it is
possible to tune the scattering length to large values bringing the
system into a universal regime, where their properties are
essentially model-independent. In this regime the properties of the system
are defined by the knowledge of only few physical low-energy
observables that the short-ranged potentials should produce. The
possibility of manipulating the interaction between trapped cold
atoms also opened new avenues to probe few-body physics as, for
example, by studying systems restricted to two dimensions
\cite{martiyanovPRL2010,frohlichPRL2011,dykePRL2011}. Mostly, the theoretical
background in few-body physics was built for systems in  three
dimensions. However, the experimental possibility to squeeze one of
the dimensions, forming trapped atomic systems in layers, asks for
deeper and larger theoretical investigations of lower dimensional
few-particle systems.

The number of spatial dimensions plays an important role in quantum
systems. For instance, let us consider the kinetic energy operator
written in angular coordinates. The dependence on the angle
variables comes through the centrifugal barrier operator, that in
three dimensions has eigenvalues always zero or positive, while in
two dimensions, for zero angular momentum, it is negative. This means
that a minimum amount of attraction is necessary to bind a three
dimensional system, while any infinitesimal attractive potential
produce a $s-$wave bound state in a two dimensional system
\cite{nielsenPRA1997,nielsenFS1999,nielsenPR2001}. In fact this was already pointed out a long
time ago. The Landau criterion says that potentials with negative
volume integral will produce a bound state for any value of the
strength in two-dimensions (see, e.g., Ref.~\cite{landau1977}). This topic
continues to be of interest and it was found that when the volume
integral is exactly zero a bound state is still present
\cite{simonAoP1976,armstrongEL2010,volosnievPRL2011}.

All this recent effort towards the two dimensional (2D)  physics is
supported by the relevance of the field for several different
applications like, e.g., high-temperature
superconductors, localization of atoms on surfaces, in
semi-conducting micro-cavities, for carbon nanotubes and organic
interface. There is an interest among the ultracold atomic gas
laboratories to produce quantum degenerate gases in low dimensions,
with the aim to probe the two-dimensional physics of quantum
systems. Early experiments already produced quasi-2D samples of
$^{133}$Cs \cite{vuleticPRL1998,morinagaPRL1999,hammesPRL2003}, $^{23}$Na
\cite{gorlitzPRL2001}, and $^{87}$Rb \cite{burgerEL2002}. Two\=/dimensional
gases with mixtures of $^{40}$K and $^{87}$Rb have been produced \cite{modugnoPRA2003,gunterPRL2005} and two\=/component
gases of $^{6}$Li \cite{dykePRL2011,martiyanovPRL2010} and $^{40}$K~\cite{frohlichPRL2011} have also been studied. Quasi\=/2D pancakes of trapped
heteronuclear diatomic molecules of $^{40}$K$^{87}$Rb were produced
in stacked layers~\cite{NPP2011}. 

Theoretically one can define precisely the dynamics of quantum
systems in two-dimensions, while in a real experiment the
confinement to 2D is typically done using an optical lattice. This
introduces a transverse energy scale, $\hbar\omega_0$, and below it
the physics is effectively 2D, while it becomes 3D at or above
$\hbar\omega_0$. To produce a 2D sample of trapped atoms in an
experimental setup, one starts with a three dimensional system.
Therefore, it is important to find observables that make possible to
distinguish experimentally when the system can be considered really
in two dimensions.

The work consists in the study of two- and three- dimensional (2D and 3D) few-body systems close to the universal regime, where the world ``universal'', which is extensively used along the work, means that the discussed properties of the quantum systems are independent of the model utilized to describe the interaction between two particles, namely, the weakly bound system is much larger than the size of the interaction. A natural way to study such properties is to  describe the pairwise interaction with Dirac$-\delta$ potentials, since the condition for universality $|a| / r_0 \gg 1$, where $a$ is the scattering length and $r_0$ the range of the potential, is  always fulfilled.

The introduction/motivation to the work 
given in Chapter~\ref{ch1} are followed by the derivation of the equations that describe the 2D and 3D dynamics of two- and three-body systems interacting through zero-ranged potentials in Chapter~\ref{ch2}. Notice that the problem consists basically in the solution of an eigenvalue\=/eigenvector problem, where the energy and the wave function of the three-body system must be determined. However, the complexity of the three-body problem, which does not have a closed solution even at the classical level, leads the problem to be described for an elaborated set of homogeneous coupled integral equations. 

The behavior of three-boson systems changes remarkably from two to three 
dimensions, since the dynamics and properties of quantum systems drastically 
change when the system is restricted to different dimensions. Two important examples 
illustrating the influence of the spatial dimension in the three-body sector are 
the Efimov effect \cite{efimovYF1970}  and the Thomas collapse \cite{thomasPR1935}. 
The Efimov states (see the beginning of Chapter~\ref{ch1}), which were predicted
and observed for three identical bosons in 3D systems \cite{kraemerNP2006,ferlainoPOJ2010}, 
are absent in 2D even in the most favorable scenario of mass\=/imbalanced
systems \cite{limZfPAHaN1980,adhikariPRA1988}.  Similarly, 
Thomas found in 1935 that the energy of a three identical bosons system subjected 
to short\=/range pairwise interactions in 3D grows without 
boundaries (collapses) when the range of the interaction approaches zero ($r_0 \to 0$).
Nevertheless, this effect was not observed in 2D systems yet. 
It is shown in Ref.~\cite{adhikariPRA1988} that both the Thomas collapse and the
Efimov effect are mathematically related to the same anomaly in the kernel of the 
three\=/body equations and they take place whenever $|a| / r_0 \to \infty$.
For instance, starting with finite and non\=/null values of $|a|$ and $r_0$, the finite and 
well\=/behaviored three\=/body spectrum will collapse when $r_0 \to 0$. On the
other hand, infinitely many weakly bound states will appear for $|a| \to \infty$.
Notice that the condition $|a| / r_0 \to \infty$ is fulfilled in both cases.

The sparseness of information about 2D three-bosons system has motivated the systematic investigation of the universal properties of mass-imbalanced systems using zero-range interactions in momentum space \cite{bellottiJoPB2011,bellottiPRA2012} and the results are shown in Chapter~\ref{ch3}. The focus is particularly  on the dependence of the three-body binding energy with masses and two-body binding energies. The critical values of these parameters (masses and two-body binding energies) allowing a given number of three-particle bound states with zero total angular momentum are determined in a form of boundaries in the multidimensional parametric space. Besides the dependence of the three-particle binding energy on the parameters be highly non-trivial, even in the simpler case of two identical particles and a distinct one,  this dependence is parametrized for the ground and first excited state in terms of {\itshape supercircles} functions  in the most general case of three distinguishable particles, as also presented in Chapter~\ref{ch3}.

The study of the universal properties of 2D three-body systems has shown an increasing number of bound states for the decreasing mass of one of the particles \cite{bellottiJoPB2011,bellottiPRA2012}. The situation where one particle is much lighter than the other two is suitably handled in the adiabatic approximation, namely the Born-Oppenheimer (BO) approximation, which is presented in Chapter~\ref{ch4}. The adiabatic potential between the heavy particles due to the light one found as the solution of a transcendental equation is mass-dependent and reveals an increasing number of bound states by decreasing the mass of one of the particles \cite{bellottiJoPB2013}. Besides, an asymptotic expression for the adiabatic potential is derived and is shown that this analytic expression faithfully corresponds to the numerically calculated adiabatic potential, even in the non-asymptotic region. The number of bound states for a heavy-heavy-light system is estimated as a function of the light-heavy mass ratio and infinitely many bound states are expected as this ratio approaches zero. However, for finite masses only finite number of bound states is always present. 

While Chapters~\ref{ch2} and \ref{ch3} are focused in the eigenvalue of the three\=/body Hamiltonian problem, Chapters~\ref{ch5} and \ref{ch6} are related to the eigenstate in momentum space, constructed from the spectator function and giving the momentum density. A key result presented in Chapter~\ref{ch5} is the derivation of an analytical expression for the asymptotic behavior of the spectator function large momentum of three-body systems in two dimensions \cite{bellottiPRA2013,bellottiNJoP2014}. This asymptotic behavior defines the one-body large momentum density, which is a strong candidate as observable quantity able to unequivocally determine whether the quantum system is restricted to two or three dimensions\cite{bellottiPRA2013}. The two- and three\=/dimensional one\=/body momentum densities are discussed respectively in Chapters~\ref{ch5} and \ref{ch6}. Besides, the one-body density defines the two- and three-body contact parameters, which relate few- and many-body properties of quantum atomic gases \cite{tanAoP2008}. 
It is shown in Chapter~\ref{ch5} that the two-body contact parameter,  which is the coefficient in the leading order in the large momentum expansion of the one\=/body density, of specific 2D systems is a universal constant, in the sense that it does not depend on the quantum level considered \cite{bellottiPRA2013,bellottiNJoP2014}. 
The three-body contact parameter, which is the coefficient in the sub\=/leading order in the large momentum expansion of the one\=/body density is not found to be universal, but the sub\=/leading functional form  is independent of the mass of the constituents in 2D. The same does not happen in 3D systems, where the functional form of the sub-leading term in the one-body momentum density depends on the mass \cite{yamashitaPRA2013}, as shown in Chapter~\ref{ch6}. Furthermore, the discussion of how the one\=/body momentum  density can be used to determine the dimensionality of the system is also made in Chapter~\ref{ch5}.

As current experimental set ups are able to continuously squeeze one dimension in order to build 2D experiments (see for instance Ref.~\cite{dykePRL2011}), it is interesting to find theoretical methods which are able to produce a continuous squeezing of one of the dimensions. In Chapter~\ref{ch7} it is presented a method that allows to study the dimensional crossover transitions of strongly interacting two- and three-bosons systems by continuously ``squeezing'' one of the dimensions \cite{yamashitaAe2014}. The particles are placed in a flat surface plus a transverse direction (compact dimension), which imposes the discretization of the momentum accordingly to the chosen type of boundary conditions.
Employing periodic boundary conditions in the compact dimension, it is shown that a sharp transition occurs in the energy spectrum  of three-body system as the system is squeezed from 3D to 2D. However, more studies are still necessary in order to relate the parameter which dials between the different situations to real experiments.


Summary and outlook are presented in Chapter~\ref{ch9}. In order to motivate the reading, the beginning of Chapters~\ref{ch2} to \ref{ch7} brings a brief motivation/introduction to the topic that will be discussed. Further details are given in Appendices~\ref{revisionst} to \ref{residues}.


\chapter{Two- and Three-body dynamics} \label{ch2}

The surprisingly fast and ongoing technological advances, which permeate our daily life, has also given tools for an extraordinary grow on the experimental studies of cold atomic quantum gases. However, the cornerstone in the study of quantum systems in laboratories is still the same: these systems are probed through collision experiments. 

In this way, the background tools for the theoretical understanding of such experiments are given by the scattering theory and some concepts of this theory are presented in the following. While, even for low density gases, the experiments are taken with several thousands of particles, it turns out that some macro properties of the systems are driven by two- and three-body correlations. This work is focused on the {\it universal} properties of three-body quantum systems, i.e., when the size of the system is much larger than the range of the interaction between the particles. Such problem is already challenging and interesting in itself, since there is no classical equivalent. 

 A brief presentation of the quantum theory of scattering for two\=/body zero\=/range potentials is given in Appendix~\ref{revisionst} and the focus here is only on the concepts and equations that are needed in order to make the reading of this thesis easier. More details are given in Appendices~\ref{revisionst},
\ref{jacobi} and \ref{melements}. Complete and formal descriptions of the scattering theory in the three-body quantum problem are given, for instance, in Refs.~\cite{schmidQMTB1974,mitraAiNP1969}. The main point here is the derivation of the integral equations for the two- and  three-body transition matrix ($T-$matrix) when the particles are assumed to interact through zero-range potentials.
Although these potentials are not realistic, their importance in the study of two- and three-body quantum systems is explained in the next section.

\section{Zero-range model and Renormalization} \label{tzmr}

The $s-$wave zero-range potentials have a separable operator form (see Eq. \eqref{eq.c2-42})
\begin{equation}
V=\lambda\left|\chi\right\rangle\left\langle\chi\right|
\end{equation}
and will be used to solve the two-body $T-$matrix (see Eq.~\eqref{eq.c2-38a})
\begin{equation}
t= V + V g_0 t \; ,
\end{equation}
as shown in Appendix~\ref{scatteringTmatrix}. For a $s-$wave separable potential, the transition matrix is  (see Eqs.~\eqref{eq.c2-46} and \eqref{eq.c2-48})
\begin{equation}
t(E)=\left|\chi\right\rangle\tau(E)\left\langle\chi\right|
\end{equation}
with 
\begin{equation}
\tau(E)=\left(\lambda^{-1}-\int{d^D p\frac{g(p)^2}{E-\frac{p^2}{2m_{red}}+\imath \epsilon}}\right)^{-1} \; .
\end{equation} 
Despite of the fact that Eq.~\eqref{eq.c2-48} holds for any generic separable potential that has the operator form given in Eq.~\eqref{eq.c2-42}, no local potential has this form besides the zero\=/range one.

Zero-range potentials are very interesting. Although they do not correspond any realistic interaction, they allow to study the phenomenology and to understand the driven physics, which dominates the properties of large quantum systems, namely systems with size much larger than the range of the potential. They guide our intuition on the expected behavior of the quantum few-body systems, since any realistic short-range potential must reproduce the results obtained with zero-range potential when the system is very large.

The $s-$wave zero-range model is introduced through a Dirac$-\delta$ interaction which is also called contact interaction. This means that the particles only interact when they touch each other. Besides, the Dirac$-\delta$ potential has the operator form given by Eq.~\eqref{eq.c2-42}. In configuration space, the matrix element of a local potential $V$ is written as
\begin{equation}
\left\langle\mathbf{R'}\right|V\left|\mathbf{R}\right\rangle=V(\mathbf{R})\delta(\mathbf{R'-R}) \ .
\label{eq.c2-42a}
\end{equation}
The Dirac$-\delta$ potential is local and $V(\mathbf{R})=\lambda\delta(\mathbf{R})$. So, Eq.~\eqref{eq.c2-42a} becomes
\begin{align}
\left\langle\mathbf{R'}\right|V\left|\mathbf{R}\right\rangle=\lambda\delta(\mathbf{R})\delta(\mathbf{R'-R}) 
=\lambda\delta(\mathbf{R})\delta(\mathbf{R'}) \ ,  \label{eq.c2-42b}
\end{align}
meaning that this potential is also separable.

In momentum space, the matrix element of the Dirac$-\delta$ potential for a $n-$dimensional system is
\begin{equation}
\left\langle\mathbf{p'}\right|V\left|\mathbf{p}\right\rangle=\frac{\lambda}{(2\pi)^D}\int{d^DR\int{d^DR' e^{\imath \mathbf{p'}\cdot\mathbf{R'}}e^{-\imath \mathbf{p} \cdot \mathbf{R}}\delta(\mathbf{R'})\delta(\mathbf{R})}} =\frac{\lambda}{(2 \pi)^D}\ .
\label{eq.c2-42c}
\end{equation}

It is possible to redefine $\left|\chi\right\rangle\ \equiv \left( 2\pi \right)^{n/2} \left|\tilde{\chi}\right\rangle$ so that Eq.~\eqref{eq.c2-42c} is equal to $\lambda$. In this way, the form factor $\left\langle\chi\right|\left.\mathbf{p}\right\rangle=\left\langle\mathbf{p}\right|\left.\chi\right\rangle=g(p)$ is equal to one for the Dirac$-\delta$ potential, as can be seen below
\begin{equation}
g(p)=\left\langle\mathbf{p}\right|\left.\chi\right\rangle= \left(2 \pi \right)^{D/2} \int{d^D R \frac{e^{-\imath \mathbf{p} \cdot \mathbf{R}}}{\left(2 \pi \right)^{D/2}}\delta(\mathbf{R})}=1 \ .
\label{eq.c2-42d}
\end{equation}

The form factor of the Dirac$-\delta$ potential in Eq.~\eqref{eq.c2-42d} introduces a divergence in the momentum integration of Eq.~\eqref{eq.c2-48}. In 2D and 3D the divergence can be treated by introducing a physical scale in the problem \cite{fredericoPiPaNP2012}, but another way to render finite the integral could be done by introducing a cut-off. It was shown in Ref.~\cite{yamashitaTHESIS2004} that both methods are equivalent when the momentum cut-off is let to be infinite. 

The scale is introduced by defining a physical value for the two-body $T-$matrix , $\lambda_R$, in a subtracted energy point defined by $E=-\mu^2$. The $T-$matrix becomes
\begin{equation}
\tau_R(-\mu^2)=\lambda_R(-\mu^2) \ ,
\label{eq.c2-49}
\end{equation} 
where the subscript $R$ means renormalized, and $\lambda_R(-\mu^2)$ is given by a physical condition. 

Inserting the condition from Eq.~\eqref{eq.c2-49} in the matrix element given by Eq.~\eqref{eq.c2-48} gives
\begin{align}
&\tau_R(-\mu^2)=\left(\lambda^{-1}-\int{d^Dp\frac{1}{-\mu^2-\frac{p^2}{2m_{red}}}}\right)^{-1}=\lambda_R(-\mu^2) \ ,& \label{eq.c2-50a}
\end{align}
which allows to express the bare strength $\lambda$ as
\begin{align}
&\lambda^{-1}=\lambda_R^{-1}(-\mu^2)+\int{d^Dp\frac{1}{-\mu^2-\frac{p^2}{2m_{red}}}} \ .& \label{eq.c2-50b}
\end{align}

A finite expression for the scattering amplitude is found by replacing $\lambda$, as given in Eq.~\eqref{eq.c2-50b}, into the matrix element in Eq.~\eqref{eq.c2-48}. The result is 
\begin{equation}
\tau_R(E)^{-1}=\lambda_R^{-1}(-\mu^2)+\int{d^Dp\left(\frac{1}{-\mu^2-\frac{p^2}{2m_{red}}}-\frac{1}{E-\frac{p^2}{2m_{red}}+\imath \epsilon}\right)} \ .
\label{eq.c2-51}
\end{equation}

\subsection{Two-body T-matrix in 2D}

Considering only bound states, i.e., $E<0$, the integral on the right-hand-side of Eq.~\eqref{eq.c2-51} for two-dimensional systems ($D=2$) is
\begin{align}
I(E)=\int{d^2p\left(\frac{1}{-\mu^2-\frac{p^2}{2m_{red}}}-\frac{1}{E-\frac{p^2}{2m_{red}}}\right)} 
=-4\pi m_{red}\ln\left(\sqrt{\frac{-E}{\mu^2}}\right) \ , \label{eq.a2-07}
\end{align}
and from Eqs.~\eqref{eq.c2-51} and \eqref{eq.a2-07}, the renormalized two-body $T-$matrix is given by
\begin{equation}
\tau_R(E)^{-1}=\lambda_R^{-1}(-\mu^2)-4\pi m_{red}\ln\left(\sqrt{\frac{-E}{\mu^2}}\right) \ .
\label{eq.c2-52}
\end{equation}

For positive energies, i.e., $E>0$ (scattering states), the scattering amplitude is obtained from the analytic continuation of Eq.~\eqref{eq.c2-52} in the upper complex semi\=/plane of $E$, as shown below:
\begin{align}
\tau_R(E)^{-1}&=\lambda_R^{-1}(-\mu^2)-4\pi m_{red}\ln\left(\sqrt{\frac{-E}{\mu^2}}\right) \; ,  \nonumber\\
&=\lambda_R^{-1}(-\mu^2)-4\pi m_{red} \ln\left(\sqrt{\frac{E}{\mu^2}}\right)+2\pi^2\;\imath\; m_{red} \; , \label{eq.c2-52b}
\end{align}
where the choice $-1=e^{-\imath \pi}$ is used due to the analytic continuation in the upper half semi\=/plane of the energy.

The matrix elements of the transition operator in Eq.~\eqref{eq.c2-47} are expressed as
\begin{equation}
\left\langle\mathbf{p'}\right|t_R(E)\left|\mathbf{p}\right\rangle=\tau_R(E) \ ,
\label{eq.c2-63}
\end{equation}
and for the sake of notation simplicity, the subscript $R$ will be suppressed in the equations from now on, i.e.,  $\tau_R(E) \equiv \tau(E)$, $\lambda_R(-\mu^2) \equiv \lambda(-\mu^2)$ and $t_R(E) \equiv t(E)$. 

Looking at the matrix element in Eq.~\eqref{eq.c2-52}, it is not straightforward to identify the $s-$wave scattering phase-shift and cross-section for the zero-range model, as they were presented in Ref.~\cite{adhikariPRA1993}. 
In units of $\hbar=2m_{red}=1$, Eq.~\eqref{eq.c2-52} becomes
\begin{align}
&\tau(E)^{-1}=\lambda^{-1}(-\mu^2)-\pi \ln\left(\frac{-E}{\mu^2}\right) \ ,&
\label{eq.c2-64}
\end{align}
where the respective analytic continuation for $E>0$ is given in Eq.~\eqref{eq.c2-52b}. Using that $\left\langle p'\right|t(E)\left|p\right\rangle=2\pi \left\langle\mathbf{p'}\right|t(E)\left|\mathbf{p}\right\rangle$, the matrix elements in Eq.~\eqref{eq.c2-63} are written as
\begin{align}
\left\langle p'\right|t(E)\left|p\right\rangle =\frac{2\pi}{\lambda^{-1}-\pi\ln\left(\frac{E}{\mu^2}\right)+\imath\pi^2} 
=\frac{2}{\pi \left(-\cot\delta_2+\imath\right)} \; ,  \label{eq.c2-65}
\end{align}
where the $s-$wave phase-shift for the zero-range model is defined as
\begin{equation}
\cot\delta_2=-\frac{1}{\pi^2\lambda(-\mu^2)}+\frac{1}{\pi}\ln\left(\frac{E}{\mu^2}\right) \; .
\label{eq.c2-66}
\end{equation}
Then, the two-dimensional scattering length, $a_2$, is found to be
\begin{align}
\overline{a}_2=-\frac{1}{\pi^2\lambda(-\mu^2)}+\frac{1}{\pi}\ln(\mu^2) 
			  =a_2 +\frac{1}{\pi}\ln(\mu^2) \; .  \label{eq.c2-67}
\end{align} 
Notice that the logarithmic term, which appears in the low energy expansion, leads to an ambiguity in the definition of the scattering length in 2D, which depends on the scale used to measure the energy. So, the binding energy of the pair, $E_B$, is chosen as the  physical scale in the problem. The bound state energy $(E<0)$ is the pole of Eq.~\eqref{eq.c2-65}, i.e.,
\begin{equation}
\ln\left(\frac{-E}{\mu^2}\right)=\frac{1}{\pi^2\lambda(-\mu^2)} \; ,
\label{eq.c2-67b}
\end{equation}
which gives
\begin{equation}
E=-\mu^2 e^{-a_2}=e^{-\overline{a}_2}=E_B \; .
\label{eq.c2-68}
\end{equation}

Remember that $\lambda(-\mu^2)$ is the physical information which was introduced in the two-body $T-$matrix integral equation to handle the ultraviolet divergence. Then, the subtraction point $\mu^2$ can be choose as the physical scale of the problem, i.e., $\mu^2=-E_B$, where   the binding energy of the pair is the zero of Eq.~\eqref{eq.c2-52}. This choice also fixes the value of $\lambda(-\mu^2)$, namely  
\begin{align}
\tau(E)^{-1}=\lambda^{-1}(E_B)-4\pi m_{red}\ln\left(\sqrt{\frac{E}{E_B}}\right)  = 0 \; ,
\end{align}
at the bound\=/state pole and therefore
\begin{align}
&\lambda^{-1}(E_B) = 0 \; . \label{eq.c2-52a}
\end{align}

Finally, the renormalized 2D two-body $T-$matrix for the zero-range model is
\begin{equation}
\tau(E)^{-1}=-4\pi m_{red}\ln\left(\sqrt{\frac{-E}{E_B}}\right) \; , 
\label{eq.c2-52c}
\end{equation}
which will be used in the calculation of the properties of three\=/body systems in 2D.

\subsection{Two-body T-matrix in 3D}
The 3D equivalent of Eq.~\eqref{eq.c2-51} is given by
\begin{equation}
\tau_R(E)^{-1}=\lambda_R^{-1}(-\mu^2)+\int{d^3p\left(\frac{1}{-\mu^2-\frac{p^2}{2m_{red}}}-\frac{1}{E-\frac{p^2}{2m_{red}}+\imath \epsilon}\right)} \ ,
\label{eq.c2-51app}
\end{equation}
where, as before, $E$ is the energy, $\mu^2$ the subtraction point, $m_{red}$ the reduced mass and the subscript $R$ means renormalized.
For $E<0$ (bound states), the integral on the right-hand-side of Eq.~\eqref{eq.c2-51app} is
\begin{align}
I(E)=\int{d^3p\left(\frac{1}{-\mu^2-\frac{p^2}{2m_{red}}}-\frac{1}{E-\frac{p^2}{2m_{red}}}\right)} 
=-4 \pi^2  m_{red}  \left( \sqrt{2m_{red} |E|} - \sqrt{2m_{red} \mu^2 }\right)   \; .
\label{Ieres3d1}
\end{align}

As in the 2D case, the subtraction point is chosen as the two-body binding energy, i.e., $-\mu^2=E_B$ and Eq.~\eqref{eq.c2-52a} also holds in the 3D case, namely $\lambda_R^{-1}(E_B)=0$. Then, dropping the subscript $R$, the two-body $T-$matrix for 3D systems is given by
\begin{equation}
\tau(E)^{-1}=-2 \pi^2  \left( 2 m_{red}\right)^{3/2}  \left( \sqrt{|E|} - \sqrt{ |E_B| }\right)   \; ,
\label{tau2b3d}
\end{equation}
which will be used in the calculation of the properties of three\=/body systems in 3D.

\section{Notation and three\=/body dynamics}
The system consists of three distinguishable particles of masses $m_\alpha$, momenta $\mathbf{k}_\alpha$ and pairwise interactions $v_\alpha$, where $\alpha=a,b,c$ labels the particles ($a,b,c$) and the notation of the potential is such that $v_a$ is the interaction between particles $b$ and $c$. 
The eigenvalue equation for the Hamiltonian  
\begin{equation}
\left( H_{0} + V \right) \Psi = E  \Psi \; ,
\label{eq.c2-69}
\end{equation}
is fulfilled by states in the discrete ($E<0$) and continuum ($E>0$) regions. The  potential given by two\=/body terms is $V=v_{a}+v_{b}+v_{c}$ and the free and full propagators are respectively given by
\begin{equation}
G_0(Z) \equiv \frac{1}{Z-H_0} \;\;\; \text{and} \;\;\; G(Z) \equiv \frac{1}{Z-H} \; ,
\end{equation} 
with $H=H_0+V$. The free Hamiltonian, in frame of the laboratory, is given by the sum over the individual kinetic energies of the particles and is written as  
\begin{equation}
H_0=\sum_{\alpha=a,b,c}\frac{k_\alpha^2}{2m_\alpha} \ .
\label{eq.c2-70}
\end{equation}

\begin{figure}[!htb]
\centering
\includegraphics[width=0.9\textwidth]{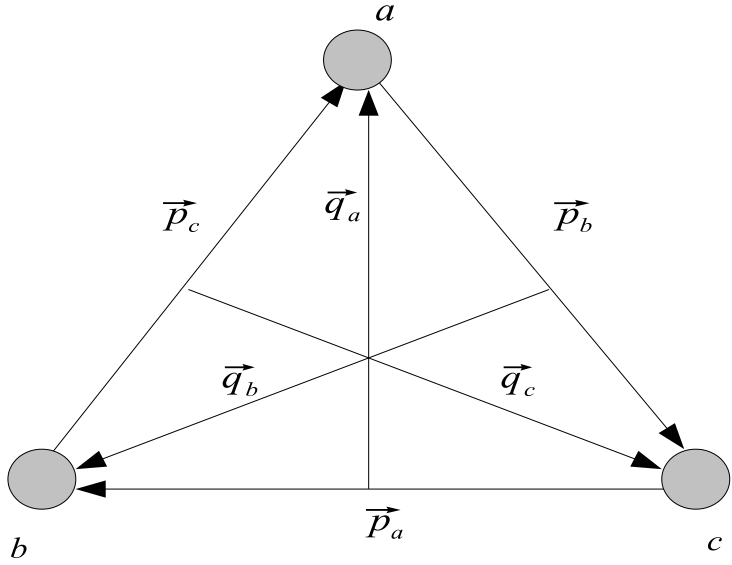}
\caption{Jacobi relative momenta}
\label{coord_jacobi}
\end{figure}

A set of Jacobi coordinates and the canonical conjugate momenta, which are shown in Fig.~\ref{coord_jacobi}, are useful when dealing with three-body problems,  since the CM motion is separated out. In this case the free Hamiltonian becomes 
\begin{equation}
H_{0}=\frac{p_{\alpha }^{2}}{2m_{\beta\gamma}}+\frac{q_{\alpha }^{2}}{2 m_{\beta\gamma,\alpha}} + \frac{Q^2}{m_{\alpha}+m_{\beta}+m_{\gamma}}\; ,  
\label{eq.c2-71}
\end{equation}%
where $\mathbf{Q}=\sum_{\alpha} \mathbf{k}_\alpha$ is the CM momentum.  Taking into account the frame of particle $\alpha$ with respect to the CM of the pair ($\beta,\gamma$), $\mathbf{q}_{\alpha }$ is the momentum of particle $\alpha$ with respect to the CM of the pair, $\mathbf{p}_{\alpha }$ is the relative momentum of the pair, $m_{\beta\gamma}$ is the reduced mass of the pair and $m_{\beta\gamma,\alpha}$ is the three\=/body reduced mass. The relative momenta and reduced masses are given by
\begin{align}
&\mathbf{q}_{\alpha }=\frac{ m_{\beta }+m_{\gamma } }{m_{\alpha }+m_{\beta }+m_{\gamma }}\left[\mathbf{k}_{\alpha }-\frac{m_{\alpha }}{m_{\beta}+m_{\gamma}}\left(\mathbf{k}_{\beta }+\mathbf{k}_{\gamma }\right)\right] \ ,& \label{eq.c2-72a} \\
&\mathbf{p}_{\alpha }=\frac{m_{\gamma }\mathbf{k}_{\beta }-m_{\beta }\mathbf{k}_{\gamma }}{m_{\beta }+m_{\gamma }} \ ,& \label{eq.c2-72b} \\
&m_{\beta\gamma}=\frac{m_{\beta }m_{\gamma }}{m_{\beta }+m_{\gamma }} \ ,& \label{eq.c2-72c} \\
&m_{\beta\gamma,\alpha }=\frac{m_{\alpha }\left( m_{\beta }+m_{\gamma}\right) }{m_{\alpha }+m_{\beta }+m_{\gamma }} \ ,& \label{eq.c2-72d}
\end{align}
with ($\alpha,\beta,\gamma$) as cyclic permutations of ($a,b,c$) (see Appendix~\ref{jacobi} for more details about Jacobi relative momenta).
It is also useful to specify an operator notation, where all two-body operators are represented with small letters, i.e., $v_\alpha,g_0$ and three-particle operators are represented by capital letters, i.e., $H,V$.

\subsection{Three-body T-matrix}
The three-body transition operator is written as 
\begin{equation}
T\left( E \right) = V + V G\left( E+\imath \epsilon\right) V  \; ,
\label{eq.c2-99}
\end{equation}%
that is the formal analogue of the two-body $T-$matrix in Eq.~\eqref{eq.c2-32}. Besides that the operator in Eq.~\eqref{eq.c2-99} is not directly related to the scattering cross section as in the two-body case, 
the relations in Eqs.~\eqref{eq.c2-38a} and \eqref{eq.c2-38b} also hold and read
\begin{equation}
T(E)=V+VG_0(E+\imath \epsilon)T(E)=V+T(E)G_0(E+\imath \epsilon)V \; .
\label{eq.c2-101}
\end{equation}

The Faddeev components of the three-body $T-$matrix (see Ref.~\cite{faddeevMATB1965,schmidQMTB1974}) are given by
\begin{equation}
T_a(E)=v_a+v_a G_0(E+\imath \epsilon)T(E) \ 
\label{eq.c2-102}
\end{equation}
and since that $V=v_a+v_b+v_c$, the transition operator from Eq.~\eqref{eq.c2-101} can be written in term of the components given by Eq.~\eqref{eq.c2-102} as
\begin{align}
T(E)=T_a(E)+T_b(E)+T_c(E) \; .
\label{eq.c2-103}
\end{align}

Inserting Eq.~\eqref{eq.c2-103} back into Eq.~\eqref{eq.c2-102} results in a system of coupled equations, which are written in matrix form as
\begin{equation}
\left( 
\begin{array}{c}
T_{a} \\ 
T_{b} \\ 
T_{c}%
\end{array}%
\right) =\left( 
\begin{array}{c}
v_{a} \\ 
v_{b} \\ 
v_{c}%
\end{array}%
\right) +\left( 
\begin{array}{ccc}
v_{a} & v_{a} & v_{a} \\ 
v_{b} & v_{b} & v_{b} \\ 
v_{c} & v_{c} & v_{c}%
\end{array}%
\right) G_0\left( 
\begin{array}{c}
T_{a} \\ 
T_{b} \\ 
T_{c}%
\end{array}%
\right) \ .  
\label{eq.c2-104}
\end{equation}%
Isolating the component $T_a$ in Eq.~\eqref{eq.c2-104} results in
\begin{align}
\left(1-v_a G_0\right)T_a=v_a+v_a G_0\left(T_b+T_c\right) \; , \label{eq.c2-105}
\end{align}
which multiplied by $\left(1-v_a G_0\right)^{-1}$ from the left gives
\begin{equation}
T_a=t_a+t_a G_0\left(T_b+T_c\right)  \; ,  \label{eq.c2-106}
\end{equation}
where the relation $t_a=\left[1-v_a G_0\right]^{-1}v_a$
was used in third line. The renormalized two\=/body $T-$matrix $t_a$ in the $abc$ system is given by
\begin{equation}
t_a \equiv t_a(E)=\left|\chi_a\right\rangle\tau_a(E)\left\langle\chi_a\right| \;\;\; \text{with} \;\;\; \tau_a(E)^{-1}=-4\pi m_{bc}\ln\left(\sqrt{\frac{-E}{E_{bc}}}\right) \; .
\label{t23b}
\end{equation}

Finally, the set of coupled equations for the Faddeev components of the three-body transition operator are written in matrix form as
\begin{equation}
\left( 
\begin{array}{c}
T_{a} \\ 
T_{b} \\ 
T_{c}%
\end{array}%
\right) =\left( 
\begin{array}{c}
t_{a} \\ 
t_{b} \\ 
t_{c}%
\end{array}%
\right) +\left( 
\begin{array}{ccc}
0 & t_{a} & t_{a} \\ 
t_{b} & 0 & t_{b} \\ 
t_{c} & t_{c} & 0%
\end{array}%
\right) G_0\left( 
\begin{array}{c}
T_{a} \\ 
T_{b} \\ 
T_{c}%
\end{array}%
\right) \ .  
\label{eq.c2-107}
\end{equation}%

These equations have the advantage to contain the only the two-body $T-$matrix, and consequently two-body energies, instead of the potential. The power of such formulation is better explained and explored in Chapter~\ref{ch3}. Besides, Eq.~\eqref{eq.c2-107} shows how the two-body scattering amplitude connects with the three-body scattering. In detail, the equation for one Faddeev component of the transition operator is given by
\begin{equation}
T_a(E_3)=t_a\left(E_3-\frac{q_a^2}{2m_{bc,a}}\right)\Biggl\{  1+ G_0(E_3+\imath \epsilon)\biggl[T_b(E_3)+T_c(E_3)\biggr]\Biggr\} \; ,
\label{eq.c2-110}
\end{equation}
where $E_3$ is the three-body energy and the other components are found by cyclic permutation of the particle labels.

Notice that the argument of the two-body $T-$matrix in Eq.~\eqref{eq.c2-52b} is the relative two-body energy, $E_2^R$, which was replaced by $E_3-\frac{q^2}{2 m_{bc,a}}$ in Eq.~\eqref{eq.c2-110}. The relative two-body energy connects with the total energy, $E_2^T$, through $E_2^T=E_2^R+\frac{q_2^2}{2 \left( m_b+m_c \right) }$,
where $q_2$ is the total momentum of the pair. At the frame of the CM in a three-body system, i.e., $Q=0$, the total energy of the pair is the difference between the three-body energy $E_3$ and the kinetic energy of the third particle, namely $E_2^T=E_3-\frac{q_1^2}{2 m_a }$.
Moreover, if $Q=0$ the momentum of the pair is exactly the momentum of the third particle. In other words, $\left|\mathbf{q}_1\right|=\left|\mathbf{-q}_2\right|=q$ and 
the relative two-body energy as function of the three-body energy is written as
\begin{align}
E_2^R=E_3-\frac{q^2}{2 m_a } -\frac{q^2}{2 \left( m_b+m_c \right) }=E_3-\frac{q^2}{2 m_{bc,a}} \; , & \label{e2r}
\end{align}
which is exactly the argument of the two-body $T-$matrix in Eq.~\eqref{eq.c2-110}.

\section{Three-body bound state equation in 2D} \label{tbsiq2d}
The three-body $T-$matrix in Eq.~\eqref{eq.c2-107} describes the three-body scattering process with pairwise short-range potentials. The transition operator generally contains all the strong interaction properties of the three-body system or, in other words, it allows to construct the resolvent of the interacting model. Therefore, the $T-$matrix  gives information about both bound ($E_3<0$) and scattering ($E_3>0$) states. The focus in the following Chapters is on three-body bound states, then the coupled homogeneous integral equations for the bound state are derived here, starting from the transition operator. It is possible to instead use directly the Faddeev decomposition for the bound\=/state wave function \cite{mitraAiNP1969}.

The completeness relation is defined as
\begin{equation}
\hat{1}=\sum_{B}\left\vert \Phi _{B}\right\rangle \left\langle \Phi
_{B}\right\vert + \int d^{2}k\left\vert \Psi _{c}^{(+)}\right\rangle \left\langle
\Psi _{c}^{(+)}\right\vert ,  \label{eq.c2-111}
\end{equation}%
where $\left\vert \Phi _{B}\right\rangle$ and $\left\vert \Psi _{c}^{(+)}\right\rangle$ represent the wave functions of bound and scattering states, respectively.  The $T-$matrix \eqref{eq.c2-99}, written in terms of the interacting resolvent decomposed in eigenstates of $H$, is written as
\begin{align}
&T\left( E_3\right)=V+\sum_{B}\frac{V\left\vert \Phi _{B}\right\rangle
\left\langle \Phi _{B}\right\vert V}{E_3-E_{B}+\imath \epsilon}+ \int d^{2}k\frac{V\left\vert
\Psi _{c}^{(+)}\right\rangle \left\langle \Psi _{c}^{(+)}\right\vert V}{E-\bar{E}_{c}+\imath \epsilon},&
\label{eq.c2-112}
\end{align}%
where the bound-state poles of the transition operator appear explicitly. When the three-body system is close to a bound state ($E_3 \approx E_{B}$), the second term on the right-hand-side of Eq.(\ref{eq.c2-112}) is dominant, due to the pole, and the part concerning to the scattering states can be neglected. Defining the bound state vertex function by $\left\vert \Gamma _{B}\right\rangle =V\left\vert \Phi_{B}\right\rangle$, the three-body $T-$matrix (\ref{eq.c2-112}) near the pole becomes
\begin{equation}
T\left( E_3\right) \approx \frac{\left\vert \Gamma _{B}\right\rangle
\left\langle \Gamma _{B}\right\vert }{E_3-E_{B}}=\frac{\left\vert \Gamma
_{B}\right\rangle \left\langle \Gamma _{B}\right\vert }{E_3+\left\vert
E_{B}\right\vert }  \ .
\label{eq.c2-113}
\end{equation}%
Then,  Eq.~\eqref{eq.c2-113} is decomposed in three Faddeev components, as in Eq.~\eqref{eq.c2-102}, which reads
\begin{equation}
T_a\left( E_3\right) \approx \frac{\left\vert \Gamma
_{a}\right\rangle \left\langle \Gamma _{B}\right\vert }{E_3+\left\vert
E_{B}\right\vert }  \ ,
\label{eq.c2-114}
\end{equation}%
where $\left\vert \Gamma_{a}\right\rangle = v_a \left\vert \Phi_{B}\right\rangle$ and $\left\langle \Gamma _{B}\right\vert=\left\langle \Phi _{B}\right\vert V$. Inserting the $T-$matrix (\ref{eq.c2-114}) in Eq.~\eqref{eq.c2-110} gives
\begin{equation}
\frac{\left\vert \Gamma _{a}\right\rangle \left\langle \Gamma_B \right\vert }{
E_3 +\left\vert E_{B}\right\vert}\approx t_{a}\left( E_3-\frac{q_{a}^{2}}{m_{bc,a}}\right) \left[ 1+G_{0}\left( E_3 \right) \left( \frac{
\left\vert \Gamma _{b}\right\rangle \left\langle \Gamma_B \right\vert }{
E_3 +\left\vert E_{B}\right\vert }+\frac{\left\vert \Gamma
_{c}\right\rangle \left\langle \Gamma_B \right\vert }{E_3 +\left\vert
E_{B}\right\vert }\right)\right] .  \label{eq.c2-115}
\end{equation}%

When the three-body system is bound, $E_3 \rightarrow-\left\vert E_{B}\right\vert$ and in this limit Eq.~\eqref{eq.c2-115} becomes a homogeneous equation, which reads
\begin{equation}
\left\vert \Gamma _{a}\right\rangle =t_{a}\left( E_3-\frac{q_{a}^{2}}{m_{bc,a}}%
\right)  G_{0}\left( E_3 \right) \left(
\left\vert \Gamma _{b}\right\rangle +\left\vert \Gamma _{c}\right\rangle
\right) \ .
 \label{eq.c2-116}
\end{equation}

Writing the two-body $T-$matrix for the one term separable potential, as in Eq.~\eqref{eq.c2-46}, gives
\begin{equation}
\left\vert \Gamma _{a}\right\rangle =\left\vert \chi_a \right\rangle \tau_{a}
\left( E_3-\frac{q_{a}^{2}}{m_{bc,a}}\right) \left\langle \chi_a \right\vert
G_{0}\left( E_3 \right) \left( \left\vert \Gamma _{b}\right\rangle
+\left\vert \Gamma _{c}\right\rangle \right) \; ,  \label{eq.c2-117}
\end{equation}
and the projection of Eq.~\eqref{eq.c2-117} in states $\left\vert \mathbf{p}_{a},\mathbf{q}_{a}\right\rangle$ results in
\begin{equation}
\left\langle \mathbf{p}_{a},\mathbf{q}_{a}\right.\left\vert \Gamma _{a}\right\rangle =\left\langle \mathbf{p}_{a}\right.\left\vert \chi_a \right\rangle \tau_{a}
\left( E_3-\frac{q_{a}^{2}}{m_{bc,a}}\right) \left\langle \chi_a,\mathbf{q}_a \right\vert
G_{0}\left( E_3 \right) \left( \left\vert \Gamma _{b}\right\rangle
+\left\vert \Gamma _{c}\right\rangle \right) \; .   \label{eq.c2-118}
\end{equation}

For Dirac$-\delta$ potentials, $\left\langle \mathbf{p}_{a},\mathbf{q}_{a}\right\vert \left. \Gamma _{a}\right\rangle =\left\langle 
\mathbf{p}_{a}\right\vert \left. \chi _{a}\right\rangle \left\langle \mathbf{q}_{a}\right\vert \left. f_{a}\right\rangle =g_{a}\left( \mathbf{p}_{a}\right) f_{a}\left( \mathbf{q}_{a}\right) =f_{a}\left( \mathbf{q}_{a}\right)$ and the $i^{th}$ Faddeev component of the three-body bound state vertex, which satisfies an homogeneous integral equation, is given by
\begin{equation}
f_{a}\left( \mathbf{q}_{a}\right) =\tau_{a} \left( E_3-\frac{q_{a}^{2}}{2m_{bc,a}}\right) \left\langle \chi_{a},\mathbf{q}_{a}\right\vert G_{0}\left(E_3 \right) \Bigl( \left\vert \chi _{b}\right\rangle \left\vert
f_{b}\right\rangle +\left\vert \chi _{c}\right\rangle \left\vert f_{c}\right\rangle \Bigr) ,  \label{eq.c2-119}
\end{equation}%
where $f_{a}$ is the spectator function, which describes the interaction of each spectator particle with the corresponding two-body subsystem. The spectator functions $f_{b}$ and $f_{c}$ are easily found by cyclic permutation of the labels ($a,b,c$) in Eq.~\eqref{eq.c2-119}.

The components $f_{a}$, $f_{b}$ and $f_{c}$ satisfy a set of three coupled homogeneous integral equations, in the case where the interaction between particles is described for zero-range potentials.  For three identical bosons, only one homogeneous integral equation has to be solved, since $f_{a}\left( \mathbf{q}_{a}\right)=f_{b}\left( \mathbf{q}_{b}\right)=f_{c}\left( \mathbf{q}_{c}\right)$. In the same way, for two identical bosons plus a distinct particle, there is a set of two coupled homogeneous integral equations, since $f_{a}\left( \mathbf{q}_{a}\right)=f_{b}\left( \mathbf{q}_{b}\right) \neq f_{c}\left( \mathbf{q}_{c}\right)$. In the most general case of three distinguishable particles, the set of coupled equations reads
\begin{align}
f_{a}\left( \mathbf{q}_{a}\right) =&\tau_{a} \left( E_3-\frac{q_{a}^{2}}{2m_{bc,a}}\right) \left\langle \chi_{a},\mathbf{q}_{a}\right\vert G_{0}\left( E_3 \right) \Bigl( \left\vert \chi _{b}\right\rangle \left\vert
f_{b}\right\rangle+\left\vert \chi _{c}\right\rangle \left\vert f_{c}\right\rangle \Bigr)  \; , &  \label{eq.c2-120a} \\
&&\nonumber\\
f_{b}\left( \mathbf{q}_{b}\right) =&\tau_{b} \left( E_3-\frac{q_{b}^{2}}{2m_{ca,b}}\right) \left\langle \chi_{b},\mathbf{q}_{b}\right\vert G_{0}\left(E_3 \right) \Bigl( \left\vert \chi _{a}\right\rangle \left\vert
f_{a}\right\rangle +\left\vert \chi _{c}\right\rangle \left\vert f_{c}\right\rangle \Bigr) \; ,& \label{eq.c2-120b} \\
&& \nonumber\\
f_{c}\left( \mathbf{q}_{c}\right) =&\tau_{c} \left( E_3-\frac{q_{c}^{2}}{2m_{ab,c}}\right) \left\langle \chi_{c},\mathbf{q}_{c}\right\vert G_{0}\left(E_3 \right) \Bigl( \left\vert \chi _{a}\right\rangle \left\vert
f_{a}\right\rangle +\left\vert \chi _{b}\right\rangle \left\vert f_{b}\right\rangle \Bigr) \; .& \label{eq.c2-120c} 
\end{align}
The matrix elements in Eqs.~\eqref{eq.c2-120a} to \eqref{eq.c2-120c}  have the same structure, namely
\begin{equation}
\left\langle \chi _{a},\mathbf{q}_{a}\right\vert
G_{0}\left( E_3 \right) \left\vert \chi _{b}\right\rangle \left\vert
f_{b}\right\rangle .
\label{eq.c2-121}
\end{equation}
These matrix elements are derived in detail in Appendix~\ref{melements} and the result is used to finally write the set of three coupled homogeneous integral equations for the  bound state of an $abc$ system as
\begin{align}
f_{a}\left( \mathbf{q}_{a}\right)  &=\left[ -4\pi \frac{m_b m_c}{m_b+m_c}\ln \left( 
\sqrt{\frac{\frac{m_a+m_b+m_c}{2m_a\left( m_b+m_c\right) }q_{a}^{2}-E_{3}}{E_{bc}}}%
\right) \right] ^{-1}\times& \nonumber\\*
&\times\left[ \int d^{2}k\frac{f_{b}\left( \mathbf{k}%
\right) }{E_{3}-\frac{m_a+m_c}{2 m_a m_c}q_{a}^{2}-\frac{m_b+m_c}{2 m_b m_c}k^{2}-\frac{1%
}{m_c}\mathbf{k}\cdot \mathbf{q}_{a}}+\right.& \nonumber  \\*
&\left. +\int d^{2}k\frac{f_{c}\left( \mathbf{k}%
\right) }{E_{3}-\frac{m_a+m_b}{2m_am_b}q_{a}^{2}-\frac{m_b+m_c}{2m_bm_c}k^{2}-\frac{1%
}{m_b}\mathbf{k}\cdot \mathbf{q}_{a}}\right] ,& \label{eq.c2-122a}
\end{align}%
\bigskip
\begin{align}
f_{b}\left( \mathbf{q}_{b}\right)  &=\left[ -4\pi \frac{m_am_c}{m_a+m_c}\ln \left( 
\sqrt{\frac{\frac{m_a+m_b+m_c}{2m_b\left( m_a+m_c\right) }q_{b}^{2}-E_{3}}{E_{ac}}}%
\right) \right] ^{-1}\times& \nonumber\\*
&\times\left[ \int d^{2}k\frac{f_{a}\left( \mathbf{k}
\right) }{E_{3}-\frac{m_b+m_c}{2m_b m_c}q_{b}^{2}-\frac{m_a+m_c}{2m_am_c}k^{2}-\frac{1%
}{m_c}\mathbf{k}\cdot \mathbf{q}_{b}}+\right.& \nonumber  \\*
&\left. +\int d^{2}k\frac{f_{c}\left( \mathbf{k}\right) 
}{E_{3}-\frac{m_a+m_b}{2m_am_b}q_{b}^{2}-\frac{m_a+m_c}{2m_am_c}k^{2}-\frac{1}{m_a}\mathbf{%
k\cdot \mathbf{q}_{b}}}\right] ,& \label{eq.c2-122b}
\end{align}
\bigskip
\begin{align}
f_{c}\left( \mathbf{q}_{c}\right)  &=\left[ -4\pi \frac{m_a m_b}{m_a+m_b}\ln \left( 
\sqrt{\frac{\frac{m_a+m_b+m_c}{2m_c\left( m_a+m_b\right) }q_{c}^{2}-E_{3}}{E_{ab}}}%
\right) \right] ^{-1}\times& \nonumber\\*
&\times\left[ \int d^{2}k\frac{f_{a}\left( \mathbf{k}%
\right) }{E_{3}-\frac{m_b+m_c}{2m_b m_c}q_{c}^{2}-\frac{m_a+m_b}{2m_a m_b}k^{2}-\frac{1%
}{m_b}\mathbf{k}\cdot \mathbf{q}_{c}}+\right.& \nonumber  \\*
&\left. +\int d^{2}k\frac{f_{b}\left( \mathbf{k}%
\right) }{E_{3}-\frac{m_a+m_c}{2m_a m_c}q_{c}^{2}-\frac{m_a+m_b}{2m_a m_b}k^{2}-\frac{1%
}{m_a}\mathbf{k}\cdot \mathbf{q}_{c}}\right] .& \label{eq.c2-122c}
\end{align}
The particles $a$, $b$ and $c$ have masses $m_a,m_b,m_c$, respectively. Also, the two-body bound state energy of each pair, defined as the scale factor of the two-body system for Dirac$-\delta$ potentials (see Eq.~\eqref{eq.c2-68}), is specifically labeled as $E_{ab}$, $E_{bc}$ and $E_{ac}$. 

The spectator functions in Eqs.~\eqref{eq.c2-122a} to \eqref{eq.c2-122c} compose the three-body bound-state wave function. Using the vertex function defined before Eq.~\eqref{eq.c2-113}, $\left\vert \Gamma _{B}\right\rangle =V\left\vert \Phi_{B}\right\rangle$ , it is possible to write that $\left(v_a+v_b+v_c\right)\left|\Psi_B\right\rangle= \left|\Gamma_a\right\rangle+\left|\Gamma_b\right\rangle+\left|\Gamma_c\right\rangle $. 
Multiplying both sides by the free resolvent results in
\begin{align}
\left|\Psi_{abc}\right\rangle=\left|\Psi_a\right\rangle+\left|\Psi_b\right\rangle+\left|\Psi_c\right\rangle= G_0(E_3)\bigl[\left|\Gamma_a\right\rangle+\left|\Gamma_b\right\rangle+\left|\Gamma_c\right\rangle\bigr] \; , \label{eq.c2-145b}
\end{align} 
where $\left|\Psi_a\right\rangle=G_0(E_3)v_a\left|\Psi_B\right\rangle$ is one of the so-called Faddeev components of the wave function.
It is possible to choose any one of the set of Jacobi momenta to project Eq.~\eqref{eq.c2-145b}. The set  ($\mathbf{q}_a$,$\mathbf{p}_a$) gives
\begin{align}
&\left\langle\mathbf{q}_a,\mathbf{p}_a\right|\left.\Psi_{abc}\right\rangle=\left\langle\mathbf{q}_a,\mathbf{p}_a\right|G_0(E_3) \Bigl(\left|\Gamma_a\right\rangle+\left|\Gamma_b\right\rangle+\left|\Gamma_c\right\rangle \Bigr) \ . \label{eq.c2-145c}
\end{align}
The matrix elements on the right-hand-side of Eq.(\ref{eq.c2-145c}) can be handled in a similar way as it is done in Appendix~\ref{melements}. Finally, the three-body  bound-state wave function is written in term of the spectator functions as
\begin{equation}
\Psi_{abc}(\mathbf{q}_a,\mathbf{p}_a)=\frac{f_a(\mathbf{q}_a)+f_b \Bigl(\mathbf{q}_b(\mathbf{q}_a,\mathbf{p}_a)\Bigr)+f_c \Bigl(\mathbf{q}_c(\mathbf{q}_a,\mathbf{p}_a)\Bigr)}{E_3-\frac{m_a+m_b+m_c}{2m_a(m_b+m_c)}\mathbf{q}_a^2-\frac{m_b+m_c}{2m_b m_c}\mathbf{p}_a^2} \ ,
\label{eq.c2-145}
\end{equation}
where the Jacobi momenta $\mathbf{q}_b$ and $\mathbf{q}_c$ are linearly related to $\mathbf{q}_a$ and $\mathbf{p}_a$ through the relations given in Appendix~\ref{jacobi}.

In a compact notation, ($\alpha,\beta,\gamma$) are introduced as cyclic permutations of the labels ($a,b,c$) and the wave function is written taking into account the momentum of particle $\alpha$ with respect to the CM of the $\beta\gamma$ subsystem as
\begin{equation}
\Psi\left(\mathbf{q}_\alpha,\mathbf{p}_\alpha\right)=\frac{f_{\alpha}\left(q_\alpha\right)+f_{\beta}\left(\left| \mathbf{p}_\alpha- \frac{m_\beta}{m_\beta+m_\gamma}\mathbf{q}_\alpha\right|  \right)+f_{\gamma}\left(\left| \mathbf{p}_\alpha+ \frac{m_\gamma}{m_\beta+m_\gamma}\mathbf{q}_\alpha\right| \right)}{-E_{3}+\frac{q_\alpha^{2}}{2m_{\beta \gamma,\alpha}}+\frac{p_\alpha^{2}}{2m_{\beta \gamma}}} , 
\label{wave}
\end{equation}
where $\mathbf{q}_\alpha,\mathbf{p}_\alpha$ are the Jacobi momenta of particle $\alpha$ with the shifted arguments given in Eqs.~\eqref{jacobiab} and \eqref{jacobiac}  and $m_{\beta \gamma,\alpha}= m_\alpha(m_\beta+m_\gamma)/(m_\alpha+m_\beta+m_\gamma)$ and $m_{\beta \gamma}= (m_\beta+m_\gamma)/(m_\beta+m_\gamma)$ are the reduced masses. 
In the same way, the spectator functions in Eq.~\eqref{wave}, i.e., $f_{\alpha,\beta,\gamma}(\mathbf{q})$, fulfill the set of three coupled homogeneous integral equations for the bound state, which in the compact notation are written as
\begin{align}
&f_{\alpha}\left( \mathbf{q}\right)  =\left[ 4\pi m_{\beta \gamma}\ln \left( 
\sqrt{\frac{\frac{q^{2}}{2m_{\beta \gamma,\alpha} }-E_{3}}{E_{\beta\gamma}}}
\right) \right] ^{-1}& \label{spec}\\*
&\times\int d^{2}k\left( \frac{f_{\beta}\left(\mathbf{k}\right) }{-E_{3}+\frac{q^{2}}{2 m_{\alpha \gamma}}+\frac{k^{2}}{2 m_{\beta \gamma}}+\frac{1}{m_\gamma}\mathbf{k}\cdot \mathbf{q}}+
\frac{f_{\gamma}\left( \mathbf{k} \right) }{-E_{3}+\frac{q^{2}}{2m_{\alpha \beta }}+\frac{k^{2}}{2m_{\beta \gamma}}+\frac{1
}{m_\beta}\mathbf{k}\cdot \mathbf{q}}\right) .& \nonumber
\end{align}
As the interaction between particles is described for $s-$waves potentials and the focus is on states with total zero angular momentum, the spectator functions do not depend on the angle, i.e., $f_{\alpha}(\mathbf{q}) \equiv f_{\alpha}(q)$. Then, the angular integration in Eq.~\eqref{spec} is solved using that
\begin{equation}
\int_{0}^{2 \pi} \frac{d \theta}{1- z \cos \theta}= \frac{1}{\sqrt{1-z^2}} \; ,
\label{angular}
\end{equation}
where the constant $z$ satisfies $|z|<1$. The result is
\begin{multline}
f_{\alpha}\left( q \right)  =2 \pi \left[ 4\pi m_{\beta \gamma}\ln \left( 
\sqrt{\frac{\frac{q^{2}}{2m_{\beta \gamma,\alpha} }-E_{3}}{E_{\beta\gamma}}}
\right) \right] ^{-1} \label{spec1} \\
\times\int_0^\infty dk\left( \frac{ k \;f_{\beta}\left( k \right) }{\sqrt{\left(-E_{3}+\frac{q^{2}}{2 m_{\alpha \gamma}}  +\frac{k^{2}}{2 m_{\beta \gamma}}\right)^2-\left(\frac{k \; q} {m_\gamma}\right)^2}} \right. \\ \left. +
\frac{k \; f_{\gamma}\left( k\right) }{\sqrt{\left(-E_{3}+\frac{q^{2}}{2m_{\alpha \beta }}+\frac{k^{2}}{2m_{\beta \gamma}}\right)^2-\left(\frac{k \; q}{m_\beta}\right)^2}}\right) \; , 
\end{multline}
which together with Eq.~\eqref{wave} build the $L_{total}=0$ bound eigenstate of the Hamiltonian with the zero\=/range force.

The study of the three-body bound states, in what follows, is based on the numerical solution of the coupled homogeneous integral equations for the spectator functions in Eq.~\eqref{spec1}. Details about the numerical methods are given in Appendix~\ref{numerical}.

\section{Three-body bound state equation in 3D} \label{tsie3d}
The naive attempt to write the three-body bound state integral equation in 3D only by changing the phase factor and the two-body $T-$matrix in Eq.~\eqref{spec} fails, since the kernel of such equation is non-compact when the interaction between particles is described for Dirac$-\delta$ potentials \cite{adhikariPRA1988}. This means that the three-body equations must be renormalized, as it was done for the two-body $T-$matrix in Sec.~\ref{tzmr}. A complete discussion about the renormalization method is given in Refs.~\cite{adhikariPRL1995a,adhikariPRL1995,fredericoPiPaNP2012}, where a discussion of the equivalent method within effective field theory can be found. 

\subsection{Renormalization of the 3B transition operator} \label{r3bto}

The Lippmann-Schwinger equation for the transition operator is
\begin{equation}
T(E) = V + V G_0(E) T(E) =  V + T(E) G_0(E) V\; ,
\label{LS}
\end{equation}
which for the sake of the notation the energy is dropped.

The subtraction point is chosen as $-\mu_{(3)}^2$ and the transition matrix in this point is $T(-\mu_{(3)}^2)$. The potential $V$ can be expressed as
\begin{equation}
V =  \left[ 1 + T\left(-\mu_{(3)}^2\right) G_0\left(-\mu_{(3)}^2\right) \right]^{-1} T\left(-\mu_{(3)}^2\right) \; ,
\label{vmu}
\end{equation}
where $T\left(-\mu_{(3)}^2\right)$ is defined as the sum over the two\=/body transition matrices in the subtraction point \cite{adhikariPRL1995a}, namely
\begin{equation}
T\left(-\mu_{(3)}^2\right)=\sum_{n=a,b,c} t_n \left( -\mu_{(3)}^2 \right) \; ,
\label{t2sum3d}
\end{equation}
with $t_n(E)$ given in Eq.~\eqref{t23b}.
Inserting the renormalized potential~\eqref{vmu} in Eq.~\eqref{LS} gives
\begin{align}
&T(E) = \left[ 1 + T\left(-\mu_{(3)}^2\right) G_0\left(-\mu_{(3)}^2\right) \right]^{-1} T\left(-\mu_{(3)}^2\right) \left[1+ G_0(E) T(E) \right] \; , \nonumber\\*
&\left[ 1 + T\left(-\mu_{(3)}^2\right) G_0\left(-\mu_{(3)}^2\right) \right]T(E) = T\left(-\mu_{(3)}^2\right) + T\left(-\mu_{(3)}^2\right) G_0(E) T(E)  \; , \nonumber\\
&T(E) + T\left(-\mu_{(3)}^2\right) G_0\left(-\mu_{(3)}^2\right)T(E)  = T\left(-\mu_{(3)}^2\right) + T\left(-\mu_{(3)}^2\right) G_0(E) T(E)  \; , \nonumber\\*
&T(E) = T\left(-\mu_{(3)}^2\right) + T\left(-\mu_{(3)}^2\right) G_1\left(E,-\mu_{(3)}^2\right) T(E)   \; ,  \label{T3D}
\end{align}
where
\begin{align}
G_1\left(E,-\mu_{(3)}^2\right)=G_0(E)-G_0\left(-\mu_{(3)}^2\right)=
-\left(\mu_{(3)}^2+E\right)G_0(E)G_0\left(-\mu_{(3)}^2\right) \; .
\label{g13d}
\end{align} 
Notice that the matrix form of Eq.~\eqref{T3D} is given in Eq.~\eqref{eq.c2-107}, meaning that each component of the renormalized three-body transition matrix is given by
\begin{align}
T_a(E_3)=t_a\left(E_3-\frac{q_a^2}{2m_{bc,a}}\right)\Biggl\{  1+ G_1\left(E_3,-\mu_{(3)}^2\right) \biggl[T_b(E_3)+T_c(E_3)\biggr]\Biggr\} \; ,
\label{T3Di}
\end{align}
which is analogous to Eq.~\eqref{eq.c2-110}, where the only difference arises from the three-body propagator.

\subsection{Three-body bound state integral equation in 3D} \label{tbsie3d}
Since Eqs.~\eqref{eq.c2-110} and \eqref{T3Di} are equivalent, the procedure to obtain the three-body bound state equation in 3D is exactly the same followed in Sec.~\ref{tbsiq2d}, only replacing $G_0(E_3) \to G_1\left(E_3,-\mu_{(3)}^2\right)$. Then, the homogeneous coupled equations for the spectator function to get the bound state energy are given by
\begin{align}
f_{a}\left( \mathbf{q}_{a}\right) =&\tau_{a} \left( E_3-\frac{q_{a}^{2}}{2m_{bc,a}}\right) \left\langle \chi_{a},\mathbf{q}_{a}\right\vert G_1\left(E_3,-\mu_{(3)}^2\right) \Bigl( \left\vert \chi _{b}\right\rangle \left\vert
f_{b}\right\rangle+\left\vert \chi _{c}\right\rangle \left\vert f_{c}\right\rangle \Bigr)  \; , &  \label{fam3d} \\
&&\nonumber\\
f_{b}\left( \mathbf{q}_{b}\right) =&\tau_{b} \left( E_3-\frac{q_{b}^{2}}{2m_{ca,b}}\right) \left\langle \chi_{b},\mathbf{q}_{b}\right\vert G_1\left(E_3,-\mu_{(3)}^2\right) \Bigl( \left\vert \chi _{a}\right\rangle \left\vert
f_{a}\right\rangle +\left\vert \chi _{c}\right\rangle \left\vert f_{c}\right\rangle \Bigr) \; ,& \label{fbm3d} \\
&& \nonumber\\
f_{c}\left( \mathbf{q}_{c}\right) =&\tau_{c} \left( E_3-\frac{q_{c}^{2}}{2m_{ab,c}}\right) \left\langle \chi_{c},\mathbf{q}_{c}\right\vert G_1\left(E_3,-\mu_{(3)}^2\right)\Bigl( \left\vert \chi _{a}\right\rangle \left\vert
f_{a}\right\rangle +\left\vert \chi _{b}\right\rangle \left\vert f_{b}\right\rangle \Bigr) \; .& \label{fcm3d} 
\end{align}
Notice that the matrix elements in Eqs.~\eqref{fam3d} to \eqref{fcm3d}  have the same structure as the ones in Eqs.~\eqref{eq.c2-120a} to \eqref{eq.c2-120c} , namely
\begin{equation}
\left\langle \chi _{a},\mathbf{q}_{a}\right\vert
G_1\left(E_3,-\mu_{(3)}^2\right) \left\vert \chi _{b}\right\rangle \left\vert
f_{b}\right\rangle .
\label{me3d}
\end{equation}
Since the term $G_1\left(E_3,-\mu_{(3)}^2\right)$ can be separated in two terms, as in Eq.~\eqref{g13d}, it turns out that each element in Eqs.~\eqref{fam3d} to \eqref{fcm3d}  is identical to the corresponding one in Eqs.~\eqref{eq.c2-120a} to \eqref{eq.c2-120c}, which are derived in detail in Appendix~\ref{melements}. The set of three coupled homogeneous integral equations for the  bound state of an $abc$ system is written in a compact form as 
\begin{align}
&f_{\alpha}\left( \mathbf{q}\right)  = \left[  2 \pi^2  \left( 2 m_{\beta \gamma} \right)^{3/2} \left( \sqrt{ \left(\frac{q^{2}}{2m_{\beta \gamma,\alpha} }-E_{3} \right)} - \sqrt{E_{\beta \gamma} }\right) \right]^{-1}   \label{spec3d}\\*
&\times \int d^{3} k \left[ \left( \frac{1}{-E_{3}+\frac{q^{2}}{2 m_{\alpha \gamma}}+\frac{k^{2}}{2 m_{\beta \gamma}}+\frac{1}{m_\gamma}\mathbf{k}\cdot \mathbf{q}} - \frac{1 }{\mu^2+\frac{q^{2}}{2 m_{\alpha \gamma}}+\frac{k^{2}}{2 m_{\beta \gamma}}+\frac{1}{m_\gamma}\mathbf{k}\cdot \mathbf{q}} \right) f_{\beta}\left(\mathbf{k}\right) \right. \nonumber\\* 
& \hskip 1.2cm \left. +\left(\frac{1}{-E_{3}+\frac{q^{2}}{2m_{\alpha \beta }}+\frac{k^{2}}{2m_{\beta \gamma}}+\frac{1}{m_\beta}\mathbf{k}\cdot \mathbf{q}} - \frac{1}{\mu^2+\frac{q^{2}}{2m_{\alpha \beta }}+\frac{k^{2}}{2m_{\beta \gamma}}+\frac{1}{m_\beta}\mathbf{k}\cdot \mathbf{q}} \right) f_{\gamma}  \left( \mathbf{k} \right) \right] .  \nonumber
\end{align}

The solutions of Eq.\eqref{spec3d} with zero total angular momentum are studied, and as the interaction between particles is described for $s-$waves potentials, the spectator functions only depends on the momentum modules, i.e., $f_{\alpha}(\mathbf{q}) \equiv f_{\alpha}(q)$. Then, the angular integration in Eq.~\eqref{spec3d} is solved by using that
\begin{equation}
\int_{-\pi}^{\pi} \frac{d\theta  \sin \theta }{1- z \cos \theta}= \ln \left( \frac{1+z}{1-z} \right) \; ,
\label{angular3d}
\end{equation}
where the constant $z$ satisfies $|z|<1$. The result is
\begin{multline} \label{spec13d}
f_{\alpha}\left( \mathbf{q}\right)  = \left[  \pi  \left( 2 m_{\beta \gamma} \right)^{3/2} \left( \sqrt{ \left(\frac{q^{2}}{2m_{\beta \gamma,\alpha} }-E_{3} \right)} - \sqrt{E_{\beta \gamma} }\right) \right]^{-1}   \\*
\times \int_0^{\infty} dk \;k^2 \left[  \left(\ln \frac{-E_{3}+\frac{q^{2}}{2 m_{\alpha \gamma}}+\frac{k^{2}}{2 m_{\beta \gamma}}+\frac{k\;q}{m_\gamma} }{-E_{3}+\frac{q^{2}}{2 m_{\alpha \gamma}}+\frac{k^{2}}{2 m_{\beta \gamma}}-\frac{k\;q}{m_\gamma} } - \ln \frac{\mu^2+\frac{q^{2}}{2 m_{\alpha \gamma}}+\frac{k^{2}}{2 m_{\beta \gamma}}+\frac{k\;q}{m_\gamma} }{\mu^2+\frac{q^{2}}{2 m_{\alpha \gamma}}+\frac{k^{2}}{2 m_{\beta \gamma}}-\frac{k\;q}{m_\gamma} } \right) f_{\beta}\left(\mathbf{k}\right) \right.\\* 
\left. +\left(\ln \frac{-E_{3}+\frac{q^{2}}{2m_{\alpha \beta }}+\frac{k^{2}}{2m_{\beta \gamma}}+\frac{k\;q}{m_\beta}}{-E_{3}+\frac{q^{2}}{2m_{\alpha \beta }}+\frac{k^{2}}{2m_{\beta \gamma}}-\frac{k\;q}{m_\beta}} - \ln \frac{\mu^2+\frac{q^{2}}{2m_{\alpha \beta }}+\frac{k^{2}}{2m_{\beta \gamma}}+\frac{k\;q}{m_\beta} }{\mu^2+\frac{q^{2}}{2m_{\alpha \beta }}+\frac{k^{2}}{2m_{\beta \gamma}}-\frac{k\;q}{m_\beta}} \right) f_{\gamma}  \left( \mathbf{k} \right) \right] .  
\end{multline}
The 3D wave function is also given by Eq.~\eqref{wave} and ($\alpha,\beta,\gamma$) are cyclic permutations of the labels ($a,b,c$).
\chapter{Universal 2D three-body bound states} \label{ch3}

The behavior of three-boson systems changes remarkably from two (2D) to three dimensions (3D), since the dynamics and properties of quantum systems drastically change when the system is restricted to different dimensions. For example, the scattering-length is defined within a constant for 2D systems \cite{adhikariAJoP1986} and, as it was already pointed out in Chapter~\ref{ch1}, the kinetic energy operator gives a negative (attractive) centrifugal barrier for 2D systems with zero total angular momentum, while the centrifugal barrier is always non-negative (zero or repulsive) for 3D systems. This means that any infinitesimal amount of attraction produce a bound state in 2D \cite{landau1977}, while a finite amount of attraction is necessary for binding a 3D system.

Another important difference between 2D and 3D systems is the occurrence of the Thomas collapse \cite{thomasPR1935} and the Efimov effect \cite{efimovYF1970}. These effects were predicted and measured for three identical bosons in 3D systems, but are absent in 2D. While in 3D the Efimov effect produces an infinite sequence of states when the scattering length diverges, previous studies have shown that the spectrum of three identical bosons in 2D contains exactly one two-body and two three-body bound states in the limit where the range of the force goes to zero \cite{bruchPRA1979,adhikariPRA1988}. Furthermore, the ratio between the three and two-body energies and radii attain universal values, no matter the detail of the short-range potential used \cite{nielsenPRA1997,nielsenPR2001}.

Starting from the well-known case of three identical bosons, universal properties of mass-imbalanced three-body systems in 2D are systematically studied using the zero-range interaction in momentum space \cite{bellottiJoPB2011,bellottiPRA2012}. In this Chapter, the focus is particularly on the dependence of the three-body binding energy with masses and two-body binding energies. The critical values of these parameters (masses and two-body binding energies) allowing a given number of three-particle bound states with zero total angular momentum are determined in a form of boundaries in the multidimensional parametric space. Moreover, it is shown that in extreme asymmetric mass systems, when one of the particles is much lighter than the other two, no bound in the number of weakly three-body bound states is found, as the light particle mass goes to zero \cite{bellottiJoPB2013}. This topic is discussed in detail in Chapter \ref{ch4}.

The dependence of the three-particle binding energy on the parameters is highly non-trivial even in the simpler case of two identical particles and a distinct one.  This dependence is parametrized for the ground and first excited state in terms of {\itshape supercicles} functions  \cite{lame1818}, even for the most general case of three distinguishable particles \cite{bellottiPRA2012}.

\section{Symmetry relations} \label{sec3.2}
An advantage in the use of two-body energies instead of interaction strengths in the homogeneous integral coupled equations for the bound state Eq.~\eqref{spec} is that only mass and energy ratios enter these equations. This means that the three-body energy divided by one of the two-body energies can be expressed as a function of four dimensionless parameters, i.e.,
\begin{align}
 \epsilon_3 = F_n\left(\frac{E_{\beta \gamma}}{E_{\alpha \beta}},
\frac{E_{\alpha \gamma}}{E_{\alpha \beta}}, \frac{m_\beta}{m_\alpha}, \frac{m_\gamma}{m_\alpha} \right)
 \equiv F_n\left(\epsilon_{\beta \gamma},\epsilon_{\alpha \gamma},\frac{m_\beta}{m_\alpha}, \frac{m_\gamma}{m_\alpha} \right),
\label{e40}
\end{align}
where $\epsilon_3=E_3/E_{\alpha \beta}$ is the scaled three-body energy, $\epsilon_{\beta \gamma}=E_{\beta \gamma}/E_{\alpha \beta}$ and $\epsilon_{\alpha \gamma}=E_{\alpha \gamma}/E_{\alpha \beta}$ are the scaled two-body energies, where $(\alpha,\beta,\gamma)$ are cyclic permutations of the particle labels $(a,b,c)$. The universal functions $F_n$ are labeled with the subscript $n$ to distinguish between ground, $n=0$, and excited states, $n>0$. Interchanging the particle labels, all the
universal functions $F_n$ must obey the symmetry relations
\begin{align} 
 F_n\left(\epsilon_{bc},\epsilon_{ac},\frac{m_b}{m_a}, \frac{m_c}{m_a} \right) &= 
 F_n\left(\epsilon_{ac},\epsilon_{bc}, \frac{m_c}{m_a},\frac{m_b}{m_a} \right)  =&  \nonumber \\
 \epsilon_{bc} 
 F_n\left(\frac{1}{\epsilon_{bc}},
 \frac{\epsilon_{ac}}{\epsilon_{bc}},
 \frac{m_a}{m_b}, \frac{m_c}{m_b}\right) &=
 \epsilon_{bc}
 F_n\left(\frac{\epsilon_{ac}}{\epsilon_{bc}},\frac{1}{\epsilon_{bc}},
 \frac{m_c}{m_b},\frac{m_a}{m_b}\right)= &  \nonumber \\   
 \epsilon_{ac}
F_n\left(\frac{1}{\epsilon_{ac}},\frac{\epsilon_{bc}}{\epsilon_{ac}},
\frac{m_a}{m_c},\frac{m_b}{m_c} \right)
 &= \epsilon_{ac}
 F_n\left(\frac{\epsilon_{bc}}{\epsilon_{ac}},\frac{1}{\epsilon_{ac}},
 \frac{m_b}{m_c},\frac{m_a}{m_c}\right).&
 \label{e57}
\end{align}

The energy and mass scaling reduces the number of unknown parameters from six to four. A straightforward advantage is that the symmetry relations in Eq.~\eqref{e57} allow investigations of $F_n$ to be taken in smaller regions of this four-parameter space, as explained in detail in the following sections.

\section{Survey of mass dependence}  \label{sec3.3}

\begin{figure}[htb!]
\centering
\includegraphics[width=0.9\textwidth]{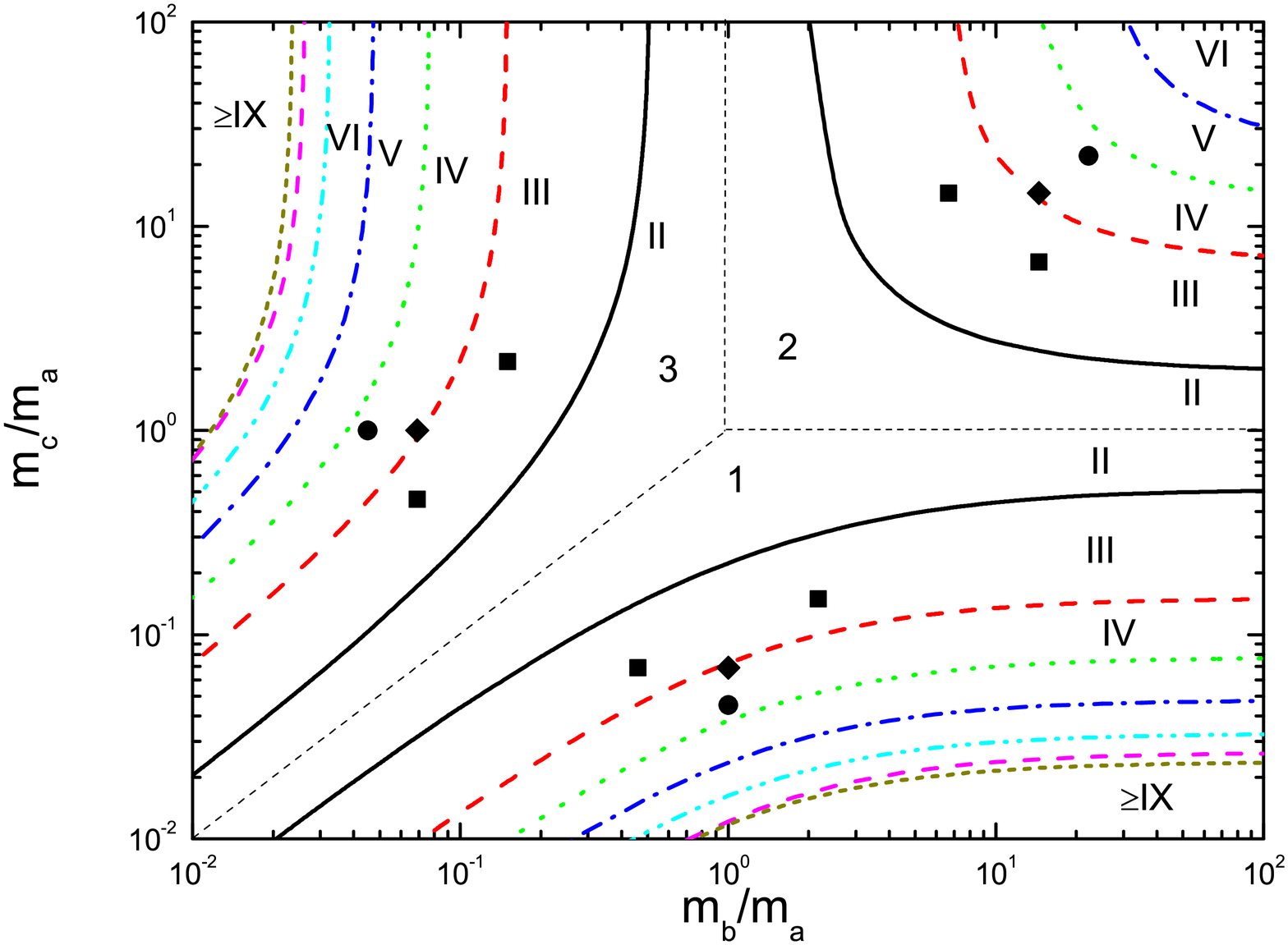}
\caption[Mass diagram of the number of three-body bound states as
  functions of two mass ratios, $\frac{m_b}{m_a}$ and $\frac{m_c}{m_a}$.  The
  three two-body energies are equal, i.e., $E_{ab}=E_{bc}=E_{ac}$.]
  {Mass diagram of the number of three-body bound states as
  functions of two mass ratios, $\frac{m_b}{m_a}$ and $\frac{m_c}{m_a}$.  The
  three two-body energies are equal, i.e., $E_{ab}=E_{bc}=E_{ac}$.
  The Roman numerals indicate the number of bound states in each region. The systems investigated are represented by
  square ($^6$Li-$^{40}$K-$^{87}$Rb), circular ($^{6}$Li-$^{133}$Cs-$^{133}$Cs) and diamond ($^{6}$Li-$^{87}$Rb-$^{87}$Rb) points. The three sets of points are related by the symmetries in Eq.~\eqref{e57}.}
\label{fig1}
\end{figure}

The mass dependence of the number of bound states for a three-body system where all the two-body subsystems have the same energy of interaction, i.e., $E_{ab}=E_{ac}=E_{bc}$ is shown in Fig.~\ref{fig1}.
In the central region around equal masses only two three-body bound states are available \cite{bruchPRA1979}. This region,
labeled II, extends in three directions corresponding to one heavy and
two rather similar light particles, that is either $\frac{m_c}{m_a} \ge 1$ and $\frac{m_b}{m_a} \ge 1$, or $m_a \simeq m_b \le m_c$.
Moving away from these regions in Fig.~\ref{fig1}, the number of stable
bound states increases in all directions. As an example, consider
$\frac{m_b}{m_a} =10$ and vary $\frac{m_c}{m_a}$ from small to large value in Fig.~\ref{fig1}.
Perhaps surprisingly, along this line
the number of bound states decreases to a minimum of two and 
subsequently increases again. The reason is that a decreasing mass asymmetry in the three-body system implies less attraction in the effective potential experienced for the light particle due to the two heavy ones (see Chapter \ref{ch4}) and consequently the disappearing states merge into the two-body continuum.  A similar
behavior is found in three dimensions 
when the attractive strength is increased and happens the disappearance of the infinitely many
Efimov states\cite{yamashitaPRA2002}. 
As the mass asymmetry increases again, the strength of effective potential also increases giving room for more bound states.

\begin{figure}[!htb]
\centering
\includegraphics[width=0.9\textwidth]{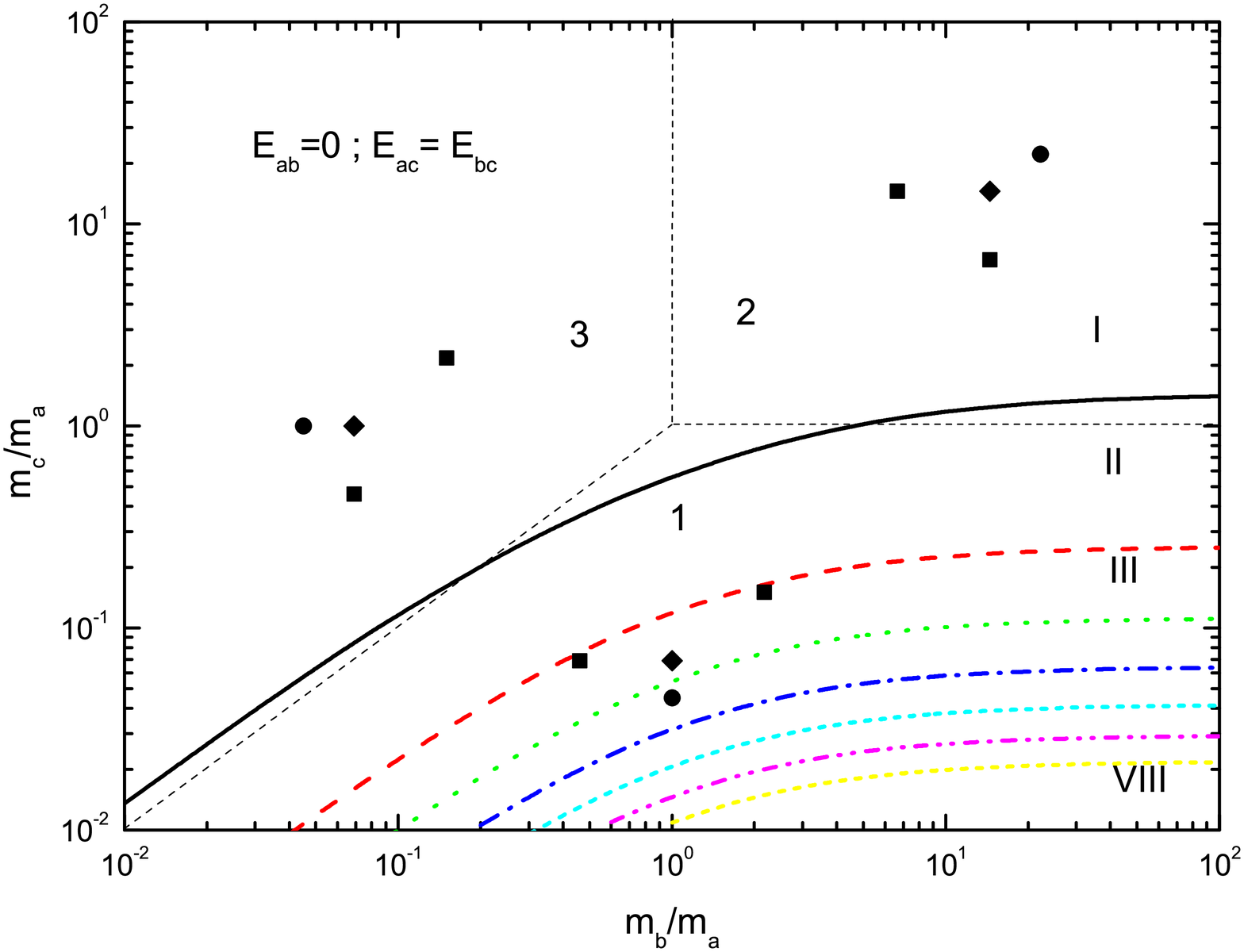}
\caption[Mass diagram of the number of three-body bound states as
  functions of two mass ratios, $\frac{m_b}{m_a}$ and $\frac{m_c}{m_a}$. The two-body energies are  $E_{ab}=0$ and $E_{ac}=E_{bc}$.]
{Mass diagram of the number of three-body bound states as
  functions of two mass ratios, $\frac{m_b}{m_a}$ and $\frac{m_c}{m_a}$. The two-body energies are  $E_{ab}=0$ and $E_{ac}=E_{bc}$.   The Roman numerals indicate the number of bound states in each region. The systems investigated are represented by
  square ($^6$Li-$^{40}$K-$^{87}$Rb), circular ($^{6}$Li-$^{133}$Cs-$^{133}$Cs) and diamond ($^{6}$Li-$^{87}$Rb-$^{87}$Rb) points. The numbers $1,2,3$ label three different sectors.} 
\label{Graph06}
\end{figure}

\begin{figure}[!htb]
\centering
\includegraphics[width=0.9\textwidth]{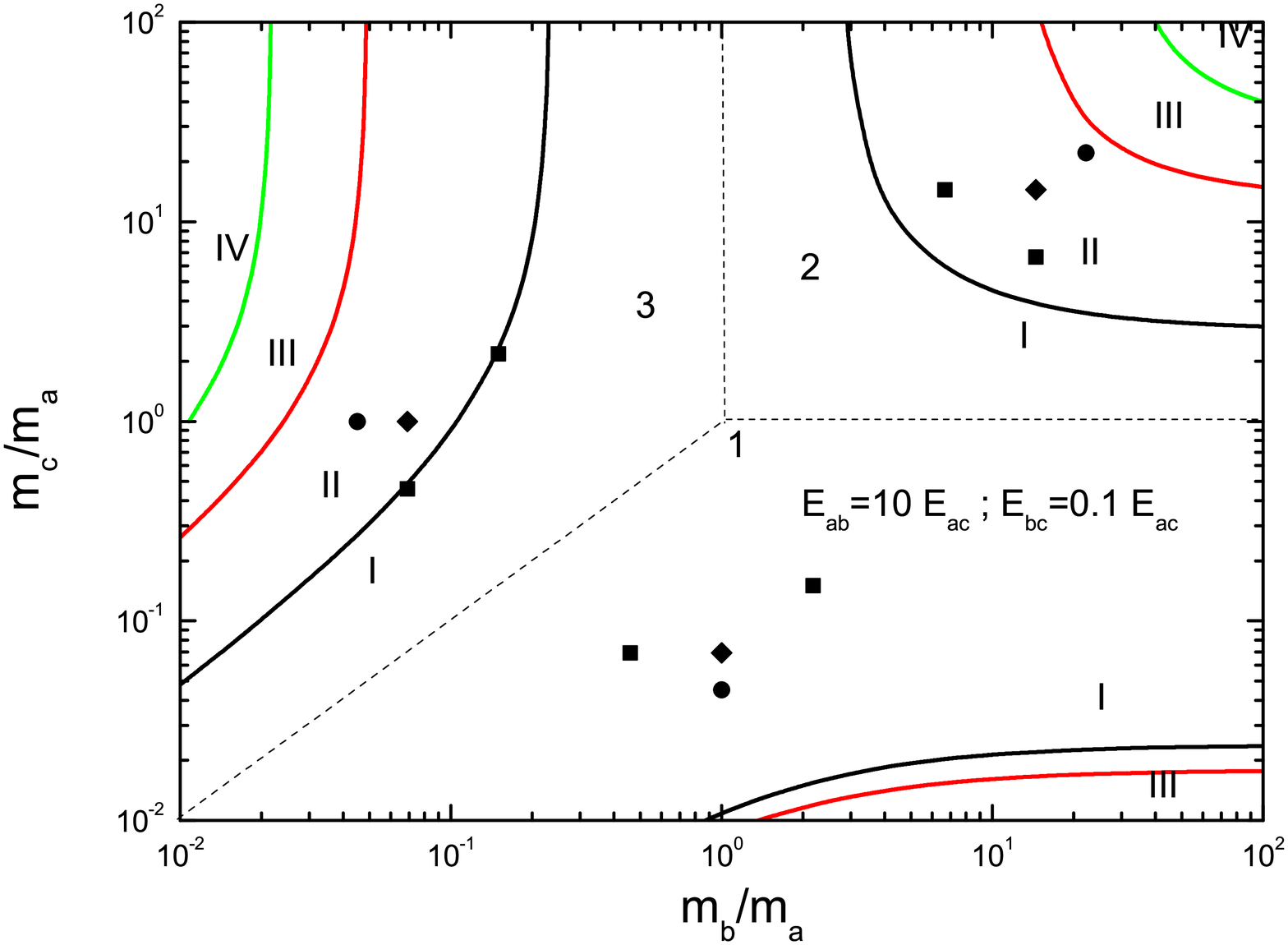}
\caption[Mass diagram of the number of three-body bound states as
functions of two mass ratios, $\frac{m_b}{m_a}$ and $\frac{m_c}{m_a}$. The two-body energies are $E_{ab}=10 E_{ac}$ and $E_{bc}=0.1 E_{ac}$.]
{Mass diagram of the number of three-body bound states as
functions of two mass ratios, $\frac{m_b}{m_a}$ and $\frac{m_c}{m_a}$. The two-body energies are $E_{ab}=10 E_{ac}$ and $E_{bc}=0.1 E_{ac}$.   The Roman numerals indicate the number of bound states in each region. The systems investigated are represented by
  square ($^6$Li-$^{40}$K-$^{87}$Rb), circular ($^{6}$Li-$^{133}$Cs-$^{133}$Cs) and diamond ($^{6}$Li-$^{87}$Rb-$^{87}$Rb) points. The numbers $1,2,3$ label three different sectors.} 
\label{Graph07}
\end{figure}

Variation of the two-body energies from all being equal leads to a
distortion of the boundaries in Fig.~\ref{fig1}, but the main structure remains. The central
region still has the smallest number of stable bound states and any
variation in the two-body energies expands region II, pushing the
other lines away from the center. This result is illustrated in Figs.~\ref{Graph06} and \ref{Graph07} and indicates that the maximum number of bound states supported for any three-body system, no matter the masses, is achieved when all the two-body subsystems interact with the same energy.  Therefore, Fig.~\ref{fig1} gives the
maximum number of stable bound states for any three-body system composed for particles with different masses. 

Although presenting the richest energy spectrum, the scenario of three distinct two-body subsystems interacting with the same energy seems hard to be implemented experimentally. However, it was recently reported in Ref.~\cite{reppPRA2013} that mixtures of $^{133}$Cs and $^{6}$Li were successfully trapped with a diverging scattering length of the $^{133}$Cs-$^{133}$Cs subsystem. A system composed of two-heavy particles and a light one is described for instance in the region where $m_c<m_a$ and $m_c<m_b$. Besides, if two particles do not interact in 2D, their energy can be set null. A mass diagram which includes such situaion (the $^{133}$Cs$^{133}$Cs$^{6}$Li system is represented as circular points in Figs.~\ref{fig1} to \ref{Graph07}) is constructed taking $E_{ab}=0$ and keeping $E_{ac}=E_{bc}$ and is shown in Fig.~\ref{Graph06}. Region I emerges in the middle of the figure, pushing the other lines away from the center. Excited states are only present in sector $1$, where the two non-interacting particles are heavier than the third one (this configuration is studied in detail in Chapter \ref{ch4}).

The symmetries in Eq.~\eqref{e57}, which clearly appear in Fig.~\ref{fig1}, can not be seen in Fig.~\ref{Graph06}, but this does not mean that symmetry was broken. This apparent contradiction comes from the way that the mass-diagram is built. In sector $1$ the light particle is $m_c$, i.e., $m_c<m_a$ and $m_c<m_b$, so that $E_{ab}=0$ means that the two heaviest particles are not interacting. Starting in sector $1$ of Fig.~\ref{Graph06} and moving towards sector $2$, after crossing the horizontal dashed line $E_{ab}=0$ does not mean that the two heaviest particles are not interacting any more, since in this region the particles $b$ and $c$ are the heaviest, i.e., $m_a<m_b$ and $m_a<m_c$ and the effective interaction between the heavy particles is mediated by light one, namely particle $a$. The same happens in region $3$, where particles $a$ and $c$ are the heaviest. In fact, a mass-diagram for imbalanced two-body energies shows information for three different systems. Therefore, each sector in Fig.~\ref{Graph06} obey the symmetries in Eq.~\eqref{e57} itself. For instance, the configuration showed in Fig.~\ref{Graph06} is also described for $E_{bc}=0$ with $m_a<m_b$ and $m_a<m_c$ or $E_{ac}=0$ with $m_b<m_a$ and $m_b<m_c$. These choices lead to two other plots, where the boundary lines in Fig.~\ref{Graph06} rotates to sector $2$ and $3$, respectively. Notice that the symmetries in Eq.~\eqref{e57} are not well defined for a non-interacting two-body subsystem, however, it is not hard to extend them for this case. 

What about a system where all two-body energies are different from each other? This scenario is shown in Fig.~\ref{Graph07}.  In sector $1$, where the energy between the two heaviest particles is greater than the other ones, the three distinct systems which are shown in Figs.~\ref{fig1} and \ref{Graph06} have only one bound state each. In this energy-configuration, region $2$ should be the most similar to the previous case, where the heavy-heavy system is not as bound as the other ones. However, sectors $2$ and $3$ seems to be almost symmetric in Fig.~\ref{Graph07}, showing that both $^{6}$Li-$^{133}$Cs-$^{133}$Cs and $^{6}$Li-$^{87}$Rb-$^{87}$Rb systems have two bound states each. A small difference is seem for $^6$Li-$^{40}$K-$^{87}$Rb, which has two bound states in sector $2$, but only one in sector $3$. The similarity between sector $1$ in Fig.~\ref{Graph06} and sector $2$ in Fig.~\ref{Graph07} is not clear enough because the strongly bound heavy-light system changes the threshold of binding the three-body system, cutting-out the most weakly three-body bound states. More details about these mass-diagrams and discussion about bound states are found in \cite{bellottiJoPB2011,bellottiPRA2012,bellottiJoPB2013}.

\section{Three-body energies for given masses}
Realistic scenarios correspond to given particles (atoms or molecules)
with known masses where in contrast, the interactions are
variable through Feshbach resonances \cite{chinRMP2010}. Once the constituents are chosen, the diagrams from the last section can be powerful guides in the experimental search for 2D three-body bound states, as they indicate how many bound states are expected to exist. However, this number vary for different two-body energies. The dependence in the number of bound states with the two-body energies for a specific system is  discussed in this section. 
  
\begin{figure}[htb!]
\centering
\includegraphics[width=0.9\textwidth]{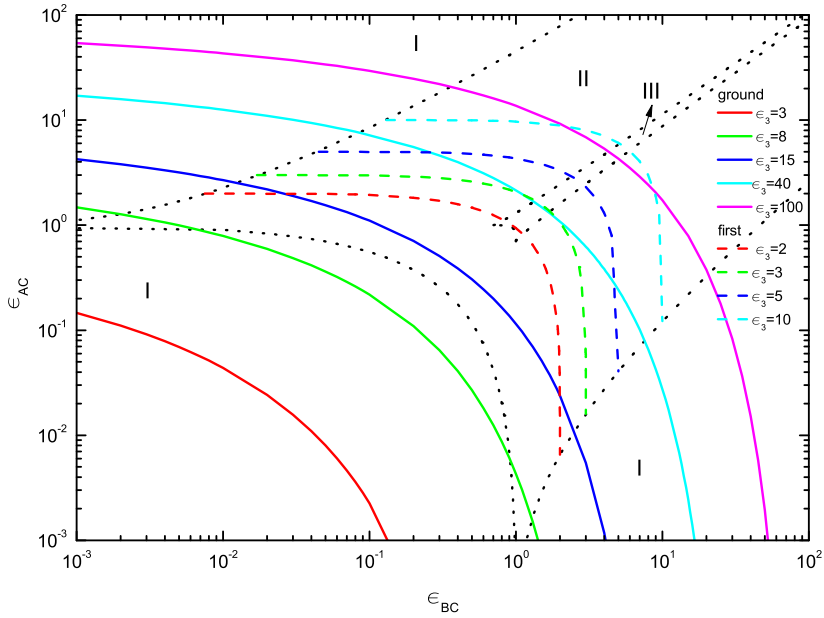}
\caption[Contour diagrams with lines of fixed $\epsilon_3$ values as
  function of the two-body energies $\epsilon_{ac}$ and $\epsilon_{bc}$.]
{Contour diagrams with lines of fixed $\epsilon_3$ values as
  function of the two-body energies $\epsilon_{ac}$ and $\epsilon_{bc}$.
  The solid and dashed curves are for ground and excited states,
  respectively. Here $a$ is $^{87}$Rb, $b$ is $^{40}$K, and $c$ is 
  $^{6}$Li.
  The dotted curves show where the number of stable bound
  states change from one (regions most asymmetric for small energies),
  to two (comparable size of the two-body energies), and to three (equal
  and large energies). The roman labels I, II, and III are as in Fig.~\ref{fig1}. }
\label{fig2}
\end{figure}

Assuming masses corresponding to the alkali
atoms $^{87}$Rb, $^{40}$K and $^{6}$Li, the two ratios of two-body
energies are left as variables where each set uniquely specifies
the three-body energies of ground and possibly excited states.
A contour diagram of the scaled three-body
energies for the two lowest stable bound states of the chosen system is shown in Fig.~\ref{fig2}.  The log-log plot can
be deceiving and on a linear scale the curves of equal scaled
three-body energy would be concave for the ground state and almost
linear or slightly convex for the first excited state in contrast to
the convex contours in Fig.~\ref{fig2}.  The chosen set of masses only
allow one, two, or three stable bound states, depending on the two-body
energies. The corresponding regions are shown by dotted curves in
Fig.~\ref{fig2}. The true extent of the regions cannot be seen.  Both
region II and III are closed, namely, region II continues along region III
up to energy ratios of about $10^{\pm5}$, and the narrow region III is
entirely embedded in region II. Other sets of mass ratios, as $^{6}$Li-$^{133}$Cs-$^{133}$Cs or $^{6}$Li-$^{87}$Rb-$^{87}$Rb, for instance, could open
region III and allow regions inside with more than three stable bound
states.

In the log-log plot of Fig.~\ref{fig2} the two-body
energies vary by five orders of magnitude, whereas the scaled
three-body energies for a stable system must be larger than all
two-body energies.  The three-body energy contours connect
minimum and maximum two-body energies, that is zero and maximum
two-body energies for the ground state and thresholds boundaries for
existence of the excited states \cite{bellottiJoPB2011}.  The contours appear in regular
intervals with larger values for increasing two-body energies.  Large
three-body energies reflect small spatial extension and therefore are less
interesting as it presumably is unreachable in the universal limit.
The contours pass continuously through the boundaries of the different regions
since the ground state exists without knowledge of the excited states. 
The contours in Fig.~\ref{fig2} for the excited state can only appear
in the regions with two or more stable bound states.  These contours
therefore must connect points of the boundaries between regions I and
II. They may cross continuously through region III precisely as the
ground state would cross through boundaries between regions I and II.
Similar contours exist within region III but are not exhibited in
this narrow strip where they are allowed.  The scaled three-body
energies are often substantially larger than the initial two-body
energies, although both arise from the same two-body interactions.

\section{Parametrization of three-body energies}
The universal functions $F_n$ defined in Eq.~\eqref{e40} are not
easily found. Their dependence on masses and two-body energies are highly non-trivial, even in a simpler scenario of two identical particles and a distinct one \cite{bellottiJoPB2011}. However, the contour diagrams in Fig.~\ref{fig2} suggest
a simple implicit dependence in terms of an extended Lam{\'e} curves
or {\itshape super ellipses} \cite{lame1818}. Note that, despite of the 
log-log scale in Fig.~\ref{fig2}, the parametrization in terms
of Lam{\'e} curves is done with the energies on a linear scale. 
The three-body energies can be written
indirectly by {\itshape super circles}, i.e.,
\begin{equation}
 \epsilon_{ac}^{t_{n}} +  \epsilon_{bc}^{t_{n}} = R_n^{t_{n}} \;,
\label{e70}
\end{equation}
where the radius, $R_n$, and the power, $t_{n}$, are functions
of $\epsilon_3$ and both depend on the two mass ratios. The term
{\itshape super circle} has been adopted since Eq.~\eqref{e70} only differs from a circumference of radii $R_n$ and coordinates ($\epsilon_{ac},\epsilon_{bc}$) in the power $t_n$, which is not equal two in general. The smallest value of $\epsilon_3$ is unity corresponding to the
two-body threshold of the $ab$ system used as the energy unit. 

\begin{figure}[!htb]
\centering
\includegraphics[width=0.9\textwidth]{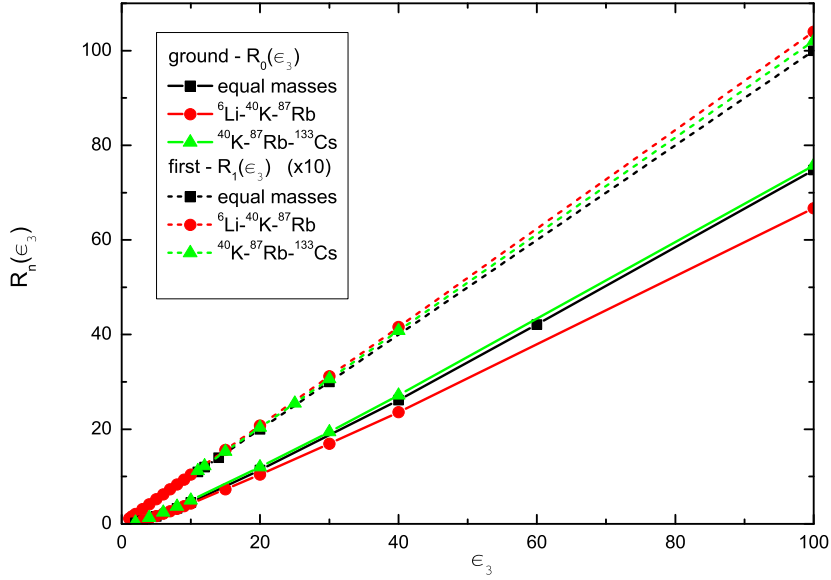}
\caption[The functions, $R_0$ and $R_1$, of the three-body
  energy $\epsilon_3$ in the super-ellipse fit for three sets of mass ratios.]
{The functions, $R_0$ and $R_1$, of the three-body
  energy $\epsilon_3$ in the super-ellipse fit for three sets of mass ratios, which are
  $(m_{\alpha},m_{\beta})= (1,1)$, $(40/87,6/87)$, $(87/133,40/133)$
  corresponding to three identical mo\-le\-cu\-les, $^{6}$Li-$^{40}$K-$^{87}$Rb, and $^{40}$K-$^{87}$Rb-$^{133}$Cs, respectively. Both
  axis are scaled up by a factor of $10$ for the excited state, where
  the maximum energy of $\epsilon_{3}=10$ corresponds to $R_1 \approx
  10$.  }
\label{fig3}
\end{figure}

\begin{figure}[!htb]
\centering
\includegraphics[width=0.9\textwidth]{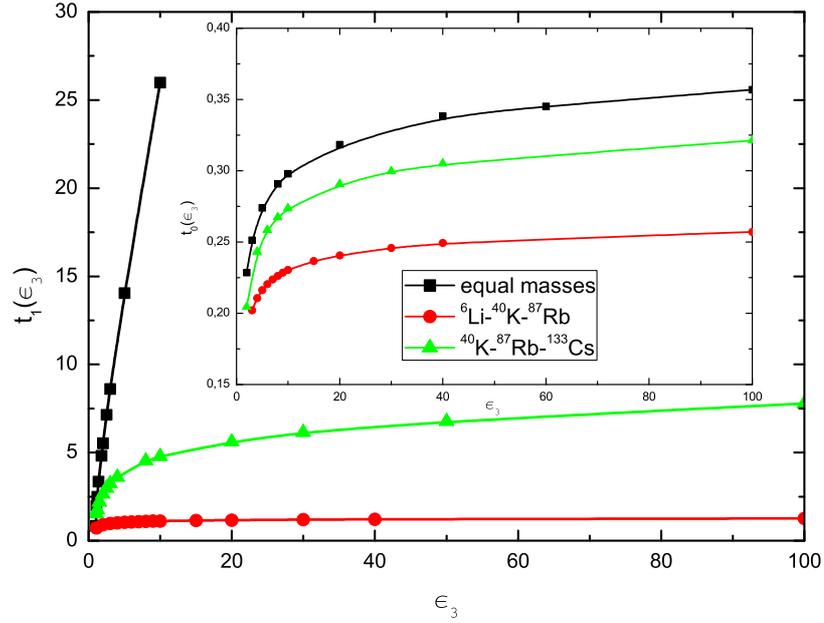}
\caption[The exponents, $t_0$ and $t_1$, in the super-ellipse fit as
  functions of the three-body energy $\epsilon_3$ for three sets of mass
  ratios.]
{The exponents, $t_0$ and $t_1$, in the super-ellipse fit as
  functions of the three-body energy $\epsilon_3$ for three sets of mass
  ratios, which are $(m_{\alpha},m_{\beta})=(1,1)$, $(40/87,6/87)$ and
  $(87/133,40/133)$ corresponding to identical molecules, $^{6}$Li-$^{40}$K-$^{87}$Rb, and $^{40}$K-$^{87}$Rb-$^{133}$Cs,
  respectively.} 
\label{fig4}
\end{figure}

Two sets of alkali atoms ($^{6}$Li-$^{40}$K-$^{87}$Rb and $^{40}$K-$^{87}$Rb-$^{133}$Cs) and a system of three identical particles are used to validate the parameterization. The fitted radius and exponent functions are respectively shown in Figs.~\ref{fig3} and \ref{fig4} for both ground and first excited states. 

The radius functions turn out to be surprisingly simple, that is
linear functions of $\epsilon_3$, which are essentially
independent of the masses.  For the ground state a slight
increase of slope with increasing three-body energy is found.  Average
estimates are
\begin{align}
 R_0(\epsilon_{3}) \approx 0.74 \epsilon_{3}-2.5 \;\;,\;\;
R_1(\epsilon_{3}) \approx \epsilon_{3} \label{e80} \;.
\end{align}

The increasing functions reflect how the contours in Fig.~\ref{fig2}
are moving to larger two-body energies with increasing
$\epsilon_{3}$. This simple
linear dependence implies that the three-body energy increases
linearly with a kind of average of the two two-body energy ratios.
Notice that the symmetric system, where all
particles are identical, has this property where two- and three-body
energies are proportional in the universal limit. To approach this limit it is assumed firstly that $a=b$, and Eqs.~\eqref{e70} and \eqref{e80} imply for
the ground states that $0.74 \epsilon_3 \approx 2.5 + \epsilon_{ac} 2^{1/t_0}$. 
When $a=b=c$, the known ratios
$\epsilon_3 \approx (2.5 + 2^{1/t_0})/0.74 = 16.52$ for the ground state and  $\epsilon_3 \approx 2^{1/t_1} =
1.267$ for the excited state are recovered. This is achieved with $t_0 \approx 0.30$ and $t_1 \approx 2.93$ and both $t_0$ and $t_1$ agree with the ones calculated in Fig.~\eqref{fig4}.

The exponents $t_{n}$ are crucial to obtain the correct
curvature of the energy contours in Fig.~\ref{fig2}. In
Fig.~\ref{fig4} are shown the functions obtained by fitting results
for the same sets of masses as in Fig.~\ref{fig3}.  These exponents
increase monotonously with $\epsilon_{3}$ from small values and the curves bend
over at some point and continue to increase linearly with a smaller
slope. Eventually the curves would stop when the states reach a
two-body threshold and become unstable. In most cases this only
happens for excited states at large energies where the universal properties are, in
practice, probably much more unlikely to realize.

The absolute sizes increase by about an order of magnitude from
ground to first excited state. As said before, the role of the exponents in
Eq.~\eqref{e70} is to adjust to the curvature of the contours in
Fig.~\ref{fig2}. Thus, large $t$ is necessary for strongly bending
curves. This immediately explain the difference between ground and
first excited state, but also the overall increase with
$\epsilon_{3}$.  This is especially pronounced for the excited
states which are squeezed in between boundaries defined by stability
towards decay to bound two-body subsystems.

The behavior of the exponents is also surprisingly simple for each
set of masses.  The relatively fast increase at small energies in
Fig.~\ref{fig4} slows down and both $t_0$ and $t_1$ approach constants
at large energy.  For the ground state, this can be accurately captured by
\begin{align}
t_0(\epsilon_{3}) \approx  \alpha_0 \frac{\epsilon_{3}^{p_0}+\beta_0}
{\epsilon_{3}^{p_0} + \gamma_0},\label{e90}
\end{align}
where $\alpha_0$ is the mass dependent constant
approached at large energy (see Fig.~\ref{fig4}). The parameters,
$(p_0,\alpha_0) \simeq (0.04-0.06,0.3-0.5)$, exhibits a small mass
dependence, whereas 
$(\beta_0,\gamma_0) \simeq -(0.93-0.95),-(0.82-0.87)$ 
are slightly negative but almost mass independent.  
Remember that stability requires $\epsilon_3 > 1$.
The value of $t_0$ for small $\epsilon_3 \approx 1$ is then in the
range of $t_0 \simeq 0.2-0.5$ as required to give the limiting value
of $\epsilon_3=16.52$. A similar parametrization for the exponent
corresponding to the excited state can be found.

Combining Eqs.~\eqref{e70}, \eqref{e80} and \eqref{e90}, the parameterized results cannot be distinguished from
the computed curves in Fig.~\ref{fig2}.

\chapter{Adiabatic approximation} \label{ch4}

Why does the number of bound states increase as one particle becomes lighter than the other two? This question arises after looking the mass-diagrams in Chapter \ref{ch3}, where an increasing number of bound states was found for decreasing the mass of one of the particles. The situation where one particle is much lighter than the other two is suitably handled in the adiabatic approximation, namely the Born-Oppenheimer (BO) approximation.

The BO approximation considers a system composed of two heavy particles and a light one, where the terms {\itshape heavy} and {\itshape light} have relative meaning: two particles are heavier than the third one. In this approximation the heavy particles move very slowly while the light particle orbits around them. In fact, for the BO approximation be valid it is enough to consider the kinetic energy of the  heavy particles is much smaller than the light particle one.

A successful implementation of the BO approximation is presented in Ref.~\cite{fonsecaNPA1979}, where the Efimov problem is solved in an analytic model. It is shown that the Efimov effect is related to a long-range effective force and it can occur even when the individual pair forces have zero range. This is an example of how long-range forces can arise in the three-body problem in a way unpredictable by two-body intuition. The BO approximation was also implemented in Ref.~\cite{limZfPAHaN1980}, looking for the Efimov effect in 2D mass-imbalanced three-body systems, however the mass-dependence of such systems was not addressed.

The BO approximation of 2D three-body systems is re-visited under the mass\=/dependence perspective \cite{bellottiJoPB2013}. As in the previous Chapter, a 2D three-body system with short-range interactions for general masses and interaction strengths is considered. The expressions for the adiabatic approximation are derived using separable zero-range potentials and yield a concise adiabatic potential between the two heavy particles in the heavy-heavy-light system when the light particle coordinate is integrated out. 

The adiabatic potential, which is found as the solution of a transcendental equation, is mass-dependent and reveals an increasing number of bound states by decreasing the mass of the light. An asymptotic expression for the adiabatic potential is derived and it is shown that this analytic expression faithfully corresponds to the numerically calculated adiabatic potential, even in the non-asymptotic region. The number of bound states for a heavy-heavy-light system is estimated as a function of the light-heavy mass ratio. Infinitely many bound states are expected as this ratio approaches zero. However, for finite masses a finite number of bound states is always expected. 

\section{Adiabatic potential}
\begin{figure}[!htb]
\centering
\includegraphics[width=0.9\textwidth]{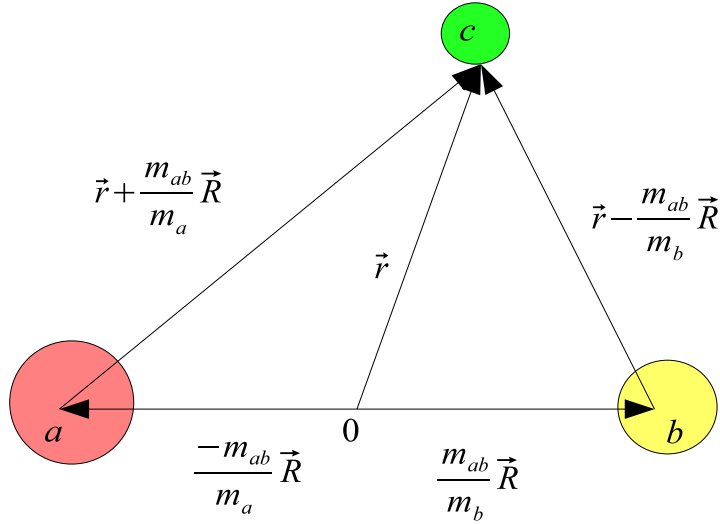}
\caption{Three-body relative coordinates used in the adiabatic approximation.} 
\label{Graph01}
\end{figure}

An $abc$ system where the two heavy particles have masses $m_a$ and $m_b$ is considered. These particles are fixed and their centers are separated out by a distance $\mathbf{R}$. The light particle has mass $m_c$ and coordinate $\mathbf{r}$ relative to the CM of the heavy-heavy subsystem. The interaction between particles is described by zero-range pairwise potentials. The notation for the potential is that $v_c$ means the interaction between particles $a$ and $b$, with $v_a,v_b$ analogously defined. The configuration of the three-body system is shown in Fig.~\ref{Graph01}. 

The Schr{\"o}dinger eigenvalue equation of the $abc$ system is $H\Psi(\mathbf{r,R})=E\Psi(\mathbf{r,R})$. The Hamiltonian $H$ is written in the relative coordinates $\mathbf{R},\mathbf{r}$ in the three-body CM frame as
\begin{align}
&H=-\frac{\hbar^2}{2m_{ab}}\nabla^2_R-\frac{\hbar^2}{2m_{ab,c}}\nabla^2_r
+v_a\left(\mathbf{r}-\frac{m_{ab}}{m_b}\mathbf{R}\right)
+v_b\left(\mathbf{r}+\frac{m_{ab}}{m_a}\mathbf{R}\right)
+v_c(\mathbf{R}),& 
\label{eq.02}
\end{align}
where the reduced masses are $m_{ab}=m_a m_b/(m_a+m_b)$ and $m_{ab,c}=m_c (m_a+m_b)/(m_a+m_b+m_c)$.

The adiabatic approximation says that it is possible to split the three-body eigenvalue equation into the solution of two two-body problems: the light particle motion is considered with respect to the heavy-heavy system and the heavy-heavy system motion is separated out. These eigenvalue equations are valid whenever the motion of the light particle is rapid compared to the motion of the heavy ones, so that the light particle dynamics can be solved while the heavy particles are instantaneously at rest. The wave function is decomposed as
\begin{equation}
\Psi(\mathbf{r,R})=\psi(\mathbf{r,R})\phi(\mathbf{R}) \ ,
\label{eq.03}
\end{equation}
where $\psi(\mathbf{r,R})$ is the wave function describing the state of the light particle for fixed $\mathbf{R}$ and $\phi(\mathbf{R})$ is the heavy subsystem wave function. The approximation is valid when the kinetic energy term, $-\frac{\hbar^2}{2m_{ab}}\nabla^2_R \psi(\mathbf{r,R})$, is small compared to the other terms in Eq.(\ref{eq.02}). Using the wave function from Eq.~\eqref{eq.03}, the eigenvalue equation becomes
\begin{align}
&H\psi(\mathbf{r,R})\phi(\mathbf{R})=E\psi(\mathbf{r,R})\phi(\mathbf{R}) \; , & \nonumber\\
&  \phi(\mathbf{R})\left(-\frac{\hbar^2}{2m_{ab,c}}\nabla^2_r+v_a\left(\mathbf{r}-\frac{m_{ab}}{m_b}\mathbf{R}\right)+v_b\left(\mathbf{r}+\frac{m_{ab}}{m_a}\mathbf{R}\right)\right)\psi(\mathbf{r,R})    \nonumber\\
& \hskip 3cm + \psi(\mathbf{r,R})\left(-\frac{\hbar^2}{2m_{ab}}\nabla^2_R+v_c(\mathbf{R})\right)\phi(\mathbf{R}) =E\psi(\mathbf{r,R})\phi(\mathbf{R}) \; ,  \nonumber \\
&\frac{\left(-\frac{\hbar^2}{2m_{ab,c}}\nabla^2_r+v_a\left(\mathbf{r}-\frac{m_{ab}}{m_b}\mathbf{R}\right)+v_b\left(\mathbf{r}+\frac{m_{ab}}{m_a}\mathbf{R}\right)\right)\psi(\mathbf{r,R})}{\psi(\mathbf{r,R})}  \nonumber\\
& \hskip 5cm + \frac{\left(-\frac{\hbar^2}{2m_{ab}}\nabla^2_R+v_c(\mathbf{R})\right)\phi(\mathbf{R})}{\phi(\mathbf{R})} = E \ . 
\label{eq.04}
\end{align}
The first term on the left-hand-side of Eq.~\eqref{eq.04} is a separation constant, $\epsilon(R)$, which does not depend on $\mathbf{r}$. Therefore, light particle equation is
\begin{equation}
\left[-\frac{\hbar^2}{2m_{ab,c}}\nabla^2_r+v_a\left(\mathbf{r}-\frac{m_{ab}}{m_b}\mathbf{R}\right)+v_b\left(\mathbf{r}+\frac{m_{ab}}{m_a}\mathbf{R}\right)\right]\psi(\mathbf{r,R})=\epsilon(R)\psi(\mathbf{r,R}) \; ,
\label{eq.05}
\end{equation}  
and the eigenvalue, $\epsilon(R)$, plays the role of an effective potential in the equation for the heavy-heavy system. From Eq.~\eqref{eq.04}, this equation is
\begin{equation}
\left(-\frac{\hbar^2}{2m_{ab}}\nabla^2_R+v_c(\mathbf{R})+\epsilon(R)\right)\phi(\mathbf{R})=E\phi(\mathbf{R}) \ .
\label{eq.06}
\end{equation}

Assuming that the potentials in Eq.(\ref{eq.05}) are separable and have the same strength, i.e., $v_\alpha=\lambda \left\vert \chi_\alpha \right\rangle \left\langle \chi_\alpha \right\vert$, the wave function of the light particle in momentum space reads
\begin{equation}
\tilde{\psi}(\mathbf{p})=\lambda \frac{g(\mathbf{p})}{\epsilon(R)-\frac{\hbar^2}{2m_{ab,c}}p^2} \left[ e^{\imath \frac{m_{ab}}{m_a}\frac{\mathbf{p \cdot R}}{\hbar}} A_{+}+ e^{-\imath \frac{m_{ab}}{m_b}\frac{\mathbf{p \cdot R}}{\hbar}} A_{-} \right] \ , \label{eq.12}
\end{equation}
where
\begin{align}
A_\pm &=\int d^{2}r^{\prime }\tilde{g}^{\dagger}\left(\mathbf{r^{\prime}} \pm \frac{m_{ab}}{m_a}\mathbf{R}\right) \Psi \left( \mathbf{r}^{\prime}\right)=\int d^{2}p^{\prime }g^{\dagger}\left(\mathbf{p^{\prime}}\right) \frac{e^{\mp \imath\frac{m_{ab}}{m_a} \frac{\mathbf{p^{\prime } \cdot R}}{\hbar}}}{2\pi}\tilde{\psi} \left( \mathbf{p}^{\prime}\right)   & \label{eq.11}
\end{align}
and $g(\mathbf{p})$ is the form factor of the potential.

The formulation of Eq.~\eqref{eq.12} in terms of $A_\pm$ leads to the system of equations
\begin{equation}
A_\pm=\lambda\int d^2p \frac{\left\vert g(\mathbf{p})\right\vert^2}{\epsilon(R)-\frac{\hbar^2}{2m_{ab,c}}p^2}\left( e^{\mp\imath \frac{\mathbf{p \cdot R}}{\hbar}} A_\mp+ A_\pm \right) \ . \label{eq.13}
\end{equation}
The non-trivial solution of Eq.(\ref{eq.13}), i.e., $A_\pm \neq 0$, gives a transcendental equation for the effective potential, which reads
\begin{equation}
\frac{1}{\lambda}=\int d^2p \frac{|g(\mathbf{p})|^2}{\epsilon(R)-\frac{\hbar^2}{2m_{ab,c}}p^2 }\left[1+ \cos \left(\frac{\mathbf{p \cdot R}}{\hbar}\right) \right] \ . \label{eq.15}
\end{equation}
Using the binding energy of the heavy-light subsystem, $E_2$, to parameterize $\lambda$ \cite{adhikariPRL1995}, Eq.~\eqref{eq.15} is rewritten as
\begin{equation}
\int d^2p |g(\mathbf{p})|^2\left[\frac{1+  \cos (\frac{\mathbf{p \cdot R}}{\hbar})} {\epsilon(R)-\frac{\hbar^2}{2m_{ab,c}}p^2}+\frac{1}{|E_{2}|+\frac{\hbar^2}{2m_{ab,c}}p^2}\right]=0 \ . \label{eq.16}
\end{equation}

Model-independent results are naturally obtained with the use of zero-range potentials and the form factor of such potential  in momentum space is a constant, i.e., $g(\mathbf{p})=1$. In this case, Eq.~\eqref{eq.16} is finite and the integration of the two terms leads to the transcendental equation for the adiabatic potential
\begin{align}
&\ln \frac{\left\vert \epsilon(R)\right\vert }{\left\vert E_2 \right\vert} =2 K_0 \left(\sqrt{\frac{2m_{ab,c}|\epsilon(R)|}{\hbar^2}}R \right)  \ ,& \label{eq.51}
\end{align}
where $K_0$ is the modified Bessel function of second kind of order zero.

The effective potential $\epsilon(R)$ is exactly defined as the solution of Eq.~(\ref{eq.51}) and is a powerful tool in understanding mass-imbalanced three-body systems in two dimensions. However, a transcendental equation involving a logarithm and a modified Bessel function of second kind is not intuitive at all. In the next section, two limiting expressions are found by expanding both sides of Eq.~(\ref{eq.51}) for small and large $R$.

\section{Asymptotic expressions}
The asymptotic form for $|\mathbf{R}| \to 0$ is found inserting the asymptotic 
form of $K_0$  for small arguments \cite{abramowitz1965} into Eq.~\eqref{eq.51}. The result is a Coulomb-like potential, which up to $2^{nd}$ order reads
\begin{equation}
\frac{|\epsilon_{asymptotic}(R)|}{|E_2|} 
\to \frac{2 e^{-\gamma}}{s(R)} \left(1-\frac{e^{-\gamma}}{2} s(R)\left[(1-\gamma)- \frac{1}{2} \ln\left(\frac{e^{-\gamma}}{2}s(R)\right) \right]  \right)^{-1} \; , \label{adpot1-subnum} 
\end{equation}
where $\gamma$ is the constant of Euler and  $s(R)=\sqrt{\frac{2m_{ab,c}\vert E_2\vert}{\hbar^2}} \ R$.

When the distance $\mathbf{R}$ between the two heavy particles is large, i.e., $\left\vert\mathbf{R}\right\vert \to \infty$, the light particle feels only the interaction from one of the heavy particles. In this limit the three-body problem becomes a two-body problem and is expected that $\left\vert E\right\vert =\left\vert E_2\right\vert$. Therefore, defining $\frac{\left\vert \epsilon(R) \right\vert }{\left\vert E_2\right\vert} =1+V(R)$, this condition is fulfilled when $V\rightarrow 0$ for $\vert\mathbf{R}\vert\rightarrow \infty$. Replacing $\left\vert E\right\vert/\left\vert E_2\right\vert$ by $1+V(R)$ in Eq.~\eqref{eq.51} and  expanding both sides up to first order in $R$ results in
\begin{equation}
V(R)=\frac{2  K_0 \left(s(R) \right)}{1+ s(R) \ K_1 \left(s(R)\right)} ,
\label{eq.35}
\end{equation}
 The asymptotic expression of the adiabatic potential for large $\mathbf{R}$ is then
\begin{equation}
\frac{\left\vert \epsilon_{asymptotic}(R) \right\vert }{\left\vert E_2\right\vert} \to 1+\frac{2  K_0 \left(s(R) \right)}{1+ s(R) \ K_1 \left(s(R)\right)} \to 1+ \sqrt{2 \pi } \frac{e^{-s(R)}}{\sqrt{s(R)}}\; .
\label{adpot2-subnum}
\end{equation}

\begin{figure}[!htb]
\centering
\includegraphics[width=0.9\textwidth]{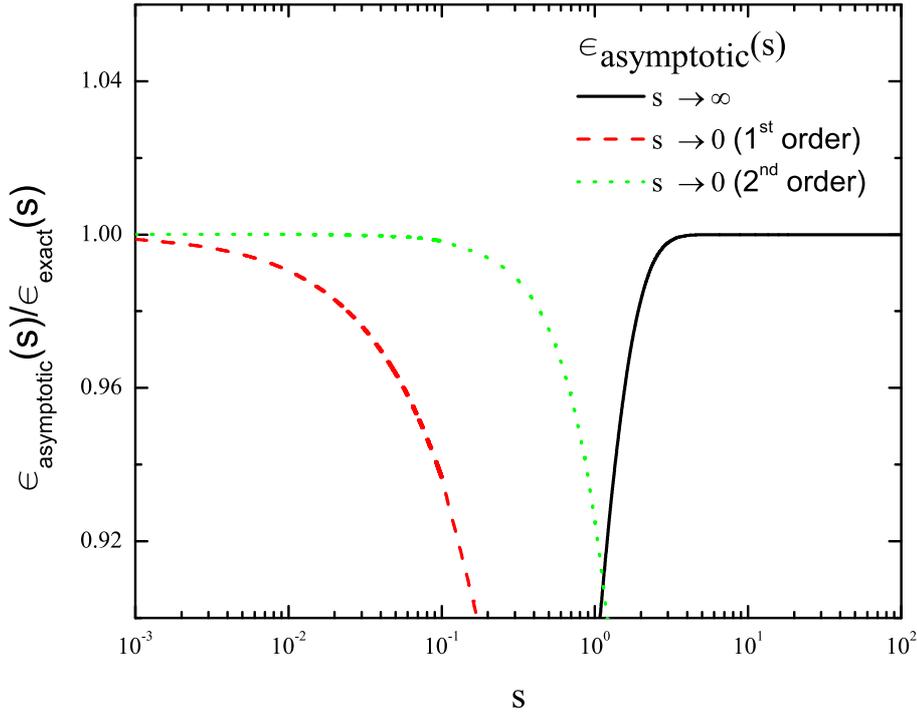}
\caption[Ratio $\epsilon_{asymptotic}(s)/\epsilon_{exact}(s)$ as
  function of the dimensionless coordinate $s(R)$, showing the validity
  of asymptotic expressions.]
{Ratio $\epsilon_{asymptotic}(s)/\epsilon_{exact}(s)$ as
  function of the dimensionless coordinate $s$, showing the validity
  of asymptotic expressions in Eqs.~\eqref{adpot1-subnum} and
  \eqref{adpot2-subnum}.  The black\=/solid and red\=/dashed curves are
  the first order expansion of the adiabatic potential at small and large distances,
  respectively.  The green\=/dotted curve is the second order expansion of the adiabatic potential at small distances.  } 
\label{Graph03}
\end{figure}

Notice that the approximation accuracy increases when higher
order terms are included in the expansions. It is possible to go to more
precise adiabatic potential representations taken higher order
expansions of Eq.~\eqref{eq.51}.  However, the results of the
approximations (\ref{adpot1-subnum}) and (\ref{adpot2-subnum}) and the
adiabatic potential (\ref{eq.51}) are almost indistinguishable in
practice. The largest deviations, shown in Fig.~\ref{Graph03},
are found in the region $0.3<s<3$, where the difference between
$\epsilon_{asymptotic}(s)$ and $\epsilon_{exact}(s)$ never exceeds $9\%$.

\begin{figure}[!htb]
\flushleft
\includegraphics[width=0.75\textwidth,angle=-90]{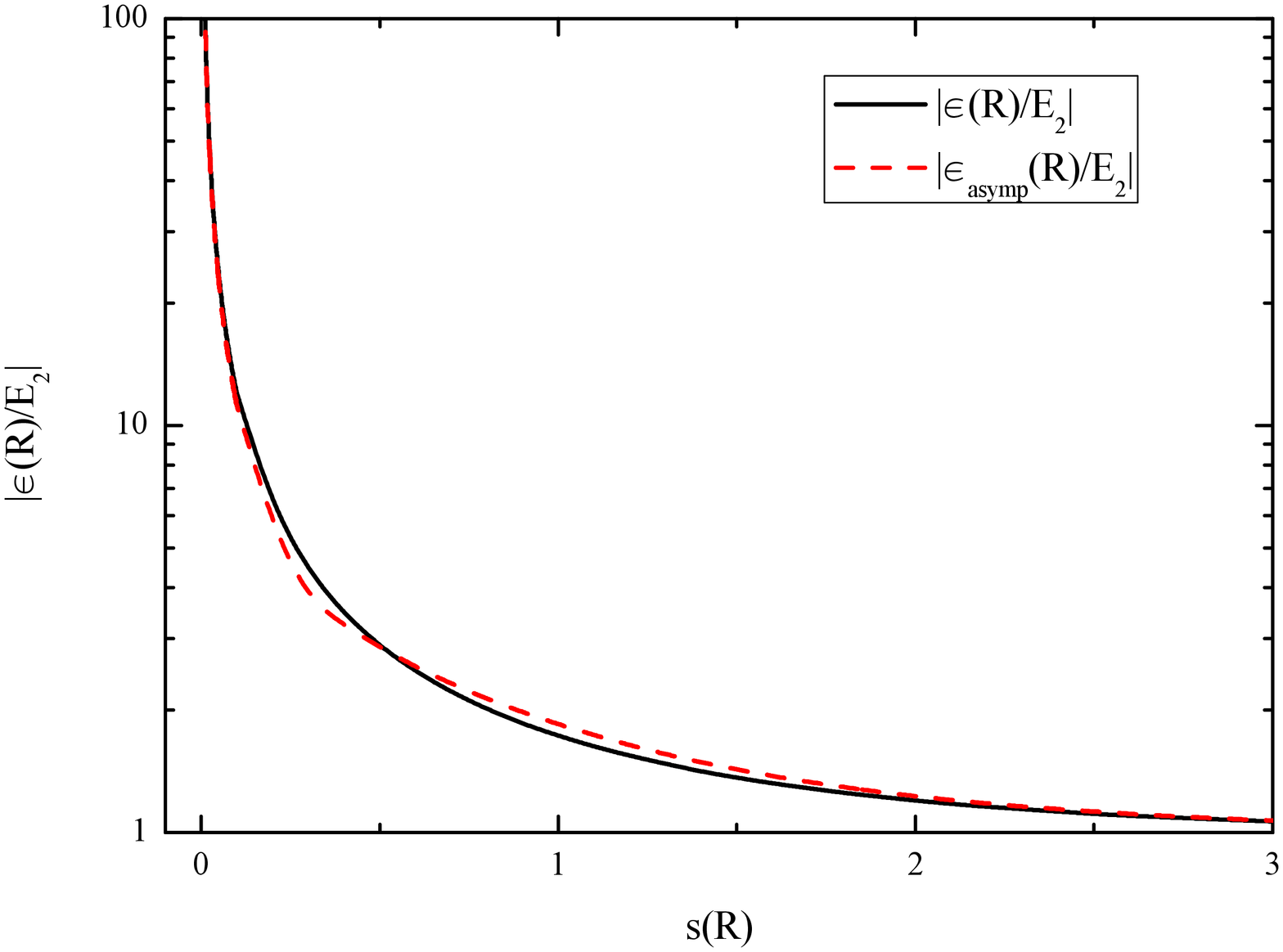}
\caption[Adiabatic potential $\left\vert  \epsilon_{asymptotic}(R)/E_2 \right\vert$ as function of the dimensionless coordinate $s(R)$.]
{Adiabatic potential $\left\vert  \epsilon_{asymptotic}(R)/E_2 \right\vert$ as function of the dimensionless coordinate $s(R)$. The solid\=/black line is the numerical solution of Eq.~\eqref{eq.51} and the dashed-red line is the asymptotic expression in Eqs.~\eqref{adpot1-subnum} and \eqref{adpot2-subnum}. The limiting expression for $R \to 0$ in Eq.~\eqref{adpot1-subnum} is plotted in the interval $0 < s(R) \leq 0.3$ and  the expression for $R \to \infty$ in Eq.~\eqref{adpot2-subnum} is plotted for $s(R) \geq 0.3$.} 
\label{adpotnumasy}
\end{figure} 

In spite of the fact that the asymptotic potential in Eqs.~\eqref{adpot1-subnum} and \eqref{adpot2-subnum} is valid respectively in the extreme limits $R\to 0$ and $R\to \infty$, it perfectly reproduces the effective potential in almost all the range of the scaled coordinate $s(R)$, since its difference to the potential numerically calculated from Eq.~\eqref{eq.51} is almost imperceptible. These features are shown in Fig.\ref{adpotnumasy}.

For $R \to 0$ the first order expansion of the effective potential in Eq.~\eqref{adpot1-subnum} resembles a hydrogen atom in 2D, where the pre-factor $1/\sqrt{m_c}$ makes the energy of the deepest states grow without boundaries when $m_c \to 0$. Furthermore, for $R \to \infty$, the potential in Eq.~\eqref{adpot2-subnum} is long-ranged and screened by a factor $\sqrt{m_c}$, which becomes less important for $m_c \to 0$. Therefore, an increasing number of bound states is expected when particle $c$ is much lighter than the other ones, i.e., $m_c \to 0$, since the adiabatic potential becomes more attractive and less screened in this limit. Still, these states will accumulate both at $R\to 0$, as the strength of the Coulomb-like potential increases, and at $R\to \infty$, where more states are allowed because the exponential moves to larger distances. However, for finite $m_c$, still the number of bound states is finite.

One might argue that the limit $|E_2| \to 0$ must produce the same effect as $m_c \to 0$ in the asymptotic form of the adiabatic potential in Eqs.~\eqref{adpot1-subnum} and \eqref{adpot2-subnum}. However, the limit where all subsystems interacting through zero-range interactions are unbound does not support three-body  bound states in 2D \cite{bruchPRA1979,limZfPAHaN1980,bellottiPRA2012}.

\section{Numerical results} \label{sec4.3}
The increasing number of three-body bound states as function of the mass of the particles, presented in Chapter~\ref{ch3}, was qualitatively explained in the last section with the asymptotic expressions of the adiabatic potential. In this section, the analytic properties of the asymptotic expressions in Eqs.~\eqref{adpot1-subnum} and \eqref{adpot2-subnum} are used in the numerical survey of bound states in the adiabatic limit. 

In the following, the analysis is done for a system of two identical heavy particles of masses $m_a=m_b=M$ in
units that $\hbar^2=M=E_{ac}=E_{bc}=E_2=1$.
The mass ratio between light and heavy particles is defined
$m=\frac{m_c}{M}$.  In this case, the
reduced mass $m_{ab,c}$ is written as
\begin{equation}
m_{ab,c}=\frac{2m}{m+2} \;\;\; \text{and} \;\;\; m_{ab,c} \to m \ \text{for} \ m \to 0 \ .
\label{eq.25}
\end{equation}
With these units, the asymptotic expression for the effective potential becomes
\begin{align}
&\epsilon(R)\to -\frac{2 e^{-\gamma}}{\sqrt{\frac{4m}{m+2}} \ R} \left(1-\frac{e^{-\gamma}}{2} \sqrt{\frac{4m}{m+2}}  \ R\left[(1-\gamma)- \frac{1}{2} \ln\left(\frac{e^{-\gamma}}{2} \sqrt{\frac{4m}{m+2}} \ R\right) \right]  \right)^{-1}  , 
\label{adpot1a-subnum} 
\end{align}
for $ \sqrt{\frac{4m}{m+2}}  \ R \leq 1.15$ and
\begin{align}
&\epsilon(R)\to -1- \sqrt{2 \pi } \frac{e^{-\sqrt{\frac{4m}{m+2}} R}}{\sqrt{\left(\sqrt{\frac{4m}{m+2}}\right)^{\frac{1}{2}}R}} \ , &  \label{adpot2a-subnum} 
\end{align}
for $\sqrt{\frac{4m}{m+2}} R\geq1.15$.
Notice that this approximation is very accurate even when
$2R \approx 1.15 \sqrt{(1+2/m)}$ where the largest deviation of $9\%$
is reached. 

The Schr{\"o}dinger equation for the heavy-heavy system in Eq.~\eqref{eq.06} is transformed in a Sturm-Liouville eigenvalue equation in a $L_z=0$ state. The wave function $\phi(R)$ is replaced by $ \frac{\chi(R)}{\sqrt{R}}$ giving
\begin{equation}
\left[-\left(\frac{d^{2}}{d R^{2}} +\frac{1}{4R ^{2}}\right)+v_c(R) +\epsilon \left( R \right) \right] \chi(R) = E_3 \ \chi(R) \; ,
\label{eq.43}
\end{equation}
where $E_3$ is the three-body energy and $\epsilon \left( R \right)$ is given in Eqs.~\eqref{adpot1a-subnum} and \eqref{adpot2a-subnum}. Genuinely bound states are present when $E_3-E_2\leq0$,
or equivalently $|E_3|\geq|E_2|$, since bound states have negative
energies.

The differential equation (\ref{eq.43}) is numerically solved to estimate the number of bound states ($N_B$) for a system with mass
ratio $m$ when the heavy particles do not interact with each other. Due to the attractive centrifugal barrier in 2D, all the two-body subsystems interact with finite energy. This means that $E_{ab}=0$ implies in a non-interacting $ab$ subsystem. Once the heavy-heavy is not interacting, $E_{ab}=0$ is translated to $v_c=0$ in Eq.~\eqref{eq.43}. If $v_c$
is attractive and able to support bound states, the three-body system
would effectively be reduced to the lightest particle moving around a
heavy-heavy dimer. The corresponding additional much deeper-lying
bound states are, however, not interesting in the present context.

The method to solve Eq.~\eqref{eq.43} numerically consists in writing this eigenvalue equation in matrix form. The operators, potential, wave function and the radial coordinate in this equation are discretized, which leads to a tridiagonal matrix form. This tridiagonal matrix is then diagonalized to give the eigenvalues of the problem. It is also possible to calculate the number of bound states ($N_B$) by solving the set of homogeneous integral equations (\ref{spec}).

\begin{figure}[!htb]
\centering
\includegraphics[width=0.9\textwidth]{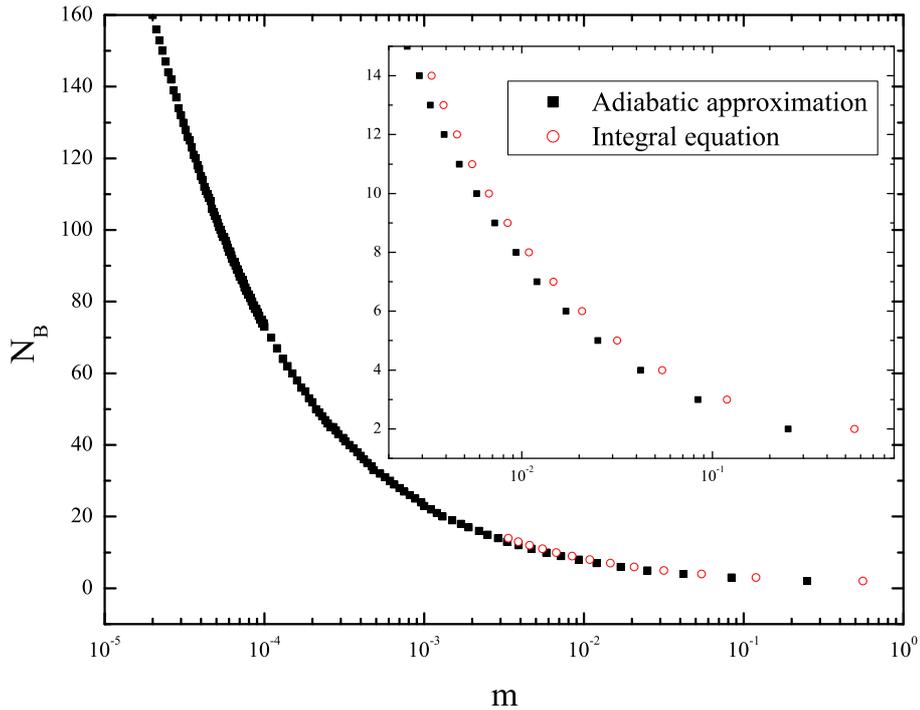}
\caption[Number of possible bound states ($N_B$) for a system with mass ratio $m$ and $E_{ab}=0$. Comparison between the adiabatic approximation and the full solution of the set of coupled homogeneous integral equations.]
{Number of possible bound states ($N_B$) for a system with masses $(1,1,m)$ and $E_{ab}=0$. The (black) squares represent the mass ratio $m$ from where $N_B$ states are bound, calculated through the adiabatic approximation (\ref{eq.43}). The (red) circles represent the solutions of the set of coupled homogeneous integral equations (\ref{spec}).} 
\label{Graph02}
\end{figure}

Counting the number of bound states, a critical mass ratio ($m_t$) is introduced, above which $N_B$ bound states are available. These critical values are shown in Fig.~\ref{Graph02}, where a comparison between the solutions of the differential equation (\ref{eq.43}) and homogeneous coupled integral equations (\ref{spec}) is made.

The results in Fig. \ref{Graph02} show that the adiabatic approximation picks up the small mass behavior very well, even for mass ratios up towards 1. There is a small error in the threshold for the number of available bound states for $0 \leq N_B \leq 14$,  but it decreases as $m \to 0$. The adiabatic approximation has an accuracy better than $10\%$ for $m=0.01$, as it can be seen in the inset of Fig.~\ref{Graph02}. Due to the numerical difficulties, it is very hard to count the number of bound states for $N_B>14$ by solving the set of coupled homogeneous integral equation (\ref{spec}). Fortunately, it is very easy to do it with the differential equation (\ref{eq.43}). 
It is clearly possible to see in Fig.~\ref{Graph02}, that $N_B \to \infty$ for $m \to 0$ as it was pointed out in the last section.

\section{Estimate of the number of bound states} \label{sec4.4}
A fit to the results presented in Fig.~\ref{Graph02} shows that the dependence of the number of bound states, $N_B$, with the mass ratio, $m$, is rather well described by
\begin{equation}
N_B \approx \frac{0.731}{\sqrt{m}} \ .
\label{numberbs}
\end{equation}
This behavior can be explained by the old quantum mechanics. An usual way to estimate the number of bound states in a semi\=/classical approximation of the one\=/dimensional quantum problem is
\begin{equation}
\int{p \ dq}=N \pi \hbar \ .
\label{BS}
\end{equation} 
Taking into account the effective potential in Eq.(\ref{eq.43}) and proper units, the number of bound states with energy up to $E_3=0$ is estimated as
\begin{equation}
N=\frac{1}{\pi \sqrt{2 m}} \int_0^\infty{dx \ \sqrt{\frac{m}{2x^2}-V(x)}}=\frac{0.733}{\sqrt{m}} \ ,
\label{estimate}
\end{equation}
where $V(x)$ is the adiabatic potential (\ref{adpot1a-subnum}) and (\ref{adpot2a-subnum}) with $x=\sqrt{\frac{4m}{m+2}}R$. One could argue that the integral in Eq.(\ref{estimate}) diverges in both limits and can not be performed. Introducing a lower and an upper cut-off in the integral, which are the same used in the numerical calculation ($10^{-2}$ and $10^5$ respectively), the result is $N=\frac{0.766}{\sqrt{m}}$. This result approaches the estimative given in Eq.~\eqref{numberbs} as the diverging term on Eq.(\ref{estimate}) becomes less important when $m$ becomes smaller. The integral in Eq.(\ref{estimate}) is practically $m-$independent for $m\leq 0.001$ with the $10^{-2}$ cut-off,  implying that the term $m/x^2$ is negligibly small by itself. The apparent divergences are due to the semi-classical estimate, and accurately removed by cut-off at both small and large $x$. The true quantum mechanical number of states are then recovered.

\begin{figure}[!htb]
\centering
\includegraphics[width=0.9\textwidth]{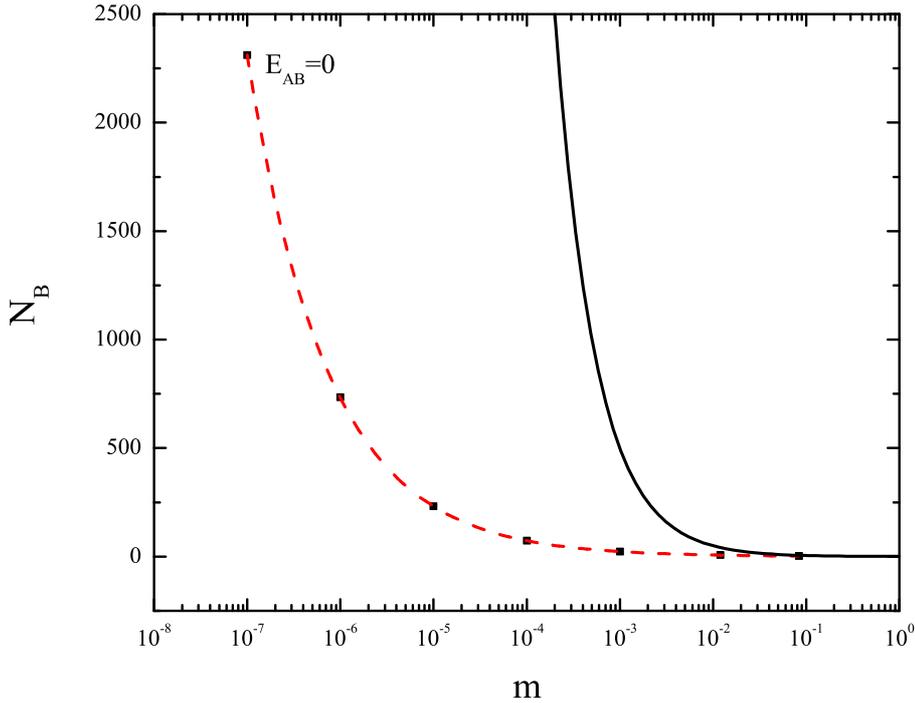}
\caption[Number of possible bound states ($N_B$) for a system with mass ratio $m$ for $E_{ab}=0$ and $E_{ab}=1$.]
{Number of possible bound states ($N_B$) for a system with masses $(1,1,m)$ for $E_{ab}=0$ (red-dashed line) given in Eq.~\eqref{estimate} and $E_{ab}=1$ given in Ref.~\cite{khuriFS2002} (black-full line). The black squares represent some of the points from Fig. \ref{Graph02}.} 
\label{Graph04}
\end{figure} 

The estimate of the number of bound states in (\ref{numberbs}) and (\ref{estimate}) nicely agree. Besides that, this estimative is less than the upper limit for a two-dimensional system with total angular momentum equal to zero, which is given in \cite{khuriFS2002}. For the adiabatic potential (\ref{adpot1a-subnum}) and (\ref{adpot2a-subnum}), this upper limit is given by $N=\frac{0.5}{m}$. The difference between both estimates is shown in Fig. \ref{Graph04}. It was shown in Chapter~\ref{ch3} that any three-body system in two dimensions will achieve its maximum number of bound states when all subsystem are bound with the same energy \cite{bellottiPRA2012} (notice that the richest energy spectrum in 2D requires a large mass asymmetry, but on the other hand the energies have to be symmetric). So, it is expected that the estimate given by the solid-black curve in Fig. \ref{Graph04} will hold for the adiabatic potential (\ref{adpot1a-subnum}) and (\ref{adpot2a-subnum}) when $E_{ab}=E_2$, since this situation gives the upper limit in the number of bound states of 2D three-body systems. Also, the number of bound states for a system with $0\leq |E_{ab}| \leq E_2$ is in the window between the black-solid curve and the red-dashed curve shown in Fig. \ref{Graph04}. 

As expected, the results confirm that the bound states accumulate
in both $R \to 0$ and $R \to \infty$ as $m \to 0$. The energy of the
lowest states seems to increase without boundaries in this limit and
the wave function vanishes slower at large distances, allowing more
bound states. This can be interpreted as an Efimov-like effect for the
two dimensional case, however, an important distinction between the 2D
and 3D case must be done. While the Efimov effect says that three
identical particles can have infinitely many bound states when $E_2
\to 0$, in 2D this limit leads to an unbound three-body system. Infinitely many bound states are only expected in 2D when $m=0$. 
Therefore, finite $m_c$ leads to a finite number of bound states.

\chapter{Momentum distribution in 2D} \label{ch5}

Another important theoretical prediction for cold atomic systems, which was reported in Ref.~\cite{tanAoP2008}, is a parameter that emerges in the study of two-component Fermi gases. This parameter, which is often called Tan's contact or two-body contact parameter and represented by $C_2$, connects universal two-body correlations and many-body properties. For instance, the variation in the energy of a two-component Fermi gas of momentum $k_F$ with the interaction strength (scattering length $a$) is directly proportional to this parameter \cite{tanAoP2008}, as it can be seen in 
\begin{equation}
2 \pi \frac{d E}{d\left[-1/(k_F a) \right]}= C_2 \; .
\label{energyc2}
\end{equation}
Furthermore, the virial theorem for this atomic gas also relates with $C_2$ through
\begin{equation}
E-2V=-\frac{C_2}{4 \pi k_F a} \; .
\label{virialc2}
\end{equation} 
These relations were confirmed in experiments with two-component Fermi gases \cite{kuhnlePRL2010}, where each side of Eqs.~\eqref{energyc2} and \eqref{virialc2} were measured independently and after compared to each other. A later experiment showed that these relations also hold for bosons \cite{wildPRL2012}. 

The quantities in the left-hand-side of Eqs.~\eqref{energyc2} and \eqref{virialc2} are defined through the many\=/body properties of the gas, while the contact parameter is defined in the few\=/body sector. A way to determine this parameter is to find the coefficient in the leading order of the asymptotic one-body large momentum density ($n(q)$) of few-body systems, namely
\begin{equation}
\lim_{q \to \infty} n(q) \rightarrow \frac{C_2}{q^4}+C_3 F(q) \; .
\end{equation} 
The next order in this expansion defines the three-body contact parameter, $C_3$, which may be important only for bosonic systems, since the Pauli principle suppresses the short-range correlations for two-component Fermi gases. Notice that the momentum dependence of the leading order term in this expansion is the same for 1D, 2D and 3D systems \cite{valientePRA2012}, but the function $F(q)$, which is strongly related to the spectator functions, depends on the the dimensionality of the system \cite{bellottiPRA2013}.

The spectator functions are the key ingredients for understanding the asymptotic one-body momentum densities of few-body systems. While the large momentum asymptotic behavior of such functions is well-known for 3D systems \cite{danilovZETF1961}, a striking result presented in this Chapter is the derivation of asymptotic expressions for the spectator functions of three-distinguishable bosons in 2D \cite{bellottiPRA2013,bellottiNJoP2014}. Using the expression for the large momentum behavior, this asymptotic equation is extended to the full range of the momentum and used to calculate an analytic expression for $C_2$ in the ground state.   

For three identical bosons, the two-body contact parameter is found to be a universal constant, in the sense that $\frac{C_2}{E_3}$ is the same for both states, each one described for a three-body energy $E_3$ \cite{bellottiPRA2013}. Furthermore, the three-body contact parameter has a very different behavior in 2D, when compared to 3D systems (the 3D system is discussed in Chapter \ref{ch6}).

It was showed in Chapters~\ref{ch3} and \ref{ch4} that, in 2D, mixed-species systems have a richer energy spectrum than symmetric mass systems. So, it is important to understand how the asymptotic one-body momentum density changes when dealing with mixed-species systems in 2D. In this case, $C_2$ is not a universal constant anymore, however the universality is recovered in at least one special case of a three-body system composed for two identical non-interacting particles. The sub-leading order in the asymptotic momentum density presents the same functional form for both equal masses and mixed-species systems \cite{bellottiNJoP2014}. 

\section{Asymptotic spectator function} \label{sec5.2}
\begin{figure}[h]
\centering
\includegraphics[width=0.9\textwidth]{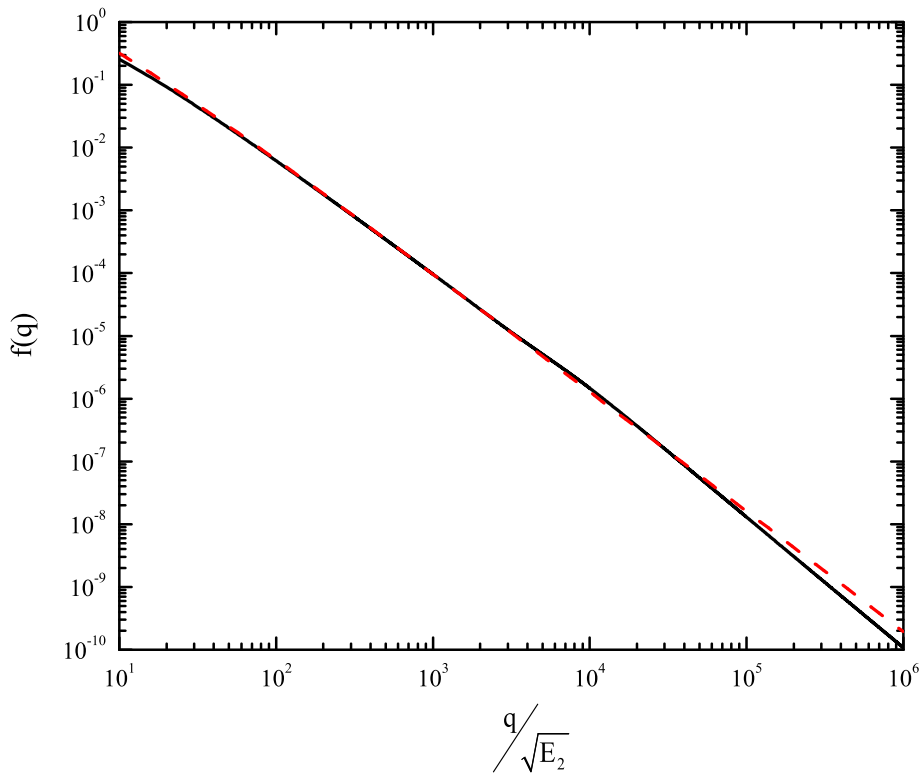}
\caption[Spectator function, $f(q)$, for the ground state calculated numerically and using the ansatz  $f(q)=A_0\frac{\ln q}{q^2}$.]
{Spectator function, $f(q)$, for the ground state calculated numerically (black solid line) and using the ansatz  $f(q)=A_0\frac{\ln q}{q^2}$ (red dashed line).
The solid (black) line tends to oscillates around the dashed (red) one as $q\rightarrow\infty$ due to finite numerical precision.}
\label{numasymf}
\end{figure}

Exhaustive numerical analysis of the spectator function in Eq.~\eqref{spec1} for three\=/identical bosons strongly suggests that the large momentum asymptotic behavior of such functions is given by
\begin{equation}
\lim_{q \to \infty} f(q) \to \Gamma \frac{\ln q}{q^2} \; ,
\label{eqch5.A00}
\end{equation}
where $\Gamma$ is a constant of normalization, as it can be seen in Fig.~\ref{numasymf}.

The ansatz in Eq.~\eqref{eqch5.A00} are, in fact, the exact large momentum asymptotic expression of the spectator function, even for mass-imbalanced systems (within a constant). A system composed of three-distinguishable particles, when all pairs are bound, have three distinct spectator functions in the wave function expressed by Eq.~\eqref{wave} . However, their large-momentum asymptotic behavior all
remain identical, except for individual proportionality factors. To prove that, Eq.\eqref{spec1} is rewritten as
\begin{multline}
f_{\alpha}\left( \mathbf{q}\right)  = 2\pi   \tau_\alpha(q,E_3)  \left[ \int_0^\infty {dk \frac{k f_{\beta}\left( k \right) }{\left(-E_{3}+\frac{q^{2}}{2 m_{\alpha \gamma}}+\frac{k^{2}}{2 m_{\beta \gamma}}\right)\sqrt{1-\frac{k^2 q^2/m_\gamma^2} {\left(-E_{3}+\frac{q^{2}}{2 m_{\alpha \gamma}}+\frac{k^{2}}{2 m_{\beta \gamma}}\right)^2}}}}+ \right. \\*
 \left. + \int_0^\infty{ dk\frac{k f_{\gamma}\left(k\right)}{\left(-E_{3}+\frac{q^{2}}{2m_{\alpha \beta}} +\frac{k^{2}}{2m_{\beta \gamma}}\right)\sqrt{1-\frac{k^2 q^2/m_\beta^2} {\left(-E_{3}+\frac{q^{2}}{2m_{\alpha \beta }}+\frac{k^{2}}{2m_{\beta \gamma}}\right)^2}}}}\right] ,  \label{eqch5.A01}
\end{multline}
with 
\begin{equation}
\tau_\alpha(q,E_3)=\left[ 4\pi m_{\beta \gamma}\ln \left( 
\sqrt{\frac{\frac{q^{2}}{2m_{\beta \gamma,\alpha} }-E_{3}}{E_{\beta\gamma}}} 
\right) \right] ^{-1} \; ,
\end{equation}
where $m_{\beta \gamma,\alpha}= m_\alpha(m_\beta+m_\gamma)/(m_\alpha+m_\beta+m_\gamma)$ and $m_{\beta \gamma}= (m_\beta+m_\gamma)/(m_\beta+m_\gamma)$ are reduced masses, $E_{\beta \gamma}$ the two-body energy and ($\alpha,\beta,\gamma$) are cyclic permutations of the particle labels ($a,b,c$).
The two terms on the right-hand-side of Eq.\eqref{eqch5.A01} have the same form, and one can be
obtained from the other by interchanging labels $\beta$ and $\gamma$.  Therefore it suffices to calculate the first integral in Eq.\eqref{eqch5.A01}.

The contribution for large $q$ can, in principle, be collected from
$k$-values ranging from zero to infinity.  Separating small and large
$k$-contributions, the integration is divided into two intervals, that
is from zero to a large ($q$-independent) momentum
$\Lambda\gg\sqrt{E_3}$, and from $\Lambda$ to infinity. Thus, Eq.~\eqref{eqch5.A01} reads
\begin{multline}
f_{\alpha}\left( \mathbf{q}\right)  = \tau_\alpha(q,E3)  \left[ \int_0^\Lambda {dk \frac{k f_{\beta}\left(k\right)}{\left(E_{3}+\frac{q^{2}}{2 m_{\alpha \gamma}}+\frac{k^{2}}{2 m_{\beta \gamma}}\right)\sqrt{1-\frac{k^2 q^2/m_\gamma^2} {\left(E_{3}+\frac{q^{2}}{2 m_{\alpha \gamma}}+\frac{k^{2}}{2 m_{\beta \gamma}}\right)^2}}}}+ \right. \\*
 \left. + \int_\Lambda^\infty {dk \frac{k f_{\beta}\left( k\right) }{\left(\frac{q^{2}}{2 m_{\alpha \gamma}}+\frac{k^{2}}{2 m_{\beta \gamma}}\right)\sqrt{1-\frac{k^2 q^2/m_\gamma^2} {\left(E_{3}+\frac{q^{2}}{2 m_{\alpha \gamma}}+\frac{k^{2}}{2 m_{\beta \gamma}}\right)^2}}}}+...\right] , \label{eqch5.A03}
\end{multline}
where the dots indicate that the second term in Eq.\eqref{eqch5.A01} should be added.

For $q\to\infty$ the first term, $f_{\alpha,1}$, on the
right-hand-side of Eq.\eqref{eqch5.A03} goes to zero as
\begin{equation}
\lim_{q \to \infty} f_{\alpha,1}\left( q\right)  \to  \frac{m_{\alpha \gamma}/m_{\beta \gamma}}{q^2 \ln(q)}\int_0^\Lambda {dk \frac{k f_{\beta}\left(k\right) }{\sqrt{1-\frac{k^2 q^2/m_\gamma^2} {\left(E_{3}+\frac{q^{2}}{2 m_{\alpha \gamma}}+\frac{k^{2}}{2 m_{\beta \gamma}}\right)^2}}}} \ ,
\label{eqch5.A04}
\end{equation} 
where $\lim_{q \to \infty} \tau_\alpha(q,E3) \to \left[ 2 m_{\beta \gamma}\ln
  q \right] ^{-1}$, and both $E_3$ and $\frac{k^{2}}{2 m_{\beta
    \gamma}}$ are much smaller than $\frac{q^{2}}{2 m_{\alpha
    \gamma}}$.
The integral in Eq. \eqref{eqch5.A04} is finite and only weakly $q-$dependent for large $q \gg \Lambda$.

The asymptotic spectator function in Eq.\eqref{eqch5.A00} can be inserted in
the second term on the right-hand-side of
Eq.\eqref{eqch5.A03}, $f_{\alpha,2}$, because in the asymptotic limit $k>\Lambda$.  In the limit of large momentum, i.e., $q\to\infty$, $f_{\alpha,2}$ is
\begin{align} 
\lim_{q \to \infty} f_{\alpha,2}\left(q\right)  &\to  \frac{\Gamma_\beta}{2 m_{\beta \gamma} \ln q} \int_\Lambda^\infty {dk \frac{ \ln k} {k \left(\frac{q^{2}}{2 m_{\alpha \gamma}}+\frac{k^{2}}{2 m_{\beta \gamma}}\right)\sqrt{1-\frac{k^2 q^2/m_\gamma^2} {\left(E_{3}+\frac{q^{2}}{2 m_{\alpha \gamma}}+\frac{k^{2}}{2 m_{\beta \gamma}}\right)^2}}}},&  \nonumber\\* 
&\to \frac{\Gamma_\beta}{q^2 \ln q} \int_{\Lambda/q}^\infty {dy \frac{ \ln y+\ln q} {y \left(\frac{m_{\beta \gamma}}{m_{\alpha \gamma}}+y^2 \right)}} ,& \label{eqch5.A06}
\end{align}
with $k=q y$ in the last expression.  Carrying out the two integrals in Eq.~\eqref{eqch5.A06} results in
\begin{align}
\int_{\Lambda/q}^\infty {dy \frac{ \ln y} {y \left(\frac{m_{\beta \gamma}}{m_{\alpha \gamma}}+y^2\right)}}&=\left. \frac{1}{2} \frac{\ln^2 y}{(\frac{m_{\beta \gamma}}{m_{\alpha \gamma}}+y^2)}\right\vert_{\Lambda/q}^\infty + \int_{\Lambda/q}^\infty {dy \frac{y \ln^2 y} { \left(\frac{m_{\beta \gamma}}{m_{\alpha \gamma}}+y^2\right)^2}} & \nonumber\\
&\to -\frac{m_{\alpha \gamma}}{2m_{\beta \gamma}}\ln^2\left(\frac{\Lambda}{q}\right)\to -\frac{\ln^2q}{2\frac{m_{\beta \gamma}}{m_{\alpha \gamma}}} ,& \label{eqch5.A07}\\
\int_{\Lambda/q}^\infty {dy \frac{1} {y \left(\frac{m_{\beta \gamma}}{m_{\alpha \gamma}}+y^2\right)}}&= \left. \frac{\ln y}{(\frac{m_{\beta \gamma}}{m_{\alpha \gamma}}+y^2)}\right\vert_{\Lambda/q}^\infty  + 2 \int_{\Lambda/q}^\infty {dy \frac{y \ln y} { \left(\frac{m_{\beta \gamma}}{m_{\alpha \gamma}}+y^2\right)^2}} & \nonumber\\
&\to -\frac{m_{\alpha \gamma}}{m_{\beta \gamma}}\ln \left(\frac{\Lambda}{q}\right)\to \frac{\ln q}{\frac{m_{\beta \gamma}}{m_{\alpha \gamma}}} ,& \label{eqch5.A08}
\end{align}
where it is assumed that the integrals in the right-hand-side of
Eqs.~\eqref{eqch5.A07} and \eqref{eqch5.A08} are finite and their
contributions can be neglected when $q \to \infty$ in comparison with
the terms carrying the log's.

Inserting Eqs.~\eqref{eqch5.A07} and \eqref{eqch5.A08} in Eq.\eqref{eqch5.A06}, the asymptotic expression of $f_{\alpha,2}$ reads
\begin{align} 
\lim_{q \to \infty} f_{\alpha,2}\left(q\right)  &\to  \frac{\Gamma_\beta}{q^2 \ln q} \int_{\Lambda/q}^\infty {dy \frac{ \ln y+\ln q} {y \left(\frac{m_{\beta \gamma}}{m_{\alpha \gamma}}+y^2 \right)}}  \; , & \nonumber\\
&\to  \frac{\Gamma_\beta}{q^2 \ln q} \left( -\frac{m_{\alpha \gamma}}{2 m_{\beta \gamma}} \ln^2q + \ln q \frac{m_{\alpha \gamma}}{m_{\beta \gamma}} \ln q\right)   \; , & \nonumber\\
&\to  \frac{m_{\alpha \gamma}}{ m_{\beta \gamma}} \Gamma_\beta \frac{\ln q}{q^2}  \; . &
\label{eqch5.A06a}
\end{align}

The missing term from Eq.~\eqref{eqch5.A01} is recovered by interchanging the labels $\gamma$ by $\beta$ in Eq.\eqref{eqch5.A06a}. The large-momentum behavior of the spectator function is therefore
\begin{equation}
\lim_{q \to \infty} f_{\alpha}\left(q\right) \to   \left(\frac{m_{\alpha \gamma}}{2m_{\beta \gamma}}\Gamma_\beta +\frac{m_{\alpha \beta}}{2m_{\beta \gamma}}\Gamma_\gamma\right) \frac{\ln q}{q^2} \ .
\label{eqch5.A09}
\end{equation} 
Replacing $f_\alpha(q_\alpha)$ in Eq.~\eqref{eqch5.A09} by its
conjectured asymptotic form, Eq.~\eqref{eqch5.A00}, results in a system of
three linear equations for the three unknown, $\Gamma_\alpha =
\frac{m_{\alpha \gamma}}{2m_{\beta \gamma}}\Gamma_\beta
+\frac{m_{\alpha \beta}}{2m_{\beta \gamma}}\Gamma_\gamma$, which can
be rewritten as $ m_{\beta \gamma} \Gamma_\alpha = m_{\alpha \gamma}
\Gamma_\beta = m_{\alpha \beta} \Gamma_\gamma $. The large-momentum asymptotic behavior for the three distinct spectator functions
are then
\begin{equation}
\lim_{q \to \infty} f_\alpha(q) \to  \frac{\Gamma}{m_{\beta \gamma}}  \frac{\ln q}{q^2} \;.
\label{eqch5.A10}
\end{equation}
This result relates the asymptotic behavior of the three spectator
functions for any state.  The remaining constant $\Gamma$ still
depends on which excited state is considered, and also
on two-body masses, energies and normalization.

\begin{figure}[!htb]%
\centering
\includegraphics[width=0.9\textwidth]{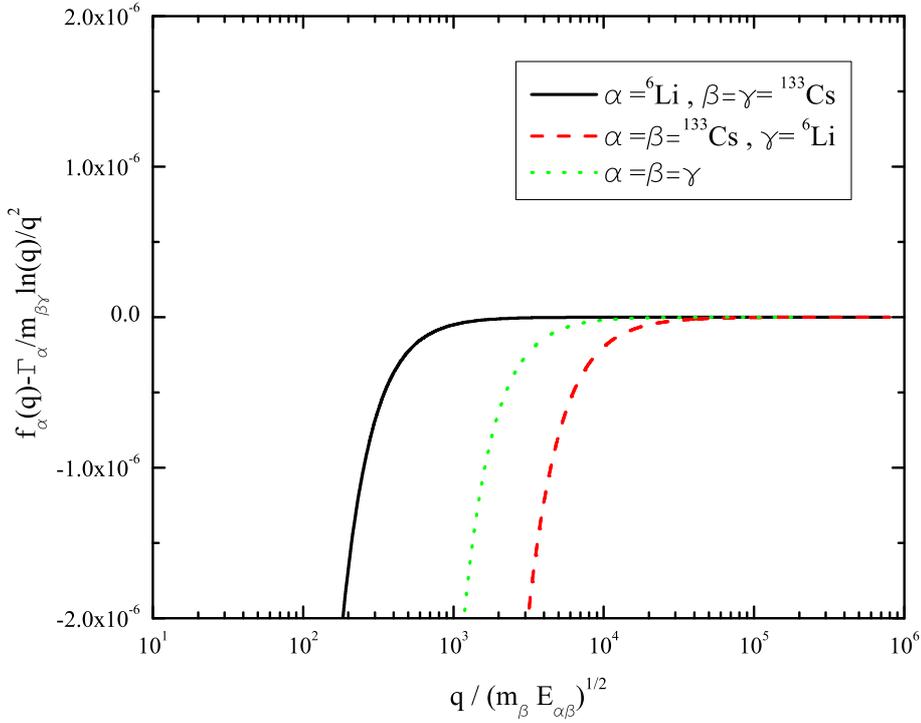}%
\caption[The difference $f_{\alpha}\left(q\right) -  \frac{\Gamma}{m_{\beta \gamma}} \frac{\ln q}{q^2}$ as a function of the momentum $q$ for three different systems.]
{The difference $f_{\alpha}\left(q\right) -  \frac{\Gamma}{m_{\beta \gamma}} \frac{\ln q}{q^2}$ as a function of the momentum $q$. The solid (black) and dash (red) lines are respectively the spectator function of $^{6}$Li and $^{133}$Cs in a $^{133}$Cs-$^{133}$Cs-$^{6}$Li system. The dot (green) line is the spectator function of a system composed for three identical bosons. Notice that Eq.~\eqref{eqch5.A10} exactly describes the asymptotic spectator function, up to the numerical accuracy.}%
\label{fig.A01}%
\end{figure} 

The derived large momentum asymptotic expression and the coefficients in
Eq.\eqref{eqch5.A10} beautifully agree with the numerical calculation.
In Fig.~\ref{fig.A01}, the difference
$f_{\alpha}\left(q\right) - \frac{\Gamma}{m_{\beta \gamma}} \frac{\ln
  q}{q^2}$ is plotted as function of the momentum $q$ for the two different
spectator functions in the $^{133}$Cs$^{133}$Cs$^{6}$Li system. The same difference for a three-body system composed of
identical bosons is also shown. The nice agreement between the analytic derivation and the numerical calculation
demonstrate that the large momentum asymptotic
behavior is always $\ln q/q^2$ for any spectator function in $2D$ three-body systems.

\begin{figure}[!htb]%
\centering
\includegraphics[width=0.9\textwidth]{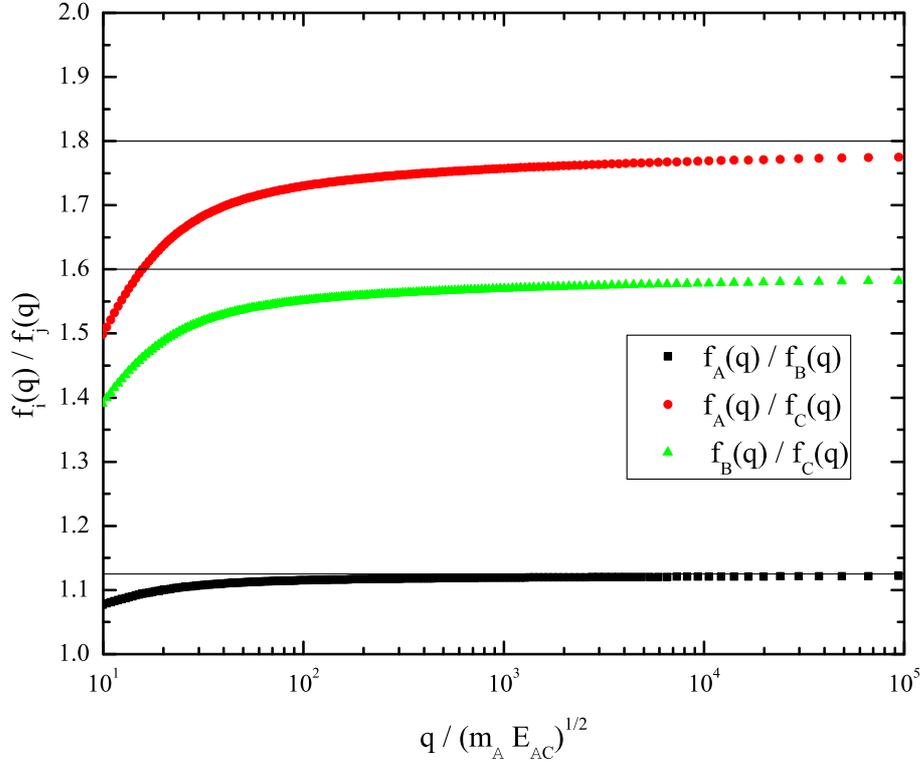}%
\caption[Ratios between the three distinct spectator function for a generic case of three distinct particles.]
{Ratios between the three distinct spectator function for a generic case of three distinct particles. Discrete points are the ratios between spectator functions numerically calculated from Eq.~\eqref{spec1} and full lines are ratios between coefficients in Eq.\eqref{eqch5.A10}.}%
\label{fig.A02}%
\end{figure} 

The general behavior of the large-momentum asymptotic form of the spectator function is further demonstrated in Fig.~\ref{fig.A02} for a system of three distinct particles. The numerically
calculated points are compared to the full lines obtained from
Eq.~\eqref{eqch5.A10}. This comparison is again consistent with the
derived asymptotic behavior, and furthermore exhibit the rate and
accuracy of the convergence.  The limit is reached within $10$\% and
$1$\% already for $q \approx 50$ and $q \approx 10^{4}$, respectively.

\subsection{Parameterizing from small to large momenta}
The asymptotic spectator function in Eq.~\eqref{eqch5.A10} seems to be a
good approximation even for moderate values of $q$, e.g., $q \approx 3
\sqrt{E_3}$.  Information about the large-distance
behavior for a given binding energy is also available, that is $\exp(-\kappa \rho)$,
where $\kappa$ is related to the binding energy and $\rho$ is the
hyper-radius.  Fourier transformation then relates to the small
momentum limit with an overall behavior of $(D+q^2)^{-1}$, where $D$
is a constant related to the energy.  This perfectly matches
Eq.~\eqref{wave} when two Jacobi momenta are present as in the
three-body system.  Therefore, a parametrization combining
the expected small momenta with the known large-momentum behavior is attempted. The result is
\begin{equation}
f_\alpha(q)=f_\alpha(0)\frac{E_3}{\ln\sqrt{E_3}} \frac{\ln\left(\sqrt{\frac{q^2}{2 m_{\beta\gamma,\alpha}}+E_3}\right)}{\frac{q^2}{2 m_{\beta\gamma,\alpha}}+E_3} \ ,
\label{eqch5.A11}
\end{equation}   
where $f_\alpha(0)$ is a normalization constant of the one-body momentum density, $n(q_\alpha)$, which satisfies 
$\int{d^2 q_\alpha\;n(q_\alpha)}=1$.
\begin{figure}[!htb]%
\centering
\includegraphics[width=0.9\textwidth]{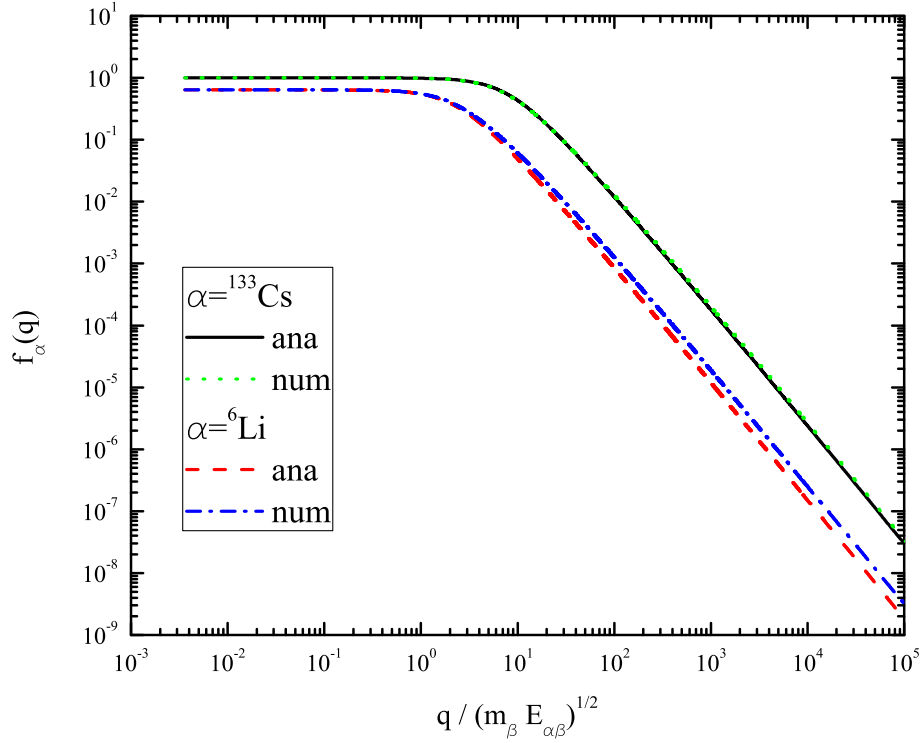}%
\caption[Comparison between the analytic spectator function estimated for the ground state given and the numeric solution of the set of coupled homogeneous integral equations.]
{Comparison between the analytic spectator function estimated for the ground state given in Eq.~\eqref{eqch5.A11} and the numeric solution of Eq.~\eqref{spec1}, for a $^{133}$Cs-$^{133}$Cs-$^{6}$Li system. The solid (black) and dot (green) lines are the analytic estimative and the numeric result for the $^{133}$Cs spectator function. The dash (red) and dash\=/dot (blue) lines are the analytic estimative and the numeric result for the $^{6}$Li spectator function. }%
\label{fig.A03}%
\end{figure} 

Notice that excited states with the same angular
structure must have a different number of radial nodes. Therefore the focus here is only on the ground state.  The expression in Eq.\eqref{eqch5.A11} parametrizes the small momentum behavior of the ground state spectator functions in general cases. As an example, a system composed of two-identical bosons and a distinct particle is shown in  Fig.~\ref{fig.A03},
where the numerical and parametrized solutions are compared.  However, when
small momenta are reproduced the large-momentum limit deviates in
overall normalization, although with the same $q$-dependence.
Surprisingly, the analytic expression is most successful for the
spectator function related to the heaviest particle in the three-body
system.  This large-momentum mismatch is due to the normalization
choice in Eq.\eqref{eqch5.A11}, which is chosen to exactly reproduce the
$q=0$ limit.

\section{Asymptotic one-body densities} \label{asymptotic2D}
The one-body density functions are observable quantities and the most
directly measurable part is the limit of large momenta, which has already been observed in experiments using time-
of-flight and the mapping to momentum space \cite{stewartPRL2010}, Bragg spectroscopy \cite{kuhnlePRL2010}
or momentum-resolved photo-emission spectroscopy \cite{frohlichPRL2011}.
The one-body momentum density of the particle $\alpha$ is defined as
$n(q_\alpha)=\int{d^2 p_\alpha
  |\Psi(\mathbf{q}_\alpha,\mathbf{p}_\alpha)|^2}$, where
$\Psi(\mathbf{q}_\alpha,\mathbf{p}_\alpha)$ is given in
Eq. \eqref{wave}. From now on, the normalization is $\int{d^2
  q_\alpha\;n(q_\alpha)}=1$.  The nine terms in $\int{d^2
  p_\alpha |\Psi(\mathbf{q}_\alpha,\mathbf{p}_\alpha)|^2}$ are then grouped into four
components with distinctly different integrand structure.  The
one-body momentum density is expressed as a sum of four terms, i.e.,
$n(q_\alpha)=\sum_{i=1}^4{n_i(q_\alpha)}$ \cite{bellottiPRA2013,bellottiNJoP2014}.

A general system of three distinguishable particles, presents three
distinct one-body momentum density distributions, each one corresponding to a
different particle.  The four terms for particle $\alpha$ are expressed as
\begin{align}
n_1(q_\alpha)&=\left|f_{\alpha}\left(q_\alpha\right)\right|^2  \int{d^2\;p \frac{1}{\left(-E_{3}+\frac{q_\alpha^{2}}{2m_{\beta \gamma,\alpha}}+\frac{p^{2}}{2m_{\beta \gamma}}\right)^2}} = \frac{2 \pi m_{\beta \gamma} \left|f_{\alpha}\left(q_\alpha\right)\right|^2}{-E_3+\frac{q_\alpha^{2}}{2m_{\beta \gamma,\alpha}}} \; , &
\label{eqch5.03a}\\
n_2(q_\alpha)&= \int{d^2\;k\frac{\left|f_{\beta}(k)\right|^2}{\left(-E_{3}+\frac{q_\alpha^{2}}{2m_{\alpha \gamma}}+\frac{k^{2}}{2m_{\beta \gamma}}+\frac{\mathbf{k}\cdot \mathbf{q_\alpha}}{m_\gamma}\right)^2}} & \nonumber\\* 
& \hskip 3cm + \int{d^2\;k\frac{\left|f_{\gamma}(k)\right|^2}{\left(-E_{3}+\frac{q_\alpha^{2}}{2m_{\alpha \beta}}+\frac{k^{2}}{2m_{\beta \gamma}}-\frac{\mathbf{k}\cdot \mathbf{q_\alpha}}{m_\beta}\right)^2}} \; , &
\label{eqch5.03b}\\
n_3(q_\alpha)&= 2 f_{\alpha}\left(q_\alpha\right) \left[\int{d^2\;k\frac{f_{\beta}(k)}{\left(-E_{3}+\frac{q_\alpha^{2}}{2m_{\alpha \gamma}}+\frac{k^{2}}{2m_{\beta \gamma}}+\frac{\mathbf{k}\cdot \mathbf{q_\alpha}}{m_\gamma}\right)^2}} \right. & \nonumber\\* 
& \left. \hskip 3cm +
\int{d^2\;k\frac{f_{\gamma}(k)}{\left(-E_{3}+\frac{q_\alpha^{2}}{2m_{\alpha \beta}}+\frac{k^{2}}{2m_{\beta \gamma}}-\frac{\mathbf{k}\cdot \mathbf{q_\alpha}}{m_\beta}\right)^2}} \right] \; ,  &
\label{eqch5.03c}\\
n_4(q_\alpha)&=  \int{d^2\;k\frac{f_{\beta}(k)f_{\gamma}(|\mathbf{k+q_\alpha}|)}{\left(-E_{3}+\frac{q_\alpha^{2}}{2m_{\alpha \gamma}}+\frac{k^{2}}{2m_{\beta \gamma}}+\frac{\mathbf{k}\cdot \mathbf{q_\alpha}}{m_\gamma}\right)^2}} & \nonumber\\* 
& \hskip 3cm +
 \int{d^2\;k\frac{f_{\gamma}(k)f_{\beta}(|\mathbf{k+q_\alpha}|)}{\left(-E_{3}+\frac{q_\alpha^{2}}{2m_{\alpha \beta}}+\frac{k^{2}}{2m_{\gamma \beta}}+\frac{\mathbf{k}\cdot \mathbf{q_\alpha}}{m_\beta}\right)^2}} \; . &
\label{eqch5.03d}
\end{align}
where the integration variable originating from Eq.~\eqref{wave} are
properly redefined to simplify the arguments of the spectator
functions in the integrands.  Only $n_4$ is then left with an angular
dependence through the spectator functions.  The
distributions for the other particles are obtained by cyclic
permutations of $(\alpha,\beta,\gamma)$ in these expressions.

The large-momentum limit of the four terms in
Eqs.~\eqref{eqch5.03a} to \eqref{eqch5.03d} is considered separately.  In three dimensions (3D), the similar problem is solved by inserting
the correspondent asymptotic spectator function into each of the
four terms in Eqs.~\eqref{eqch5.03a} to \eqref{eqch5.03d}, and evaluating
the corresponding integrals \cite{castinPRA2011,yamashitaPRA2013}.  This procedure
is not guaranteed to work in 2D  because smaller than asymptotic momentum values may
contribute in the integrands.  However, for 3D it was shown that the
leading order in the integrands is sufficient to provide both leading
and next to leading order of the one-body momentum distributions.  The
details of these calculation in 3D can be found in \cite{castinPRA2011} for
three identical bosons and in \cite{yamashitaPRA2013} for mass-imbalanced
systems. The 3D momentum distributions  are discussed in Chapter~\ref{ch6}.

The large-momentum behavior of the spectator functions changes a lot with dimension, going from $\sin(\ln(q))/q^2$ in 3D to $\ln(q)/q^2$ in
2D. Naively proceeding in 2D as successfully done in 3D,
the integrals in
Eqs.~\eqref{eqch5.03a} to \eqref{eqch5.03d} diverge.  This divergence problem is circumvented by
following the procedure used in the derivation of the asymptotic
spectator functions.  In the following,  each of the four
momentum components defined in Eqs.~\eqref{eqch5.03a} to \eqref{eqch5.03d} are worked out.
In addition, the next-to-leading order
term arising from the dominant $n_2$-term must be simultaneously considered, since it has the same order as the leading order of $n_3$- and $n_4$-terms.

\subsection{Asymptotic contribution from $n_1(q_\alpha)$}
This term is straightforward to calculate.
The argument of the spectator function in Eq.~\eqref{eqch5.03a} does not
depend on the integration variable.  The large-momentum limit is then
found by replacing the spectator function by its asymptotic form and
taking the large $q$ limit after a simple integration, resulting in   
\begin{align}
\lim_{q_\alpha \to \infty} n_1(q_\alpha)& \to  4 \pi m_{\beta \gamma,\alpha} m_{\beta \gamma} \frac{\left|f_{\alpha}\left(q_\alpha\right)\right|^2} {q_\alpha^{2}} \to  4 \pi \frac{m_{\beta \gamma,\alpha}}{m_{\beta \gamma}} \Gamma^2 \frac{\ln^2 (q_\alpha)}{q_\alpha^6}. & 
\label{eqch5.04a}
\end{align}

\subsection{Asymptotic contribution from $n_2(q_\alpha)$}
Integrating the two terms in
Eq.~\eqref{eqch5.03b} over the angle is possible, once the integrand has a simple structure
where the spectator function is angle independent.  The result 
\begin{align}
n_2(q_\alpha)&= 2 \pi \int_0^\infty{dk \frac{k \left|f_{\beta}(k)\right|^2 \left( -E_{3}+\frac{q_\alpha^{2}}{2m_{\alpha \gamma}}+\frac{k^{2}}{2m_{\beta \gamma}}\right)}{\left[\left( -E_{3}+\frac{q_\alpha^{2}}{2m_{\alpha \gamma}}+\frac{k^{2}}{2m_{\beta \gamma}}\right)^2-\frac{k^2\;q_\alpha^2}{m_\gamma^2}\right]^{3/2}}} &\nonumber \\
&\hskip 3cm + 2 \pi \int_0^\infty{dk \frac{k \left|f_{\gamma}(k)\right|^2 \left( -E_{3}+\frac{q_\alpha^{2}}{2m_{\alpha \beta}}+\frac{k^{2}}{2m_{\beta \gamma}}\right)}{\left[\left( -E_{3}+\frac{q_\alpha^{2}}{2m_{\alpha \beta}}+\frac{k^{2}}{2m_{\beta \gamma}}\right)^2-\frac{k^2\;q_\alpha^2}{m_\beta^2}\right]^{3/2}}} &
\label{eqch5.B01}
\end{align}    
is then expanded for large $q$.  Since
$\int_0^\infty{dk\;k\;|f_\alpha(k)|^2}$ is finite, the large momentum
expansion becomes
\begin{align}
\lim_{q_\alpha \to \infty} n_2(q_\alpha)& \to  \frac{8 \pi}{q_\alpha^4} \left(m_{\alpha \gamma}^2 \int_0^\infty {dk\;k\;\left|f_{\beta}(k)\right|^2}+m_{\alpha \beta}^2 \int_0^\infty {dk\;k\;\left|f_{\gamma}(k)\right|^2} \right)+n_5(q_\alpha) \; , & \nonumber\\
&\equiv  \frac{C_{\beta \gamma}}{q_\alpha^4}  + n_5(q_\alpha) \ , &
\label{eqch5.04b}
\end{align}
where $C_{\beta \gamma}$ is the so\=/called two\=/body contact parameter.

As mentioned before, the second term on the right-hand-side of Eq.~\eqref{eqch5.04b}, $n_5(q_\alpha)$, which is
sub-leading term in the expansion of $n_2(q_\alpha)$ has the same asymptotic behavior as $n_3(q_\alpha)$ and
$n_4(q_\alpha)$. This term is kept and derived later.

It is important to emphasize that the one-body large-momentum leading order term comes only
from $n_2(q_\alpha)$.  The spectator function can not be replaced
by its asymptotic expression, because the main contribution to
$\int_0^\infty{dk\;k\;|f_\alpha(k)|^2}$ arises from small $k$.  This
replacement would therefore lead to a completely wrong result.
However, this is not always the case, as later shown for
$n_5(q_\alpha)$.

\subsection{Asymptotic contribution from $n_3(q_\alpha)$}
The structure of $n_3(q_\alpha)$ in
Eq.~\eqref{eqch5.03c} is similar to $n_2(q_\alpha)$ in
Eq.~\eqref{eqch5.03b}. The only difference is that the spectator function
under the integration sign is not squared anymore. This functional
difference leads to a completely different result. The angular integration, which only
involves the denominator, can still be carried out as in the previous
case. Integrating Eq.~\eqref{eqch5.03c} over the
angle gives
\begin{multline}
n_3(q_\alpha)= 4 \pi f_{\alpha}(q_\alpha)\left(\int_0^\infty {dk \frac{k f_{\beta}(k) \left( -E_{3}+\frac{q_\alpha^{2}}{2m_{\alpha \gamma}}+\frac{k^{2}}{2m_{\beta \gamma}}\right)}{\left[\left( -E_{3}+\frac{q_\alpha^{2}}{2m_{\alpha \gamma}}+\frac{k^{2}}{2m_{\beta \gamma}}\right)^2-\frac{k^2\;q_\alpha^2}{m_\gamma^2}\right]^{3/2}}} \right. \\*
 \left. \hskip 3cm +  \int_0^\infty{dk \frac{k f_{\gamma}(k) \left( -E_{3}+\frac{q_\alpha^{2}}{2m_{\alpha \beta}}+\frac{k^{2}}{2m_{\beta \gamma}}\right)}{\left[\left( -E_{3}+\frac{q_\alpha^{2}}{2m_{\alpha \beta}}+\frac{k^{2}}{2m_{\beta \gamma}}\right)^2-\frac{k^2\;q_\alpha^2}{m_\beta^2}\right]^{3/2}}} \right) \; . 
\label{eqch5.B02}
\end{multline}

Here, the difference between $n_2$ and $n_3$ becomes important. Since the integral
$\int_0^\infty{dk\;k\;f(k)}$ is divergent, Eq.~\eqref{eqch5.B02} can not be expanded
 as done for Eq.~\eqref{eqch5.B01}.  Instead, the trick is to
proceed as done in obtaining the asymptotic spectator function.  The integration in Eq.~\eqref{eqch5.B02} is separeted a large, but finite,
momentum, $\Lambda \gg \sqrt{E_3}$, and each term on the
right-hand-side is then split in two others.  The two terms only differ by
simple factors, and therefore details are only given  for the first term.
Changing variables to $k=q_\alpha y$, Eq.~\eqref{eqch5.B02} becomes
\begin{multline}
\lim_{q_\alpha \to \infty} n_3(q_\alpha) \to  16 \pi m_{\beta\gamma}^2 \frac{f_{\alpha}(q_\alpha)}{q_\alpha^2} \int_0^{\Lambda/q_\alpha}{dy \frac{y f_{\beta}(q_\alpha y) \left( -\frac{2 m_{\beta \gamma} E_{3}}{q_\alpha^{2}}+\frac{m_{\beta \gamma}}{m_{\alpha \gamma}}+y^2 \right)}{\left[\left( -\frac{2 m_{\beta \gamma} E_{3}}{q_\alpha^{2}}+\frac{m_{\beta \gamma}}{m_{\alpha \gamma}}+y^2 \right)^2-\frac{4 m_{\beta \gamma}^2}{m_\gamma^2}y^2\right]^{3/2}}} \\* 
+ 16 \pi m_{\beta\gamma}^2 \frac{f_{\alpha}(q_\alpha)}{q_\alpha^4} \frac{\Gamma}{m_{\alpha\gamma}} \int_{\Lambda/q_\alpha}^\infty{dy \frac{ [\ln(q_\alpha)+\ln(y)] \left( \frac{m_{\beta \gamma}}{m_{\alpha \gamma}}+y^2 \right)}{y\;\left[\left(\frac{m_{\beta \gamma}}{m_{\alpha \gamma}}+y^2 \right)^2-\frac{4 m_{\beta \gamma}^2}{m_\gamma^2}y^2\right]^{3/2}}} + ...\ ,
\label{eqch5.B03}
\end{multline}
where $f_\beta(k)$ is replaced by its asymptotic form and $E_3$ is
neglected in the second term on the right-hand-side, where $\sqrt{E_3}
\ll \Lambda$ and $k>\Lambda$.  In the limit $q_\alpha \to \infty$, the
integral in the first term on the right-hand-side of Eq.~\eqref{eqch5.B03} vanishes and therefore does not
contribute to the large-momentum limit. The integrals in the second
term are
\begin{align}
\int_{\Lambda/q_\alpha}^\infty{dy \frac{ \ln(y)\; h(y)}{y}}&=\left.\frac{1}{2} \ln^2(y)\;h(y) \right\vert_{\Lambda/q_\alpha}^\infty -\frac{1}{2} \int_{\Lambda/q_\alpha}^\infty {dy  \ln^2(y) g(y)} \nonumber\\
\to & -\frac{m_{\alpha \gamma}^2}{2m_{\beta \gamma}^2}\ln^2\left(\frac{\Lambda}{q_\alpha}\right) \to -\frac{m_{\alpha \gamma}^2}{2m_{\beta \gamma}^2} \ln^2(q_\alpha) ,& \label{eqch5.B04}\\
\int_{\Lambda/q_\alpha}^\infty{dy \frac{h(y)}{y}}&=\left. \ln(y) h(y) \right\vert_{\Lambda/q_\alpha}^\infty -\int_{\Lambda/q_\alpha}^\infty {dy  \ln(y) g(y)} & \nonumber\\
\to &-\frac{m_{\alpha \gamma}^2}{m_{\beta \gamma}^2}\ln\left(\frac{\Lambda}{q_\alpha}\right) \to \frac{m_{\alpha \gamma}^2}{m_{\beta \gamma}^2} \ln(q_\alpha) ,& \label{eqch5.B05}
\end{align}
where 
\begin{align}
h(y)= \left(\frac{m_{\beta \gamma}}{m_{\alpha
    \gamma}}+y^2\right)\left[\left(\frac{m_{\beta \gamma}}{m_{\alpha
      \gamma}}+y^2 \right)^2-\frac{4 m_{\beta
      \gamma}^2}{m_\gamma^2}y^2\right]^{-3/2} , \\
g(y)=\frac{d\;h(y)}{dy},\;\; \lim_{y \to 0} \ln^2(y) g(y) \to 0\;,\;\;
\lim_{y \to \infty} \ln^2(y) g(y) \to 0 \;.
\end{align}
The function $g(y)$ and its limits ensure that the integrals on the
right-hand-side of Eqs. \eqref{eqch5.B04} and \eqref{eqch5.B05} are finite
and their contributions to the momentum distribution can be neglected
when $q_\alpha \to \infty$.

Finally, inserting the results given in Eqs.~\eqref{eqch5.B04} and
\eqref{eqch5.B05} into Eq.~\eqref{eqch5.B03} and replacing the spectator
function $f_\alpha(q_\alpha)$ by its asymptotic form, the
$n_3(q_\alpha)$ leading order term is given by
\begin{equation}
\lim_{q_\alpha \to \infty} n_3(q_\alpha) \to  8 \pi \left(\frac{m_{\alpha\gamma}+m_{\alpha \beta}}{m_{\beta\gamma}}\right) \Gamma^2 \frac{\ln^3(q_\alpha)}{q_\alpha^6} \ ,
\label{eqch5.04c}
\end{equation}
where the second term in the right-hand-side of Eq.~\eqref{eqch5.B02} is
recovered and added by the interchange of $m_{\alpha \gamma} \to
m_{\alpha \beta}$ in Eqs.~\eqref{eqch5.B03} to \eqref{eqch5.B05}.

Although $n_2(q_\alpha)$ and $n_3(q_\alpha)$ have rather similar form,
their contributions to the one-body large momentum density are quite
different. The sub-leading order,
$n_5(q_\alpha)$, of $n_2(q_\alpha)$ is comparable to the
$n_3(q_\alpha)$ leading order, given in Eq.~\eqref{eqch5.04c}.

\subsection{Asymptotic contribution from $n_4(q_\alpha)$}
This is the most complicated of the four
additive terms in the one-body momentum density. The angular
dependence in both spectator arguments can not be removed
simultaneously by variable change.  The formulation in
Eq.~\eqref{eqch5.03d} has the advantage that the argument in
$f_\gamma(|\mathbf{k+q_\alpha}|)$ (or in
$f_\beta(|\mathbf{k+q_\alpha}|)$) is never small in the limit of large
$q_\alpha$.  This is in contrast with the choice of variables where the
numerator in the first term of Eq.~\eqref{eqch5.03d} would be
$f_{\gamma}(k)f_{\beta}(|\mathbf{k-q_\alpha}|)$, and the argument in
$f_\beta$ would consequently be small as soon as $k$ is comparable to
$q_\alpha$.

The main contribution to the integrals in Eq.~\eqref{eqch5.03d} arise
from small $k$.  For large $q_\alpha$, the
approximation, $f_\gamma(|\mathbf{k+q_\alpha}|) \approx
f_\gamma(q_\alpha)$ (or $f_\beta(|\mathbf{k+q_\alpha}|) \approx
f_\beta(q_\alpha)$) is used and the integrals are then identical to the
terms of $n_3$ in Eq.~\eqref{eqch5.03c}. Keeping track of the slightly
different mass factors immediately leads the asymptotic
limit to be
\begin{equation}
\lim_{q_\alpha \to \infty} n_4(q_\alpha) \to  4 \pi \Big(\frac{m_{\alpha\gamma}}{m_{\alpha \beta}} 
+ \frac{m_{\alpha\beta}}{m_{\alpha \gamma}} \Big)
\Gamma^2 \frac{\ln^3(q_\alpha)}{q_\alpha^6} \ .
\label{eqch5.04d}
\end{equation}

\subsection{Asymptotic contribution from $n_5(q_\alpha)$}
This is the next-to-leading order
contribution from the $n_2(q_\alpha)$ term.  It turns out that this
term has the same large-momentum behavior as the leading orders of
both $n_3(q_\alpha)$ and $n_4(q_\alpha)$. By definition 
\begin{equation} \label{eqch5.n5a}
n_5(q_\alpha) =  n_2(q_\alpha) - \lim_{q_\alpha \to \infty} n_2(q_\alpha)
 = n_2(q_\alpha) -\frac{C_{\beta \gamma}}{q_\alpha^4}
\end{equation}
which can be rewritten in detail as
\begin{multline}
n_5(q_\alpha) = \lim_{q_\alpha \to \infty}
 2 \pi \int_0^\infty{dk k \left|f_{\beta}(k)\right|^2} \\*\times
 \left(\frac{
\left( -E_{3}+\frac{q_\alpha^{2}}{2m_{\alpha \gamma}}+\frac{k^{2}}{2m_{\beta \gamma}}\right)}{\left[\left( -E_{3}+\frac{q_\alpha^{2}}{2m_{\alpha \gamma}}+\frac{k^{2}}{2m_{\beta \gamma}}\right)^2-\frac{k^2q_\alpha^2}{m_\gamma^2}\right]^{3/2}}  - \frac{4 m_{\alpha \gamma}^2}{q_\alpha^4}\right) + .... \;,
\label{eqch5.n5b}
\end{multline}    
where the dots denote that the last term in Eq.~\eqref{eqch5.B01} is obtained by
interchange of labels $\beta$ and $\gamma$.  The tempting procedure is now to
expand the integrand around $q_{\alpha} = \infty$ assuming that
$q_{\alpha}$ overwhelms all terms in this expression.  This
immediately leads to integrals corresponding to the cubic momentum multiplying 
the spectator function which however is not converging.  On the other
hand Eq.~\eqref{eqch5.n5b} is perfectly well defined due to the large-$k$
cut-off from the denominator.  In fact, the spectator function is
multiplied by $k^3$ and $1/k^{3}$ at small and large $k$-values,
respectively.  The integrand therefore has a maximum where the main
contribution to $n_5$ arises. This peak in $k$ moves towards infinity
proportional to $q$.

Computing $n_5(q_\alpha)$, the integration is divided into two
intervals, that is from zero to a finite but very large $k$-value,
$\Lambda_s$, and from $\Lambda_s$ to infinity.  The small momentum
interval, $k/q_{\alpha} \ll 1$, allows an expansion in $k/q_{\alpha}$
leading to the following contribution $n_{5,1}(q_{\alpha})$:
\begin{equation}
n_{5,1}(q_\alpha) = 8 \pi \frac{m_{\alpha \gamma}^2}{q_\alpha^6} 
 \left( 3 \frac{m^{2}_{\alpha \gamma}}{m_\gamma^2}
- \frac{m_{\alpha \gamma}}{m_{\beta \gamma}}\right)
\int_0^{\Lambda_s}dk k^3 \left|f_{\beta}(k)\right|^2 
+ \frac{\omega}{q_{\alpha}^8} .... \; ,
\label{eqch5.n5c}
\end{equation}    
where $\omega$ is  constant. Thus, the contribution from this small momentum integration vanish with
the $6'th$ power of $q_{\alpha}$, which is faster than the sub-leading orders of the other terms kept.

Choosing $\Lambda_s$ sufficiently large such that the spectator function 
reaches its asymptotic behavior in Eq.~\eqref{eqch5.A10},  the momentum integration over larges values can now be performed by omitting the small
$E_3$-terms and changing the integration variable to $y$, i.e., $k^2=y q_{\alpha}^2$,  results in
\begin{align}
n_{5,2}(q_\alpha) & = \frac{ \pi \Gamma^2}{q_{\alpha}^6}
\int_{\Lambda_s^2/q_{\alpha}^2}^{\infty} \frac{dy}{y^2} 
\bigl[\ln^2(y) + \ln^2(q_{\alpha}^2) + 2 \ln y \ln (q_{\alpha}^2)\bigr] &\nonumber\\
& \hskip 3cm \times\left(\frac{1 + y \; m_{\alpha \gamma}/ m_{\beta \gamma}}
{\left[(1 + y \; m_{\alpha \gamma}/ m_{\beta \gamma})^2 -
 4 y \; m^2_{\alpha \gamma}/ m^2_{\gamma}\right]^{3/2}} -1\right) + .... \;, &
\label{eqch5.n5d}
\end{align}
where the large $y$-limit behaves like $\ln^2 y/y^4$ and therefore
assuring rapid convergence, whereas the integrand for small $y$
behaves like $(\ln^2(y) + \ln^2(q^2) + 2 \ln y \ln (q^2))/y$. Integration from an arbitrary minimum value, $y_L$ (independent of $q_\alpha$), of $y>\Lambda_s^2/q_{\alpha}^2$ gives a $q_{\alpha}$-independent value
except for the logarithmic factors and $q_{\alpha}$ in the
numerator. Thus the large-$q_{\alpha}$ dependence is found from very
small values of $y$ close to the lower, and vanishing, limit.  Expanding  around small $y$, the limit for large
$q_{\alpha}$ approach zero as
\begin{multline}
\lim_{q_\alpha \to \infty} n_{5,2}(q_\alpha) \to  \frac{16 \pi \Gamma^2}{q_{\alpha}^6}
 \left( 3 \frac{m^{2}_{\alpha \gamma}}{m_\gamma^2}
- \frac{m_{\alpha \gamma}}{m_{\beta \gamma}}\right) \\* \times
\int_{\Lambda_s^2/q_\alpha^2}^{y_L} \frac{dy}{y} 
\bigl[ \ln^2(y)  + \ln^2(q_{\alpha}^2) + 2 \ln y \ln (q_{\alpha}^2) \bigr] \; .
\label{eqch5.n5e}
\end{multline}
Together with the missing term from Eq.~\eqref{eqch5.n5d}, which comes from the interchange of labels $\beta$ and $\gamma$ the final result is
\begin{equation}
\lim_{q_\alpha \to \infty} n_{5}(q_\alpha) \to  \frac{16 \pi}{3}
 \left[ 3 \left( \frac{m^{2}_{\alpha \gamma}}{m_\gamma^2}+\frac{m^{2}_{\alpha \beta}}{m_\beta^2} \right)
- \frac{m_{\alpha \gamma} + m_{\alpha \beta}}{m_{\beta \gamma}}\right]  \Gamma^2 \frac{\ln^3(q_\alpha)}{q_\alpha^6}\; .
\label{eqch5.n5f}
\end{equation}

\section{Contact parameters}

The expressions for the asymptotic one\=/body densities, which were analytically derived in Eqs.~\eqref{eqch5.04a}, \eqref{eqch5.04b}, \eqref{eqch5.04c}, \eqref{eqch5.04d} and \eqref{eqch5.n5f} are collected and then compared to numerical calculations.

\subsection{Analytic expressions}

Two- and three-body contact parameters are defined via the
large-momentum one-body density. The two-body contact parameter,
$C_{\beta\gamma}$, is the proportionality constant of the leading
order $q_\alpha^{-4}$ term, which arises solely from $n_2(q_\alpha)$
in Eq.~\eqref{eqch5.04b}.  A system of three distinguishable particles have
three contact parameters related to the momentum distribution of
each particle. 
The two-body contact parameter, $C_{\beta \gamma}$, is defined in Eq.~\eqref{eqch5.04b}, where the momentum distribution of the particle $\alpha$ is considered with respect to the CM of the ($\beta,\gamma$) subsystem. This parameter reads
\begin{align}
C_{\beta \gamma} &= 8 \pi m_{\alpha \gamma}^2 \int_0^\infty {dk\;k\;\left|f_{\beta}(k)\right|^2} + 8 \pi m_{\alpha \beta}^2 \int_0^\infty {dk\;k\;\left|f_{\gamma}(k)\right|^2} \; .
\label{c2alpha}
\end{align} 
In the same way, the two-body parameters related to the momenta of particles $\beta$ and $\gamma$ are given by
\begin{align}
C_{\alpha \gamma} &= 8 \pi m_{\beta \gamma}^2 \int_0^\infty {dk\;k\;\left|f_{\alpha}(k)\right|^2} + 8 \pi m_{\alpha \beta}^2 \int_0^\infty {dk\;k\;\left|f_{\gamma}(k)\right|^2} \; , \label{c2beta} \\
C_{\alpha \beta} &= 8 \pi m_{\alpha \gamma}^2 \int_0^\infty {dk\;k\;\left|f_{\beta}(k)\right|^2} + 8 \pi m_{\beta \gamma}^2 \int_0^\infty {dk\;k\;\left|f_{\alpha}(k)\right|^2} \; , \label{c2gamma}
\end{align}  
and a relation between the three independent parameters is found, from  Eqs.~\eqref{c2alpha}, \eqref{c2beta} and \eqref{c2gamma}, to be
\begin{equation}
C_{\alpha \beta} + C_{\alpha \gamma} = C_{\beta \gamma} + 16 \pi m_{\beta \gamma}^2 \int_0^\infty {dk\;k\;\left|f_{\alpha}(k)\right|^2} \; .
\label{eqch5.06}
\end{equation}

For a specific system, where two of the particles are non-interacting, both the corresponding two-body energy and spectator function vanish, i.e., $E_{\beta \gamma}=0$ leads to $f_\alpha(q)=0$ in Eq.~\eqref{spec1} \cite{bellottiJoPB2011,bellottiPRA2012}. In this case, Eq.~\eqref{eqch5.06} becomes a simple relation between the three two-body contact parameters, that is
\begin{equation}
C_{\alpha \beta} + C_{\alpha \gamma} = C_{\beta \gamma} \;\;\text{for}\;\; E_{\beta\gamma}=0 \; .
\label{eqch5.08}
\end{equation}
Notice that this relation between different two-body
parameters does not depend on the system dimension.  Although the
calculations in this chapter are for 2D systems, the relation in Eq.~\eqref{eqch5.08} applies as
well for 3D systems with a non-interacting subsystem.  Note that 
a non-interacting system and a vanishing two-body energy is not
the same in 3D, where attraction is required to provide a state
with zero binding energy.

The three-body contact parameter expressed by $C_{\beta \gamma,\alpha}$, is the coefficient of the next-to-leading order term
in the one-body large-momentum density distribution given by $\ln^3(q_\alpha)/q_\alpha^6$ (see Ref.~\cite{bellottiPRA2012} and references therein).  For
distinguishable particles there are again  three of these parameters,
each one related to the momentum distributions of the different particles.  The
asymptotic behavior, $\ln^3(q_\alpha)/q_\alpha^6$, receives
contributions from the three terms in
Eqs.~\eqref{eqch5.04c}, \eqref{eqch5.04d} and \eqref{eqch5.n5f}.  In total
\begin{equation}
C_{\beta \gamma,\alpha} = 16 \pi \left( \frac{m_{\alpha \gamma}+m_{\alpha \beta}}{6 m_{\beta \gamma}}+\frac{m_{\alpha \gamma}}{4m_{\alpha \beta}} +
\frac{m_{\alpha \beta}}{4m_{\alpha \gamma}}
 + \frac{m_{\alpha \gamma}^2}{m_\gamma^2}+\frac{m_{\alpha \beta}^2}{m_\beta^2}\right) \Gamma^2 \ .
\label{eqch5.07}
\end{equation}

It is worth emphasizing that only a logarithmic factor distinguishes
the behavior of the three-body contact term from the next order one,
$\ln^2(q_\alpha)/q_\alpha^6$ which arises from the leading order of $n_1$ (see Eq.~\eqref{eqch5.04a}), the next-to-next order of 
 $n_2$ (see Eq.~\eqref{eqch5.04b}) as well as from next order of $n_3$, and $n_4$ (see Eqs.~\eqref{eqch5.04c} and \eqref{eqch5.04d}) .  In practice, it must be a
 huge challenge to distinguish between terms differing by only one
 power of $\ln(q_\alpha)$ in experiments.

If one of the two-body subsystems is non\=/interacting, the three-body contact parameter in Eq.~\eqref{eqch5.07} becomes
\begin{equation}
C_{\beta \gamma,\alpha} = 16 \pi \left( - \frac{m_{\alpha \gamma}+m_{\alpha \beta}}{3 m_{\beta \gamma}}+\frac{m_{\alpha \gamma}}{4m_{\alpha \beta}} +
\frac{m_{\alpha \beta}}{4m_{\alpha \gamma}}
+ \frac{m_{\alpha \gamma}^2}{m_\gamma^2}+\frac{m_{\alpha \beta}^2}{m_\beta^2}\right) \Gamma^2 \ ,
\label{eqch5.07a}
\end{equation}
which is obtained by collecting contributions only from  the 
non-vanishing $n_4$ and $n_5$ terms (since $f_\alpha(q)=0$, $n_1$ and $n_3$ do not contribute). Cyclic permutations of the indices in Eqs.~\eqref{eqch5.07} and \eqref{eqch5.07a} show that the three different
three-body contact parameters are related by the mass factors in
Eqs.~\eqref{eqch5.07} and \eqref{eqch5.07a}.  This conclusion holds for all
excited states. 

\subsection{Identical Bosons} \label{secib}
\begin{figure}[ht!]
\centering
\includegraphics[width=0.9\textwidth]{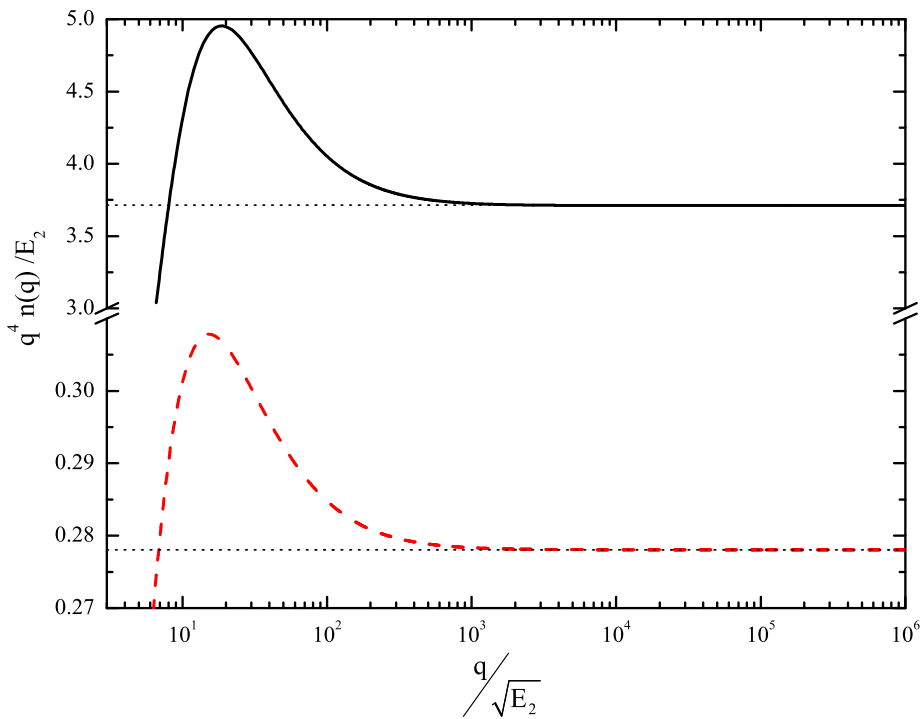}
\caption[LO momentum distribution tail, $q^4 n(q)$, for ground 
and excited three-body states.]
{LO momentum distribution tail, $q^4 n(q)$, for ground 
(upper solid black line) and excited (lower dashed red line) 
three-body states. Note that the vertical axis is not uniform. 
The asymptotic dashed lines are the analytical results given by Eq.~\eqref{LOtail}.}
\label{fig1ch5}
\end{figure}

For three-identical bosons, all the two-body contact parameters in Eq.~\eqref{eqch5.06} are identical. Introducing the label $n$ to distinguish between ground, $n=0$ and excited $n>0$ states, the parameter reads 
\begin{equation}
C_2^n= 4\pi \int_{0}^{\infty}dk k \left|f_{n}(k) \right|^{2} \; .
\label{c2equ}
\end{equation}  
The leading order (LO) behavior of the one-body large-momentum density in Eq.~\eqref{eqch5.04b}, which is characterized by $C_2$, can be seen in Fig.~\ref{fig1ch5} for both ground and first excited states and reads 
\begin{align}
n_{3}^{0}(q)\to \frac{3.71E_2}{q^4} \;\;\; \textrm{and} \;\;\; n_{3}^{1}(q)\to \frac{0.28E_2}{q^4}.
\label{LOtail}
\end{align}

Each one of the two states is defined exclusively by its corresponding three-body energy $E_3$. Scaling $C_2$ with $E_3$ results in  $3.71E_2/16.52E_2=0.224$ and $0.28E_2/1.270E_2=0.219$ for ground and excited state respectively. This striking result demonstrates the state-independence of the LO term in 2D to within the numerical accuracy of about 2\%. The two-body contact for a bosonic system in 2D with short-range attractive interactions in the limit of zero range is 
\begin{align}
C_2/E_3=0.222\pm 0.0025 \; ,
\label{c2equ1}
\end{align}
where $E_3$ is the trimer energy. 

The universal behavior of the tail of the momentum distribution is far from trivial. 
In 3D and at unitarity, the discrete
scale invariance induced by the ultraviolet sensitivity  of the three-body dynamics, 
implies that the system should behave similarly irrespective of which 
trimer state is considered.
This does not occur in 2D and the universal trimer energies are in some sense 
magic numbers multiplying the only scale available, $E_2$. The above
result show that in spite of this major difference, the 2D momentum 
tail displays universal behavior, i.e., $C_2/E_3$ has the same 
value for both ground and excited states. This should be compared to the 2D relation for the trimer energy $\frac{dE}{d \ln a}= \pi N C_2$ derived on general grounds in Ref.\cite{wernerPRA2012}. The factor N appears due a different normalization and this result nicely agrees with Eq.~\eqref{c2equ1}.

The same does not happen to the next-to-leading (NLO) order term, where  $C_3^0/E_3^0 \neq C_3^1/E_3^1$. The three-body contact parameters in Eq.~\eqref{eqch5.07} for the two states of a system of identical particles are
\begin{equation}
C_3^{0}=52.07 \;\;\; \textrm{and} \;\;\; C_{3}^{1}=1.01.
\label{c3equ}
\end{equation} 

While the (LO) behavior of the momentum distribution exhibits the same 
$\frac{C_2}{k^{4}}$ tail in 1D, 2D, and 3D, since it derives solely from 
two-body physics \cite{valientePRA2012}, $C_2$ depends on
what system is addressed and whether few-body bound states are present. On the other hand, the functional form of the NLO term also changes when the system is confined to different dimensions. Collecting results from Eqs.~\eqref{c2equ1} and \eqref{c3equ}, the 2D tail is
\begin{align}
n_{2D}(k)\to \frac{1}{k^4}C_2+\frac{\textrm{ln}^3(k)}{k^6}C_3 \; ,
\label{n2d}
\end{align}
while for  bosons in 3D, the tail reads \cite{castinPRA2011,braatenPRL2011}
\begin{align}
n_{3D}(k)\to \frac{1}{k^4}C_2+\frac{\cos[2s_0 \textrm{ln}(\sqrt{3}k/\kappa_*)+\phi]}{k^5}C_3 \; ,
\label{n3d}
\end{align}
where $s_0=1.00624$ and $\phi=-0.87280$ are constants that can be 
determined from a full solution of the three-bosons problem in 3D at 
unitarity (see Chapter \ref{ch6}) with trimer energy $E_3=\kappa^{2}_{*}$. The log-periodic
three-body NLO term derives from the Efimov effect, whose
solution can be used to determine $C_2=53.097/\kappa_*$ and $C_3=-89.263/\kappa_{*}^{2}$
\cite{castinPRA2011}. 

\begin{figure}[ht!]
\centering
\includegraphics[width=0.9\textwidth]{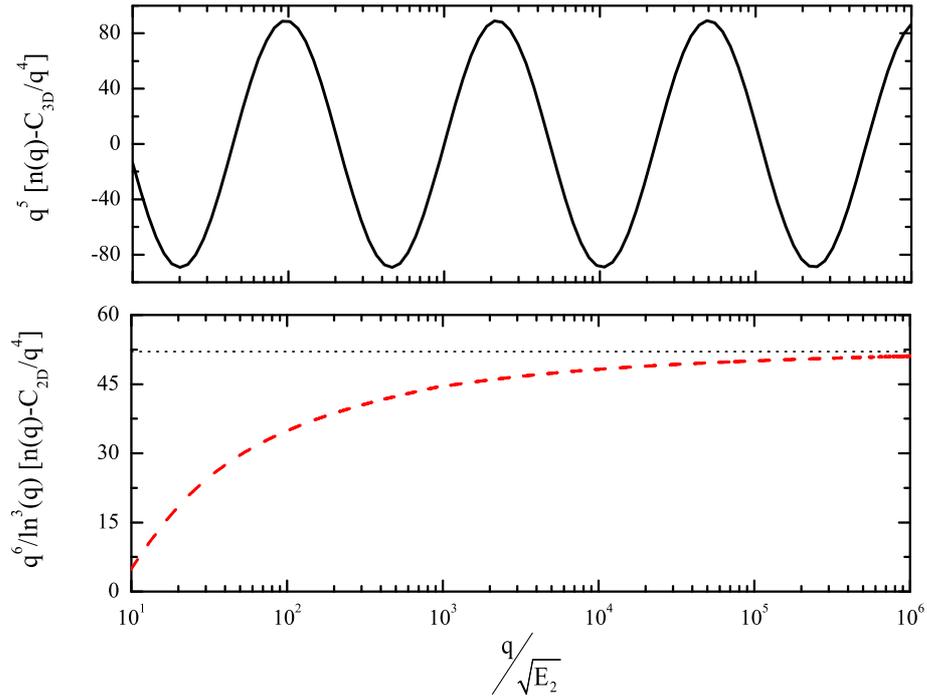}
\caption[NLO momentum distribution comparison of 3D and 
2D.]
{NLO momentum distribution comparison of 3D (upper panel) and 
2D (lower panel). The 2D momentum distribution
is the one of the ground state, but the result is similar for the 
excited state.}
\label{fig2ch5}
\end{figure}

Expressions in Eqs.~\eqref{n2d} and \eqref{n3d} have the same and expect LO behavior, but
vastly different NLO term, as shown in Fig.~\ref{fig2ch5}. The oscillations seen in Eq.~\eqref{n3d}
can be traced directly to the discrete scaling symmetry and are independent of the state considered . It is 
known that the condition on the dimension, $D$, for this behavior
is $2.3<D<3.8$ \cite{nielsenPRA1997,nielsenPR2001}. Imagining an interpolation
between 2D and 3D \cite{yamashitaAe2014}, the log-periodic terms would be expected only
in this range of $D$. The NLO term is therefore a 
tell-tale sign of effective dimensionality of the system. The dimensional crossover is discussed in Chapter~\ref{ch7}.

In experiments that study cold 2D quantum gases, a tight
transverse optical lattice potential is used to reduce the motion in this
direction (see Ref.~\cite{blochRoMP2008}). The strength of the transverse optical lattice 
can be used to interpolate between 2D and 3D behavior of 
fermionic two-component systems \cite{dykePRL2011,sommerPRL2012}. 

Results here are for bosonic systems, and
demonstrate how the NLO 
part of the momentum distribution can be used as a measure of the effective
dimensionality felt by the particles in the system by identifying 
the presence of log-periodic behavior. 
The extreme cases of 2D and 3D are shown in Fig.~\ref{fig2ch5} where the 
log-periodic oscillations are clearly seen in the latter, 
while the former has a smooth behavior. The form of the tail at the crossover is still unknown.

A measurement of the overall functional form of the NLO term is
thus enough to determine the effective dimensionality of the 
squeezed bosonic gas. Since experiments have shown that it is 
possible to reach both the extreme 2D and the 3D regime, there must
necessarily be a dimensional crossover that can be seen in the 
NLO behavior. A theoretical formulation of how the dimensionality can be smoothly changed is presented in Chapter \ref{ch7}.

\subsection{Mass-imbalanced systems} \label{secmis}

\subsubsection{Two-body contact parameters}
The analytical results in this Chapter hold for any mass-imbalanced
three-body system. Such a system has six independent parameters, which
are reduced to four by choosing one mass and one energy as units \cite{bellottiPRA2012} (shown in Chapter \ref{ch3}, see Eq.~\eqref{e57}).
This simply implies that all results can be expressed
as ratios of masses and energies, and in this way provides very
useful scaling relations. However, results depending on four independent
parameters are still hard to display and digest.

A systems composed of two identical particles, $a$, and a distinct one, $c$, has from the beginning four independent parameters, which are reduced to two after the choice of units.  From now on, $E_{ac}$ and $m_a$ are the energy and mass units, and to simplify the notation, the mass ratio $m=\frac{m_c}{m_a}$ is defined.  In this case, the two-body contact  parameters in Eqs.\eqref{c2alpha} to \eqref{c2gamma} are given by
\begin{align}
&C_{aa}=16 \pi \left(\frac{m}{1+m}\right)^2 \int_0^\infty {dk\;k\;\left|f_{a}(k)\right|^2} \ ,& \label{eqch5.09a} \\
&C_{ac}= \frac{C_{aa}}{2}+ 2 \pi \int_0^\infty {dk\;k\;\left|f_{c}(k)\right|^2} \ ,& \label{eqch5.09b}
\end{align}

As shown in the previous section, for three identical particles where all masses and interactions are
the same, $C_{aa}=C_{ac} = C_2$, and the quantity $\frac{C_2}{E_3}$ is a universal constant in 2D \cite{bellottiPRA2013}, since it does not depend on the quantum state considered.  Maintaining universal conditions for all excited states in mass-imbalanced systems, which have more excited states \cite{bellottiPRA2012,bellottiJoPB2013}, must be more demanding.

Detailed investigations reveal that when the mass-energy symmetry is
broken, meaning that particles and two\=/body energies are not identical, the universality of $\frac{C_2}{E_3}$ does not hold anymore.
The two two-body contact parameters defined in Eqs.~\eqref{eqch5.09a} and
\eqref{eqch5.09b} divided by the three-body energy are not the same for
all possible bound states in this general case.  However, at least
in one special case of two identical non-interacting particles,
$E_{aa}=0$, the universality is recovered.  This condition leads to
$f_c=0$ in the set of coupled homogeneous integrals
equations \eqref{spec1} and two universal two-body contact
parameters are related by
\begin{align}
&C_{ac}= \frac{C_{aa}}{2} \;\;\text{for}\;\; E_{aa}=0 \ .& \label{eqch5.10}
\end{align}

\begin{figure}[!htb]%
\centering
\includegraphics[width=0.9\textwidth]{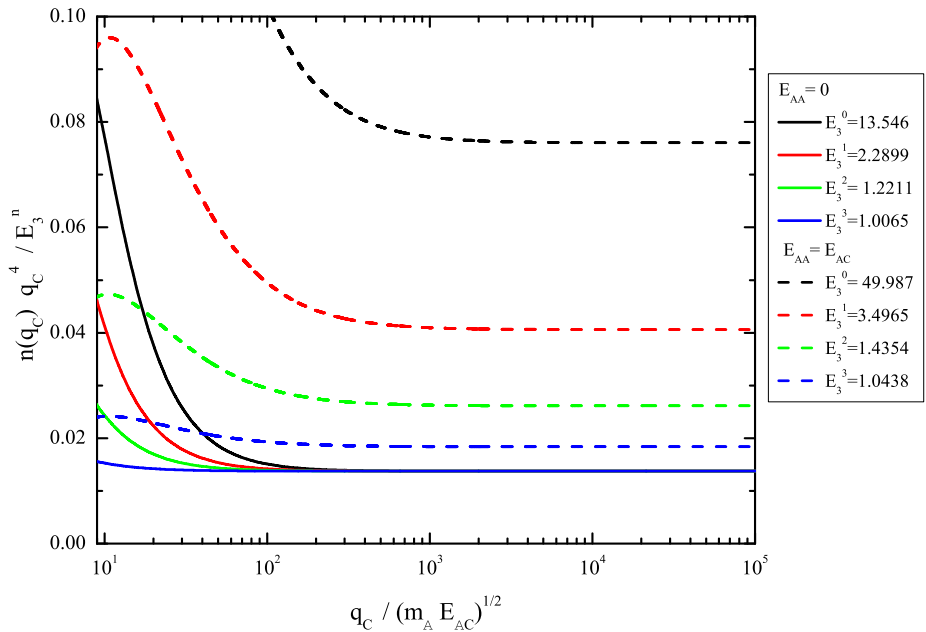}%
\caption[The leading order term of the one-body momentum density divided by $E_3^n$ for each
bound state labeled as $n$.]
{The leading order term of the one-body momentum density divided by $E_3^n$ for each
bound state labeled as $n$ in a system composed of two identical ($a=^{133}$Cs) particles
and a distinct one ($c=^{6}$Li) as a function of the momentum q for both $E_{aa}=E_{ac}$ and $E_{aa}=0$.}
\label{fig.07}%
\end{figure}

The effect of the two-body energy on the contact parameter is shown in Fig.~\ref{fig.07} for the $^{133}$Cs$^{133}$Cs$^{6}$Li system, where $a=^{133}$Cs and $c=^{6}$Li.  This system has four excited states in both cases of $E_{aa}=E_{ac}$ and $E_{aa}=0$ and the coefficients (two\=/body contact) of the large-momentum limit reach constants in all cases.  For $E_{aa}=0$, universality is observed, since
all two-body contacts ratios, $C_{ac}/E_3$, are equal in units of the
three-body energy.  This case is rather special because two particles
do not interact and the three-body structure is determined by the
identical two-body interactions in the identical subsystems.  In
other words the large-momentum limit of the one\=/body density for particle $a$ is determined
universally by the properties of the $ac$ subsystem.  The
other contact parameter, $C_{aa}/E_3$, is also universal and follows
from Eq.~\eqref{eqch5.10}.

This picture changes when $E_{aa}=E_{ac}$, as seen in
Fig.~\ref{fig.07}. Now, in the large-momentum limit, the coefficients $C_{aa}$ and $C_{ac}$ of the one\=/body densities change with the excitation energy.
The systematics is that both $C_{aa}/E_3$ and $C_{ac}/E_3$ as function
of excitation energy move towards the corresponding values for
$E_{aa}=0$, one from below and the other from above.  First the
non-universality of the ratios with the two\=/body energies is understandable, since the interaction of the two
identical particles now must affect the three-body structure at small
distances, and hence at large momenta.  However, as the three-body
binding energy decreases, the size of the system increases and details
of the short-distance structure becomes less important. 

The quantities $\frac{C_{aa}}{E_3}$ and $\frac{2 \pi}{E_3}
\int_0^\infty {dk\;k\;\left|f_{c}(k)\right|^2}$ are defined by the
limit of large-$q$ in  $n_2$ in Eq.~\eqref{eqch5.04b}.  Plotting
the corresponding pieces of $n_2(q) q^4$ as function of $q$ lead to
figures similar to Fig.~\ref{fig.07}, where the different bound state excitations
show distinct results for $E_{aa}=E_{ac}$, while they all coincide for
$E_{aa}=0$.  The ratio of the coefficients $\frac{C_{ac}}{E_3}$,
$\frac{C_{aa}}{E_3}$ and $\frac{2 \pi}{E_3} \int_0^\infty
{dk\;k\;\left|f_{c}(k)\right|^2}$ are
shown in table~\ref{tab1}. The results are presented for two different two\=/body energies and two
different systems represented by $c=^{6}$Li,  $a=^{133}$Cs or
$a=^{40}$K.  These numerical calculations confirm the
systematics described above in complete agreement with
Eqs.~\eqref{eqch5.09b} and \eqref{eqch5.10}.



\begin{table}[!htb]
\centering
\caption[The coefficients $\frac{C_{ac}}{E_3}$,
  $\frac{C_{aa}}{E_3}$ and $\frac{2 \pi}{E_3} \int_0^\infty
  {dk\;k\;\left|f_{c}(k)\right|^2}$.]
{The coefficients $\frac{C_{ac}}{E_3}$,
  $\frac{C_{aa}}{E_3}$ and $\frac{2 \pi}{E_3} \int_0^\infty
  {dk\;k\;\left|f_{c}(k)\right|^2}$ defined by Eqs.~\eqref{c2alpha} to \eqref{c2gamma}, are
  shown for two different interactions and two
  different systems represented by $c=^{6}$Li and $a=^{133}$Cs or
  $a=^{40}$K.  Values in the fifth column are plotted in
  Fig.~\ref{fig.07}.}
\begin{tabular}{cccccc}
\hline
system & $\frac{E_{aa}}{E_{ac}}$ & state  & $\frac{C_{aa}}{E_3}$ & $\frac{C_{ac}}{E_3}$ & $\frac{2 \pi}{E_3} \int_0^\infty {dk\;k\;\left|f_{c}(k)\right|^2}$ \\ \hline
\multirow{5}{*}{$a=^{133}$Cs $c=^{6}$Li}  
	& \multirow{4}{*}{1} 
		& Ground & 0.02210  & 0.07625  & 0.06503 \\ 
		&& First  & 0.02495  & 0.04062  & 0.02812 \\ 
		&& Second & 0.02616  & 0.02612  & 0.01305 \\ 
		&& Third  & 0.02718  & 0.01837  & 0.00478 \\ \cline{2-6}
	&\multirow{1}{*}{0}
		& all & 0.02748  & 0.01374  & 0 \\ \hline 

\multirow{4}{*}{$a=^{40}$K  $c=^{6}$Li}
	&\multirow{3}{*}{1}
		& Ground & 0.06337  & 0.11499  & 0.08372 \\ 
		&& First  & 0.07438  & 0.08256  & 0.04727 \\ 
		&& Second & 0.07934  & 0.05369  & 0.01840 \\ \cline{2-6}
	&\multirow{1}{*}{0}
		& all & 0.08304  & 0.04152  & 0 \\ 
\hline		
\end{tabular}
\label{tab1}
\end{table}

\subsubsection{Mass-dependence of the two-body contacts}
\begin{figure}[!htb]%
\centering
\includegraphics[width=0.9\textwidth]{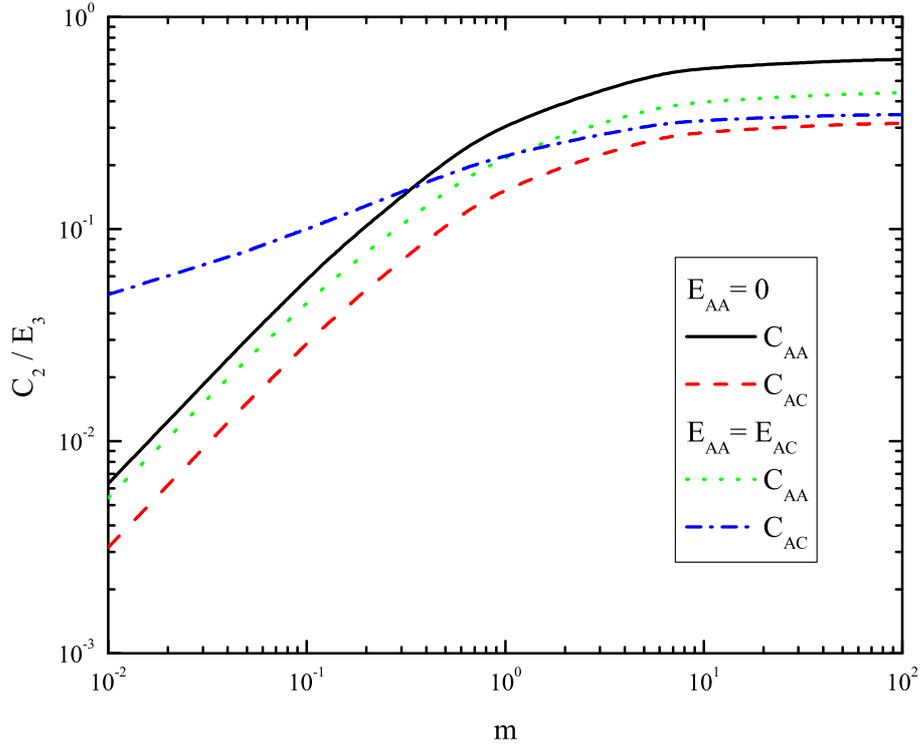}%
\caption[The two-body parameters $C_{aa}$ and $C_{ac}$ as function of the mass ratio $m$.]
{The two-body parameters $C_{aa}$ and $C_{ac}$ defined in Eqs.~\eqref{eqch5.09a} and \eqref{eqch5.09b} as function of the mass ratio $m=\frac{m_c}{m_a}$ for an $aac$ system in both cases where $E_{aa}=0$ and $E_{aa}=E_{ac}$.}%
\label{fig.10}%
\end{figure}

In general, for two identical particles, the two-body contact
parameters divided by the three-body energy depend on the mass ratio
$m$.  The dependence change from universal for $E_{aa}=0$ to
non-universal for $E_{aa}=E_{ac}$.  The mass dependence for ground states
are shown in Fig.~\ref{fig.10}, where the ratio 
$\frac{C_{aa}}{C_{ac}}=2$ in Eq.~\eqref{eqch5.10} is shown to hold for $E_{aa}=0$ in
the entire mass interval investigated.  It is also possible to see how the second term
on the right-hand-side of Eq.~\eqref{eqch5.09b} affects the relation
between the two two-body contact parameters.  Fig.~\ref{fig.10} shows
that the values rapidly increase from small $m$ up to
$1$ and become almost constant above $m \approx 5$.  This behavior is
similar to mass-imbalanced system in 3D \cite{yamashitaPRA2013}.

\subsubsection{Estimate of the two-body contact in the ground state}
\begin{figure}[!htb]%
\centering
\includegraphics[width=0.9\textwidth]{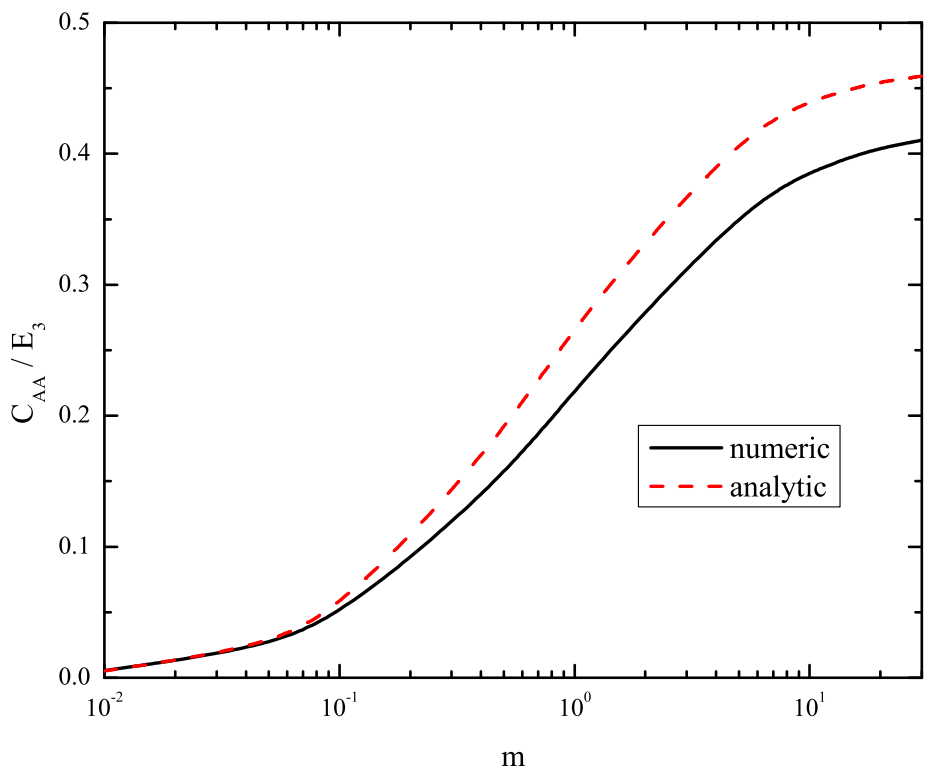}%
\caption[Comparison between the analytic estimative and numerical calculation of $C_{aa}$.]
{Comparison between the analytic estimative of $C_{aa}$, given by Eq.~\eqref{eqch5.11} and the numerical calculation from Eq.~\eqref{eqch5.09a}.}%
\label{fig.01}%
\end{figure}

The parametrization of the ground state spectator function in Eq.~\eqref{eqch5.A11} is used to estimate the dependence of the two-body contact parameter on the three-body energy. Inserting it in Eq.~\eqref{eqch5.09a} gives
\begin{equation}
\frac{C_{aa}}{E_3}= 16 \pi \frac{m^2}{(1+m)(2+m)} f_a^2(0) \left(1+ \frac{2}{\ln(E_3)}+\frac{2}{\ln^2(E_3)}\right) .
\label{eqch5.11}
\end{equation}  
A comparison between this approximation and the numerical results is
shown in Fig. \ref{fig.01}.  Notice that Eq.~\eqref{eqch5.11} provides a
fairly good estimate, which is accurate within $5\%$ for small $m$,
around $10\%$ for $m>1$, and within about $20\%$ deviation in the
worst case of $m =1$.  The divergence in Eq.~\eqref{eqch5.11} for $E_3
\rightarrow 1$ means that the two-body contact parameters diverge when
the three-body system approaches the threshold of binding.  This does
not reveal the full energy dependence since the normalization factor,
$f_a^2(0)$, also is state and energy dependent.

\subsubsection{Three-body contact parameters}
The non-universality of the two-body contact parameters, and even of the three-body one for identical particles (see Eq.~\eqref{c3equ}), does not encourage to check the universality of the three-body contact
parameter in mass-asymmetric systems. However, at least
the system with two non-interacting identical particles turned out to
be universal and may lead to an interesting large-momentum three-body
structure.  As before, inserting $E_{aa}=0$ in the set of coupled
integral equations \eqref{spec1} gives $f_c(q_c)=0$.  Then
Eqs.~\eqref{eqch5.03a} to \eqref{eqch5.03d} show directly that $n_1(q_c)$
and $n_3(q_c)$ vanish when $f_c(q_c)=0$, leaving only
contributions from $n_4(q_c)$ and $n_5(q_c)$.

\begin{figure}[!htb]%
\centering
\includegraphics[width=0.9\textwidth]{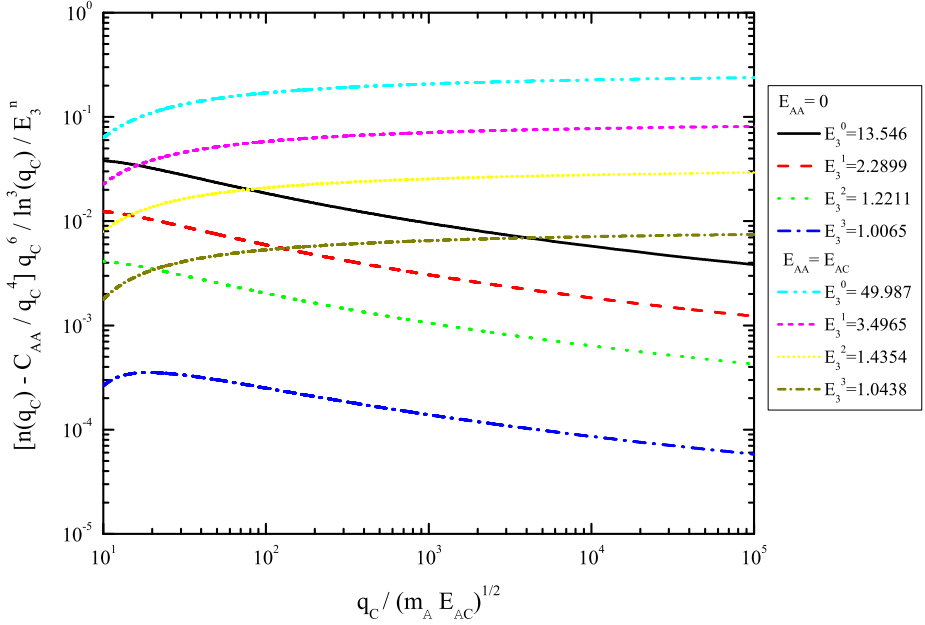}%
\caption[The sub-leading order of the one-body momentum density divided by $E_3^n$
for each bound state labeled as $n$.]
{The sub-leading order of the one-body momentum density divided by $E_3^n$
for each bound state labeled as $n$ in a system composed of two identical 
particles ($a=^{133}$Cs) and a distinct one ($c=^{6}$Li) as a function of the momentum q for both $E_{aa}=E_{ac}$ and $E_{aa}=0$.}%
\label{fig.02}%
\end{figure}

The sub-leading order of the
large-momentum distribution multiplied by $q_c^6/\ln^3(q_c)$ is shown in Fig.~\ref{fig.02} , that is
$C_{aa,c}$, as functions of $q_c$ for the four bound states of the
system $a=^{133}$Cs and $c=^{6}$Li for both $E_{aa}=E_{ac}$ and
$E_{aa}=0$.  Only one of these three-body contact parameters is shown,
since the other one, $C_{ac,a}$, is related state-by-state through the
mass factors in Eqs.~\eqref{eqch5.07} and \eqref{eqch5.07a}.  The momentum
dependence flattens at much larger $q_c$ is not shown in figure.  The values are divided by the three-body energy and no simple energy scaling were obtained.  Not
surprisingly, a more complicated and non-universal behavior is
present.

However, it is striking to see that this sub-leading order in the
large-momentum limit is negligibly small for $E_{aa}=0$ compared
to the interacting case with $E_{aa}=E_{ac}$.  This suggests that a negligible
three-body contact parameter combined with a universal two-body
contact parameter can be taken as a signature of a two-body
non-interacting subsystem within a three-body system in 2D.

\section{Discussion about possible experiments}
As demonstrated in Sec.~\ref{secmis}, the NLO term in the momentum 
distribution can be used to distinguish whether the two-body subsystems are interacting when three-body systems are being taken into account. Maybe more important, the NLO term also carries a tell-tale signature of the dimensionality of 
the quantum system under study. The 2D to 3D crossover 
was studied in Ref.~\cite{dykePRL2011,sommerPRL2012}, 
and it has been shown that both the 3D and the 
strict 2D limits are accessible in experiments. The crossover was here discussed by using formalism applicable to 
either pure 2D or pure 3D without explicit consideration of the
external confinement. The results predict that a proof-of-principle
experiment is possible by going to the two strict limits. However, 
the full crossover including the intermediate regime (quasi-2D)
where the transverse confinement must be taken explicitly into
account, which is experimentally addressable, should be 
explored theoretically in the future. A first step in this direction is taken in \cite{yamashitaAe2014} and discussed in Chapter \ref{ch7}.

The units (the dimer binding energy $E_2$) and the 
effects of the transverse confinement on this two-body bound state need to be considered when connecting the results to the experiments.
The interaction is controlled by Feshbach resonances \cite{chinRMP2010}. 
However, under the confinement, the dimer energy is modified and 
becomes $E_2=B\hbar\omega_z\exp(-\sqrt{2\pi}l_z/|a|)/\pi$ \cite{petrovPRA2001}.
Here, $\omega_z$ is the transverse harmonic confinement frequency, $l_z=\sqrt{\hbar/m\omega_z}$ 
the trapping length, $a$ the 3D scattering length associated with the 
Feshbach resonance and $B=0.905$ is a constant. This formula holds for $a<0$
and $|a| \ll l_z$, while on resonance, $|a|\to\infty$ and $E_2=0.244\hbar\omega_z$.
Corrections also arise from the non-harmonic optical lattice, 
but they are not essential for the discussion which follows. 
The dimer energy scale can be 
converted into a momentum scale, $k_0$, 
through $E_2=\hbar^2k_{0}^{2}/2m$. Accessing the tail behavior and
the 2D-3D crossover, the range $k\sim 10^1-10^3 k_0$ is enough, as shown in Figs.~\ref{fig1ch5} and \ref{fig2ch5}. 
Some 2D 
Bose gas experiments \cite{hungNP2011,yefsahPRL2011} use $l_z\sim 3800a_0$,
where $a_0$ is the Bohr radius, which implies that $k_0\sim 10^{-4}a_{0}^{-1}$
when $|a|=\infty$. For the momentum distribution measurements \cite{stewartPRL2010,kuhnlePRL2010},
the maximum momentum reported is about $k\sim 10^{-3}a_{0}^{-1}$. This 
implies that an order of magnitude or two beyond the reported capabilities
is necessary. However, if $a$ is tuned away from resonance to the $a<0$ side, 
$E_2$ will decrease rapidly according to the formulas above, inducing
a corresponding rapid decrease of $k_0$ which should render the 
physics discussed here within reach of current experimental setups.  
Notice that the van der Waals length scale of about $100a_0$ is 
in the deep tail, so there is no conflict with the universal 
zero-range description employed here.



\chapter{Momentum distribution in 3D} \label{ch6}

A key result in the study of the two-dimensional (2D) one-body density is the analytic expression of the large-momentum asymptotic behavior for the spectator functions. In Chapter~\ref{ch5}, mass-imbalanced systems were addressed and it was shown that the spectator functions have the same asymptotic behavior, i.e., $\ln(q)/q^2$, each one having specific normalization constants, irrespective of the quantum state considered, with a relation among them.


Three-dimensional (3D) three-body systems have geometric scaling between consecutive three-body states for $|a| \to \infty$, as predicted in the seventies by V. Efimov \cite{efimovYF1970} and firstly observed in cold atomic gases around 35 years latter \cite{kraemerNP2006}. The experimental verification of the effect has opened a new research direction dubbed Efimov physics \cite{ferlainoPOJ2010}. In practice, the Efimov effect occurs when the size of the three-body system, given for the scattering  length $|a|$ is much larger than interaction range $r_0$, i.e., $|a|/r_0 \to \infty$. In this limit, a sequence of three-body bound states
occurs wherein two successive states always have the same fixed
ratio of their binding energies. The scaling of the energy levels implies that some of properties of three-body 3D systems are universal, in the sense that are independent of the state considered once properly rescaled. The focus here is in the two- and three- body contact parameters, defined via the one-body momentum density.

The two- and three-body contact parameters were determined for a 3D system of three identical bosons in Ref.~\cite{castinPRA2011} and for mixed-species systems in Ref.~\cite{yamashitaPRA2013}. Unlike the 2D analogue, the results show that the influence of non-equal masses in three-body systems goes beyond changing the contact parameters values, i.e., the sub-leading term in the one-body momentum distribution, which defines $C_3$, has a different and mass-dependent functional form in each case.

The study is taken for three-body bound Efimov-like states
for systems that contain two identical bosons
and a third distinguishable particle. The contact parameters of such systems are addressed when the masses are
different and for different strengths of the interaction
parameters. For that aim the one-body
momentum distributions are computed and its asymptotic behavior is studied. Only the universal regime is considered, since all two-body potentials are described by zero-range interactions. Results are shown for the experimentally relevant cases $^6$Li$^{133}$Cs$^{133}$Cs and $^6$Li$^{87}$Rb$^{87}$Rb, as in Chapter \ref{ch5}.


\section{Formalism and definitions}\label{formal}

The $AAB$ system is constituted by  two identical bosons $A$ and a third  particle $B$ of different kind. The universal limit $|a|\gg r_0$, where the range of the two-body potentials can be neglected, is naturally achieved  by introducing zero-range interactions between the particles. 

The set of 3D homogeneous integral equations are given by Eq.~\eqref{spec13d} for a general system of three distinguishable particles. For the $AAB$ system, the $s-$wave coupled subtracted integral equations for the
spectator functions, $\chi$, corresponding to a bound state, where $|E_3|$ is the absolute value of the 
three-body energy, is written  in units of $\hbar=m_A=1$ as
\begin{align}
\chi_{A A}(y)=2\tau_{A A}(y;|E_3|) \int_0^\infty dx
\frac{x}{y} & G_1(y,x;|E_3|)\chi_{A B}(x) \; ,& \label{chi1} \\
\chi_{A B}(y)=\tau_{A B}(y;|E_3|) \int_0^\infty dx
\frac{x}{y} & \left[G_1(x,y;|E_3|) \chi_{A A}(x) \right. & \nonumber\\
&+ \left. {\cal A}
G_2(y,x;|E_3|) \chi_{A B}(x)\right] \; , &\label{chi2}
\end{align}
with
\begin{align}
\tau_{A A}(y;|E_3|)&\equiv \frac{1}{\pi}
\left[\sqrt{|E_3|+\frac{{\cal A}+2}{4{\cal A}} y^2} \mp \sqrt{E_{AA}}
\right]^{-1} \; , &  \label{tau1}
\\
\tau_{A B}(y;|E_3|)&\equiv
\frac{1}{\pi}\left(\frac{{\cal A}+1}{2{\cal A}}\right)^{3/2}
\left[\sqrt{|E_3|+\frac{{\cal A}+2}{2({\cal A}+1)} y^2} \mp
\sqrt{E_{AB}}\right]^{-1} \; , & \label{tau2}
\end{align}
\begin{align}
G_1(y,x;|E_3|)\equiv \ln &
\frac{2{\cal A}(|E_3|+x^2+xy)+y^2({\cal A}+1)}{2{\cal A}(|E_3|+x^2-xy)+y^2({\cal A}+1)} & \nonumber\\*
&-\ln\frac{2{\cal A}(\mu^2+x^2+xy)+y^2({\cal A}+1)}{2{\cal A}(\mu^2+x^2-xy)+y^2({\cal A}+1)} \; , &
\label{G1} \\
G_2(y,x;|E_3|)\equiv  \ln &\frac{2({\cal A}|E_3|
+xy)+(y^2+x^2)({\cal A}+1)}{2({\cal A} |E_3|-xy)+(y^2+x^2)({\cal A}+1)} & \nonumber\\*
&-\ln \frac{2({\cal A}\mu^2 +xy)+(y^2+x^2)({\cal A}+1)}{2({\cal A}\mu^2 -xy)+(y^2+x^2)({\cal A}+1)} \; , &
\label{G2}
\end{align}
where $x$ and $y$ denote dimensionless momenta and ${\cal A}$ is the mass ratio ${\cal A}=m_B/m_A$. Notice that a slightly different notation is introduced for the study of the 3D system, in order to avoid confusion with the previous 2D case. 

The interaction strengths of the $AA$ and $AB$ subsystems are parametrized by the energies $E_{AA}$ and $E_{AB}$, and the plus and minus signs in Eqs.~\eqref{tau1} and \eqref{tau2} refer to virtual and bound two-body subsystems, respectively \cite{yamashitaPRA2002,bringasPRA2004,yamashitaFS2008}. 

The universal regime of the $AB$ system is studied with $|a_{AB}|\to \infty$ and/or $E_{AB}\to 0$.
In light of the fact that experimental information about mixed systems of the
$AAB$ type is still sparse, the two extreme cases of  $i)$ $E_{AA}=0$ and $ii)$ a non-interacting $AA$ subsystem are considered. Notice that conditions $i)$ and $ii)$ are equivalent in 2D systems \cite{bellottiNJoP2014}, while they are distinct in 3D. 

As before, $\mathbf{q_\alpha}$ is the Jacobi momentum from $\alpha$ particle to the
center-of-mass of the pair ($\beta\gamma$) and $\mathbf{p_\alpha}$ the relative
momentum of the pair. The four terms in the one-body momentum density for each constituent are defined in the most general case of three distinguishable particles in Eqs.~\eqref{eqch5.03a} to \eqref{eqch5.03d}. The $AAB$ system has only two distinct distributions, namely of type $\alpha=\beta=A$ and type $\gamma=B$.  Following the notation and units defined in this Chapter, the wave function in Eq.~\eqref{wave} is written in terms of the spectator
functions in the basis $|\mathbf{q}_B\mathbf{p}_B\rangle$ as
\begin{align}
\langle\mathbf{q}_B\mathbf{p}_B|\Psi\rangle&= \frac{\chi_{AA}(q_B)+\chi_{AB}(q_A)+\chi_{AB}(q_A')}{|E_3|+H_0} \; , & \nonumber \\
&=\frac{\chi_{AA}(q_B)+\chi_{AB}(|\mathbf{p}_B-\frac{\mathbf{q}_B}{2}|)+\chi_{AB}(|\mathbf{p}_B+\frac{\mathbf{q}_B}{2}|)}{|E_3|+H_0} \; , \label{psiqb}
\end{align}
or in the basis $|\mathbf{q}_A\mathbf{p}_A\rangle$ as
\begin{align}
\langle\mathbf{q}_A\mathbf{p}_A|\Psi\rangle&=\frac{\chi_{AA}(|\mathbf{p}_A-\frac{{\cal A}}{{\cal A}+1}\mathbf{q}_A|)+ \chi_{AB}(|\mathbf{p}_A+\frac{1}{{\cal A}+1}\mathbf{q}_A|)+\chi_{AB}(q_A)} {|E_3|+H_0^\prime} \; , & \end{align}
where $H_0=\frac{p_B^2}{2m_{AA}}+\frac{q_B^2}{2m_{AA,B}}$ and
$H_0^\prime=\frac{p_A^2}{2m_{AB}}+\frac{q_A^2}{2m_{AB,A}}$. The reduced masses
are given by $m_{AA}=\frac12$, $m_{AA,B}=\frac{2{\cal A}}{{\cal A}+2}$,
$m_{AB}=\frac{{\cal A}}{{\cal A}+1}$ and $m_{AB,A}=\frac{{\cal A}+1}{{\cal A}+2}$.

The momentum distributions for the particles $A$ and $B$ are
\begin{equation}
\label{nqaqb}
n(q_B)=\int d^3p_B |\langle\mathbf{q}_B\mathbf{p}_B|\Psi\rangle|^2,
\hspace{1cm} n(q_A)=\int d^3p_A |\langle\mathbf{q}_A\mathbf{p}_A|\Psi\rangle|^2
\end{equation}
and they are normalized such that $\int d^3q \; n(q)=1$. 

Since the results here are compared to Ref.~\cite{castinPRA2011}, note that the definition
of momentum distributions as well as their normalizations differ for factor of $1/(2\pi)^3$ multiplying the definition of
$n(q)$, which is normalized to 3, the number of particles, in that reference.

\section{Asymptotic spectator function}\label{asymp}
The asymptotic behavior of the spectator function is used in
deriving some analytic formulas and compare to corresponding numerical
results. The large momentum regime $\sqrt{|E_3|}\ll
q$ is accessed by taking the limit $\mu\to\infty$ and
$|E_3|=E_{AA}=E_{AB}\to 0$. The coupled equations
for the spectator functions in Eqs.~\eqref{chi1} and \eqref{chi2} consequently simplify and become
\begin{align}
\chi_{A A}(y)&=\frac{2}{\pi} \left[y\sqrt{\frac{{\cal A}+2}{4{\cal A}}}
\right]^{-1} \int_0^\infty dx \frac{x}{y} G_{1a}(y,x)\chi_{A B}(x) \; , & \label{chi1a} \\
\chi_{A B}(y)&=
\frac{1}{\pi}\left(\frac{{\cal A}+1}{2{\cal A}}\right)^{3/2}
\left[y\sqrt{\frac{{\cal A}+2}{2({\cal A}+1)}}\right]^{-1} \times & \nonumber\\
& \int_0^\infty dx \frac{x}{y}
\left[G_{1a}(x,y) \chi_{A A}(x) + {\cal A} G_{2a}(y,x) \chi_{A B}(x)\right] \; , & \label{chi2a}
\end{align}
where
\begin{align}
G_{1a}(y,x)&\equiv \ln
\frac{2{\cal A}(x^2+xy)+y^2({\cal A}+1)}{2{\cal A}(x^2-xy)+y^2({\cal A}+1)} \; ,&
\label{G1a} \\
G_{2a}(y,x)&\equiv  \ln \frac{
(y^2+x^2)({\cal A}+1)+2xy}{(y^2+x^2)({\cal A}+1)-2xy} \; . \label{G2a}
\end{align}

The coupled equations in Eqs.~\eqref{chi1a} and \eqref{chi2a} are solved by using the ansatz
\begin{equation}
\chi_{A A}(y)=c_{AA}\; y^{-2+\imath s} ~~~{\text{ and }} ~~~~ \chi_{A
B}(y)=c_{AB}\;y^{-2+\imath s}, \label{sol}
\end{equation}
where $y$ once again denotes a dimensionless momentum.
Inserting the functions \eqref{sol} in the set of coupled
equations and performing the scale transformation $x=y\;z$ in the
integrand of Eqs.~\eqref{chi1a} and \eqref{chi2a}, it is derived the set of equations for the constants $c_{AA}$ and $c_{AB}$, given by
\begin{align}
c_{AA} = & c_{AB}\frac{2}{\pi}\sqrt{\frac {4{\cal A}}{{\cal A}+2}}  \int_0^\infty
dz\; z^{-2+1+\imath s} \; \ln \frac{2{\cal A}(z^2+z)+({\cal A}+1)}{2{\cal A}(z^2-z)+({\cal A}+1)} \; , & \label{chi1a1} \\
c_{AB} = &
\frac{1}{\pi}\left(\frac{{\cal A}+1}{2{\cal A}}\right)^{3/2} \sqrt{\frac
{2({\cal A}+1)}{{\cal A}+2}}
\int_0^\infty dz\; z^{-2+1+\imath s}
\left[c_{A A}\;\ln
\frac{2{\cal A}(1+z)+z^2({\cal A}+1)}{2{\cal A}(1-z)+z^2({\cal A}+1)} \right.  \nonumber \\*
& \hspace{4cm} +\left. {\cal A} \;
 c_{A B}\;
\ln \frac{(1+z^2)({\cal A}+1)+2z}{(1+z^2)({\cal A}+1)-2z}
 \right] . & \label{chi2a1}
\end{align}

Returning to Eqs.~\eqref{chi1a} and \eqref{chi2a}, there are two
solutions which are complex conjugates of each other, i.e., $z^{\pm\imath s}$.
Apart from an overall normalization, there is still a relative phase between 
these two independent solutions, which is determined by requiring 
the wave function to be zero at a certain momentum denoted $q^*$. 
This parameter is known as the three-body parameter 
\cite{nielsenPR2001,braatenPR2006}. This is the 
momentum-space equivalent of the coordinate-space three-body parameter
which is now believed to be simply related to the van der Waals 
two-body interaction of the atoms in question 
\cite{berningerPRL2011,schmidtEPJB2012,sorensenJoPBAMP2013}.
In this case the
asymptotic form of the spectator functions becomes
\begin{equation}
\chi_{AA}(q)=c_{AA}\; q^{-2} \sin(s\;\ln q/q^*) ~~~~{\text{and}}
~~~\chi_{AB}(q)=c_{AB}\; q^{-2} \sin(s\;\ln q/q^*)\; ,
\label{chiasymp}
\end{equation}
where $q$ denotes momentum and the boundary condition $\chi(q^*)=0$ is fulfilled.
\begin{figure}[!htb]%
\subfigure[\ Sixth excited state, $E_3=-8.6724\times10^{-12}$ for a Rb-Rb-Li molecule. ]
{\includegraphics[width=0.49\textwidth]{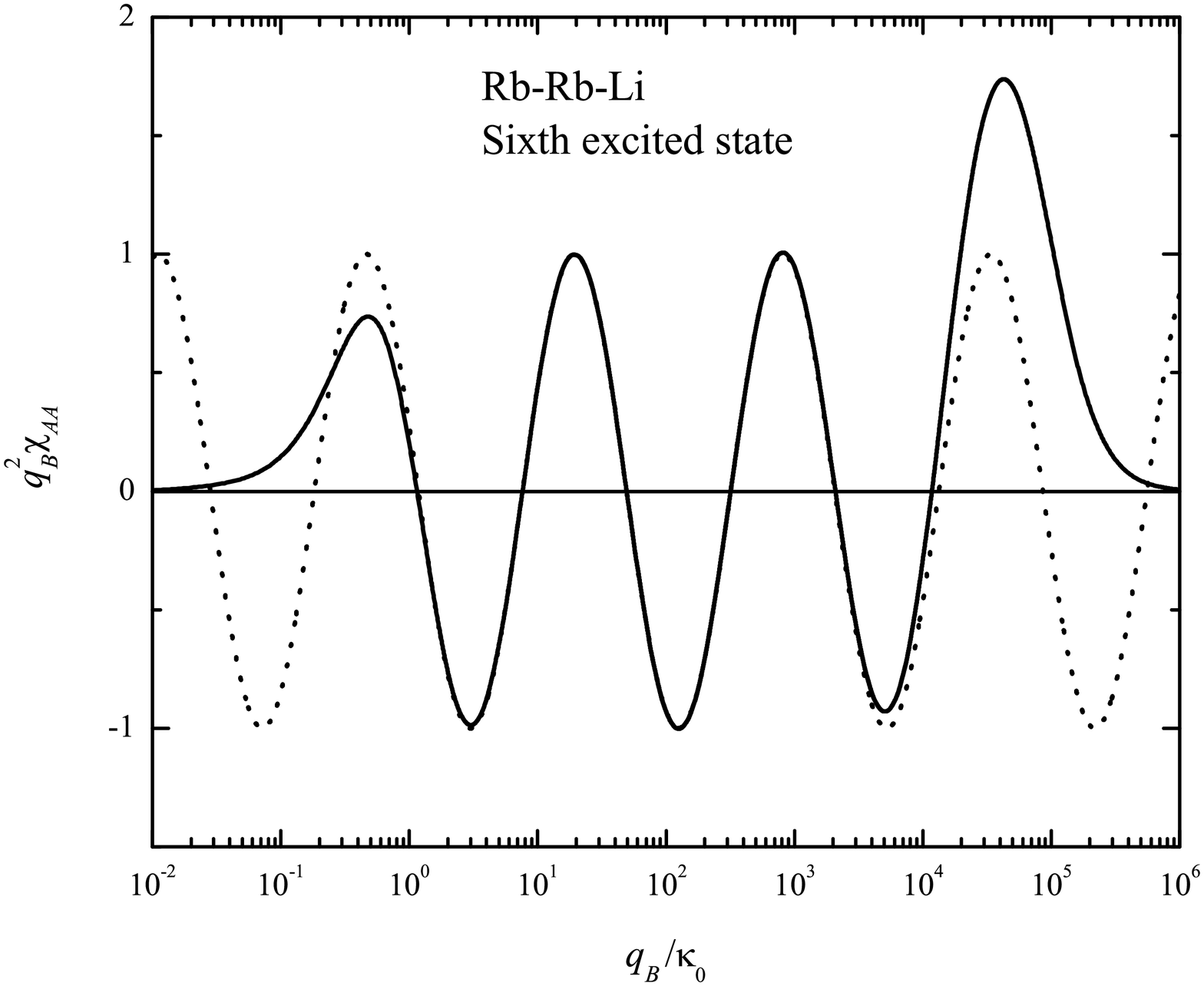}}
\subfigure[\ Eighth excited state, $E_3=-8.9265\times10^{-13}$ for a Cs-Cs-Li molecule.]
{\includegraphics[width=0.49\textwidth]{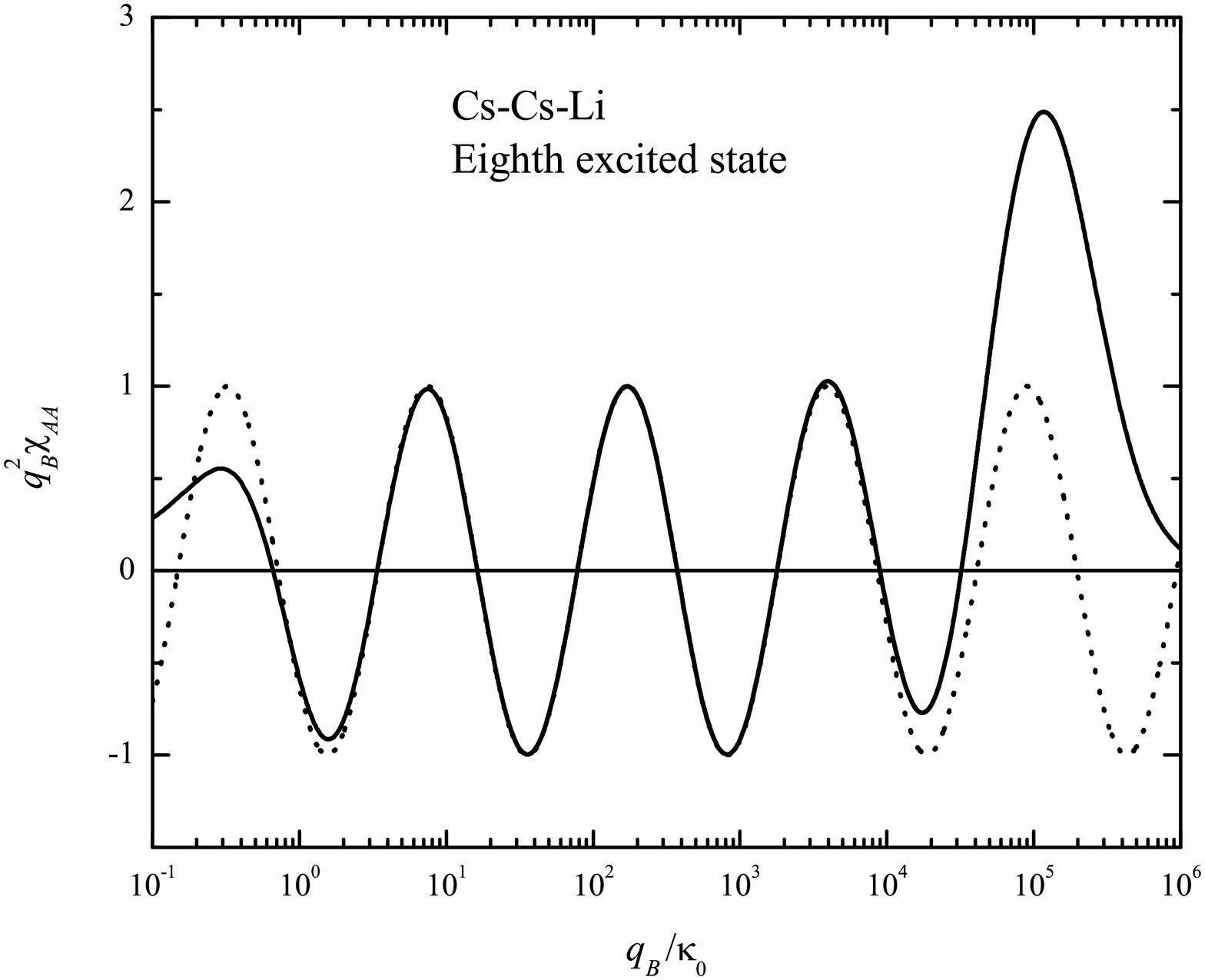}}
\caption[Comparison between the numerical solution of the set of coupled homogeneous integral equations and the asymptotic formula of fhe spectator function $\chi_{AA}(q)$.]
{The spectator function $\chi_{AA}(q)$ of a high excited state for $E_{AA}=E_{AB}=0$, solution of the
coupled equations (\ref{chi1}) and (\ref{chi2}) (solid line),
compared with the asymptotic formula (\ref{sol}) (dotted line). $E_3$ energies are 
given in arbitrary units.} \label{chirb}
\label{figasym}
\end{figure}

The asymptotic form of the spectator function should be compared
with the solutions of the subtracted equations in the limit of large
momentum, constrained by the window $\kappa_0<<q_B<<\mu$, where 
$\kappa_0\equiv\sqrt{|E_3|}$. 
The spectator functions $\chi_{AA}(q)$ for Rb-Rb-Li and Cs-Cs-Li 
compared to the respective asymptotic formula are 
shown in Fig.~\ref{figasym}. 
The difference between the numerical calculation and the analytic behavior is easily understood. While the limit $\mu\to\infty$ in taken in the analytical derivations, $\mu^2=1$ is used for the subtraction point in the numerical calculation  (see for instance Ref.~\cite{fredericoPiPaNP2012} for a detailed discussion and references therein). Notice that this subtraction method is basically equivalent to the procedure employed by Danilov \cite{danilovZETF1961} to regularize the original three-body Skorniakov-Ter-Martirosian equation \cite{skorniakovZETF1956}. A very detailed discussion of these issues is given by Pricoupenko \cite{pricoupenkoPRA2010,pricoupenkoPRA2011}. Therefore, the two curves would coincide in the idealized limit where $\kappa_0=0$ and $\mu\to \infty$ and the effect of finite  value of these two quantities is seen on each end of both plots. The window of validity for the use of the asymptotic formulas, i.e., $\sqrt{|E_3|}<<q<<\mu$ can be clearly seen in these figures.

\subsection{Scaling parameter}
Although the asymptotic expression in Eq.~\eqref{chiasymp} is already known, the procedure of using it to solve the coupled integral equations in Eqs.~\eqref{chi1a} and \eqref{chi2a} leads to an expression for the scaling parameter $s$ of an $AAB$ system. Inserting Eq.~\eqref{chi1a1} into Eq.~\eqref{chi2a1},  the set of coupled equations can be written as a single transcendental equation
\begin{align}
\frac{1}{\pi}\left(\frac{{\cal A}+1}{2{\cal A}}\right)^{3/2}  \sqrt{\frac
{2({\cal A}+1)}{{\cal A}+2}}\left({\cal A}I_1(s)
+\frac{2}{\pi}\sqrt{\frac {4{\cal A}}{{\cal A}+2}}I_2(s)I_3(s)\right)
=1 \; , \label{chi2a2}
\end{align}
where 
\begin{equation}
I_1(s)=\int_{0}^{\infty}dz z^{-1+is}\,\,\textrm{ln}\left[\frac{(z^2+1)({\cal A}+1)+2z}{(z^2+1)({\cal A}+1)-2z}\right] 
=\frac{2\pi}{s}\frac{\sinh(\theta_1 s-\frac{\pi}{2}s)}{\cosh(\frac{\pi}{2}s)} \; , \label{I1} 
\end{equation}
\begin{multline}
I_2(s)=\int_{0}^{\infty}dz z^{-1+is}\,\,\textrm{ln}\left[\frac{2{\cal A}(z^2+z)+{\cal A}+1}{2{\cal A}(z^2-z)+{\cal A}+1}\right] 
\\*
=\frac{2\pi}{s}\frac{\sinh(\theta_2 s-\frac{\pi}{2}s)}{\cosh(\frac{\pi}{2}s)}\left(\frac{{\cal A}+1}{2{\cal A}}\right)^{is/2} \; , \label{I2}
\end{multline}
\begin{multline}
I_3(s)=\int_{0}^{\infty}dz z^{-1+is}\,\,\textrm{ln}\left[\frac{2{\cal A}(1+z)+({\cal A}+1)z^2}{2{\cal A}(1-z)+({\cal A}+1)z^2}\right] 
\\*
= \frac{2\pi}{s}\frac{\sinh(\theta_2 s-\frac{\pi}{2}s)}{\cosh(\frac{\pi}{2}s)}\left(\frac{{\cal A}+1}{2{\cal A}}\right)^{-is/2} \; . \label{I3} 
\end{multline}
The angles are given by $\tan^2\theta_1={\cal A}({\cal A}+2)$ and $\tan^2\theta_2=({\cal A}+2)/{\cal A}$ with
the conditions $\pi/2<\theta_1$ and $\theta_2<\pi$. For the special case of equal masses, i.e., ${\cal A}=1$,  $\theta_1=\theta_2$, 
$I_1=I_2=I_3$ and
\begin{equation}
\left(\frac{1}{\pi}\sqrt{\frac{4}{3}}I_1(s)\right)+2\left(\frac{1}{\pi}\sqrt{\frac{4}{3}}I_1(s)\right)^2-1=0,
\end{equation}
for which the physically relevant solution is seen to be
\begin{equation}
\frac{1}{\pi}\sqrt{\frac{4}{3}}I_1(s)=\frac{1}{2}.
\label{I1a}
\end{equation}

The Efimov equation for the scaling parameter $s$ in a system of identical bosons is then recovered from Eqs.~\eqref{I1} and \eqref{I1a} \cite{efimovYF1970,nielsenPR2001}. Another very interesting and relevant case is when there is no interaction between the two $A$ particles, in which case $c_{AA}=0$ in Eq.~\eqref{chi2a1}. The equation for the scale factor, Eq.~\eqref{chi2a2}, now simplifies and gives
\begin{equation}\label{noAA}
\frac{{\cal A}}{\pi}\left(\frac{{\cal A}+1}{2{\cal A}}\right)^{3/2}
\sqrt{\frac{2({\cal A}+1)}{{\cal A}+2}}I_1(s)=1.
\end{equation}

\begin{figure}[htb!]
\centering
\includegraphics[width=0.9\textwidth]{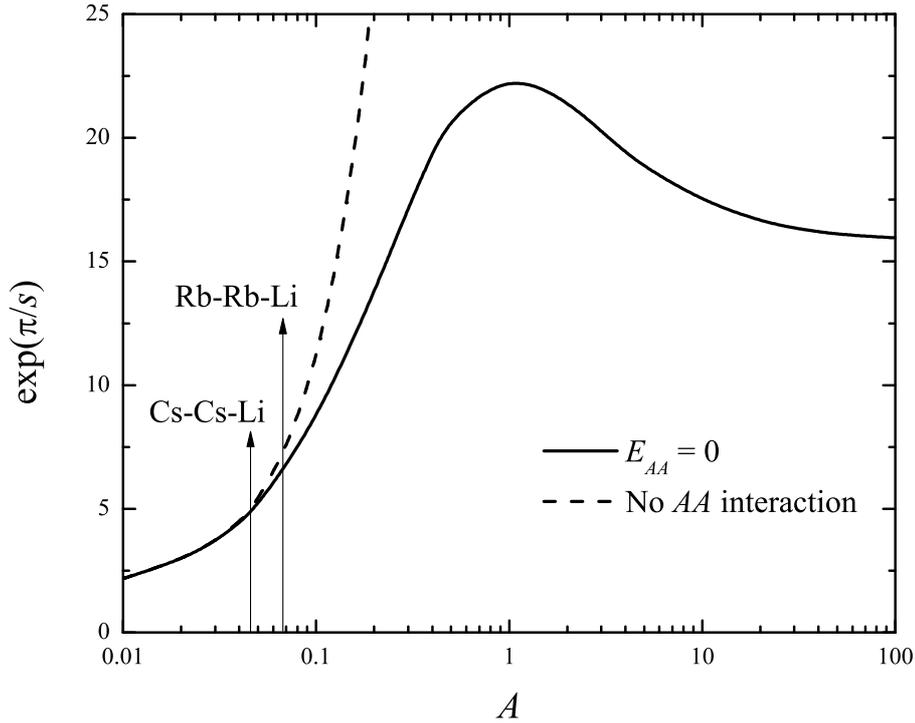}
\caption[Scaling parameter $s$ as a
function of the mass ratio ${\cal A}$ for $E_{AA}=0$ (resonant interactions) and with no interaction between AA.]
{Scaling parameter $s$ as a
function of ${\cal A}=m_B/m_A$ for $E_{AA}=0$ and $E_{AB}=0$(resonant interactions), solid line,
and for the situation
where $E_{AB}=0$ but with no interaction between AA, dashed line. The arrows
show the corresponding mass ratios for $^{133}$Cs-$^{133}$Cs-$^{6}$Li and
$^{87}$Rb-$^{87}$Rb-$^{6}$Li.} \label{s}
\end{figure}

The scaling factors, $\exp(\pi/s)$, are plotted in Fig.~\ref{s} for the
cases when all three subsystems have resonant interaction, which is the 
expression in Eq.~\eqref{chi2a2} valid for $E_{AA}=E_{AB}=0$ (solid line) and
when there is no interaction in the $AA$ subsystem, which is the 
expression in Eq.~\eqref{noAA} valid for $E_{AB}=0$ (dashed line). 

What is important to notice is that for $m_A\gg m_B$ (${\cal A}\ll 1$), the
scaling factors are very similar, and both are much smaller than the
equal mass case where ${\cal A}=1$. Therefore, it is expected that the $AAB$ system with
heavy $A$ and light $B$, would have many universal three-body
bound states ($s$ large or equivalently $e^{\pi/s}$ small) 
{\it irrespective} of whether the heavy-heavy
subsystem is weakly or strongly interacting. This feature is similar to the 2D case and recent experiments
with mixtures of $^{6}$Li and $^{133}$Cs indicate that there could be
a resonance of the $^{6}$Li-$^{133}$Cs subsystem at a point where
the scattering length in the $^{133}$Cs-$^{133}$Cs system is close to 
zero, i.e., weak interaction in the $AA$ subsystem \cite{reppPRA2013,tungPRA2013}.

\section{Asymptotic one-body densities}\label{amomentum}
The large-momentum one-body density $n(q_B)$, i.e., the single-particle momentum distribution for 
particle $B$ is calculated similarly to the procedure used for the 2D case. From Eqs.~\eqref{psiqb} and \eqref{nqaqb} it is possible to split the 
momentum density into nine terms, which can be reduced to four considering 
the symmetry between the two identical particles $A$. This simplifies 
the computation of the momentum density to the form $n(q_B)=\sum_{i=1}^4 n_i(q_B)$,
where
\begin{align}
n_1(q_B)&=\left|\chi_{AA}(q_B)\right|^2\int d^3p_B
 \frac{1}{\left(|E_3|+p_B^2+q_B^2\frac{{\cal A}+2}{4{\cal A}}\right)^2} \; 
 =\pi^2\frac{\left|\chi_{AA}(q_B)\right|^2 }{\sqrt{|E_3|+q_B^2\frac{{\cal A}+2}{4{\cal A}}}} \;  , \label{n1} 
\\
n_2(q_B)&= 2\int d^3p_B
 \frac{\left|\chi_{AB}(|\mathbf{p}_B-\frac{\mathbf{q}_B}{2}|)\right|^2}{\left(|E_3|+p_B^2+q_B^2\frac{{\cal A}+2}{4{\cal A}}\right)^2} \; , 
\label{n2} \\
n_3(q_B)&= 2\;\chi^{\ast}_{AA}(q_B)\int d^3p_B
 \frac{\chi_{AB}(|\mathbf{p}_B-\frac{\mathbf{q}_B}{2}|)}{\left(|E_3|+p_B^2+q_B^2\frac{{\cal A}+2}{4{\cal A}}\right)^2}+c.c.\label{n3} \; ,& \\
n_4(q_B)&= \int d^3p_B
 \frac{\chi^{\ast}_{AB}(|\mathbf{p}_B-\frac{\mathbf{q}_B}{2}|)\chi_{AB}(|\mathbf{p}_B+\frac{\mathbf{q}_B}{2}|)}
 {\left(|E_3|+p_B^2+q_B^2\frac{{\cal A}+2}{4{\cal A}}\right)^2}+c.c. \; . &\label{n4}
\end{align}

Going to the large momentum domain, $q \gg \sqrt{|E_3|}$, the limit $|E_3|\to 0$ is taken, i.e., the three-body energy is assumed to be negligible, since the focus is the imprint of excited Efimov states on the momentum distribution. These states are very extended and do not feel any short-range effects besides those encoded in the three-body parameter, $q^*$, discussed above. 

The asymptotic forms for the spectator functions are used 
in the integrals of Eqs.~\eqref{n1} to \eqref{n4}, where the integration is being performed from 0 to $\infty$. This may a priori 
cause problems at small momenta. However, a numerical 
check shows that the different behavior of the spectator functions at low momenta contributes 
only at next-to-next-to-leading order (NNLO). This procedure is the same as 
the one used in Ref.~\cite{castinPRA2011}.

The large momentum limit of the four terms in Eqs.~\eqref{n1} to \eqref{n4} is worked out in Appendix~\ref{derivations}, which is supplemented by Appendix~\ref{residues}. The derivation of the large momentum contributions of the terms $n_3(q_B)$ (see Eq.~\eqref{n3}) and $n_4(q_B)$ (see Eq.~\eqref{n4}) require several non\=/trivial mathematical steps, which let them too long and make the presentation lengthly in the bulk of the Chapter. However, it is worth to emphasize that the equations derived in Appendix~\ref{derivations} are the basis of the following discussion and their tricky derivation deserves to be looked out. Then, the large momentum contribution of the four terms in Eqs.~\eqref{n1} to \eqref{n4} up to next-to-leading order (NLO) is found to be
\begin{align}
\left<n_1(q_B)\right> & = \frac{\pi^2}{q_B^5}\; \left|c_{AA}\right|^2\;  \sqrt{ \frac{{\cal A}}{{\cal A}+2}} \; , \\
\left<n_2(q_B)\right> & =\frac{8{\cal A}^2}{q_B^4\;({\cal A}+1)^2}\int d^3q_A\;\left|\chi_{AB}(q_A)\right|^2 \nonumber\\*
& \hskip 4cm -\frac{8\pi^2 \left|c_{AB}\right|^2}{q_{B}^{5}} \frac{{\cal
A}^3({\cal A}+3)}{({\cal A}+1)^3 \sqrt{{\cal A}({\cal A}+2)}} \; , \label{n2asym}\\
\left<n_3(q_B)\right> & = \frac{4 \pi^2 c_{AA}\;c_{AB}}{q_B^5 \cosh\left(\frac{s\pi}{2}\right)}  \left\{\sqrt{\frac{{\cal A}}{{\cal A}+2}}\cos\left(s\ln \sqrt{\frac{{\cal A}+1}{2{\cal A}}}\right)\cosh\left[s\left(\frac{\pi}{2}-\theta_3\right)\right] \right.  \nonumber\\*
& \hskip 3cm \left. + \sin\left(s\ln \sqrt{\frac{{\cal A}+1}{2{\cal A}}}\right)\sinh\left[s\left(\frac{\pi}{2}-\theta_3\right)\right]\right\} \; , \\
\left<n_4(q_B)\right> & =\frac{ 8 \pi^2 |c_{AB}|^2 {\cal A}^2 }{s\; q_B^5 \sqrt{{\cal A}({\cal A}+2)} \cosh\left(\frac{s\pi}{2}\right)}   \left\{\sqrt{{\cal A}({\cal A}+2)}\sinh\left[s\left(\frac{\pi}{2}- \theta_4\right) \right] \right. \nonumber\\*
& \hskip 5cm \left.-\frac{s\; {\cal A} }{{\cal A}+1}\cosh\left[s\left(\frac{\pi}{2}- \theta_4\right) \right]\right\} \; ,
\end{align}
where $\tan\theta_3=\sqrt{\frac{{\cal A}+2}{{\cal A}}}$ for $0\leq\theta_3\leq\pi/2$ and $\tan\theta_4=\sqrt{{\cal A}({\cal A}+2)}$ for $0\leq\theta_4\leq\pi/2$.

\section{Leading and sub-leading terms}
As discussed in Sec.~\ref{secib}, the leading order term $\frac{C}{q_B^4}$ has the same functional form as in 2D and comes only from $n_2$, i.e., the first term on the right-hand-side of Eq.~\eqref{n2asym}. The constant $C$ is simply given by $C=\frac{8{\cal A}^2}{({\cal A}+1)^2}\int d^3q_A \left|\chi_{AB}(q_A)\right|^2$, which gives $C/\kappa_0=0.0274$ for  $^{133}$Cs-$^{133}$Cs-$^6$Li and $C/\kappa_0=0.0211$ for $^{87}$Rb-$^{87}$Rb-$^6$Li. For ${\cal A}=1$, the value $3(2\pi)^3C/\kappa_0=52.8$ is close to the exact value $53.097$, obtained in Ref.~\cite{castinPRA2011}. The factor $3(2\pi)^3$ comes from the different choice of normalization. 

The small discrepancy from both values of the contact parameter for a system composed for identical bosons arises from numerical issues. While the numerical value obtained here is calculated for the second excited state, the exact value in Ref.~\cite{castinPRA2011} is calculated for an arbitrary highly excited state.

\begin{figure}[htb!]
\centering
\includegraphics[width=0.9\textwidth]{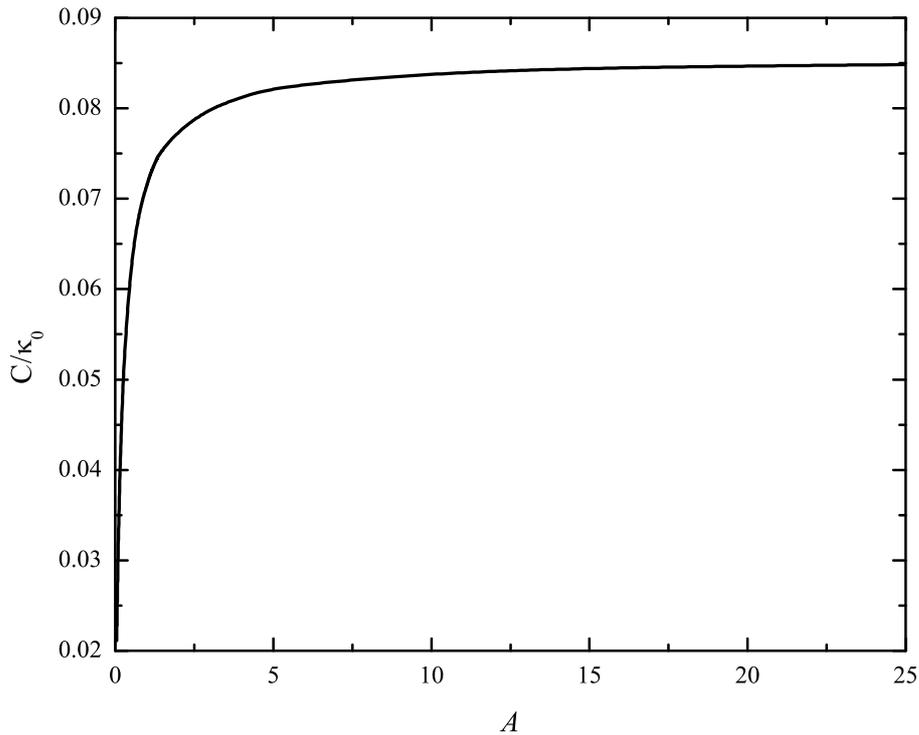}
\caption{$C/\kappa_0$ for mass ratios in the range $6/133 \leq {\cal A} \leq 25$.} \label{cmass}
\end{figure}

In Fig.~\ref{cmass} is shown the value of $C/\kappa_0$ for mass ratios in the range $6/133 \leq {\cal A} \leq 25$. The increase is very rapid until $A \approx 5$, beyond which an almost constant value is reached.  A similar behavior is found in 2D, as shown in Fig.~\ref{fig.10}.

In Ref.~\cite{castinPRA2011} it is shown that the non-oscillatory term of 
order $q_B^{-5}$ coming from $n_1$ to $n_4$ exactly cancels for ${\cal A} = 1$. However, this conclusion does not hold in the general case ${\cal A}\neq1$, as demonstrated bellow.

The four analytic expressions derived for each of the components of the one-body momentum distribution in Sec.~\ref{amomentum} are defined apart the coefficients $c_{aa}$ and $c_{ab}$.  The ratio between these coefficients is given by Eq.~\eqref{chi1a1}, which can be used to eliminate one of these factors. The other one can be determined from the overall normalization of the wave function. As the study is focused on the general behavior of the momentum distribution, the normalization is not relevant and the remaining coefficient is set to unit from now on, i.e., $c_{AB}=1$. 

\begin{figure}[htb!]
\centering
\includegraphics[width=0.9\textwidth]{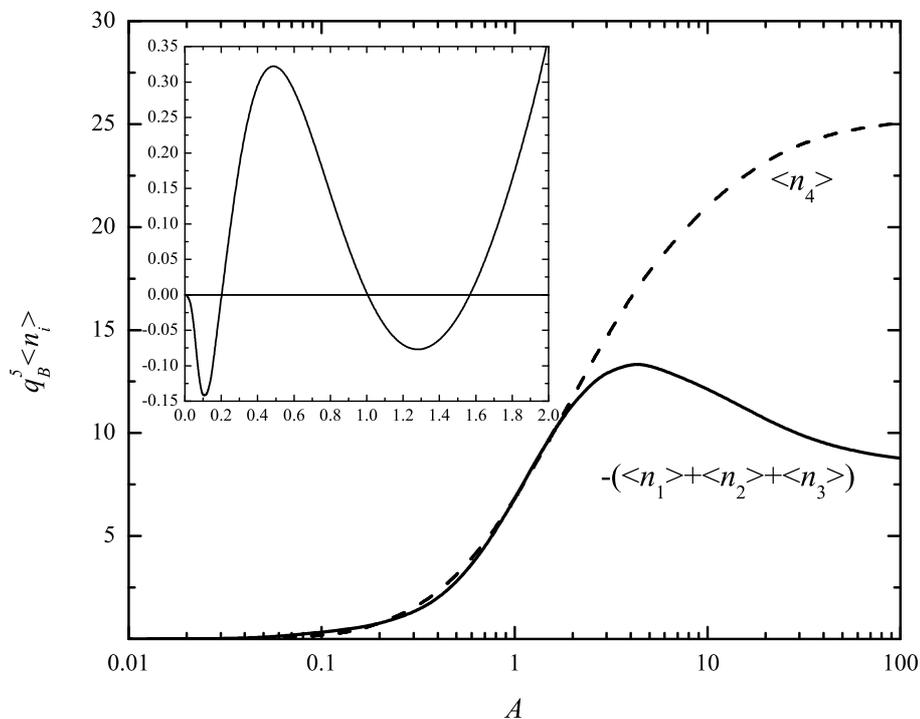}
\caption[Non-oscillatory contributions for $n_1$, $n_2$, $n_3$ and $n_4$ as a function of the mass 
ratio ${\cal A}$.]
{Non-oscillatory contributions for $n_1$, $n_2$, $n_3$ and $n_4$ as a function of the mass 
ratio ${\cal A}$. Their sum, showed in the inset, cancels exactly for ${\cal A}=$ 0.2, 1 and 1.57.}
\label{cancel_nonosc}
\end{figure}

\begin{figure}[htb!]
\centering
\includegraphics[width=0.9\textwidth]{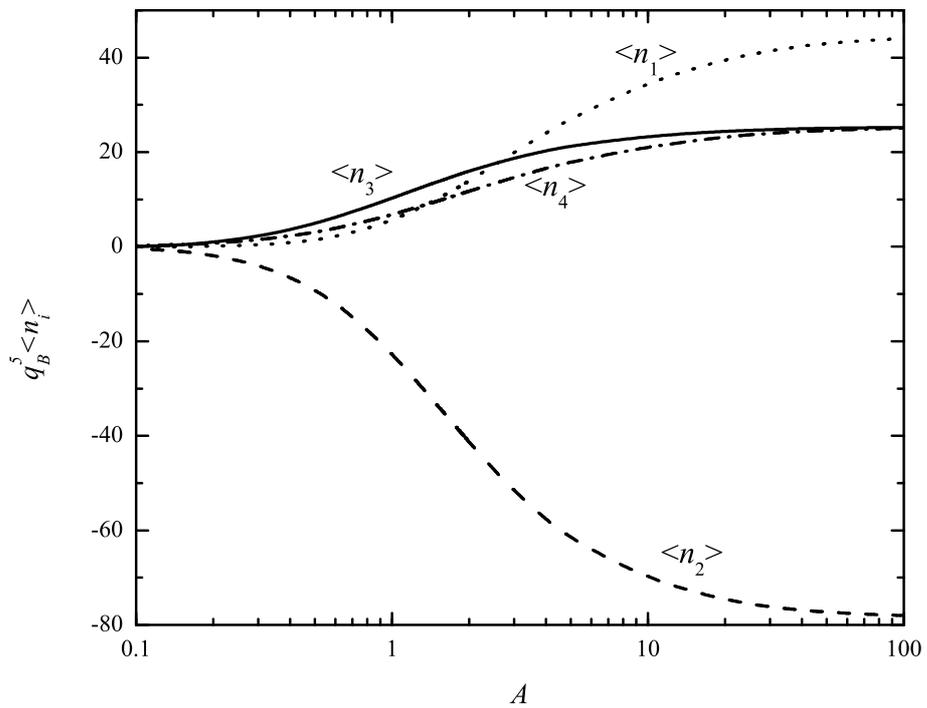}
\caption{Individual non-oscillatory contributions for $n_1$, $n_2$, $n_3$ and $n_4$ as a function of the mass 
ratio ${\cal A}$.}
\label{n1_n2_n3_n4}
\end{figure}

The contribution $-(n_1+n_2+n_3)$ and $n_4$ are shown in Fig.~\ref{cancel_nonosc} as a function of mass ratio ${\cal A}$ (each individual component $n_i$ as function of the mass ratio ${\cal A}$ is shown in Fig.~\ref{n1_n2_n3_n4}). What is immediately seen is that for ${\cal A}=1$ the result of Ref.~\cite{castinPRA2011} is reproduced, i.e., that the $q_{B}^{-5}$ non-oscillatory term cancels. However, for general ${\cal A}$ this does not hold and a $q_{B}^{-5}$ term in the asymptotic momentum distribution should also be expected for systems with two identical and a third particle.

This demonstrates that non-equal masses will generally influence not only the value of the contact parameter attributed to three-body bound states but also the functional form of the asymptotic momentum tail. Another important difference between 2D and 3D system arises here. Remember that systems of non-equal masses have the same functional form of the next-order contribution to  the momentum distribution (see Sec.~\ref{asymptotic2D}). Curiously, there is an oscillatory behavior around ${\cal A}\sim 1$ from the sum of all contributions. This is shown in the inset of Fig.~\ref{cancel_nonosc} where zero-crossings are seen at ${\cal A}=0.2$, $1$, and $1.57$. It seems quite clear that all the oscillatory terms, which depend on the scale factor $s$, are the reason for this interesting behavior, but a physical explanation for it was not found yet.

However, what makes this interesting is the fact that if ratios of typical isotopes of alkali atoms 
like Li, Na, K, Rb, and Cs are taken, then one can get rather close to 0.2 or 1.57. For instance, taking one $^{133}$Cs
and two $^{85}$Rb yields ${\cal A}=1.565$, while one $^{7}$Li atom and two $^{39}$K atoms yields
${\cal A}=0.179$. These interesting ratios are thus close to experimentally accessible species.

\section{Numerical examples}\label{results}

Some numerical examples of momentum distributions for the experimentally 
interesting systems with large mass ratios are now provided. Focus is on
$^{133}$Cs-$^{133}$Cs-$^6$Li and $^{87}$Rb-$^{87}$Rb-$^6$Li systems, where two extreme possibilities are investigated: (i) the heavy-heavy subsystems, 
i.e., $^{133}$Cs-$^{133}$Cs and $^{87}$Rb-$^{87}$Rb, 
have a two-body bound state at zero energy
and (ii) the opposite limit where they do not interact. 
In the first case the heavy atoms are 
at a Feshbach resonance with infinite scattering length, 
while in the second case they are far from resonance and a 
negligible background scattering length is assumed.
As it was recently 
demonstrated for the $^{133}$Cs-$^{6}$Li mixture, there are Feshbach resonances 
in the Li-Cs subsystem at positions where the Cs-Cs scattering length is non-resonant
\cite{reppPRA2013,tungPRA2013}. 
While this does not automatically imply that the Cs-Cs channel
can be neglected, the assumption (ii) is made here. 
The formalism can be modified in a 
straightforward manner to also include interaction in the heavy-heavy subsystem.

As before the focus is on the $AAB$ system, where $A$ refers to the identical (bosonic) atoms,
$^{133}$Cs or $^{87}$Rb, and $B$ to $^6$Li. 
By solving Eq.~\eqref{chi2a2}, the scaling factors are $s(6/133)=2.00588$ and $s(6/87)=1.68334$ when
assuming that all three subsystems have large scattering lengths (solid line in Fig.~\ref{s}).
The situation where the interaction between the two 
identical particles is turned off is shown by the dashed line in Fig.~\ref{s}. 
In this case, $s({\cal A})$ is calculated from Eq.~\eqref{chi2a1} by setting $c_{AA}=0$. This 
yields $s(6/133)=1.98572$ and $s(6/87)=1.63454$.

\begin{figure}[htb!]
\centering
\includegraphics[width=0.9\textwidth]{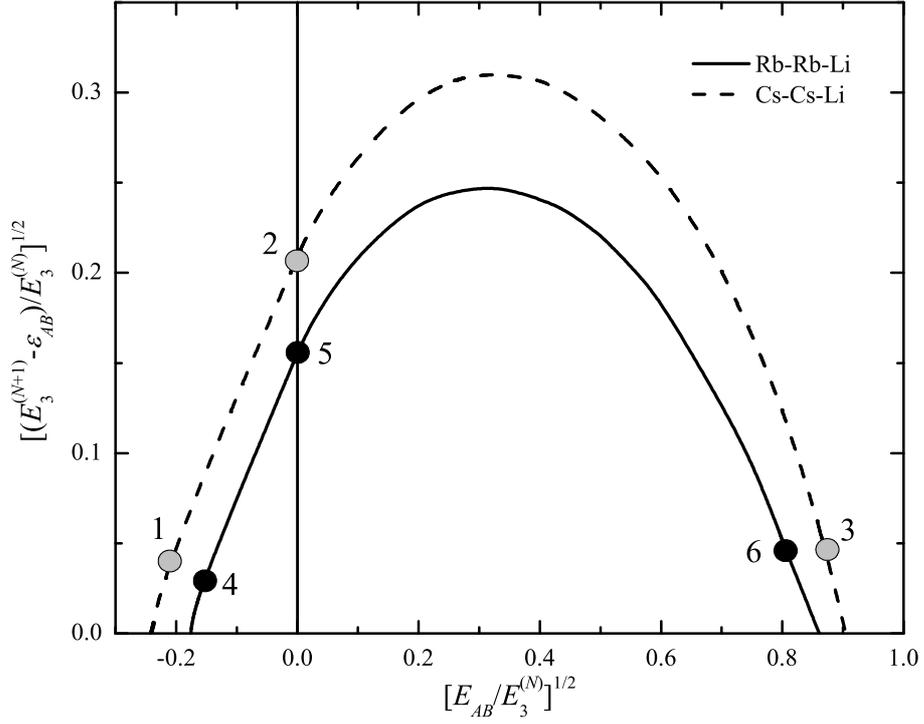}
\caption[Scaling plot of the Efimov states indicating the points where the momentum distributions
have been calculated.]
{$E_{AB}$ is the Cs-Li or Rb-Li two-body energy (Cs-Cs and Rb-Rb two-body energies are
zero). The negative and positive parts refer, respectively, to virtual and bound $AB$ states, such that 
$\epsilon_{AB}=0$ and $\epsilon_{AB}\equiv E_{AB}$, respectively, on the negative and positive sides.
The circles labeled from 1 to 6 mark the points where the momentum distributions
have been calculated.}
\label{bell}
\end{figure}

Firstly the binding energies are considered. Assuming that the Cs-Cs and Rb-Rb two-body energies 
are zero, a system satisfying the universality condition $|a| \gg r_0$ implies that any 
observable should be a function of the remaining two- and three-body scales, which can be 
conveniently chosen as $|E_3|^{(N)}$ and $E_{AB}$ (the Cs-Li or Rb-Li two-body energy). Here 
$N$ denotes the $N$th consecutive three-body bound state with $N=0$ being the lowest one. 
Thus, the energy of an $N+1$ state can be plotted in terms of a scaling function relating 
only $E_{AB}$ and the energy of previous state. The limit cycle, which should be in principle reached 
for $N\rightarrow\infty$, is achieved pretty fast such that the curve in Fig.~\ref{bell} is constructed 
using $N=2$ \cite{yamashitaPRA2002,fredericoPiPaNP2012}. In this figure, the negative and positive parts of the 
horizontal axis refer, respectively, to virtual and bound two-body $AB$ states. The circles
labeled from 1 to 6 mark the points where the momentum distributions have been calculated. 
The points 1 and 4 represent the Borromean case, the points 2 and 5 are the ``Efimov situation'' 
and in points 3 and 6 $AB$ is bound. 

\begin{figure}[!htb]%
\subfigure[\ $A=^{133}$Cs. \label{123Cs}]
{\includegraphics[width=0.49\textwidth]{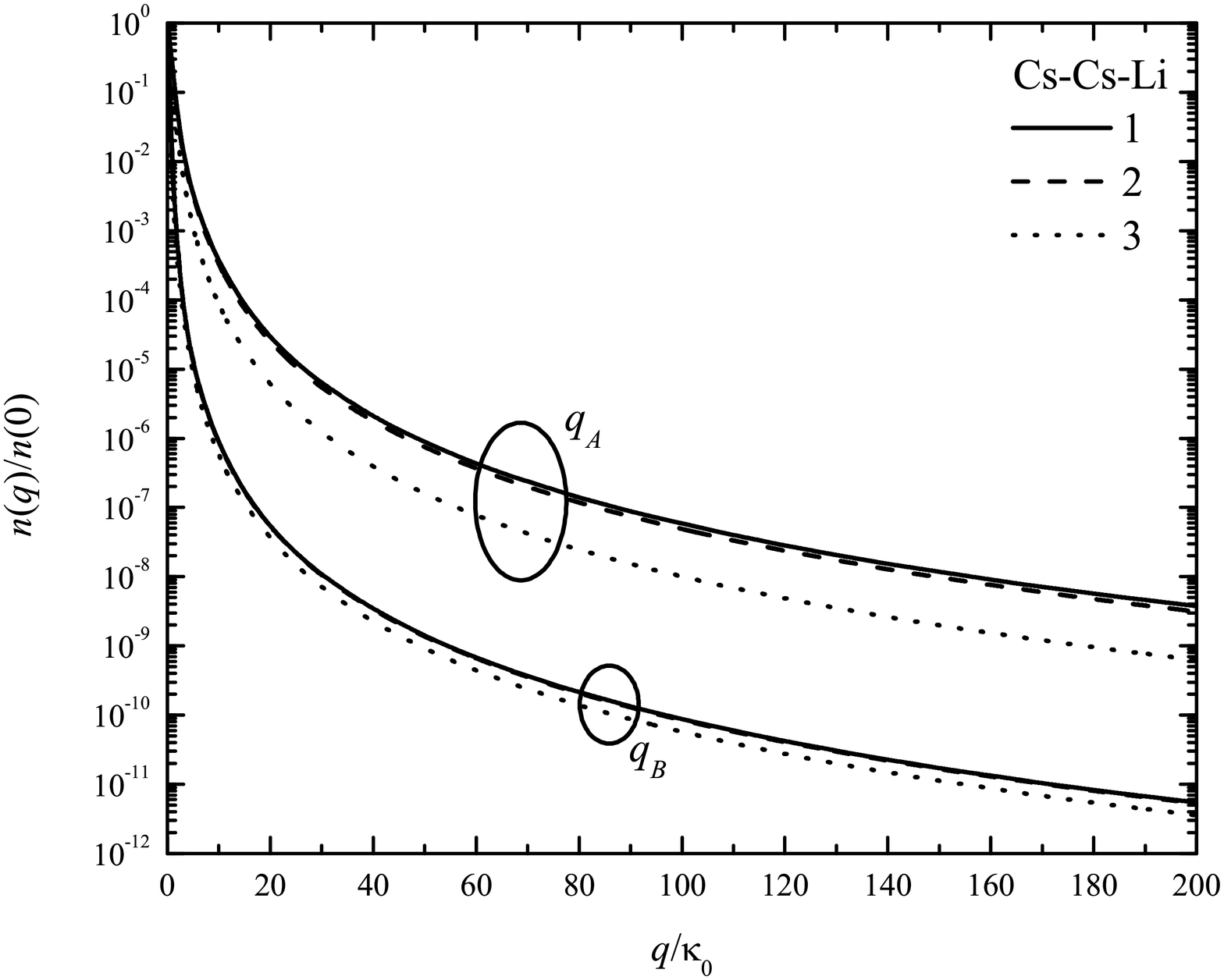}}
\subfigure[\ $A=^{87}$Rb.  \label{123Rb}]
{\includegraphics[width=0.49\textwidth]{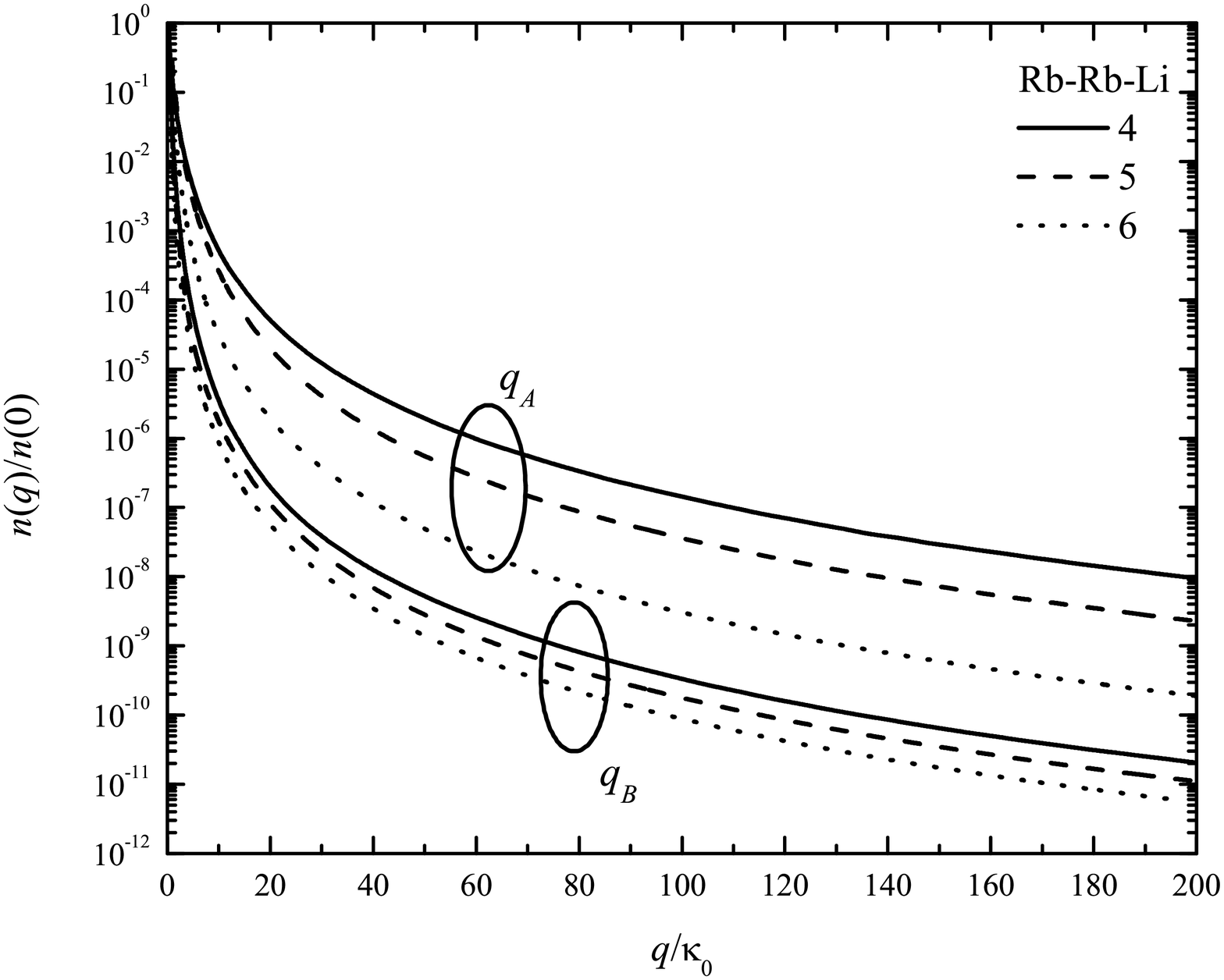}}
\caption[Momentum distribution for the second excited state as a function of the relative
momentum of one particle to the CM of the remaining pair.]
{ Momentum distribution for the second excited state as a function of the relative
momentum of one $A$-particle, $q_A$, or $^6$Li, $q_B$, to the center-of-mass of the remaining pair 
$A$-$^6$Li or $A$-$A$. The solid, dashed and dotted lines were calculated 
for the two- and three-body energies satisfying the ratios indicated by the points 1 to 3 ($A=^{133}$Cs) and 4 to 6 ($A=^{87}$Rb) in Fig.~\ref{bell}. The circles show the set of curves related to $q_A$ or $q_B$.} 
\label{123CsRb}
\end{figure}

Fig.~\ref{123Cs} and \ref{123Rb} give the momentum distributions of the second excited states
for the energy ratio $\sqrt{E_{AB}/|E_3|}$ given by the points labeled from 1 to 6 in Fig.~\ref{bell}. 
According to previous calculations \cite{yamashitaNPA2004}, for fixed 
three-body energy the size of the system increases as the number of bound 
two-body subsystems increase. Thus, it seems reasonable that the Borromean case 
decreases slower. This behavior is clearly seen in Figs.~\ref{123Cs} and \ref{123Rb}. 
The distance of one atom to the center-of-mass of the other two is much larger for $^6$Li than 
for $^{133}$Cs or $^{87}$Rb, due to the large difference of the masses. Therefore, the momentum distribution for the heavier atom, $q_A$ set, decreases much slower 
than that for the lighter one, $q_B$ set.


\begin{figure}[!htb]%
\subfigure[\ $A=^{133}$Cs. \label{scalingCs}]
{\includegraphics[width=0.49\textwidth]{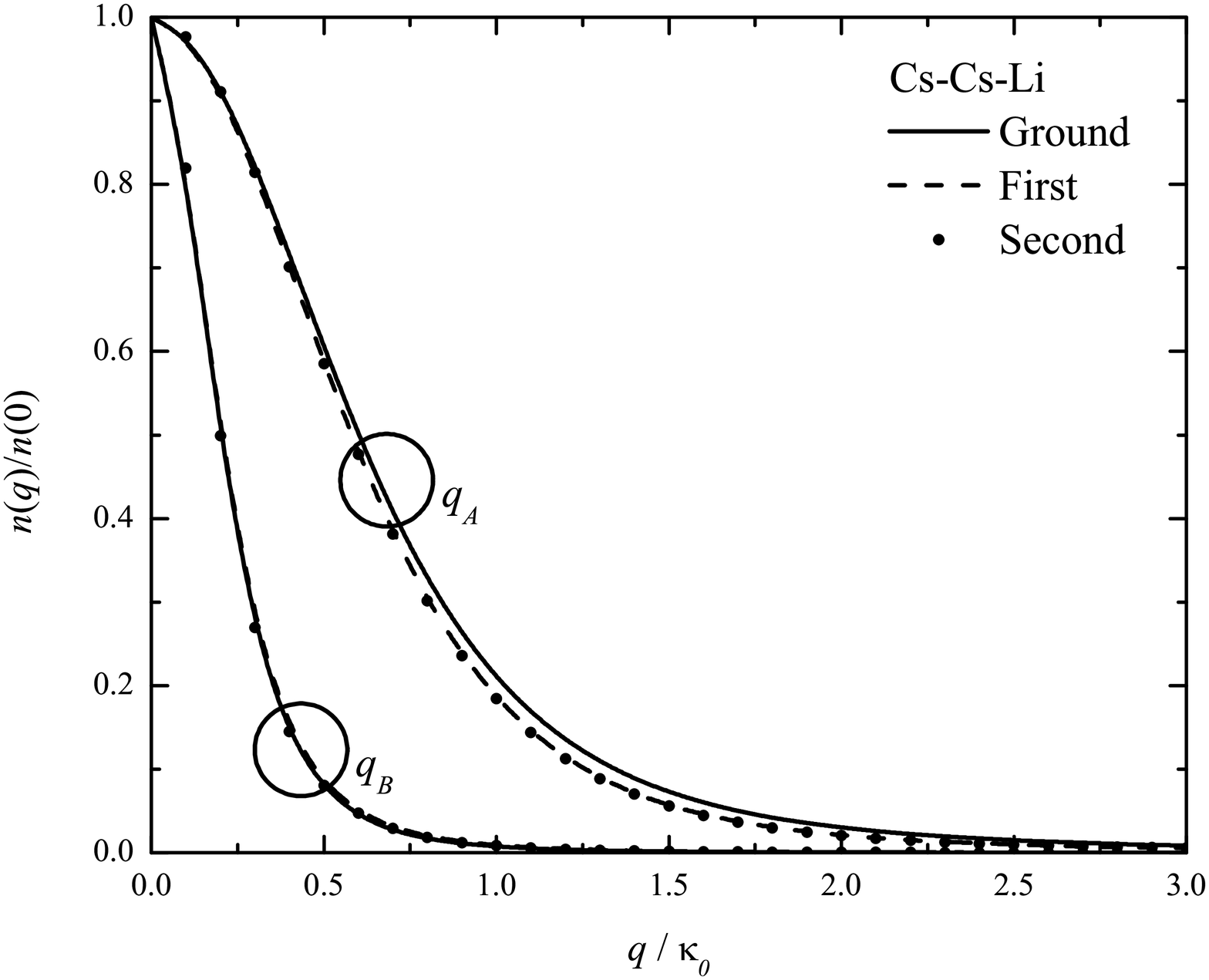}}
\subfigure[\ $A=^{87}$Rb.  \label{scalingRb}]
{\includegraphics[width=0.49\textwidth]{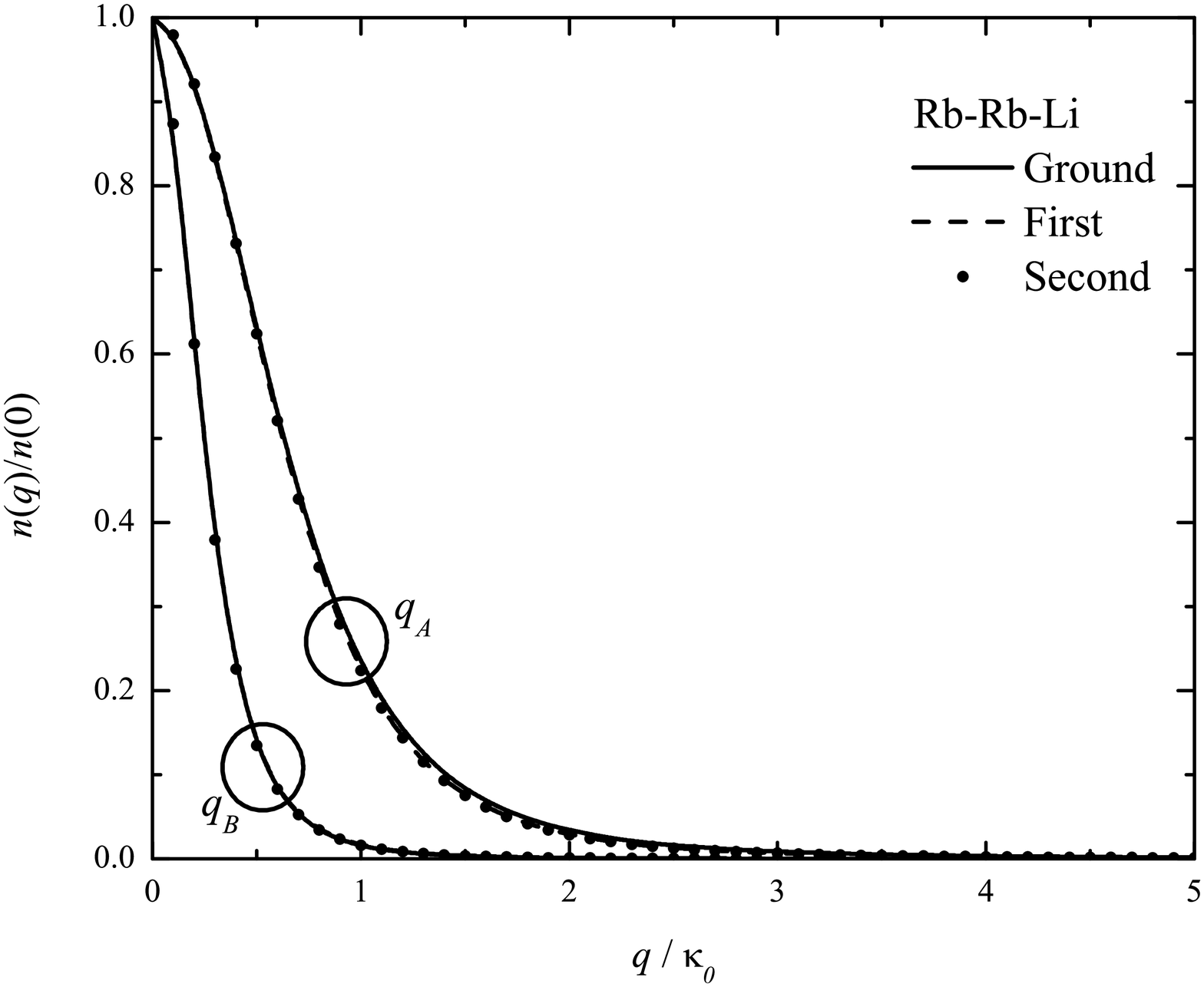}}
\caption[Rescaled momentum distribution for the ground, first and second excited
states as a function of the relative momenta of one particle to the CM of the remaining pair.]
{ Rescaled momentum distribution for the ground, first and second excited
states as a function of the relative momenta of the $A$-particle to the center-of-mass of the pair
$^{6}$Li-$A$, $q_A$, and of $^6$Li to the center-of-mass of the pair
$A$-$A$, $q_B$. The subsystem binding energies are all set to zero.
Normalization to unity at zero momentum.} 
\label{scalingCsRb}
\end{figure}

Figs.~\ref{scalingCs} and \ref{scalingRb} show the rescaled momentum distributions for the
ground, first and second excited states. In these figures, the subsystem energies 
are chosen to zero, corresponding to the transition point to a Borromean configuration. 
In this situation, the only low-energy scale is $|E_3|$ (remember that the
high-momentum scale is $\mu=1$). Therefore, in units in which
$|E_3|=1$, to achieve a universal regime, in principle, to wash-out
the effect of the subtraction scale, $\mu$, it is necessary to go to a highly excited state.  However, 
a universal low-energy regime of $n(q_B)/n(q_B=0)$ is seen for momentum of 
the order of $\sqrt{|E_3|}$, even for the ground state which is smaller 
than excited states. Then, in practice, the universal 
behavior of the momentum distribution is approached rapidly.

\chapter{Dimensional crossover} \label{ch7}

Examples of how the dimensionality  affects the properties of physical system have been  recently attracting great interest, since the possibility of probing lower dimensional systems has being continuously increasing. One recent and famous example is graphene, which is itself a effective two-dimensional (2D) structure  with good mechanical and electrical properties~\cite{novoselovS2004}.  Besides that, the experimental study of one-dimensional (1D) systems, which is an useful theoretical laboratory to study physical problems, has also been recently reported~\cite{serwaneS2011,zurnPRL2012}. Following this line, maybe the most surprisingly achievement is the study of the so-called quantum dots~\cite{ramosPRL2011}, which  are approached as zero-dimensional systems.

Dimensionality also plays an important role in the behavior of few- and many-body quantum systems. As it was discussed in Chapters~\ref{ch3} and \ref{ch6}, the Efimov effect \cite{efimovYF1970}, where a geometric series of three-body bound
states of three bosons occurs at the threshold for binding
the two-body subsystems, is present in 3D systems but absent when the dimension  is reduced to two. A straightforward consequence discussed in Chapters~\ref{ch5} and \ref{ch6} is that systems restricted to different dimensions have very distinct two- and three-body contact parameters and asymptotic forms of the momentum distribution at the next\=/to\=/leading order. The contact parameters can be defined via the one-body large momentum density and relate few- and many-body properties of quantum atomic gases \cite{tanAoP2008}.

Cold atomic gases have proven their ability to be excellent
quantum simulators due to the tunability of interactions, 
geometry and inter-particle statistical properties.
At the few-body level, three-body states linked to the
Efimov effect have been observed in three 
dimensions (see for example Ref.~\cite{kraemerNP2006,berningerPRL2011a})
using a variety of different atomic species and two-body 
Feshbach resonances \cite{chinRMP2010}.

In spite of the tunability 
of the external trapping geometry of cold atomic systems, there has
been little study of how the three-boson
bound state problem undergoes its dramatic change from displaying the 
Efimov effect in three dimensions, yet in two dimensions the systems
only holds two bound states \cite{bruchPRA1979}. A key question is 
whether it is possible to interpolate these limits in simple theoretical 
terms and subsequently explore this in simulations using 
both more involved numerical methods and experimental setups.

A model that has the ability to interpolate
geometrically between two and three spatial dimensions and thus study
this crossover for both two- and three-body bound states 
of identical bosons is proposed. A ``squeezed'' dimension whose size can be varied to interpolate the two 
limits is employed with periodic 
boundary conditions (PBC) . This model has the unique feature that it can be 
regularized analytically which is a great advantage for its numerical 
implementation allowing to go smoothly between both limits. 


The theoretical elegance and tractability of calculations in the three-body
system is itself a strong incentive for pursuing this geometry, but in spite of this elegance, a direct connection between experiments and the parameter that dials between different dimensions with PBC in this model was not found yet. On the other hand, it is also possible to formulate the problem with open boundary conditions (OBC). While
for many experimental setups in cold atoms the transverse confining geometry
is given by a harmonic trapping potential and a recent theoretical 
study \cite{levinsenAe2014} has considered the properties of three-boson states 
under strong transverse confinement,  the recent successful production of box potential traps with bosons  \cite{gauntPRL2013,schmidutzPRL2014} means that OBC (hard wall) are now accessible. Besides that, the formulation of the problem with OBC is let for a future consideration.

The method allows to study the dimensional crossover transitions of strongly interacting two- and three-bosons systems by continuously  ``squeezing'' one of the dimensions. The motion of the particles is separated into two directions, namely a flat surface plus a transverse direction (compact dimension) which has the position/momentum discretized accordingly to the chosen type of boundary conditions. Notice that, unless in the pure 2D limit, the problem is always 3D (Quasi- 2D). In the following, the case of periodic boundary condition (PBC) is considered.

\section{Renormalization with a compact dimension}
Effects of a compact dimension are investigated in a system of three-identical bosons with zero-range pairwise interactions.
The formal expression of the two- and three-body transition operators is the same as the one presented in Chapter~\ref{ch2}, which reads
\begin{equation}
T(E) = V + V G_0(E) T(E) \; .
\label{LSch7}
\end{equation}
However, the matrix elements are not the same now, due to the restriction of the momentum in the transverse direction, that arises from the boundary condition from which the system is subjected. Being $\mathbf{k}$ and $\mathbf{k}^{\prime}$ respectively the momenta of the incoming and outgoing waves, the matrix elements of Eq.~\eqref{LSch7} are given by
\begin{align}
\left\langle \mathbf{k}^{\prime} \right| T(E) \left| \mathbf{k} \right\rangle =& 
\left\langle \mathbf{k}^{\prime} \right| V \left| \mathbf{k} \right\rangle + 
\sumint d\mathbf{q} d\mathbf{q}^{\prime} \left\langle \mathbf{k}^{\prime} \right| V \left| \mathbf{q} \right\rangle 
\left\langle \mathbf{q}\right| G_0(E) \left| \mathbf{q}^{\prime} \right\rangle
\left\langle \mathbf{q}^{\prime} \right| T(E) \left| \mathbf{k} \right\rangle \; , 
\label{LSp}
\end{align}  
where the symbol $\sumint$ indicates that all momenta are being taking into account, i.e., there is an integration over the continuum momentum in the plane ($p_{\perp}$) and a sum over the discrete perpendicular momentum ($p_z$). The particular form of $\sumint$ depends on the type of boundary conditions that are being addressed.

Once more, the interaction between particles is described by Dirac$-\delta$ potentials. As it is discussed in Sec.~\ref{tzmr}, the matrix element for this kind of potential is $\left\langle \mathbf{k}^{\prime} \right| V \left| \mathbf{q} \right\rangle=V\left( \mathbf{k}^{\prime}, \mathbf{q}\right)= \lambda$ and therefore Eq.~\eqref{LSp} has also to be renormalized. 
Following the procedure from Refs.~\cite{adhikariPRL1995a,adhikariPRL1995}, which is briefly presented in Sec.~\ref{r3bto}, the renormalized transition matrix for two- and three-body systems subjected to a compact dimension are given by
\begin{align}
\left\langle \mathbf{k}^{\prime} \right| T(E) \left| \mathbf{k} \right\rangle =& T(-\mu^2) - T(-\mu^2) (\mu^2+E)
\sumint d\mathbf{q} G_0(\mathbf{q},E)G_0(\mathbf{q},-\mu^2) T\left(\mathbf{q}, \mathbf{k}, E \right) \; , 
\label{LSR}
\end{align}  
where in the two-body sector $T(-\mu^2)$ is a constant, namely $T(-\mu^2)= \lambda$ and in the three-body sector $T(-\mu^2)$ is the sum over the renormalized two-body $T-$matrix given in Eq.~\eqref{t2sum3d}.

In the two-body sector, since the right-hand-side of Eq.~\eqref{LSR} is independent of $\mathbf{k}^{\prime}$ and $\mathbf{k}$, it is possible to define $T\left(\mathbf{q}, \mathbf{k}, E \right)= \left\langle \mathbf{q} \right| T(E) \left| \mathbf{k} \right\rangle = \left\langle \mathbf{k}^{\prime} \right| T(E) \left| \mathbf{k} \right\rangle = \tau(E)$. The two-body $T-$matrix becomes
\begin{align}
\tau(E) =& \lambda - \lambda (\mu^2+E) \tau(E) \sumint d\mathbf{q} G_0(\mathbf{q},E)G_0(\mathbf{q},-\mu^2) \;.
 \nonumber\\* 
 =& \frac{\lambda}{1+ \lambda (\mu^2+E)  \sumint d\mathbf{q} G_0(\mathbf{q},E)G_0(\mathbf{q},-\mu^2)} \; . 
\label{LSR2}
\end{align}  

In the three-body sector, the renormalized $T-$matrix in Eq.~\eqref{LSR} is given by
\begin{align}
T(E) = T(-\mu^2) + T(-\mu^2) \left[G_0(E)-G_0(-\mu^2) \right] T(E)   \; ,
\label{te3d}
\end{align}
is the sum over the renormalized two-body $T-$matrix given in Eq.~\eqref{t2sum3d}. Notice that Eq.~\eqref{te3d} is identical to Eq.~\eqref{T3D}, meaning that the derivation of the coupled homogeneous integral equations for the three-body bound state is the same as in Sec.~\eqref{tbsie3d}. The only difference now is that the momentum $\mathbf{q}$ and the phase factor $d \mathbf{q}$ are restricted by the boundary conditions.

\section{3D - 2D transition with PBC} \label{32pbc}
Periodic boundary conditions (PBC) are assumed to be valid for the relative
distance between the particles in the compact dimension, chosen to
be $z$. Then, the relative momentum is given by $\mathbf
p_\perp=(p_x,p_y)$  in the flat 2D surface and by
\begin{align}
p_z=\frac{2 \pi n}{L}=\frac{n }{ R} \ , \;\;\; n=0,\pm1,\pm2, \hdots \; ,
\end{align}
in the transverse direction, with $L=2\pi R$ being the size of the compact dimension corresponding
to a radius $R$, which is the parameter that dials between two and three-dimensions.
When $R \to 0$ it selects the 2D case and in the opposite limit,
i.e.,  $R\to \infty$, the 3D case is selected. 
The momentum $\mathbf{q}$ and its corresponding phase factor $d \mathbf{q}$ in Eq.~\eqref{LSR2} are, with PBC, given by
\begin{align}
q^2= p_{\perp}^2+\frac{n^2}{R^2} \;\;\; \text{and} \;\;\; d\mathbf{q}=\frac{1}{R}d^2p_{\perp} \; .
\label{qdq}
\end{align}
The symbol $\sumint$, which indicates an integration over the continuum momentum in the plane ($p_{\perp}$) and a sum over the discrete perpendicular momentum ($p_z=\frac{n}{R}$) in this case reads
\begin{align}
\sumint d\mathbf{q} \equiv \sum_{n=-\infty}^{\infty} \int \frac{1}{R}d^2 p_{\perp} \; .
\label{sumintpbc}
\end{align}

\subsection{Two-body scattering amplitude  } \label{2bcompac}

Replacing $q^2= p_{\perp}^2+\frac{n^2}{R^2}$ and $d\mathbf{q}=\frac{1}{R}d^2p_{\perp}$, Eq.~\eqref{LSR2} becomes
\begin{align}
R^{-1} \tau(E) = \left[\lambda^{-1}R - (\mu^2+E)  \sumint d^2p_{\perp} \frac{1}{\left(E - p_{\perp}^2 - \frac{n^2}{R^2} + \imath \epsilon\right)\left(\mu^2 + p_{\perp}^2 + \frac{n^2}{R^2} \right)} \right]^{-1} \; .
\label{LSR1}
\end{align}  

The choice of $-\mu^2 = E_2$ leads to $\lambda^{-1}=0$ (see Sec.~\ref{tzmr} ) and the matrix elements in Eq.~\eqref{LSR1} are given by
\begin{align}
R^{-1} \tau_{p}(E) = \left[ I_R(E) \right]^{-1} \; , 
\end{align} 
where the subscript $p$ distinguish  between the matrix elements of systems restricted purely to 2D or 3D to the compacted dimension (quasi-2D) with PBC discussed here and the function $I_R(E)$ is given by
\begin{align}
I_R(E) =&  \sumint d^2 p_{\perp} \frac{- (E-E_2)}{\left(E - p_{\perp}^2 - \frac{n^2}{R^2} + \imath \epsilon\right)\left(-E_2 + p_{\perp}^2 + \frac{n^2}{R^2} \right)} \; .
\label{IRa}
\end{align}  

For $E<0$  the function in Eq.~\eqref{IRa} reads
\begin{align}
I_R(E) =& - \sumint d^2 p_{\perp} \left[ \frac{1}{|E_2| + p_{\perp}^2 + \frac{n^2}{R^2} } - \frac{1}{|E| + p_{\perp}^2 + \frac{n^2}{R^2}}  \right]\; , \nonumber\\
=& - \pi \sum_{n = -\infty}^{\infty} \lim_{\Lambda \to \infty} \left[ \ln \left(|E_2| + p_{\perp}^2 + \frac{n^2}{R^2} \right) \Bigr|_{p_{\perp}=0}^{\Lambda} - \ln \left(|E| + p_{\perp}^2 + \frac{n^2}{R^2} \right) \Bigr|_{p_{\perp}=0}^{\Lambda}   \right]\; , \nonumber\\
=& - \pi \sum_{n = -\infty}^{\infty} \lim_{\Lambda \to \infty} \left[ \ln \left(\frac{|E_2| + \Lambda^2 + \frac{n^2}{R^2}}{|E| + \Lambda^2 + \frac{n^2}{R^2}} \right)  + \ln \left(\frac{|E| + \frac{n^2}{R^2}}{|E_2| + \frac{n^2}{R^2}} \right) \right]\; , \nonumber\\
=& - \pi \sum_{n = -\infty}^{\infty}  \ln \left(\frac{|E| + \frac{n^2}{R^2}}{|E_2| + \frac{n^2}{R^2}}\right) = - \pi \sum_n  \ln \left(\frac{-E + \frac{n^2}{R^2}}{|E_2| + \frac{n^2}{R^2}}\right) \; .
\label{IR1}
\end{align}  

On the other hand, for $E>0$  the analytic extension of Eq.~\eqref{IR1} must be
\begin{multline}
I_R(E) = - \pi \sum_{n = -\infty}^{\infty}  \left\{ \ln \left(\frac{-E + \frac{n^2}{R^2}}{|E_2| + \frac{n^2}{R^2}}\right) \Theta\left(\frac{n^2}{R^2}-E\right) \right. \\* \left. + \left[\ln \left(\frac{E - \frac{n^2}{R^2}}{|E_2| + \frac{n^2}{R^2}}\right) -\imath \pi \right] \Theta\left(E-\frac{n^2}{R^2}\right) \right\}\; .
\label{IRb}
\end{multline}  
The sum over $n$ in Eq.~\eqref{IRb} can be performed analytically for $E<0$, noticing that
\begin{align}
\sum_{n = -\infty}^{\infty} \ln \left(\frac{a^2+n^2}{b^2+n^2}\right) = 2  \ln \left[\frac{\sinh \left(\pi a\right)}{\sinh \left(\pi b\right)}\right] \; .
\label{sumab}
\end{align}  
Then, the function $I_R(E)$ becomes
\begin{align}
I_R(E) =& - 2\pi \ln \left[\frac{\sinh \left(\pi R \sqrt{|E|}  \right)}{\sinh \left(\pi R \sqrt{|E_2|}  \right)} \right] \; ,
\label{IR}
\end{align}  
and the matrix elements of the transition operator for two-body systems restricted to a compact dimension with PBC are given by 
\begin{align}
\tau_p(E)^{-1} = R^{-1}  I_R(E) 
= -\frac{2 \pi}{R} \ln \left[\frac{\sinh \left(\pi R \sqrt{|E|}  \right)}{\sinh \left(\pi R \sqrt{|E_2|}  \right)} \right] \; .
\label{taup}
\end{align}
Notice that $\tau_p(E)$ recovers the matrix elements of 3D and 2D systems in the limits $R \to \infty$ and $R \to 0$, respectively. The first case is straightforward an reads
\begin{align}
\tau_{3D}^{-1}(E) = \lim_{R \to \infty} \tau_p^{-1}(E) 
= - 2\pi^2 \left(\sqrt{|E|} - \sqrt{|E_2|} \right)\; ,
\label{tau3dpbc}
\end{align}
which is identical to Eq.~\eqref{tau2b3d}.

Before going to the 2D limit, it is important to notice that, as it was said before, a quasi-2D system is in practice a 3D system. Then, the units of  $\tau_{3D}^{-1}(E)$ and $\tau_p^{-1}(E)$ are exactly the same and reads $[E] . [L]^3$, as it can be easily seen in from Eq.~\eqref{LSp}, where
\begin{align}
\left\langle \mathbf{k}^{\prime} \right| T(E) \left| \mathbf{k} \right\rangle = \int \frac{d^3 x}{\left(2 \pi\right)^3}  e^{\imath \mathbf{k}^{\prime} \cdot \mathbf{x}} e^{-\imath \mathbf{k}\cdot \mathbf{x}}\left\langle \mathbf{x} \right| V \left| \mathbf{x} \right\rangle \; .
\end{align}
On the other hand, the unit of $\tau_{2D}^{-1}(E)$ is $[E] . [L]^2$, which gives $\frac{\left[\tau_{2D}^{-1}(E)\right]}{\left[\tau_{p}^{-1}(E) \right]} =[L]^{-1}$. Taking into account the correct units, the 2D limit of Eq.~\eqref{IR} reads
\begin{align}
\tau_{2D}^{-1}(E) = \lim_{R \to 0}  R \; \tau_p^{-1}(E)  
= - 2\pi \ln \left(   \sqrt{\frac{|E|} {|E_2|}} \right) \; ,
\label{tau2Dpbc}
\end{align}
which is identical to Eq.~\eqref{eq.c2-52c}.

\subsection{Dimer binding energy } \label{vdecrpbc}

Up to this point, the energy of the dimer, $E_2$, was considered unchanged during the squeezing of the trap. In this subsection it is shown how $E_2$ changes as the parameter $R$, which controls the size of the trap, is continuously tuned from 3D to 2D when the squeezing is performed with PBC. 
The analysis starts with the denominator of Eq.~\eqref{LSR1} in the limit of $\mu\to \infty$, which reads
\begin{align}
R^{-1}\tau_p(E)= \left[R \lambda^{-1} (\mu \to \infty) - \sumint \frac{d^2p_\perp }{ E-\mathbf{p}_{\perp}^2-\frac{n^2}{R^2}+\imath  \epsilon}
\right]^{-1}\ . \label{rtau1}
\end{align}
Therefore, the ultraviolet divergence has to be removed by $\lambda^{-1}(\mu \to \infty)$, which can be can be chosen as
\begin{align}
\lambda^{-1}(\mu \to \infty) = \int \frac{d^3p }{ E_2^{3D}-\mathbf{p}^2}  \; ,\label{rtau2}
\end{align}
where the limit $R\to \infty$ leads to the bound-state pole at the dimer energy in 3D, namely $\tau_p^{-1}(E)=0$ at $E=E_2^{3D}$.  The renormalized scattering amplitude becomes
\begin{align}
R^{-1}\tau_p(E)=\left[R \int \frac{d^3p }{ E_2^{3D}-\mathbf{p}^2}-\sumint \frac{d^2p_\perp }{ E-\mathbf{p}_{\perp}^2-\frac{n^2}{R^2}+ \imath \epsilon } \right]^{-1}\ , \label{rtau3}
\end{align}
and to get a finite value for $\tau_p(E)$ the ultraviolet cutoffs in both divergent terms have to be chosen consistently to keep the correct 3D limit. It is enough to regularize the momentum integral in the plane, namely $d^2p_\perp$, in both terms of Eq.~\eqref{rtau3} with an UV cutoff $\Lambda$ and then perform the limit $\Lambda\to\infty$. The integration in $d^2p_\perp$ in Eq.~\eqref{rtau3} was already done in Eq.~\eqref{IR1} and changing variables to $y\equiv R\;p$, the renormalized $T-$matrix is given by
\begin{multline}
R\tau_p^{-1}(E)=\pi \lim_{\Lambda\to\infty} \left[ \int^{\infty}_{-\infty} dy  \ln \left(\frac{-E_2^{3D}R^2+y^2 }{ -E_2^{3D}R^2+y^2+(\Lambda\,R)^2 }\right) \right.  \\
 \left.  - \sum_{n=-\infty}^{\infty}  \ln \left( \frac{-E\;R^2+n^2 - \imath \epsilon}{ - E\;R^2+n^2+(\Lambda\,R)^2+ \imath \epsilon }\right)
\right] \; . \label{rtau4}
\end{multline}
The sum and the integral of the terms in the right-hand-side of Eq.~\eqref{rtau4} are respectively given by Eq.~\eqref{sumab} and by
\begin{align}
\int_{-\infty}^{\infty} dy \ln \left( \frac{a^2 + y^2}{b^2+y^2} \right) = 2 \pi \left( \frac{a^2}{|a|}-\frac{b^2}{|b|} \right) \; .
\label{intab}
\end{align}
For negative energies, Eq.~\eqref{rtau4} becomes
\begin{align}
R\tau_p^{-1}(E)&=2\pi \lim_{\Lambda\to\infty} \left\{ \pi\, R \left(\sqrt{-E_2^{3D}}- \sqrt{-E_2^{3D}+\Lambda^2}\right) 
\right. \nonumber\\*  
& \left. \hskip 5cm - \ln\left[ \frac{\sinh \left(\pi R \sqrt{-E} \right)}{ \sinh \left(\pi R \sqrt{-E + \Lambda^2}\right)} \right]  \right\} \; , \nonumber\\
&=2\pi \left\{ \pi\, R \sqrt{-E_2^{3D}} - \ln\left[\sinh \left(\pi R \sqrt{-E} \right) \right] + L\left(\Lambda\right) \right\}\; , \label{rtau5a}
\end{align}
where 
\begin{align}
L\left(\Lambda\right)& \to \lim_{\Lambda\to\infty} \left\{ - \pi\, R  \sqrt{-E_2^{3D}+\Lambda^2} + 
\ln\left[ \sinh \left(\pi R \sqrt{-E + \Lambda^2} \right)  \right]  \right\}\; , \nonumber\\
&\to \ln \lim_{\Lambda\to\infty} \frac{1}{2} e^{ - \pi\, R  \sqrt{-E_2^{3D}+\Lambda^2}} \left( e^{\pi R \sqrt{-E + \Lambda^2}} - e^{-\pi R \sqrt{-E + \Lambda^2}} \right)  \; , \nonumber\\
&\to \ln \frac{1}{2} \lim_{\Lambda\to\infty} \left( 1- e^{-2 \pi R \Lambda \sqrt{1-E/\Lambda^2 }} \right)  \to - \ln 2 \; , 
\label{l1}
\end{align}
in the limit $\Lambda \to \infty$ gives a finite result.
The two-body $T-$matrix is finally written as 
\begin{align}
R^{-1} \tau_p(E)&=(2\pi)^{-1}\left\{ \pi\, R\sqrt{-E_2^{3D}}- \ln \left[ 2 \sinh \left( \pi R \sqrt{-E}\right) \right] \right\}^{-1} \ . \label{rtau5}
\end{align}
The formula in Eq.~\eqref{rtau5} for the 2B scattering amplitude can be generalized to allow negative scattering lengths by recognizing that $\sqrt{-E_2^{3D}}\to1/a$ in the zero-range limit. Therefore, the energy of the bound dimer in quasi\=/2D for positive or negative 3D scattering lengths is the solution of
\begin{align}
 \tau_p^{-1}(E)&=(2\pi)\left\{ \pi\, a^{-1}- R^{-1} \ln \left[2  \sinh \left(\pi R \sqrt{-E}\right) \right] \right\} = 0\; , 
 \label{rtau6}
\end{align}
and can be written as
\begin{equation}
\sqrt{-E}
= \frac{1}{\pi R}\sinh^{-1} \frac{e ^{\pi\, R/a}}{2} \ ,
\label{rtau7}
\end{equation}
For $R\to 0$ it goes to $\sqrt{-E}\sim (\pi R)^{-1}\,  \sinh^{-1} \frac{1}{2}  = 0.153174\,  R^{-1}$, which does not depend on the scattering length. Therefore, for any 3D two-body subsystem - bound or virtual - a strong deformation of the trap towards 2D always binds the dimer. 

\subsection{Trimer bound state equation} \label{3bcompac}

Considering the momentum $\mathbf{q}$ and the phase factor $d \mathbf{q}$  as given in Eq.~\eqref{qdq}, where $\mathbf{q} = \mathbf{q}_{\perp} + \mathbf{q}_z$ and $d \mathbf{q} = \frac{1}{R} d^2 q_{\perp}$, the three-body free Hamiltonian (see Eq.~\eqref{eq.c2-71}) becomes
\begin{align}
H_0^p \left( \mathbf{q} , \mathbf{k} \right) &= \left( \mathbf{q}_{\perp} + \mathbf{q}_z \right)^{2} + \left( \mathbf{k}_{\perp} + \mathbf{k}_z \right)^{2} + \left( \mathbf{q}_{\perp} + \mathbf{q}_z \right) \cdot \left( \mathbf{k}_{\perp} + \mathbf{k}_z \right) \; , \nonumber\\
&= q_{\perp}^2+k_{\perp}^2+\mathbf{q}_{\perp} \cdot \mathbf{k}_{\perp} + \frac{n^2}{R^2}+\frac{m^2}{R^2} + \frac{n\;m}{R^2} \; .
\label{h0p}
\end{align}

Considering the proper two-body $T-$matrix and $H_0^p$ respectively from Eqs.~\eqref{taup} and \eqref{h0p}, the three-body bound state equation for a compact dimension with PBC is found to be
\begin{align}
f\left( \mathbf{q}\right)   &= - \frac{2}{R} \; \tau_c \left( \frac{3}{4} q^2-E_3 \right) \sumint d \mathbf{k}  \left( \frac{ f\left(  \mathbf{k}\right)} {-E_{3}+H_0^p \left( \mathbf{q} , \mathbf{k} \right) } - \frac{f\left(  \mathbf{k} \right)}{\mu^2+H_0^p \left( \mathbf{q} , \mathbf{k} \right)  }\right)  \; , 
\end{align}
which after introducing the discrete momentum is written as
\begin{multline}
f\left( \mathbf{q}_{\perp},n \right)   = - \frac{2}{R} \; \tau_c \left[ \frac{3}{4} \left(q_{\perp}^2+\frac{n^2}{R^2} \right)-E_3 \right]   \\* \times \sum_{m=-\infty}^{\infty} \int d^2k_{\perp}  \left( \frac{ f\left( \mathbf{k}_{\perp} , m\right)} {-E_{3}+H_0^p \left( \mathbf{q} , \mathbf{k} \right) }  - \frac{f\left( \mathbf{k}_{\perp} , m\right)}{\mu^2+H_0^p \left( \mathbf{q} , \mathbf{k} \right)  }\right)  \; .
\label{bsecompac}
\end{multline}

Note that the subtraction is kept even after the discretization because the Thomas collapse is always present for any finite compact radius, no matter how small.

As the interaction between the particles corresponds to $s-$waves potentials and the focus is on states with zero angular momentum, the spectator functions do not depend on the angle, i.e., $f(\mathbf{q}) \equiv f(q)$ and the angular integration in Eq.~\eqref{bsecompac} is solved using Eq.~\eqref{angular}. Then, introducing  dimensionless variables, $\epsilon_3=E_3/\mu^2$, $\epsilon_2=E_2/\mu^2$, $r=R\;\mu$, $y_\perp=q_\perp/\sqrt{\mu}$ and $x_\perp=k_\perp/\sqrt{\mu}$, the subtracted equation for the three-body bound state with a compact dimension subjected to PBC is written as 
\begin{multline}
f\left( y_{\perp},n \right) =  \left\{ \pi \ln \left[\frac{\sinh \left(\pi r \sqrt{\frac{3}{4} \left(y_{\perp}^2+\frac{n^2}{r^2} \right)-\epsilon_3}  \right)}{\sinh \left(\pi r \sqrt{\epsilon_2}  \right)} \right] \right\}^{-1} \\* 
\times \sum_{m=-\infty}^{\infty} \int_0^{\infty}  dx_{\perp} \; x_{\perp} \; f\left( x_{\perp} , m\right) \left( \frac{1} {\sqrt{\left(-\epsilon_{3}+y_{\perp}^2+x_{\perp}^2+ \frac{n^2}{r^2}+\frac{m^2}{r^2} + \frac{n\;m}{r^2} \right)^2 - y_{\perp}^2\; x_{\perp}^2} } \right. \\* \left.
- \frac{1} {\sqrt{\left(1+y_{\perp}^2+x_{\perp}^2+ \frac{n^2}{r^2}+\frac{m^2}{r^2} + \frac{n\;m}{r^2} \right)^2 - x_{\perp}^2\; y_{\perp}^2} } \right)    \; .
\label{bsecompac1}
\end{multline}

It is worthwhile to remind that Eq.~\eqref{bsecompac1} for $R \to \infty$ returns precisely Eq.~\eqref{chi1} for the spectator function in the 3D case.

\section{2D - 1D with PBC }
Periodic boundary conditions (PBC) are assumed to be valid for the relative distance between the particles in the compact dimension, chosen to be $y$. The relative momentum is given by $\mathbf{p}=(p_x)$  in the linear dimension and by
\begin{align}
p_y=\frac{\pi n}{L}=\frac{n }{ R} \ , \;\;\; n=0, \pm 1, \pm 2, \pm 3, \hdots \; ,
\end{align}
in the transverse direction, with $L=2\pi R$ being the size of the compact dimension corresponding
to a radius $R$, that is again the parameter that dials between two and three-dimensions. The limit $R \to 0$ selects the 1D case and  in the opposite limit  $R\to \infty$, the 2D one is selected. 

Now, the momentum $\mathbf{q}$ and its corresponding phase factor $d \mathbf{q}$ in Eq.~\eqref{LSR2} are, with PBC, given by
\begin{align}
q^2= p_{x}^2+\frac{n^2}{R^2} \;\;\; \text{and} \;\;\; d\mathbf{q}=\frac{1}{R}dp_{x} \; .
\label{qdq1d}
\end{align}
The symbol $\sumint$ indicates that all momenta are being taking into account, i.e., there is an integration over the continuum momentum in the linear coordinate ($p_{x}$) and a sum over the discrete perpendicular momentum ($p_y=\frac{n}{R}$). In other words
\begin{align}
\sumint d\mathbf{q} \equiv \sum_{n=-\infty}^{\infty} \int \frac{1}{R}d p_{x} \; .
\label{sumintapp1d}
\end{align}

\subsection{Two-body scattering amplitude}

Replacing $q^2= p_{x}^2+\frac{n^2}{R^2}$ and $d\mathbf{q}=\frac{1}{R}d p_{x}$, Eq.~\eqref{LSR2} becomes
\begin{align}
R^{-1} \overline{\tau}_{p}(E) = \left[\lambda^{-1}R - (\mu^2+E)  \sumint d p_{x} \frac{1}{\left(E - p_{x}^2 - \frac{n^2}{R^2} + \imath \epsilon\right)\left(\mu^2 + p_{x}^2 + \frac{n^2}{R^2} \right)} \right]^{-1} \; .
\label{LSR1p1d}
\end{align}  
which leads to the renormalized scattering amplitude written as
\begin{align}
R^{-1} \overline{\tau}_{p}(E)=\left[R \int \frac{d^2p }{ E_2^{2D}-\mathbf{p}^2}-\sumint \frac{dp }{ E-\mathbf{p}^2-\frac{n^2}{R^2}+\imath \epsilon}
\right]^{-1} \; . \label{rtau31}
\end{align}
In order to get a finite value for $\tau_{p}(E)$ the ultraviolet cutoffs in both divergent terms have to be chosen consistently to keep the correct 2D limit. Using that
\begin{align}
\sum_{n=-\infty}^{\infty} \frac{1}{n^2+ a^2} = \frac{\pi \coth\left( \pi \; a\right) }{a} \; ,
\end{align}
the $T-$matrix in Eq.~\eqref{rtau31} is worked  for negative  energies, which is enough for bound trimer calculations and reads
\begin{align}
R^{-1} \overline{\tau}_{p}(E)&=(R\pi)^{-1} \lim_{\Lambda \to \infty}\left[  -\int^{\Lambda}_{-\Lambda} dp \frac{1}{\sqrt{-E_2^{2D }+p^2}} 
\right. \nonumber \\* 
& \left. \hskip 5cm + \int^{\Lambda}_{-\Lambda}  dp \frac{\coth\left(\pi \, R\sqrt{-E+p^2}\right)}{\sqrt{-E+p^2}}
\right]^{-1} \; , \nonumber \\ 
\overline{\tau}_{p}(E)&=\pi^{-1}\left[  \int^{\infty}_{-\infty} dp \left(  \frac{\coth\left(\pi \, R\sqrt{-E+p^2}\right)}{\sqrt{-E+p^2}}-\frac{1}{\sqrt{-E_2^{2D }+p^2}} \right)
\right]^{-1} \ . \label{rtau51}
\end{align}
The energy of the bound dimer in the quasi 1D situation is given by the pole of Eq.~\eqref{rtau51} at negative energies and it is found to be
\begin{equation}
 \int^{\infty}_{-\infty} dp \left(  \frac{\coth\left(\pi \, R\sqrt{-E_2+p^2}\right)}{\sqrt{-E_2+p^2}}-\frac{1}{\sqrt{-E_2^{2D }+p^2}} \right)=0\ ,
\label{rtau71}
\end{equation}
which for $R\to \infty$ gives correctly that  $E_2\to E_2^{2D} $ as in this limit  $\coth R \to 1$. For $R\to 0$ it should happen that  $-E_2\to \infty$ as given in Ref.~\cite{delfinoIJoQC2011}.

\subsection{Trimer bound state equation}

Considering the momentum $\mathbf{q}$ and the phase factor $d \mathbf{q}$  as given in Eq.~\eqref{qdq1d}, where $\mathbf{q} = \mathbf{q}_{x} + \mathbf{q}_y$ and $d \mathbf{q} = \frac{1}{R} d q_{x}$, the three-body free Hamiltonian (see Eq.~\eqref{eq.c2-71}) becomes
\begin{align}
\overline{H}_0^{p} \left( \mathbf{q} , \mathbf{k} \right) &= \left( \mathbf{q}_x + \mathbf{q}_y \right)^{2} + \left( \mathbf{k}_x + \mathbf{k}_y \right)^{2} + \left( \mathbf{q}_x + \mathbf{q}_y \right) \cdot \left( \mathbf{k}_x + \mathbf{k}_y \right) \; , \nonumber\\
&= q_x^2+k_x^2+\mathbf{q}_x \cdot \mathbf{k}_x + \frac{n^2}{R^2}+\frac{m^2}{R^2} + \frac{n\;m}{R^2} \; .
\label{h0p1d}
\end{align}

Considering the proper two-body $T-$matrix and $H_0^p$ respectively from Eqs.~\eqref{rtau51} and \eqref{h0p1d}, the three-body bound state equation for a compact dimension with PBC is found to be
\begin{align}
f(q_x,n) =R^{-1} \overline{\tau}_{p} \left[E_3-\frac{3}{4} \left(q_x^2+\frac{n^2 }{ R^2}\right)\right]
\sum_{m=-\infty}^\infty \int_{-\infty}^\infty dk_x 
 \frac{ f( k_x,m)}{ E_3-\overline{H}_0^{p}\left( \mathbf{q} , \mathbf{k} \right)} \; . \label{zre500}
\end{align}

In this case of quasi-1D there is no need of UV regularization of the integral equations as the Thomas collapse of the trimer in quasi-2D is absent in 2D and, therefore, in quasi-1D.

\section{Trimer at 3D - 2D crossover with PBC}

\begin{figure}[!htb]
\centering
\includegraphics[width=0.9\textwidth]{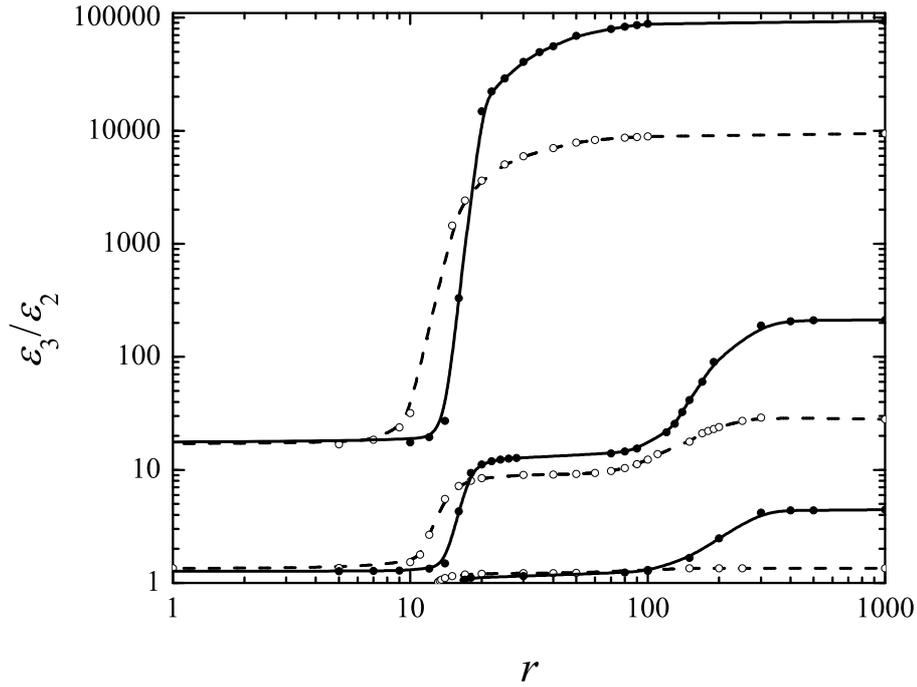}
\caption[Dimensional crossover of the three\=/body binding energy spectrum.]
{$\epsilon_3/\epsilon_2$ as a function of $r$, for $\epsilon_2=$ $10^{-7}$ 
(full circles) and $10^{-6}$ (empty circles). The solid and dashed lines are guides 
to the eye. As the 2D limit ($r\to 0$) is approached, higher 
excited states disappear and only the ground and first excited states remain.}
\label{compact}
\end{figure}

The results for the trimer binding energy shown in Fig.~\ref{compact} are obtained from the numerical solution of the 
integral bound state equation for the spectator function in quasi-2D (see Eq.~\eqref{bsecompac1}).
In order to explore the dimensional crossover transition, Fig.~\ref{compact} shows the ratios 
$\epsilon_3/\epsilon_2$ as a function of the compact dimension radius $r$, for the ground, first, and second
excited states. Note that the last state goes into the continuum before the 
2D limit is reached.

The computations were performed for two fixed two-body 
energies $\epsilon_{2}=10^{-6}$ (empty circles/dashed lines) 
and $10^{-7}$ (full circles/solid lines). In a pure 3D calculation
these parameters are on the $a>0$ side of the resonance and only three three-body bound states are found from the solution of the integral equation~\eqref{spec13d}, such that the well known Efimov ratio between two consecutive 
three-body states, $\sim 515$, is still far if the second and first 
excited states are considered. 
The points at which the energies are calculated are 
shown explicitly, while the curves are guides to the eye.
For the largest radius ($r=1000$) the energies are obtained from the pure 3D equation given in Eq.~\eqref{spec13d}.
The plot shows a very interesting dimensional crossover result, where only one sharp transition is present 
for the ground state while there are two for the first excited state. 
This behavior can be understood by
considering the size of the trimer given roughly by $\bar{r}\sim1/\sqrt{\epsilon_3}$.

For $\epsilon_2=10^{-7}$, the ground state plateau for $\epsilon_3/\epsilon_2=93330$ is placed at $\bar{r}=10.35$  and first 
excited state plateau for  $\epsilon_3/\epsilon_2=211.79$ at $\bar{r}=217.29$.
These $\bar{r}$ values give 
approximately the region of the jumps signaling that the 3D limit, represented by the plateau, 
is reached once the trimer size matches the size of the squeezed dimension, $r$. The same 
analysis can be made for $\epsilon_2=10^{-6}$ with $\bar{r}=10.27$ and $\bar{r}=188.98$, 
respectively, for the ground and first excited state. Varying $r$ from large to small 
values, the 3D$\rightarrow$2D transition occurs for $r\sim10$, where it is possible to notice the 
disappearance of the higher excited states in order to reproduce the well known 
2D results with two trimer bound state energies proportional to $\epsilon_2$ with the ratios $\epsilon_3/\epsilon_2=16.52$ 
and $\epsilon_3/\epsilon_2=1.27$ \cite{bruchPRA1979}.

From the experimental point of view it may be difficult to keep the dimer energy 
constant. However, the transition observed in Fig.~\ref{compact} 
will not disappear due to a variation of $\epsilon_2$ with $r$. The increase of the 
dimer energy will merely move the beginning of the jumps towards smaller $r$. The 
optimal way to probe these jumps is to start from a two-body energy in  
the unitary limit ($a\to \infty$) where the 2D plateaus are fixed. Larger dimer energies will cause  
the 3D plateau to move to lower $\epsilon_3/\epsilon_2$ ratio and push the beginning of the transition 
to smaller $r$, thus making the transition region broader.

In the case where the dimer energy is not tuned to be fixed and runs with $r$, it is possible to  estimate with Eq.~\eqref{rtau7} that for the small values of $r\sim 10-20$, when compared to the 3D scattering length of $a\sim$ 10$^3$, $\epsilon_2\sim 0.02346/r^2$ is  a reasonable approximation. In the case of $\epsilon_2^{3D}=10^{-6}$, one has for  $r=10$, $\epsilon_2\sim2.3\times10^{-4}$  and $\epsilon_2/\epsilon_2^{3D}=230$. For $r=20$, $\epsilon_2\sim5.1\times10^{-5}$ one get the ratio $\epsilon_2/\epsilon_2^{3D}=51$.
The results shown in Figure \ref{compact} for fixed $\epsilon_2=10^{-6}$, are expected to have some changes, as $r=10$, where  $\epsilon_3/\epsilon_2^{3D}$ is comparable to $\epsilon_2/\epsilon_2^{3D}=230$, while for $r=20$, $\epsilon_3/\epsilon_2^{3D}$  is quite large compared
to $\epsilon_2/\epsilon_2^{3D}=51$. Therefore, one expects that while some changes in the above picture will be expected for small radius, for $r > 20$ it will be quite unaffected. 

\chapter{Summary and outlook} \label{ch9}

Quantum few-body systems composed of two and three particles with attractive zero range pairwise interactions for general masses and interaction strengths were considered in two and three dimensions. In order to have the methods clearly stated, the Faddeev decomposition was used to write the set of homogeneous coupled integral equations for the bound state as well as the momentum-space wave function, which are the starting points for the analytical and numerical investigations of the universal properties in few-body systems.

The universal properties are those that any potential with similar constraints (observables) are able to describe in a model-independent form.  They occur when the system is large compared to the range of the potential and in this case the details of the basic two-body ingredients are unimportant. Then, zero-range interaction models introduced through Dirac$-\delta$ potential are useful, since all properties are determined at distances outside the potential \cite{fredericoPiPaNP2012}.


A considerable simplification in the 2D three-body problem comes by defining scaling functions and choosing appropriates energy and mass scales. This reduces the number of unknown parameters in the set of coupled homogeneous integral equations for the bound state and simplifies the presentation of the results \cite{bellottiPRA2012}.

It was shown in Chapter~\ref{ch3} that the number of three\=/body bound states in 2D varies from one and up depending on mass ratios and two-body subsystem energies, with symmetric mass system having the fewest and the most for two heavy particles and a light one. An upper limit for the number of bound states in any $abc$ system was found when all the subsystems interact with the same energy, namely $E_{ab}=E_{ac}=E_{bc}$ \cite{bellottiPRA2012}. Since this configuration seems still hard to be experimentally implemented, a feasible situation corresponds to three-body systems composed for a heavy\=/heavy non\=/interacting subsystem, i.e.  $m_c \ll m_a=m_b$ and $E_{ab}=0;E_{ac}=E_{bc}$, which has also a rich energy spectrum \cite{bellottiJoPB2013}.

For example, 2D mixtures of $^{87}$Rb-$^{87}$Rb-$^{6}$Li and $^{133}$Cs-$^{133}$Cs-$^{6}$Li are expected to have respectively 3 and 4 three-body universal bound states \cite{bellottiJoPB2013}.  It is very important to note 
that these numbers do  not depend on the exact two-body energy in the 
$^6$Li-$^{133}$Cs subsystem. These two-body energies in the 2D setup are
functions of the 3D low-energy scattering length  of 
the particular Feshbach resonance that is used in experiment to tune 
the interaction \cite{blochRoMP2008}. However, as long as there is such a resonance, 
the results should hold when the system is squeezed into a 
two-dimensional geometry. The possibility of tunning the 
binding energy of each pair and performing experiments mixing molecules 
and atoms should open new avenues for even richer two-dimensional 
three-body spectra. 

One of these promising configurations was recently reported as the 
experimental realization of $^{133}$Cs-$^{6}$Li systems \cite{reppPRA2013,tungPRA2013}, where it even looks as if three-body
bound states can be expected when the subsystem $^{133}$Cs-$^{133}$Cs is almost non-interacting. 
Therefore, the system $^{133}$Cs-$^{133}$Cs-$^{6}$Li  seems to be the 
most promising realistic combination to experimentally achieve a rich three-body 
energy spectrum in 2D. Some other mixtures with large mass imbalances under study at 
the moment  are Lithium-Ytterbium \cite{hansenPRA2013} 
and Helium-Rubidium \cite{knoopPRA2012}.  Cases of layered systems with long-range dipolar interactions are also extremely promising for 
finding bound states \cite{armstrongEPJD2012,volosnievPRA2012}, since that some of these are expected to have a universal low-energy character.

The three-body ground and first excited state energies in 2D have been successfully parameterized by universal functions that are so-called {\itshape super circles} (powers different from two) where the coordinates are the independent two-body energy ratios and the  radius parameter is the three-body energy. The latter is an approximately linear function of the three-body energy, independent of masses, while the powers of the coordinates are functions of both mass and three-body energy.  This result can be used to estimate three-body energies and the number of bound states, and as a measure of the deviation from the universal zero-range limit \cite{bellottiPRA2012}.


The interesting scenario in 2D where a particle is much lighter than the other two ($m_c \ll m_a \approx m_b$) presents a rich energy spectrum even for non\=/interacting heavy particles ($E_{ab}=0$) and was analyzed in the Born-Oppenheimer (BO) approximation, where the light particle coordinate is integrated out from the Schr{\"o}dinger equation leading to an effective adiabatic potential between the heavy particles in the heavy-heavy-light system, as presented in Chapter~\ref{ch4}.

The adiabatic potential, which was found as the solution of a transcendental equation, is mass-dependent and reveals an increasing number of bound states for the decreasing mass of one of the particles. An asymptotic expression for the adiabatic potential was derived and was shown that this analytic expression faithfully corresponds to the numerically calculated adiabatic potential, even in the non-asymptotic region where the biggest deviation is still less than $9\%$  \cite{bellottiJoPB2013}. This means that the asymptotic expression can be directly applied in 2D three-body system calculations, even in the non-asymptotic regions.

An estimate of the number of bound states as a function of the light-heavy mass ratio $m$ for a heavy-heavy-light system in 2D was done using the analytic expression in the JWKB approximation. Infinitely many bound states are expected as this ratio approaches zero since the number of states is proportional to $1/\sqrt{m}$  \cite{bellottiJoPB2013}. However, for finite masses a finite number of bound states is always expected. The explicitly mass-dependence of the asymptotic expression shows that the adiabatic potential becomes more attractive and less screened as one particle is much lighter than the other two ($m\to 0$). This behavior explains the increasing number of bound states when the  mass of the light particle is decresead. Besides, being $R$ the distance between the heavy particles, the bound states accumulate both at $R\to 0$ and $R\to \infty$.  While particles with finite mass always produce a finite number of bound states, an Efimov\=/like effect is expected only for $m=0$  \cite{bellottiJoPB2013}.

In order to experimentally observe the presence of these three-body 
bound states in 2D one should be able to use similar techniques to those
used for the study of Efimov states in 3D, i.e. loss measurements
\cite{kraemerNP2006} and photo-association \cite{lompeS2010}.
It may also be possible to use RF techniques as in  experiments that
studied two-body bound states and many-body pairing in two-dimensional 
Fermi gases \cite{frohlichPRL2011,sommerPRL2012}. 
Another possible experimental signature of 2D three-body systems is
through the momentum distribution and the two- and three-body 
contact parameters which appears as 
coefficients \cite{wernerPRA2012,bellottiPRA2012,bellottiNJoP2014}. 
These coefficients depends sensitively on the presence of 
bound two- and three-body states.


Important ingredients in the study of the momentum distribution are the Faddeev components, also called spectator functions, that compose the  wave\=/function in momentum space. 
A key result shown in Chapter~\ref{ch5} was the finding of an exact analytic expression for the spectator functions in 2D of any generic $abc$ system in the large momentum regime  \cite{bellottiPRA2013,bellottiNJoP2014}, where the normalization of three distinct spectator functions relates to each other through a constant, properly weighted by reduced masses. These analytical results are supported by accurate numerical computations, which confirmed both the asymptotic behavior and the relation between the asymptotic expressions for different spectator functions in a generic case of three distinguishable particles. 

The spectator functions and their asymptotic behavior define both the two- and the three-
body contact parameters, $C_2$ and $C_3$. The parameter $C_2$ arises from integration of the spectator
functions over all momenta, so that both small and large momenta contribute. 
The three two-body parameters in a system of three distinguishable particles in 2D are related by simple mass scaling, however they are in general not universal in the sense of being independent of the state when more than one excited state is present, or at least the energy scaling is more complicated than the corresponding one for identical bosons, where the three-body energy was used as the measuring unit \cite{bellottiNJoP2014}. This non-universal behavior was expected, since unlike for 3D systems, 2D three-body states does not present any geometric  scaling. Then, the surprisingly and interesting result also showed in Chapter~\ref{ch5} is that the two-body parameter scales with the three-body energy and becomes independent of the quantum state considered for three-body systems composed of one distinguishable and two identical and non-interacting particles, and also in the case of three-identical bosons \cite{bellottiPRA2013,bellottiNJoP2014}. 
In these cases the third particle apparently does not disturb the short-distance structure arising from
the other two particles and therefore the two-body contact parameter turn out to be universal.
This is similar to the 3D case with three identical bosons where $C_2$ is universal in the scaling
or Efimov limit where the binding energy is negligible.

The three\=/body contact parameter $C_3$ depends only on the
large\=/momentum asymptotic behavior of the spectator function and was fully determined due to the analytic expression found for the spectator function in 2D for the large momentum regime \cite{bellottiNJoP2014}. 
Unlike $C_2$, the three-body contact parameters do not turn
out to be universal in any of the investigated cases.  This parameter
is highly sensitive to the large-momentum asymptotic spectator
function.  Indeed, the proportionality coefficient determined in the
asymptotic region, appears explicitly in the expression for $C_3$.  It was found that $C_3$ drastically change from the cases of two interacting and 
non-interacting identical particles.  The values of $C_3$ for any
state in the non-interacting scenario are always less than $10\%$ of
the values obtained for interacting identical particles.  The
"absence" of a significant $C_3$, combined with universal $C_2$
parameters, is then a signature of a non-interacting subsystem within the
three-body system \cite{bellottiNJoP2014}.

It is of interest to know the spectator functions in 2D for all, both small
and large, momenta, since the two-body contact parameters are integral
quantities.  The knowledge about the asymptotic
spectator functions was used to guess an analytic structure describing
approximately the ground state.  This analytic form of the spectator function
was used to derive an expression for the two-body contact parameter between the
two identical non-interacting particles in the ground state. The mass ratio,
$m$, and energy, $E_3$, dependences then appear explicitly in addition
to a more hidden, but much weaker, dependence in a normalization
constant.  The derived expression deviates from the exact value of
$C_2$ about $10\%$ for $m \gg 1$, about $20\%$ when $m=1$ and
less than $5\%$ for $m \ll 1$.  Although the exact value is
not fully reproduced, this formula presents a powerful way to determine the
two-body contact parameter as function of $m$ within percents~\cite{bellottiNJoP2014}.

The single-particle momentum distribution was also calculated for  3D
systems consisting of two identical bosons and a third particle of a different
kind. Zero-range interaction was considered in the regime of a finite number of three-body bound states and also in the 
Efimov limit. 
Again, the asymptotic momentum distribution was analytically
calculated as a function of the mass ratio and it was shown in Chapter~\ref{ch6} 
that the corresponding functional form is sensitive to this mass ratio \cite{yamashitaPRA2013}. In the case of equal masses
the results of Ref.~\cite{castinPRA2011} were reproduced. The leading term has 
a $q^{-4}$ tail while the sub-leading contribution is $q^{-5}$ times a log-periodic 
function that is a characteristic of the Efimov limit. In particular,
it was shown that for general mass ratios there is a non-oscillatory $q^{-5}$ contribution
which happens to  vanish (and leave the oscillatory contribution behind) when 
the mass ratio is 0.2, 1, or 1.57 \cite{yamashitaPRA2013}.

Exemplifying the study above, the coefficient of the 
$q^{-4}$ tail, which is the two-body contact parameter, was numerically determined for the systems $^{133}$Cs-$^{133}$Cs-$^{6}$Li and 
$^{87}$Rb-$^{87}$Rb-$^{6}$Li. 
For these cases, the momentum distributions of excited
Efimov trimers for both the heavy and light components were also calculated. The numerical results demonstrate
that the momentum distributions of ground, first, and second excited Efimov trimers 
approach a universal form at large but also at small momentum, indicating that 
one does not need to go to highly excited (and numerically challenging) three-body states
in order to study the universal behavior of Efimov states in momentum space \cite{yamashitaPRA2013}.

As a brief parenthesis, it is worth to highlight that the steps employed in the challenging analytical derivation of the sub-leading terms $n_3$ and $n_4$ would be used as a complete and exciting example in a course of complex analysis (see Chapter~\ref{ch6} and Appendix~\ref{residues}).


In 3D systems, the two\=/body contact has been observed in experiments using time\=/
of\=/flight and the mapping to momentum space \cite{stewartPRL2010}, Bragg spectroscopy \cite{kuhnlePRL2010}
or momentum\=/resolved photo\=/emission spectroscopy 
\cite{frohlichPRL2011}. Measuring the sub-leading term and thus accessing $C_3$ requires more
precision, which has so far only produced the upper limits for the particular case of $^{87}$Rb \cite{wildPRL2012}.
In 2D systems the functional form of the sub-leading term is different from the 3D case, so it is
difficult to compare with the 3D case. However, given that the precision improves continuously
it should be possible to also probe the 2D case when tightly squeezing a 3D sample. As it was 
shown here, the mass ratio can change the values of the contact parameters significantly. It is thus
expected that mixtures of different atoms is the most promising direction to make a measurement
of a 2D contact parameter.

A first step in the study of higher-order correlations
and dimensional 3D - 2D crossover was taken by demonstrating how trimer observables in
strongly-interacting quantum gases can be used to probe dimensionality \cite{bellottiPRA2013}.
Specifically, the breakdown of scale-invariance due to the Efimov effect is directly seen in 
the functional form of the tail of the momentum distribution.
A clear direction of study is a full inclusion of the transverse
direction and the discrete spectrum that it brings, since it was shown that 
a crossover with fundamental influence on the momentum tail will 
happen. A first try to mapping the dimensional crossover out in a system that would be squeezed by optical
lattice(s) is shown in Chapter~\ref{ch7} (see also Ref.~\cite{yamashitaAe2014}). For that aim, periodic boundary conditions were used to change from 3D to quasi\=/2D the trimer physics.

Squeezing the transverse direction with periodic boundary conditions, a sharp transition was found in the energy spectrum  of the trimer as the compact dimension changes from a 3D to a 2D situation~\cite{yamashitaAe2014}. This is an ongoing project and more studies are still necessary in order to relate the parameter which dials between the different dimensions to real experiments. Only the transition in the trimer energy spectrum was still studied and another interesting topic is to follow the dimensional crossover transition of the wave function, since it was shown that the spectator function drastically changes with the dimensionality. An intriguing point is to identify what is the form of the spectator function in the sharp crossover region, where the system is clearly neither 2D or 3D.

Another ongoing project is the study of the two\=dimensional three\=/body systems containing one fermion that is identical to the ones of a Fermi sea background. So far, our studies in 2D have concerned mass\=/imbalanced three\=/body systems of different nature in the vacuum. In the Fermi sea background, the dynamics of the shallow molecules having in its components fermions have to also account for the Pauli\=/blocking effect. Furthermore, the boson self\=/energy has to be treated as it has been done for two\=/body systems in, for example, Refs.~\cite{meeraPRA2011,saschaPRA2011,schmidtPRA2012}. It is open the question on the effects of Pauli\=/Blocking and self\=/energies corrections for the three\=/body systems. The numerical solution of this problem is much more time\=/demanding and more technically challenging, but worth while to solve in view of the experimental possibility of squeezing systems with bosons immersed in cold atomic Fermi gases.

Further projects are the study of range\=/corrections in the universal properties presented in this thesis and the investigation of four\=/body systems in 2D.  
For  real applications it is necessary to care about details of
the short-range interaction, as the universal limit in which the
range of the interaction is zero is an idealization. Therefore, it
is necessary to study how the introduction of real potentials with a
given range make the few-particle properties deviate from the
results obtained in the universal regime. The need for details of
the interaction  must be significant when  the system is probed
at wavelengths close to the potential range, while for low energies
the range effect can be possibly studied by a systematic expansion,
as has been done, e.g. for three- \cite{platterPRA2009} and four-boson  \cite{hadizadehPRA2013} systems in 3D.
Therefore, the corrections in the binding and
structure of three- and four-body systems due to a finite
interaction range in 2D will be investigated for several different possibilities of
constituents, paying attention to the  implications of the four\=/body scale~\cite{hadizadehPRL2011}.

The three-particle system for large mass asymmetries can have a
large number of bound states in two-dimensions, as shown in Chapters~\ref{ch3} and \ref{ch4} (see Ref.~\cite{bellottiJoPB2011,bellottiPRA2012,bellottiJoPB2013}).
It will be interesting to find how the interwoven three- and
four-body spectra come with different possibilities of particles
masses. In such case, it is expected that  some of these four-body states
will be resonances in the atom plus triatomic molecule continuum, as
has been seen in 3D calculations in traps \cite{thogersenEPL2008} and
without traps \cite{NPP2009,deltuvaPRA2010}. 
In the case of the
formulation in momentum space, to obtain the position of the
four-body resonances a rotation to the complex momentum plane of the
integral equation will be required, as has been done in the case of
the calculation of three-boson resonances in 3D \cite{bringasPRA2004}.

\appendix

\chapter{Revision of the scattering theory} \label{revisionst}

\section{Brief presentation of the quantum scattering} \label{qst}
\begin{figure}[!htb]
\centering
\includegraphics[width=0.8\textwidth]{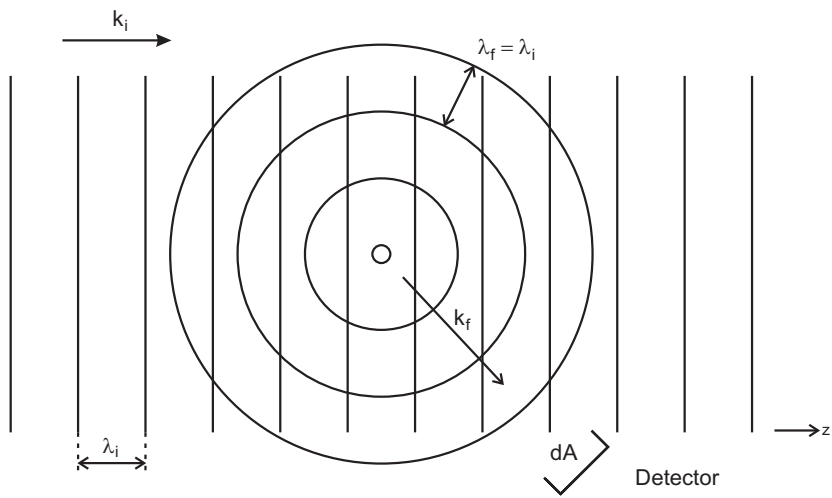}
\caption[Schematic figure showing the quantum scattering process.]
{Schematic figure showing the quantum scattering process. $\mathbf{k}$ and $\lambda$ are respectively the relative momentum and corresponding wavelength. The subscripts $i$ and $f$ means initial and final.} 
\label{scattering_waves}
\end{figure}

The two-body problem is much simpler than the three-body one, since there are only two possibilities for the two-body systems: a scattering state or a bound state with structureless constituents. In the quantum regime, the particles can be described as waves (see for example Ref.~\cite{pizaMQ2002,landau1977}). In this scenario, two colliding particles are described by an incoming  wave and a scattered one as shown in Fig.~\ref{scattering_waves}.  

The Hamiltonian for two particles that interact through a generic potential $V$ is written as
\begin{equation}
H'=\frac{\mathbf{k}_1^2}{2m_1}+\frac{\mathbf{k}_2^2}{2m_2}+V \; ,
\label{eq.c2-01}
\end{equation}
where $\mathbf{k}_i$ and $m_i$ are  respectively the momentum and mass of the particle $i$ in the laboratory frame.
If the potential $V$ is translational invariant, the Hamiltonian is given by
\begin{equation}
H=\frac{\mathbf{k}^2}{2(m_1+m_2)}+\frac{\mathbf{p}^2}{2m_{12}}+V \; ,
\label{eq.c2-02}
\end{equation}
where $\mathbf{k}=\mathbf{k}_1+\mathbf{k}_2$ is the total momentum and $\mathbf{p}=\frac{m_2\mathbf{k}_1-m_1\mathbf{k}_2}{m_1+m_2}$ the relative one. The reduced mass of the system is $m_{12}=\frac{m_1m_2}{m_1+m_2}$.
As there are no external forces, the CM motion is free and just the relative motion has to be considered. The Hamiltonian for the relative motion is
\begin{equation}
H=\frac{\mathbf{p}^2}{2m_{12}}+V \equiv H_0+V \; .
\label{eq.c2-06}
\end{equation}

The solution of the Schr\"odinger equation for the scattered outgoing wave, \\ ${\left(H_0+V \right)\psi^{+}=E \psi^{+}}$ is formulated as 
\begin{equation}
\psi^{+}= \psi_0^{+} + g_0(E) V \psi^{+} \; ,
\label{scheq}
\end{equation}
where $\psi_0^{+}$ is the solution of the homogeneous equation  and the free resolvent $g_0(E)$ is given by
\begin{equation}
g_0(E) \equiv \frac{1}{E-H_0+\imath \epsilon} \; . 
\label{eq.c2-18}
\end{equation}
Analogously, the resolvent (or Green's function) is defined as
\begin{equation}
g(E)\equiv \frac{1}{E-H+\imath \epsilon} \; .
\label{eq.c2-17}
\end{equation}

Relations between the free and full propagators, respectively in Eqs.~\eqref{eq.c2-18} and \eqref{eq.c2-17}, are found by noticing that 
\begin{equation}
V=H-H_0=g_0^{-1}(E)-g^{-1}(E) \; .
\label{eq.c2-20}
\end{equation}
Multiplying Eq.~\eqref{eq.c2-20} for $g_0$ from left and $g$ from right gives
\begin{align}
g_0^{-1}-g^{-1} = V \; , \; 
1-g_0g^{-1} = g_0V \; , \;
g-g_0 = g_0Vg \; , \;
g =g_0 + g_0Vg \; .  \label{eq.c2-21a}
\end{align}
On the other hand, multiplying Eq.~\eqref{eq.c2-20} for $g$ from left and $g_0$ from right results in
\begin{align}
g_0^{-1}-g^{-1} = V \; , \;
gg_0^{-1}-1 = gV \; , \;
g-g_0 = gVg_0 \; , \;
g=g_0 + gVg_0 \; . \label{eq.c2-21b}
\end{align}
Then, the two-body transition matrix $t(E)$ is found by inserting Eq.~\eqref{eq.c2-21a} into Eq.~\eqref{eq.c2-21b}. The result is
\begin{align}
g&=g_0+g_0V \left(g_0+gVg_0 \right) 
=g_0+g_0Vg_0+g_0VgVg_0 
=g_0+g_0\left[V+VgV\right]g_0 
  \nonumber\\
&\equiv g_0+g_0tg_0 \; , \label{eq.c2-31}
\end{align} 
where the transition matrix ($T-$matrix), $t(E)$, is defined as 
\begin{equation}
t=V+VgV \; .
\label{eq.c2-32}
\end{equation}
Notice that the $T-$matrix, as well as the free and full propagators, depends explicitly on the energy, i.e., $t \equiv t(E)$, $g \equiv g(E)$ and $g_0 \equiv g_0(E)$, however, this dependence is not shown in Eqs.~\eqref{eq.c2-21a} to \eqref{eq.c2-32} in order to let the visualization of such equations clearer.

The transition matrix is a key ingredient in the study of quantum systems, since it relates directly to the main observable in the scattering problem: the connection between theory and measurable data is made through the scattering phase\=/shifts and cross\=/section. While the study of the scattering phase-shift and cross-section is broadly made for three-dimensional systems in several text books, the analogous two-dimensional (2D) problem is beautifully described in Ref.~\cite{adhikariPRA1993}.

\section{Scattering equation for the two-body T-matrix} \label{scatteringTmatrix}
The $T-$matrix plays a central role in the scattering problem and, it is the base of the mathematical framework used to describe quantum few-body problems. An integral equation for this operator is found by combining the two expressions for the full propagator in Eqs.~\eqref{eq.c2-21a} and \eqref{eq.c2-31}. Inserting the latter in the former one gives:
\begin{align}
g=g_0 + g_0 V \left( g_0+g_0 t g_0  \right) 
 =g_0 + g_0 V g_0 + g_0 V g_0 t g_0  
 =g_0 + g_0 \left[V+V g_0 t \right] g_0 \; . 
\label{eq.c2-37}
\end{align}
Comparing the two forms of the resolvent as written in Eqs.~\eqref{eq.c2-31} and \eqref{eq.c2-37}, the scattering integral equation comes as
\begin{equation}
t= V + V g_0 t \; .
\label{eq.c2-38a}
\end{equation} 
Alternatively, it is possible to insert the resolvent in Eq.~\eqref{eq.c2-31} into Eq.~\eqref{eq.c2-21b}, which results in
\begin{equation}
t = V + t g_0 V \; .
\label{eq.c2-38b}
\end{equation}

The integral equation for the $T-$matrix, Eq.~\eqref{eq.c2-38a}, can in principle be solved for any potential $V$, however, more complex the potential more difficult to solve the equation. Hopefully, there is a class of potentials which allows the algebraic manipulation of Eq.~\eqref{eq.c2-38a}. These potentials are called separable and have the operator form
\begin{equation}
V=\lambda\left|\chi\right\rangle\left\langle\chi\right| \ ,
\label{eq.c2-42}
\end{equation}
where $\lambda$ is the potential strength. 

Inserting the potential given in Eq.~\eqref{eq.c2-42}, in the two-body scattering integral equation from Eq.~\eqref{eq.c2-38a} leads to
\begin{equation}
t(E)=\lambda\left|\chi\right\rangle\left\langle\chi\right|+\lambda\left|\chi\right\rangle\left\langle\chi\right|g_0(E)t(E) \; .
\label{eq.c2-43}
\end{equation}
In order to find the term $\left\langle\chi\right|g_0(E)t(E)$, Eq.~\eqref{eq.c2-43} is multiplied by $\left\langle\chi\right|g_0(E)$ from the left, which gives
\begin{align}
&\left\langle\chi\right|g_0(E)t(E)=\lambda\left\langle\chi\right|g_0(E)\left|\chi\right\rangle\left\langle\chi\right|+\lambda\left\langle\chi\right|g_0(E)\left|\chi\right\rangle\left\langle\chi\right|g_0(E)t(E) \ ,& \nonumber\\
&\left\langle\chi\right|g_0(E)t(E)=\frac{\lambda\left\langle\chi\right|g_0(E)\left|\chi\right\rangle\left\langle\chi\right|}{1-\lambda\left\langle\chi\right|g_0(E)\left|\chi\right\rangle} \ .& \label{eq.c2-44}
\end{align}
Inserting Eq.~\eqref{eq.c2-44} back in Eq.~\eqref{eq.c2-43}, the two-body $T-$matrix becomes
\begin{align}
t(E)
=&\left|\chi\right\rangle\left(\frac{1}{\lambda^{-1}-\left\langle\chi\right|g_0(E)\left|\chi\right\rangle}\right)\left\langle\chi\right| \ , \label{eq.c2-45}
\end{align}
which in a compact form becomes
\begin{equation}
t(E)=\left|\chi\right\rangle\tau(E)\left\langle\chi\right| \ ,
\label{eq.c2-46}
\end{equation}
where the matrix element $\tau(E)$ is given by
\begin{equation}
\tau(E)=\left(\lambda^{-1}-\left\langle\chi\right|g_0(E)\left|\chi\right\rangle\right)^{-1} \ .
\label{eq.c2-47}
\end{equation}

Introducing the identity $\mathbf{\hat{1}}=\int{d^Dp\left|\mathbf{p}\right\rangle\left\langle\mathbf{p}\right|}$ in Eq.~\eqref{eq.c2-47}, the integral form of the $T-$matrix is
\begin{align}
\tau(E)=&\left(\lambda^{-1}-\iint{d^Dp'd^Dp\left\langle\chi\right|\left.\mathbf{p'}\right\rangle\left\langle\mathbf{p'}\right|g_0(E)\left|\mathbf{p}\right\rangle\left\langle\mathbf{p}\right.\left|\chi\right\rangle}\right)^{-1} \; ,  \nonumber\\
=&\left(\lambda^{-1}-\int{d^Dp\frac{g(p)^2}{E-\frac{p^2}{2m_{red}}+\imath \epsilon}}\right)^{-1} \ , \label{eq.c2-48}
\end{align}
where $m_{red}$ is the two-body reduced mass and $g(p) \equiv \left\langle\mathbf{p}\right.\left|\chi\right\rangle$ is the form factor of the potential $V$.

\chapter{Jacobi relative momenta} \label{jacobi}

\section{Classical three-body problem}
\begin{figure}[!htb]
\centering
\includegraphics[width=0.8\textwidth]{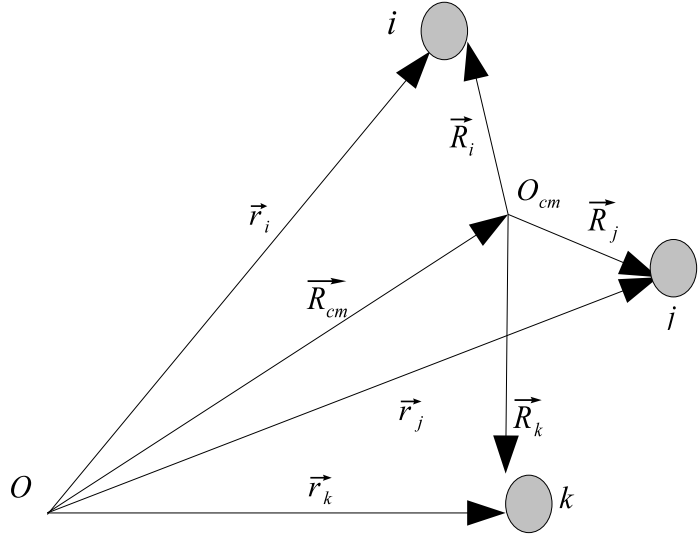}
\caption{Three-body coordinates in laboratory  frame.}
\label{coord_jacobi_three_body}
\end{figure}

The physical attributes of the three particles are labeled as $\alpha=i,j,k$.  Their masses are $m_\alpha$, and their positions and velocities in the laboratory frame are respectively given by $\mathbf{r_\alpha}$ and $\mathbf{v}_\alpha=\frac{d\mathbf{r_\alpha}}{dt}$. The momentum of each particle is $\mathbf{K_\alpha}=m_\alpha \mathbf{v}_\alpha$.
The total momentum $\mathbf{K}$ and the free Hamiltonian $H_0$ of the three-body system are written as
\begin{align}
\mathbf{K}&=m_i \mathbf{v}_i+m_j \mathbf{v}_j+m_k \mathbf{v}_k=\mathbf{K}_i+\mathbf{K}_j+\mathbf{K}_k \ ,& \label{eq.a3-24a}\\
H_0&=\frac{\mathbf{K}_i^2}{2m_i}+\frac{\mathbf{K}_j^2}{2m_j}+\frac{\mathbf{K}_k^2}{2m_k} \ .& \label{eq.a3-24b}
\end{align}
The center-of-mass (CM) position and velocity are found to be
\begin{align}
\mathbf{R}_{cm}&=\frac{m_{i}\mathbf{r}_{i}+m_{j}\mathbf{r}_{j}+m_{k}\mathbf{r}_{k}}{M} \ ,&  \label{eq.a3-01a} \\
\mathbf{V}_{cm}&=\frac{d\mathbf{R_{cm}}}{dt}=\frac{m_{i}\mathbf{v}_{i}+m_{j}\mathbf{v}_{j}+m_{k}\mathbf{v}_{k}}{M} \ ,&  \label{eq.a3-01b}
\end{align}%
where the total mass is  $M=m_{i}+m_{j}+m_{k}$. The relatives coordinate and velocity between each particle $\alpha$ and the CM of the system are shown in Fig.~\ref{coord_jacobi_three_body} and read
\begin{align}
\mathbf{R}_{\alpha }&=\mathbf{r}_{\alpha }-\mathbf{R}_{cm} \ ,&  \label{eq.a3-02}\\
\mathbf{V}_{\alpha }&=\frac{d\mathbf{R}_{\alpha }}{dt}=\mathbf{v}_{\alpha }-\mathbf{V}_{cm} \ .&  \label{eq.a3-03}
\end{align}

In the CM frame, the momentum of particle $\alpha$ is written as 
\begin{equation}
\mathbf{k}_{\alpha}=m_{\alpha }\mathbf{V}_{\alpha } \ ,  \label{eq.a3-04}
\end{equation}
where $V_\alpha$ is given in Eq.~\eqref{eq.a3-03}. Besides, at the CM frame the three momenta must fulfill 
\begin{equation}
\mathbf{k}_{i}+\mathbf{k}_{j}+\mathbf{k}_{k}=0.  \label{eq.a3-05}
\end{equation}

Inserting the CM velocity from Eq.~\eqref{eq.a3-01b} and the relative velocity of particle $\alpha$ from Eq.~\eqref{eq.a3-03} into Eq.~\eqref{eq.a3-04}, the momentum of particle $\alpha$ in the CM frame becomes
\begin{align}
\mathbf{k}_\alpha&=m_\alpha\left( \mathbf{v}_\alpha-\frac{m_\alpha\mathbf{v}_\alpha+m_\beta\mathbf{v}_\beta+m_{\gamma}\mathbf{v}_{\gamma}}{M}\right) \; ,  & \nonumber\\ 
&=\frac{m_\alpha}{m_\alpha+m_\beta+m_\gamma}\Bigl[ (m_\alpha+m_\beta+m_\gamma)\mathbf{v}_\alpha-m_\alpha\mathbf{v}_\alpha+m_\beta\mathbf{v}_\beta+m_{\gamma}\mathbf{v}_{\gamma}\Bigr] \; ,  & \nonumber\\ 
&=\frac{m_\alpha\left( m_\beta+m_\gamma\right) }{m_\alpha+m_\beta+m_\gamma}\left( \mathbf{v}_\alpha-\frac{m_\beta\mathbf{v}_\beta+m_\gamma\mathbf{v}_\gamma}{m_\beta+m_\gamma}\right)\ , &  \label{eq.a3-06b}
\end{align}

Remembering that ($\alpha,\beta,\gamma$) represent cyclic permutations of ($i,j,k$), the momentum of each particle in the CM frame is
\begin{align}
\mathbf{k}_{i}&=\frac{m_{i}\left( m_{j}+m_{k}\right) }{m_{i}+m_{j}+m_{k}}\left(\mathbf{v}_{i}-\frac{m_{j}\mathbf{v}_{j}+m_{k}\mathbf{v}_{k}}{m_{j}+m_{k}}\right) \ ,& \label{eq.a3-07} \\
\mathbf{k}_{j} &=\frac{m_{j}\left( m_{i}+m_{k}\right) }{m_{i}+m_{j}+m_{k}}%
\left( \mathbf{v}_{j}-\frac{m_{i}\mathbf{v}_{i}+m_{k}\mathbf{v}_{k}}{%
m_{i}+m_{k}}\right) \ ,&  \label{eq.a3-08} \\
\mathbf{k}_{k} &=\frac{m_{k}\left( m_{i}+m_{j}\right) }{m_{i}+m_{j}+m_{k}}%
\left( \mathbf{v}_{k}-\frac{m_{i}\mathbf{v}_{i}+m_{j}\mathbf{v}_{j}}{%
m_{i}+m_{j}}\right) \ .&  \label{eq.a3-09}
\end{align}
Notice that the momenta given in Eqs.~\eqref{eq.a3-07} to \eqref{eq.a3-09} fulfill the relation in Eq.~\eqref{eq.a3-05}. 

\section{Jacobi relative momenta}

\begin{figure}[!htb]
\centering
\includegraphics[width=0.9\textwidth]{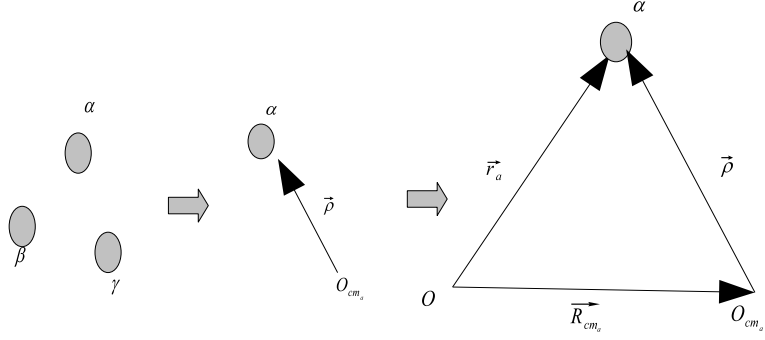}
\caption[Relation of the new coordinates with the coordinates in frame of the laboratory.]
{Relation of the new coordinates with the coordinates in frame of the laboratory for a system composed of an $\alpha$ particle and $cm_{\alpha }$. }\label{coord_jacobi_three_body_cms}
\end{figure}

As it was stated in Chapter~\ref{ch2}, an advantage in the use of the Jacobi momenta in the study of the three-body problem is that the motion of the CM can be separated out.  The Jacobi momenta are illustrated in Fig.~\ref{coord_jacobi}, where  $\mathbf{q}_{\alpha }$ is identified as the momentum of the particle $\alpha$ with respect to $cm_{\alpha }$, the CM of particles ($\beta,\gamma$).  Also, $\mathbf{p}_{\alpha }$ is the relative momentum of the two-body system ($\beta,\gamma$). In order to identify all these relative momenta, the three-body motion is split out in the motion of one particle $\alpha$ plus the motion of the CM of the remaining pair, $cm_{\alpha }$. Considering $cm_{\alpha }$ as a particle, the problem is roughly reduced to a two-body problem, as shown in Fig. \ref{coord_jacobi_three_body_cms}. 

In Fig.~\ref{coord_jacobi_three_body_cms}, the relatives coordinate $\mathbf{\rho }$ and velocity $\sigma$ are
\begin{align}
\mathbf{\rho_\alpha} &=\mathbf{r}_{\alpha }-\mathbf{R}_{cm_{\alpha }} \; , &  \label{eq.a3-10} \\
\mathbf{\sigma_\alpha} &=\mathbf{v}_{\alpha }-\mathbf{V}_{cm_{\alpha }} \; . &  \label{eq.a3-11}
\end{align}
The mass of the particle $cm_{\alpha }$ is given by
\begin{equation}
M_{cm_{\alpha }} =m_{\beta }+m_{\gamma } \; .  
\label{eq.a3-16} 
\end{equation}
In the same way, its position and velocity in the laboratory frame are
\begin{align}
\mathbf{R}_{cm_{\alpha }} &=\frac{m_{\beta }\mathbf{r}_{\beta }+m_{\gamma }\mathbf{r}_{\gamma }}{m_{\beta }+m_{\gamma }} \; ,&  \label{eq.a3-12a} \\
\mathbf{V}_{cm_{\alpha }} &=\frac{m_{\beta }\mathbf{v}_{\beta }+m_{\gamma }\mathbf{v}_{\gamma }}{m_{\beta }+m_{\gamma }} \; . &  \label{eq.a3-12b}
\end{align}

From the classical two-body problem, it must not be hard to identify the relative momentum between particles $\alpha$ and $cm_\alpha$ as
\begin{equation}
\mathbf{q}_{\alpha }=\mu_\alpha \mathbf{\sigma _\alpha} \ ,  \label{eq.a3-14}
\end{equation}%
where the reduced mass $\mu_\alpha$ reads
\begin{equation}
\mu_\alpha =\frac{m_{\alpha }M_{cm_{\alpha }}}{m_{\alpha }+M_{cm_{\alpha }}}=\frac{%
m_{\alpha }\left( m_{\beta }+m_{\gamma }\right) }{m_{\alpha }+m_{\beta
}+m_{\gamma }}.  \label{eq.a3-15}
\end{equation}
Inserting the relative velocity from Eq.~\eqref{eq.a3-11} and the reduced mass \eqref{eq.a3-15} in Eq.~\eqref{eq.a3-14}, the relative momentum between particle $\alpha$ and the CM of the ($\beta,\gamma$) subsystem is found to be
\begin{equation}
\mathbf{q}_{\alpha }=\frac{m_{\alpha }\left( m_{\beta }+m_{\gamma }\right) }{%
m_{\alpha }+m_{\beta }+m_{\gamma }}\left( \mathbf{v}_{\alpha }-\frac{%
m_{\beta }\mathbf{v}_{\beta }+m_{\gamma }\mathbf{v}_{\gamma }}{m_{\beta
}+m_{\gamma }}\right) .  \label{eq.a3-17}
\end{equation}
Notice the Jacobi momentum in Eq.~\eqref{eq.a3-17} is identical to the momentum in Eq.~\eqref{eq.a3-06b}, meaning that 
\begin{align}
&\mathbf{q}_{i}=\mathbf{k}_{i}, \ \mathbf{q}_{j}=\mathbf{k}_{j}, \ \mathbf{q}_{k}=\mathbf{k}_{k}&  . 
\label{eq.a3-18}
\end{align}
The relative momentum between particles $\beta$ and $\gamma$, $\mathbf{p}_{\alpha }$, is 
\begin{equation}
\mathbf{p}_{\alpha }=\frac{m_{\beta }m_{\gamma }}{m_{\beta }+m_{\gamma }}\mathbf{v}_{\beta \gamma }=\frac{m_{\beta }m_{\gamma }}{m_{\beta }+m_{\gamma }}\left(\mathbf{v}_{\beta }-\mathbf{v}_{\gamma }\right) =\frac{m_{\gamma }\mathbf{k}_{\beta }-m_{\beta }\mathbf{k}_{\gamma}}{m_{\beta }+m_{\gamma }} \ ,  
\label{eq.a3-19}
\end{equation}
and using Eq.~\eqref{eq.a3-18} it becomes
\begin{align}
\mathbf{p}_{i}&=\frac{m_{k}\mathbf{q}_{j}-m_{j}\mathbf{q}_{k}}{m_{j}+m_{k}} \; , & \label{eq.a3-20a} \\
\mathbf{p}_{j}&=\frac{m_{i}\mathbf{q}_{k}-m_{k}\mathbf{q}_{i}}{m_{i}+m_{k}} \; , & \label{eq.a3-20b} \\
\mathbf{p}_{k}&=\frac{m_{j}\mathbf{q}_{i}-m_{i}\mathbf{q}_{j}}{m_{j}+m_{i}} \; . & \label{eq.a3-20c}
\end{align}

Writing the momentum of each particle as function of the Jacobi momenta ($\mathbf{q}_{\alpha },\mathbf{p}_{\alpha }$), the free Hamiltonian in Eq.~\eqref{eq.a3-24b} becomes 
\begin{equation}
H_{0}=\frac{\mathbf{q}_{\alpha }^{2}}{2m_{\beta \gamma,\alpha }}
+\frac{\mathbf{p}_{\alpha }^{2}}{2m_{\beta \gamma }} 
+\frac{Q^2}{m_{\alpha}+m_{\beta}+m_{\gamma}},
\label{eq.a3-21}
\end{equation}%
where $\mathbf{Q}=\sum_{\alpha} \mathbf{k}_\alpha=\sum_{\alpha} \mathbf{q}_\alpha$ is the total momentum and the reduced masses are
\begin{align}
m_{\beta \gamma ,\alpha } &=\frac{m_{\alpha }\left( m_{\beta }+m_{\gamma}\right) }{m_{\alpha }+m_{\beta }+m_{\gamma }} \ ,&  \label{eq.a3-22} \\
m_{\beta \gamma } &=\frac{m_{\beta }m_{\gamma }}{m_{\beta }+m_{\gamma }} \ .& \label{eq.a3-23}
\end{align}

The shifted arguments of the spectator functions in the wave\=/function from Eq.~\eqref{eq.c2-145} are found by manipulating Eqs.~\eqref{eq.a3-05}, \eqref{eq.a3-18}, \eqref{eq.a3-20a}, \eqref{eq.a3-20b} and \eqref{eq.a3-20c}. The momentum $\mathbf{q}_j(\mathbf{q}_i,\mathbf{p}_i)$ reads
\begin{align}
\mathbf{q}_j(\mathbf{q}_i,\mathbf{p}_i) &= \frac{m_{j}+m_{k}}{m_{k}} \mathbf{p}_i + \frac{m_{j}}{m_{k}} \mathbf{q}_k = \frac{m_{j}+m_{k}}{m_{k}} \mathbf{p}_i + \frac{m_{j}}{m_{k}} \left(-\mathbf{q}_i - \mathbf{q}_j \right) \; , \nonumber\\
\left( 1 + \frac{m_{j}}{m_{k}} \right) \mathbf{q}_j(\mathbf{q}_i,\mathbf{p}_i) &= \frac{m_{j}+m_{k}}{m_{k}} \mathbf{p}_i - \frac{m_{j}}{m_{k}} \mathbf{q}_i  \; , \nonumber\\
\mathbf{q}_j(\mathbf{q}_i,\mathbf{p}_i) &=  \mathbf{p}_i + \frac{m_{j}}{m_{j}+m_{k}} \mathbf{q}_i   \; .
\end{align}
In the same way, momentum $\mathbf{q}_k(\mathbf{q}_i,\mathbf{p}_i)$ is found to be
\begin{align}
\mathbf{q}_k(\mathbf{q}_i,\mathbf{p}_i) &= - \frac{m_{j}+m_{k}}{m_{j}} \mathbf{p}_i + \frac{m_{k}}{m_{j}} \mathbf{q}_j = - \frac{m_{j}+m_{k}}{m_{j}} \mathbf{p}_i + \frac{m_{k}}{m_{j}} \left(-\mathbf{q}_i - \mathbf{q}_k \right) \; , \nonumber\\
\left( 1 + \frac{m_{k}}{m_{j}} \right) \mathbf{q}_k (\mathbf{q}_i,\mathbf{p}_i)&= -\frac{m_{j}+m_{k}}{m_{j}} \mathbf{p}_i - \frac{m_{k}}{m_{j}} \mathbf{q}_i  \; , \nonumber\\
\mathbf{q}_k (\mathbf{q}_i,\mathbf{p}_i) &= - \left( \mathbf{p}_i + \frac{m_{k}}{m_{j}+m_{k}} \mathbf{q}_i \right)  \; .
\end{align}

The same manipulation can be made for the momenta $\mathbf{q}_i(\mathbf{q}_j,\mathbf{p}_j)$, $\mathbf{q}_k(\mathbf{q}_j,\mathbf{p}_j)$, $\mathbf{q}_i(\mathbf{q}_k,\mathbf{p}_k)$ and $\mathbf{q}_j(\mathbf{q}_k,\mathbf{p}_k)$.  Introducing the variables ($\alpha,\beta,\gamma$) as cyclic permutation of the particle labels ($i,j,k$), the six combinations needed can be simply written as
\begin{align}
\mathbf{q}_\beta(\mathbf{q}_\alpha,\mathbf{p}_\alpha) &= \mathbf{p}_\alpha - \frac{m_{\beta}}{m_{\beta}+m_{\gamma}} \mathbf{q}_\alpha  \; , \label{jacobiab}\\
\mathbf{q}_\gamma(\mathbf{q}_\alpha,\mathbf{p}_\alpha) &=  - \left( \mathbf{p}_\alpha + \frac{m_{\gamma}}{m_{\beta}+m_{\gamma}} \mathbf{q}_\alpha \right)  \; . \label{jacobiac}
\end{align}

\chapter{Matrix elements of the three-body resolvent} \label{melements}

The six matrix elements which appear in Eqs.~\eqref{eq.c2-120a} to \eqref{eq.c2-120c} are calculated in detail in this Appendix. These elements have the same structure as given in Eq.~\eqref{eq.c2-121}, namely
\begin{equation}
ME=\left\langle \chi _{\alpha },\mathbf{q}_{\alpha}\right\vert G_{0}\left( E \right) \left\vert \chi _{\beta}\right\rangle \left\vert f_{\beta }\right\rangle \ .
\label{eq.a4-01}
\end{equation}   
Defining two resolutions of the unit as
\begin{align}
&\hat{1}=\int d^{2}q_{\beta}\left\vert \mathbf{q}_{\beta}\right\rangle \left\langle \mathbf{q}_{\beta }\right\vert \ \text{and} \ \hat{1}=\int d^{2}p_{\alpha}\left\vert \mathbf{p}_{\alpha}\right\rangle \left\langle \mathbf{p}_{\alpha}\right\vert \ ,
\label{eq.a4-02a}
\end{align}
the matrix element in Eq.~\eqref{eq.a4-01} becomes
\begin{align}
ME&=\int d^{2}q_{\beta }\left\langle \chi _{\alpha },\mathbf{q}_{\alpha }\right\vert G_{0}\left( \epsilon \right) \left\vert\mathbf{q}_{\beta }\right\rangle \left\langle \mathbf{q}_{\beta }\right. \left\vert\chi _{\beta }\right\rangle\left\vert f_{\beta }\right\rangle \ ,& \nonumber\\
&=\int d^{2}q_{\beta }d^{2}p_{\alpha }d^{2}p_{\beta }\left\langle \chi_{\alpha },\mathbf{q}_{\alpha }\right\vert \left.\mathbf{p}_{\alpha}\right\rangle \left\langle \mathbf{p}_{\alpha }\right\vert G_{0}\left(\epsilon \right) \left\vert\mathbf{p}_{\beta }\right\rangle \left\langle 
\mathbf{p}_{\beta }\right. \left\vert \chi _{\beta },\mathbf{q}_{\beta}\right\rangle f_{\beta }\left( \mathbf{q}_{\beta}\right) \ ,&   \nonumber \\
&=\int d^{2}q_{\beta }d^{2}p_{\alpha }d^{2}p_{\beta }\frac{g_{\alpha}\left( \mathbf{p}_{\alpha }\right) g_{\beta }\left( \mathbf{p}_{\beta}\right) }{E-\frac{\mathbf{q}_{\alpha }^{2}}{2m_{\beta \gamma ,\alpha }}-\frac{\mathbf{p}_{\alpha }^{2}}{2m_{\beta \gamma }}}\left\langle \mathbf{q}_{\alpha },\mathbf{p}_{\alpha }\right. \left\vert \mathbf{p}_{\beta }, \mathbf{q}_{\beta }\right\rangle f_{\beta }\left( \mathbf{q}_{\beta }\right)\ ,&  \label{eq.a4-02b}
\end{align}
where ($\alpha,\beta,\gamma$) are cyclic permutations of the particle labels ($i,j,k$).

The matrix element $\left\langle \mathbf{q}_{\alpha },\mathbf{p}_{\alpha}\right. \left\vert \mathbf{p}_{\beta},\mathbf{q}_{\beta }\right\rangle$ is written as \cite{schmidQMTB1974}
\begin{equation}
\left\langle \mathbf{q}_{\alpha },\mathbf{p}_{\alpha }\right. \left\vert 
\mathbf{p}_{\beta },\mathbf{q}_{\beta }\right\rangle =
\delta \Bigl( \mathbf{p}_{\alpha }-\mathbf{p}_{\alpha }^{\prime }\left( \mathbf{p}_{\beta },\mathbf{%
q}_{\beta } \right)\Bigr)
 \delta \Bigl( \mathbf{q}_{\alpha }-\mathbf{q}%
_{\alpha }^{\prime }\left( \mathbf{p}_{\beta },\mathbf{q}_{\beta }\right)
\Bigr) .  \label{eq.a4-03}
\end{equation}

The six possibilities  in Eqs.~\eqref{eq.c2-120a} to \eqref{eq.c2-120c} come from permutation of particles label in Eq.~\eqref{eq.a4-01}. Handling Eq.~\eqref{eq.a4-03} requires some manipulation of the Jacobi relative momenta in Eqs.~\eqref{eq.a3-18} and \eqref{eq.a3-19} and  
$\mathbf{p}_{\alpha}^{\prime }$ is written as function of the others momenta as
\begin{align}
\mathbf{p}_{\alpha}^{\prime }\left( \mathbf{p}_{\beta},\mathbf{q}_{\beta}\right)&= \frac{m_{\gamma}\mathbf{q}_{\beta}-m_{\beta}\mathbf{q}_{\gamma}}{m_{\gamma}+m_{\beta}} \ ,&  \nonumber \\
&=\frac{m_{\gamma}}{m_{\gamma}+m_{\beta}}\mathbf{q}_{\beta}-\frac{m_{\beta}}{m_{\gamma}+m_{\beta}}\left(-\mathbf{q}_{\alpha}-\mathbf{q}_{\beta}\right) \ ,   \nonumber \\
&=\mathbf{q}_{\beta}+\frac{m_{\beta}}{m_{\gamma}+m_{\beta}}\mathbf{q}_{\alpha} \ ,&  \label{eq.a4-04}
\end{align}%
and the first term on the right-hand-side of Eq.(\ref{eq.a4-03}) is
\begin{equation}
\delta \left( \mathbf{p}_{\alpha}-\mathbf{p}_{\alpha}^{\prime }\left( \mathbf{p}_{\beta},%
\mathbf{q}_{\beta}\right) \right) =\delta \left( \mathbf{p}_{\alpha}-\mathbf{q}_{\beta}-%
\frac{m_{\beta}}{m_{\gamma}+m_{\beta}}\mathbf{q}_{\alpha}\right) .  \label{eq.a4-05}
\end{equation}%
In the same way, $\mathbf{q}_{\alpha}^{\prime }$ reads
\begin{align}
\mathbf{q}_{\alpha}^{\prime }\left( \mathbf{p}_{\beta},\mathbf{q}_{\beta}\right)&=-\frac{m_{\gamma}+m_{\alpha}}{m_{\gamma}}\mathbf{p}_{\beta}+\frac{m_{\alpha}}{m_{\gamma}}\mathbf{q}_{\gamma} \ ,& \nonumber \\
&=-\frac{m_{\gamma}+m_{\alpha}}{m_{\gamma}}\mathbf{p}_{\beta}+\frac{m_{\alpha}}{m_{\gamma}}\left(-\mathbf{q}_{\alpha}-\mathbf{q}_{\beta}\right) \ ,& \label{eq.a4-06}
\end{align}
and the second term on the right-hand-side of Eq.(\ref{eq.a4-03}) is
\begin{align}
\delta \left( \mathbf{q}_{\alpha}-\mathbf{q}_{\alpha}^{\prime }\left( \mathbf{p}_{\beta},\mathbf{q}_{\beta}\right) \right)&=\delta\left(\mathbf{q}_{\alpha}+\frac{m_{\gamma}+m_{\alpha}}{m_{\gamma}}\mathbf{p}_{\beta}+\frac{m_{\alpha}}{m_{\gamma}}\left( \mathbf{q}_{\alpha}+ \mathbf{q}_{\beta}\right) \right) \ ,&   \nonumber \\
&=\delta \left( \frac{m_{\gamma}+m_{\alpha}}{m_{\gamma}}\mathbf{p}_{\beta}+\frac{m_{\gamma}+m_{\alpha}}{m_{\gamma}}\mathbf{q}_{\alpha}+\mathbf{q}_{\beta}\right) \ ,& \nonumber \\
&\equiv \delta \left( \mathbf{p}_{\beta}+\mathbf{q}_{\alpha}+\frac{m_{\alpha}}{m_{\gamma}+m_{\alpha}}\mathbf{q}_{\beta}\right) \ .&  \label{eq.a4-07}
\end{align}
The matrix element $\left\langle \mathbf{q}_{\alpha},\mathbf{p}_{\alpha}\right.
\left\vert \mathbf{p}_{\beta},\mathbf{q}_{\beta}\right\rangle $ is written as
\begin{equation}
\left\langle \mathbf{q}_{\alpha},\mathbf{p}_{\alpha}\right. \left\vert \mathbf{p}_{\beta},%
\mathbf{q}_{\beta}\right\rangle =\delta \left( \mathbf{p}_{\alpha}-\mathbf{q}_{\beta}-%
\frac{m_{\beta}}{m_{\gamma}+m_{\beta}}\mathbf{q}_{\alpha}\right) \delta \left( \mathbf{p}_{\beta}+%
\mathbf{q}_{\alpha}+\frac{m_{\alpha}}{m_{\gamma}+m_{\alpha}}\mathbf{q}_{\beta}\right) 
\label{eq.a4-08}
\end{equation}
and remembering that  it was shown in Sec.~\ref{tzmr} that the form factor of the Dirac$-\delta$ potential is $g(\mathbf{p})=1$, Eq.~\eqref{eq.a4-02b} becomes
\begin{align}
ME&=\int d^{2}q_{\beta}d^{2}p_{\alpha}d^{2}p_{\beta}\frac{\delta \left( \mathbf{p}_{\alpha}-\mathbf{q}_{\beta}-\frac{m_{\beta}}{m_{\beta}+m_{\gamma}}\mathbf{q}_{\alpha}\right) \delta \left(\mathbf{p}_{\beta}+\mathbf{q}_{\alpha} +\frac{m_{\alpha}}{m_{\alpha}+m_{\gamma}}\mathbf{q}_{\beta}\right)}{E-\frac{\mathbf{q}_{\alpha}^{2}}{2m_{\beta\gamma,\alpha}}-\frac{\mathbf{p}_{\alpha}^{2}}{2m_{\beta\gamma}}}f_{\beta}\left( \mathbf{q}_{\beta}\right) \ ,&   \nonumber \\
&=\int d^{2}q_{\beta}\frac{f_{\beta}\left( \mathbf{q}_{\beta}\right)}{E-\frac{\mathbf{q}_{\alpha}^{2}}{2m_{\beta\gamma,\alpha}}-\frac{\left(\mathbf{q}_{\beta}+\frac{m_{\beta}}{m_{\beta}+m_{\gamma}} \mathbf{q}_{\alpha}\right) ^{2}}{2m_{\beta\gamma}}} \ . &  \label{eq.a4-09}
\end{align}

Finally, the matrix element which appears in Eqs.~\eqref{eq.c2-122a} to \eqref{eq.c2-122c} are given by
\begin{align}
\left\langle \chi _{\alpha},\mathbf{q}_{\alpha}\right\vert
G_{0}\left( E \right) \left\vert \chi _{\beta}\right\rangle \left\vert
f_{\beta}\right\rangle
&=\int d^{2}q_{\beta}\frac{f_{\beta}\left( \mathbf{q}_{\beta}\right)}{E-\frac{q_{\alpha}^{2}}{2 m_{\alpha \gamma}}-\frac{q_{\beta}^{2}}{2 m_{\beta \gamma}}-\frac{1}{m_\gamma}\mathbf{q}_{\beta}\cdot \mathbf{q}_{\alpha}} \ , &  \label{eq.a4-10}
\end{align}
where ($\alpha,\beta,\gamma$) are cyclic permutations of the particle labels ($i,j,k$).
\chapter{Numerical methods} \label{numerical}

The set of integral homogeneous coupled equations in Eq.~\eqref{spec1} does not have analytic solution in general and then, it has to be numerically solved. There are several well established methods available in the literature to solve integral equations as given, for example, in Ref.~\cite{press2007}. Besides, classical techniques as Gauss-Legendre quadratures are used for the numerical discretization of the integral equations, the Newton-Raphson method is used to find zeros of functions and the Gauss decomposition method is used to find the determinant of a matrix. Optimized routines  which implement these techniques are available in the libraries of the programming languages as C, C++ and Fortran and are also described in Ref.~\cite{press2007}.

The three-body problem studied in this thesis consists, basically, of an eigenvalue - eigenvector problem, where the determination of the energy (eigenvalue) leads to the determination of the spectator, and consequently, the wave function (eigenvector).

\section{Three-body energy (eigenvalue)} \label{tbee}
In order to illustrate the methods employed in the numerical solution of Eq.~\eqref{spec1}, the symmetric mass case is considered, i.e., $m_a=m_b=m_c=m$ and $E_{ab}=E_{ac}=E_{bc}=E_2$. This means that only one integral equation has to be solved and choosing $E_2$ and $m$ as the energy and mass units, the three-body energy $E_3$ and the momenta $q$ and $k$ in Eq.~\eqref{spec1} are rewritten as $E_3 \equiv \frac{E_3}{E_2}$ and $q \equiv \frac{q}{\sqrt{m E_2}}$. In units of $E_2=m=1$, Eq.~\eqref{spec1} becomes 
\begin{align}
&f\left( q \right)  = 2\; \left[ \ln \left( \sqrt{\frac{3}{4}q^{2}-E_{3}} \right) \right] ^{-1}
\int_0^\infty dk  \frac{ k \;f\left( k \right) }{\sqrt{\left(-E_{3}+q^{2}+k^{2}\right)^2-\left(k \; q\right)^2}} ,& 
\label{eq.a7-01}
\end{align}
which in a compact form reads
\begin{equation}
f(q)=\int_0^\infty{K(E_3,q,q')f(q')dq'} \ ,
\label{eq.a7-02}
\end{equation}
where $\mathbf{k} \equiv \mathbf{q}'$ and the kernel $K(E_3,q,q')$ is defined by
\begin{equation}
K(E_3,q,q')=2\; \left[ \ln \left( \sqrt{\frac{3}{4}q^{2}-E_{3}} \right) \right] ^{-1}
\frac{ q' }{\sqrt{\left(-E_{3}+q^{2}+q'^{2}\right)^2-\left(q' \; q\right)^2}} .
\label{eq.a7-03}
\end{equation}

The Gauss-Legendre mesh-points are used for the discretization of the kernel in  Eq.~\eqref{eq.a7-03}, where the discrete momentum $q \equiv x_i$ correspond to one set of points with the respective Gauss-Legendre weights $dq \equiv \omega_i$. The Gauss\=/Legendre mesh points and weights are generated, in general, to calculate integrals in the interval $[-1,1]$. Since the kernel has to be discretized in the interval $[0,\infty[$, a possible transformation of the set of mesh points and weights is
\begin{align}
q_i&=\frac{1+x_i}{1-x_i} \,& \nonumber\\
w_i&=\frac{2}{(1-x_i)^2} \omega_i \ . \label{eq.a7-21}
\end{align}

Therefore, the discretization of the homogeneous integral equation in Eq.~\eqref{eq.a7-02} is written as
\begin{align}
&f(q_i)=\sum_{j=1}^{N}{K(E_3,q_i,q_j)f(q_j)w_j} \ ,& \nonumber\\*
&f(q_i)-\sum_{j=1}^{N}{K(E_3,q_i,q_j)f(q_j)w_j}=0 \ ,& \nonumber\\* 
&\left(\delta_{ij}-\sum_{j=1}^{N}{K(E_3,q_i,q_j)w_j}\right)f(q_j)=0 \ ,& \label{eq.a7-04}
\end{align}
where $\delta_{ij}$ is the Kronecker's delta, $1 \leq i \leq N$ and $K(E_3,q_i,q_j)$ reads
\begin{equation}
K(E_3,q_i,q_j)=2\; \left[ \ln \left( \sqrt{\frac{3}{4}q_i^{2}-E_{3}} \right) \right] ^{-1}
\frac{ q_j }{\sqrt{\left(-E_{3}+q_i^{2}+q_j^{2}\right)^2-\left(q_i \; q_j\right)^2}} \; .
\label{eq.a7-05}
\end{equation}
The matrix form of Eq.~\eqref{eq.a7-04} is given by
\begin{equation}
HF=0
\label{eq.a7-06}
\end{equation}
whit
\begin{equation} 
H=
\left(\begin{array}{cccc}
1-K(E_3,q_1,q_1)w_1 & -K(E_3,q_1,q_2)w_2 & \cdots & -K(E_3,q_1,q_N)w_N \\ 
-K(E_3,q_2,q_1)w_1 & 1-K(E_3,q_2,q_2)w_2 & \cdots & -K(E_3,q_2,q_N)w_N \\
\vdots & \vdots & \ddots & \vdots \\ 
-K(E_3,q_N,q_1)w_1 & -K(E_3,q_N,q_2)w_2 & \cdots & 1-K(E_3,q_N,q_N)w_N \\
\end{array}\right)
\label{eq.a7-06a}
\end{equation}
and
\begin{equation} 
F=
\left(\begin{array}{c}
f(q_{1}) \\ 
f(q_{2}) \\
\vdots \\ 
f(q_{N})%
\end{array}\right)
\; .
\label{eq.a7-06b}
\end{equation}

The matrix equation~\eqref{eq.a7-06} only admits non-trivial solution for
\begin{equation}
\det H = \det\left(\delta_{ij}-\sum_{j=1}^{N}{K(E_3,q_i,q_j)w_j}\right)=0 \ .
\label{eq.a7-07}
\end{equation}
The determinant of $H$ is a function of the three-body energy $E_3$, namely 
\begin{equation}
F(E_3)=\det\left(\delta_{ij}-\sum_{j=1}^{N}{K(E_3,q_i,q_j)w_j}\right) \ ,
\label{eq.a7-08}
\end{equation}
and it is calculated, for instance with the Gauss method or the QR\=/decomposition \cite{press2007}.
When $E_3$ corresponds to a three-body bound state energy, $E_3^n$, the determinant of $H$ must be null. In other words, 
\begin{equation}
F(E_3^n)=0 \ ,
\label{eq.a7-09}
\end{equation}
where the superscript $n$ labels the three-body energy for the $n^{th}$ bound state. which satisfies Eqs.~\eqref{eq.a7-01} and \eqref{eq.a7-07}. 

The Newton-Raphson method is used in order to find $E_3^n$ from Eq.~\eqref{eq.a7-08}.  Expanding Eq.~\eqref{eq.a7-09} up to first order around $E_3$ gives
\begin{align}
\centering
&F(E_3^n)=F(E_3)+(E_3^n-E_3)F'(E_3)=0 \ ,& \nonumber\\
&E_3^n=E_3-\frac{F(E_3)}{F'(E_3)} \ ,& \label{eq.a7-10}
\end{align}
where $F'(E_3)=\frac{dF(E_3^n)}{dE_3^n}\left.\right\vert_{E_3^n=E_3}$.
The three\=/body bound\=/state energy  is found by successively iterations of Eq.~\eqref{eq.a7-10}, where the output of the $m^{th}$ iteration,  $E_m$, is used as thin input of the consecutive one. This means that
\begin{align}
\centering
&E_m=E_{m-1}-\frac{F(E_{m-1})}{F'(E_{m-1})} \;\;\; \text{for} \;\;\; m=1,2,3... \; ,& \label{eq.a7-10a}
\end{align}
Notice that an appropriate guess is needed for $E_0$. A good guess can be found by plotting $F(E)$ vs. $E$ and taking $E_0$ as the point where $F(E_0) \approx 0$. However, the kernel in Eq.~\eqref{eq.a7-03} presents the so-called well of attraction, which means that even bad guesses for $E_0$ would lead to the right final result. The iterative process in Eq.~\eqref{eq.a7-10a} must be repeated until
\begin{equation}
\left|\frac{E_m-E_{m-1}}{E_{m-1}}\right| \leq acc \ ,
\label{eq.a7-11}
\end{equation} 
where $acc$ is the desired accuracy. 

Unfortunately, the functions $F$ and $F'$ in Eq.~\eqref{eq.a7-10a} does not have an analytical form. While $F(E_m)$ can be easily calculated from Eq.~\eqref{eq.a7-08}, its derivative is found from the definition of derivative 
\begin{equation}
F'(E) =  \lim_{\Delta E \to 0} \frac{F(E+\Delta E) - F(E)}{ \Delta E} \; ,
\end{equation}
where the discrete version reads
\begin{equation}
F'(E_m) =  \frac{F(E_m) - F(E_{m-1})}{ E_m-E_{m-1}} \; ,
\label{derdisc}
\end{equation}

Finally, inserting the derivative~\eqref{derdisc} in Eq.~\eqref{eq.a7-10a}, the iterative equation for the three-body energy becomes
\begin{align}
\centering
&E_m=E_{m-1}-F(E_{m-1}) \frac{ E_{m-1}-E_{m-2}}{F(E_{m-1}) - F(E_{m-2})} \;\;\; \text{for} \;\;\; m \geq 2 \; ,& 
\label{eq.a7-10b}
\end{align}
where now $E_1$ has also to be guessed. A good try is $E_1 = 1.1 E_0$.

It is straight forward to extended the method above for the case of three distinguishable particles, since only the matrix $H$ in Eq.~\eqref{eq.a7-06a} and $F$ in Eq.~\eqref{eq.a7-06b} have to be redefined.  The set of coupled integral equations is given in Eq.~\eqref{spec1}. These equations read
\begin{multline}
f_{\alpha}\left( q \right)  =2 \pi \left[ 4\pi m_{\beta \gamma}\ln \left( 
\sqrt{\frac{\frac{q^{2}}{2m_{\beta \gamma,\alpha} }-E_{3}}{E_{\beta\gamma}}}
\right) \right] ^{-1}   \label{spec1-a} \\*
\times \int_0^\infty dk\left( \frac{ k \;f_{\beta}\left( k \right) }{\sqrt{\left(-E_{3}+\frac{q^{2}}{2 m_{\alpha \gamma}}+\frac{k^{2}}{2 m_{\beta \gamma}}\right)^2-\left(\frac{k \; q} {m_\gamma}\right)^2}} \right. \\* \left.
+\frac{k \; f_{\gamma}\left( k\right) }{\sqrt{\left(-E_{3}+\frac{q^{2}}{2m_{\alpha \beta }}+\frac{k^{2}}{2m_{\beta \gamma}}\right)^2-\left(\frac{k \; q}{m_\beta}\right)^2}}\right), 
\end{multline}
or in a more compact form
\begin{align}
f_\alpha(q)&=\int_0^\infty{K_{\alpha \beta}(E_3,q,k)f_\beta(k)dk}+\int_0^\infty{K_{\alpha \gamma}(E_3,q,k)f_\gamma(k)dk} \ ,& \label{eq.a7-14d}
\end{align}
where, writing the labels of each particle explicitly results in three equations, namely
\begin{align}
f_a(q)&=\int_0^\infty{K_{12}(E_3,q,k)f_b(k)dk}+\int_0^\infty{K_{13}(E_3,q,k)f_c(k)dk} \ ,& \label{eq.a7-14a}\\
f_b(q)&=\int_0^\infty{K_{21}(E_3,q,k)f_a(k)dk}+\int_0^\infty{K_{23}(E_3,q,k)f_c(k)dk} \ ,& \label{eq.a7-14b} \\
f_c(q)&=\int_0^\infty{K_{31}(E_3,q,k)f_a(k)dk}+\int_0^\infty{K_{32}(E_3,q,k)f_b(k)dk} \ ,& \label{eq.a7-14c} 
\end{align}
with the kernels defined as
\begin{align}
K_{12}(E_3,q,k)&=  \frac{  \left[ 2 m_{bc}\ln \left(\sqrt{\frac{\frac{q^{2}}{2m_{bc,a}}-E_{3}}{E_{bc}}}\right) \right] ^{-1} k }
{\sqrt{\left(-E_{3}+\frac{q^{2}}{2 m_{ac}}+\frac{k^{2}}{2 m_{bc}}\right)^2-\left(\frac{k \; q} {m_c}\right)^2}} \; , & \label{eq.a7-15a} 
\\
K_{13}(E_3,q,k)&=\frac{  \left[ 2 m_{bc}\ln \left(\sqrt{\frac{\frac{q^{2}}{2m_{bc,a}}-E_{3}}{E_{bc}}}\right) \right] ^{-1} k }
{\sqrt{\left(-E_{3}+\frac{q^{2}}{2 m_{ab}}+\frac{k^{2}}{2 m_{bc}}\right)^2-\left(\frac{k \; q} {m_b}\right)^2}} \; , & \label{eq.a7-15b} \\
K_{21}(E_3,q,k)&=  \frac{  \left[ 2 m_{ac}\ln \left(\sqrt{\frac{\frac{q^{2}}{2m_{ac,b}}-E_{3}}{E_{ac}}}\right) \right] ^{-1} k }
{\sqrt{\left(-E_{3}+\frac{k^{2}}{2 m_{ac}}+\frac{q^{2}}{2 m_{bc}}\right)^2-\left(\frac{k \; q} {m_c}\right)^2}} \; , & \label{eq.a7-15c} \\
K_{23}(E_3,q,k)&=\frac{  \left[ 2 m_{ac}\ln \left(\sqrt{\frac{\frac{q^{2}}{2m_{ac,b}}-E_{3}}{E_{ac}}}\right) \right] ^{-1} k }
{\sqrt{\left(-E_{3}+\frac{q^{2}}{2 m_{ab}}+\frac{k^{2}}{2 m_{ac}}\right)^2-\left(\frac{k \; q} {m_a}\right)^2}} \; , & \label{eq.a7-15d} \\
K_{31}(E_3,q,k)&=\frac{  \left[ 2 m_{ab}\ln \left(\sqrt{\frac{\frac{q^{2}}{2m_{ab,c}}-E_{3}}{E_{ab}}}\right) \right] ^{-1} k }
{\sqrt{\left(-E_{3}+\frac{k^{2}}{2 m_{ab}}+\frac{q^{2}}{2 m_{bc}}\right)^2-\left(\frac{k \; q} {m_b}\right)^2}} \; , & \label{eq.a7-15e} \\
K_{32}(E_3,q,k)&=\frac{  \left[ 2 m_{ab}\ln \left(\sqrt{\frac{\frac{q^{2}}{2m_{ab,c}}-E_{3}}{E_{ab}}}\right) \right] ^{-1} k }
{\sqrt{\left(-E_{3}+\frac{k^{2}}{2 m_{ab}}+\frac{q^{2}}{2 m_{ac}}\right)^2-\left(\frac{k \; q} {m_a}\right)^2}} \; . & \label{eq.a7-15f} 
\end{align}

Therefore, the discretization of the homogeneous integral equation in Eq.~\eqref{eq.a7-14d} is written as
\begin{align}
&f_\alpha(q_i) = \sum_{j=1}^{N}{K_{\alpha \beta}(E_3,q_i,q_j)f_\beta(q_j)w_j} + \sum_{j=1}^{N}{K_{\alpha \gamma}(E_3,q_i,q_j)f_\gamma(q_j)w_j} \ ,& \nonumber\\
&f_\alpha(q_i) - \sum_{j=1}^{N}{K_{\alpha \beta}(E_3,q_i,q_j)f_\beta(q_j)w_j} - \sum_{j=1}^{N}{K_{\alpha \gamma}(E_3,q_i,q_j)f_\gamma(q_j)w_j} = 0\ ,& \nonumber\\
&\left(\delta_{ij}- \sum_{j=1}^{N}{K_{\alpha \beta}(E_3,q_i,q_j)w_j} - \sum_{j=1}^{N}{K_{\alpha \gamma}(E_3,q_i,q_j)w_j}\right) 
\left(\begin{array}{c}
f_\alpha(q_i) \\ 
f_\beta(q_i) \\
f_\gamma(q_i)
\end{array}\right)=0 \ ,& \label{eq.a7-16b}
\end{align}
where $\delta_{ij}$ is the Kronecker's delta, $1 \leq i \leq N$ and $K_{\alpha \beta}$ and $K_{\alpha \gamma}$ are defined in Eqs.~\eqref{eq.a7-15a} to \eqref{eq.a7-15f}. The matrix equation of the discretized kernel is in fact a so-called matrix by blocks and reads
\begin{equation} 
\left(\begin{array}{ccc}
\mathds{1} & H_{12} & H_{13}  \\ 
H_{21} & \mathds{1} & H_{23}  \\ 
H_{31} & H_{32} & \mathds{1} 
\end{array}\right)
\left(\begin{array}{c}
f_a \\ 
f_b \\
f_c 
\end{array}\right) = 0 \ ,
\label{eq.a7-16}
\end{equation}
where $\mathds{1}$ is the identity matrix and the matrix blocks $H_{\alpha \beta}$ and $f_\alpha$ are given by
\begin{align}
H_{\alpha \beta} &= \sum_{j=1}^{N}{K_{\alpha \beta}(E_3,q_i,q_j)w_j} \;\;\; \text{for} \;\;\; 1 \leq i \leq N \; , & \label{eq.a7-16a}\\
f_\alpha &= f_\alpha(q_{i}) \;\;\; \text{for} \;\;\; 1 \leq i \leq N \; . & \label{eq.a7-16c}
\end{align}
Or, in matrix form
\begin{equation} 
f_\alpha = 
\left(\begin{array}{c}
f_\alpha(q_{1}) \\ 
f_\alpha(q_{2}) \\
\vdots \\ 
f_\alpha(q_{N})%
\end{array}\right)
\label{eq.a7-17a}
\end{equation}
and, in the same way, the matrix blocks $H_{\alpha\beta}$ are
\begin{equation}
H_{\alpha\beta} =
\left(\begin{array}{cccc}
K_{\alpha\beta}(E_3,q_1,q_1)w_1 & K_{\alpha\beta}(E_3,q_1,1_2)w_2 & \cdots & K_{\alpha\beta}(E_3,q_1,q_N)w_N \\ 
K_{\alpha\beta}(E_3,q_2,q_1)w_1 & K_{\alpha\beta}(E_3,q_2,q_2)w_2 & \cdots & K_{\alpha\beta}(E_3,q_2,q_N)w_N \\
\vdots & \vdots & \ddots & \vdots \\ 
K_{\alpha\beta}(E_3,q_N,q_1)w_1 & K_{\alpha\beta}(E_3,q_N,q_2)w_2 & \cdots & K_{\alpha\beta}(E_3,q_N,q_N)w_N \\
\end{array}\right) \; .
\label{eq.a7-17b}
\end{equation}

Finally, Eq.~\eqref{eq.a7-16} can be written as Eq.~\eqref{eq.a7-06}, i.e., $HF=0$ and the steps between Eqs.~\eqref{eq.a7-07} and \eqref{eq.a7-10b} are the same, giving the three-body energy for a system composed for three-distinguishable particles.

\section{Spectator functions (eigenvector)}
The method is illustrated for the case of identical particles, since it allows a simpler notation. However, the procedure is general and easily extended to the case of three-distinguishable particles, as it was done in the calculation of the three-body energy in Sec.~\ref{tbee}.

Once the three-bode energy $E_3$ is calculated, it should be inserted again in Eq.~\eqref{eq.a7-04} in order to once more generate the matrix $H$, as given in Eq.~\eqref{eq.a7-06a}. Now, the matrix equation $HF=0$ (see Eq.~\eqref{eq.a7-06}) is simply a set of $N$ equations for $N$ unknown, namely, each value of the spectator function $f(q_n)$ has to be determined in a point $q_n$, with $1 \leq n \leq N$.

The set of $N$ equations is given by
\begin{equation}
\sum_{j=1}^{N} H_{ij}(E_3,q_i,q_j) f(q_j) = 0 \;\;\; \text{for} \;\;\;  1 \leq i \leq N \; .
\label{setn}
\end{equation}
Notice, however, that the three-body energy calculated in the last section is the one which fulfills $\det H = 0$. This means that one of the equations from the set is redundant and that the system in Eq.~\eqref{setn} can not be unequivocally determined. In other words, $N-1$ variables will be given in terms of an arbitrary value.
This freedom in the system is utilized to set, for instance, $f(q_1)=1$. Then, eliminating a line and a column in Eq.\eqref{eq.a7-06}, the system of equations becomes 
\begin{equation}
\sum_{j=2}^{N} H_{ij}(E_3,q_i,q_j) f(q_j) = -H_{i1} f(q_1) = -H_{i1} \;\;\; \text{for} \;\;\;  2 \leq i \leq N \; ,
\label{setn1}
\end{equation}
or in the matrix form

\setlength{\unitlength}{7cm} 
\begin{center}
\begin{picture}(1,1)
\put(-0.3,0){\line(0,1){1}}
\put(-0.3,1){\line(1,0){1}}
\put(-0.2,0.9){\line(1,0){0.9}}
\put(-0.2,0.9){\line(0,-1){0.9}}
\put(0.7,0.9){\line(0,-1){0.9}}
\put(-0.2,0){\line(1,0){0.9}}
\put(0.0,0.5){$\tilde{H}_{(N-1) x (N-1)}$}
\put(-0.4,0.5){\oval(0.1,1.1)[l]}
\put(0.8,0.5){\oval(0.1,1.1)[r]}
\put(1.0,0.5){\oval(0.1,1.1)[l]}
\put(1.15,0.5){\oval(0.1,1.1)[r]}
\put(1.05,1){$1$}
\put(1.0,0.9){$f(q_2)$}
\put(1.05,0.2){$\vdots$}
\put(1.05,0.5){$\vdots$}
\put(1.05,0.8){$\vdots$}
\put(1.0,0){$f(q_N)$}
\put(1.3,0.5){$=0$ ,}
\end{picture}
\end{center}
where all the unknown $f(q_n)$ with $2 \leq n \leq N$ are determined in term of $f(q_1)=1$. There is no problem in choosing an arbitrary value for $f(q_1)$, since the wave-function is defined unless a normalization constant. The Gauss\=/ Jordan elimination method in then used to solve the set algebraic equations in Eq.~\eqref{setn1}. Besides, this method still holds in the case of three-distinguishable particles, where the only difference arises from the matrix $H$ and $F$ which have to be respectively defined as in Eqs.~\eqref{eq.a7-16a} and \eqref{eq.a7-16c}, whose matrix form is given in Eq.~\eqref{eq.a7-16}.  

\chapter{Asymptotic one-body density in 3D} \label{derivations}

The large momentum limit of the four terms in Eqs.~\eqref{n1} to \eqref{n4} is worked out in this Appendix.

\subsection{Asymptotic contribution from $n_1(q_B)$}
As in the 2D case, this is also the simplest term. The asymptotic form of the spectator function from Eq.~\eqref{chiasymp} is inserted in Eq.~\eqref{n1}, the first out of the four terms of the momentum distribution. Taking the large momentum limit results in
\begin{align}
n_1(q_B)&\to 2\pi^2\sqrt{ \frac{{\cal A}}{{\cal A}+2}}\;\frac{\left|\chi_{AA}(q_B)\right|^2 }{q_B }
\to 2\pi^2\; \left|c_{AA}\right|^2\;  \sqrt{ \frac{{\cal A}}{{\cal A}+2}}\; \frac{\left|\sin(s\;\ln q_B/q^*)\right|^2}{q_B^5} \; , &
\nonumber\\
&\to \frac{\pi^2}{q_B^5}\; \left|c_{AA}\right|^2\;  \sqrt{ \frac{{\cal A}}{{\cal A}+2}} \; , &
\end{align}
where the $1/2$ came from the average of  the oscillating part.

\subsection{Asymptotic contribution from $n_2(q_B)$}
For large $q_B$, Eq.~\eqref{n2} becomes
\begin{align}
n_2(q_B)= &2\int d^3q_A
 \frac{\left|\chi_{AB}(q_A)\right|^2}{\left(q_A^2+\mathbf q_A\cdot \mathbf
q_B+q_B^2\;\frac{{\cal A}+1}{2{\cal A}}\right)^2} 
\; , & \nonumber\\* &
= \frac{8{\cal A}^2}{q_B^4\;({\cal A}+1)^2}\int d^3q_A\;\left|\chi_{AB}(q_A)\right|^2 + n_{2s}(q_B) \; ,
 \label{n2asymapp}
\end{align}
where a sub-leading part, $n_{2s}(q_B)$, is retained  since it is of the same order as
the leading order of the other terms. It is important to emphasize that the one-body large-momentum leading order comes only
from $n_2(q_B)$.  The spectator function can not be replaced
by its asymptotic expression, because the main contribution to
$\int_0^\infty{dq_A\;q_A^2\;|\chi_{AB}(q_A)|^2}$ arises from small $q_A$.  This
replacement would therefore lead to a completely wrong result.
However, this is not always the case, as shown below for $n_{2s}(q_{B})$.

The sub-leading term is
\begin{align}
n_{2s}(q_B) &= \int d^3q_A \left|\chi_{AB}(q_A)\right|^2\left[
 \frac{2}{\left(q_A^2+\mathbf q_A\cdot \mathbf
q_B+q_B^2\;\frac{{\cal A}+1}{2{\cal A}}\right)^2}-\frac{8{\cal
A}^2}{({\cal A}+1)^2}\frac{1}{q_B^4}\right] \; ,  \nonumber\\*
&= \left|c_{AB}\right|^2 \int \frac{d^3q_A}{q_{A}^{4}} \left[
 \frac{1}{\left(q_A^2+\mathbf q_A\cdot \mathbf
q_B+q_B^2\;\frac{{\cal A}+1}{2{\cal A}}\right)^2}-\frac{4{\cal
A}^2}{({\cal A}+1)^2}\frac{1}{q_B^4}\right] \; , \nonumber\\*
&= \frac{2\pi
\left|c_{AB}\right|^2}{q_{B}^{5}}\int_{-\infty}^{\infty}
\frac{dx}{x^2} \left[
 \frac{1}{x^4+\frac{1}{{\cal A}}x^2+(\frac{{\cal A}+1}{2{\cal A}})^2}-\frac{1}{(\frac{{\cal A}+1}{2{\cal A}})^2}\right] \; ,
\label{sublead}
\end{align}
where it was used in the second line the asymptotic form of
$|\chi_{AB}(q_A)|^2=|c_{AB}|^3q_{A}^{-4}/2$ obtained after averaging over the oscillatory term 
in Eq.~\eqref{chiasymp}. Next, the angular integral was solved and the variable $q_A=q_B x$ introduced. Since the integrand is even the integration can be extended to the entire real axis.

The function under the integral,
\begin{align}
f(x)=\frac{1}{x^2} \left[
 \frac{1}{x^4+\frac{1}{{\cal A}}x^2+(\frac{{\cal A}+1}{2{\cal A}})^2}-\frac{1}{(\frac{{\cal A}+1}{2{\cal A}})^2}\right],
\end{align}
falls off faster than $1/x$ for $|x|\to\infty$. Therefore, it is extended to the complex domain, where a contour in the
upper-half plane that includes the real axis and a semi-circle of large radius in a counterclockwise orientation is considered. The poles of $f(x)$ have to be determined in order to use the Cauchy residue theorem. Since $f(x)$ is regular at $x=0$, the four poles are out in the
complex plane and are given by
\begin{align}
x_1=r e^{i\theta_1/2},\,\,x_2=r e^{i(\pi-\theta_1/2)},\,\,x_3=r
e^{i(\pi+\theta_1/2)},\,\,x_4=r e^{-i\theta_1/2},
\end{align}
where $r=\sqrt{\tfrac{{\cal A}+1}{2{\cal A}}}$ and
$\tan^2\theta_1={\cal A}({\cal A}+2)$. Following the convention from Eqs.~\eqref{I1} to \eqref{I3}, where $\pi/2<\theta_1<\pi$, then $x_1$ and $x_2$ are the poles in the upper-half plane. The sum of the two residues is
\begin{align}
\textrm{Res}(f,x_1)+\textrm{Res}(f,x_2)=-\frac{1}{ir^3}\frac{{\cal
A}({\cal A}+3)}{({\cal
A}+1)^2}\frac{\cos(\frac{\theta_1}{2})}{\sin(\theta_1)} \; .
\end{align}
Using the residue theorem, the sub-leading term in Eq.~\eqref{sublead} becomes
\begin{align}
n_{2s}(q_B)&= -\frac{4\pi^2
\left|c_{AB}\right|^2}{q_{B}^{5} 2\sin(\frac{\theta_1}{2})}\frac{{\cal
A}({\cal A}+3)}{({\cal A}+1)^2} \left(\frac{2{\cal A}}{{\cal
A}+1}\right)^{3/2} \; ,
\end{align}
where, from the definition of $\theta_1$, $cos\theta_1=-\frac{1}{{\cal A}+1}$ and 
$\left[2\sin(\frac{\theta_1}{2})\right]^{-1}=\sqrt{\frac{{\cal A}+1}{2({\cal A}+2)}}$. Finally, the sub-leading term in $n_2$ is given by
\begin{equation}
\left<n_2(q_B)\right>=-\frac{8\pi^2 \left|c_{AB}\right|^2}{q_{B}^{5}} \frac{{\cal
A}^3({\cal A}+3)}{({\cal A}+1)^3 \sqrt{{\cal A}({\cal A}+2)}} \ ,
\end{equation}
where the special case ${\cal A}=1$ yields $\left<n_2(q_B)\right>=-4\pi^2 \left|c_{AB}\right|^2/(\sqrt{3}q_{B}^{5})$. The sub-index $s$ was dropped, since this term has the same order as the leading-order of the $n_3(q_B)$ and $n_4(q_B)$ terms.

\subsection{Asymptotic contribution from $n_3(q_B)$}
The structure of $n_3(q_B)$ in
Eq.~\eqref{n3} is similar to $n_2(q_B)$ in
Eq.~\eqref{n2}. The only difference is that the spectator function
under the integration sign is not squared anymore. This small functional
difference leads to a completely different result. 
Neglecting the three-body energy and changing variables to $\mathbf{q}_A=\mathbf{p}_B-\frac{\mathbf{q}_B}{2}$ in Eq.~\eqref{n3} results in
\begin{align}
n_3(q_B)&= 2\chi^{\ast}_{AA}(q_B)  \int d^3q_A \frac{\chi_{AB}(q_A)}{\left(q^{2}_A+q_B^2\frac{{\cal A}+1}{2{\cal A}}+{\mathbf q}_A\cdot{\mathbf{q}_B}\right)^2}+c.c. \; ,  \nonumber\\
	    &= 2\chi^{\ast}_{AA}(q_B)  \int d^3 y \; q_B^3 \frac{\chi_{AB}(q_B \; y)}{\left(y^2 q^{2}_B+q_B^2\frac{{\cal A}+1}{2{\cal A}}+ y q_B^2 \cos \theta \right)^2}+c.c. \; ,  \nonumber\\
	    &= \frac{2\chi^{\ast}_{AA}(q_B)}{q_B}  \int d^3 y  \frac{\chi_{AB}(q_B \; y)}{\left(y^2 + \frac{{\cal A}+1}{2{\cal A}}+ y  \cos \theta \right)^2}+c.c. \; ,  \nonumber\\*
   	    &=4 \pi \frac{\chi^{\ast}_{AA}(q_B)}{q_B}  \int_0^\infty dy \; y^2 \int_0^\pi  \frac{d \theta \sin \theta \chi_{AB}(q_B \; y)}{\left(y^2 + \frac{{\cal A}+1}{2{\cal A}}+ y  \cos \theta \right)^2}+c.c. \; .  \label{intne1}
\end{align}
where in the second line it was defined  ${\mathbf q}_A=q_B{\mathbf y}$. 

\begin{figure}[htb!]
\centering
\includegraphics[width=0.9\textwidth]{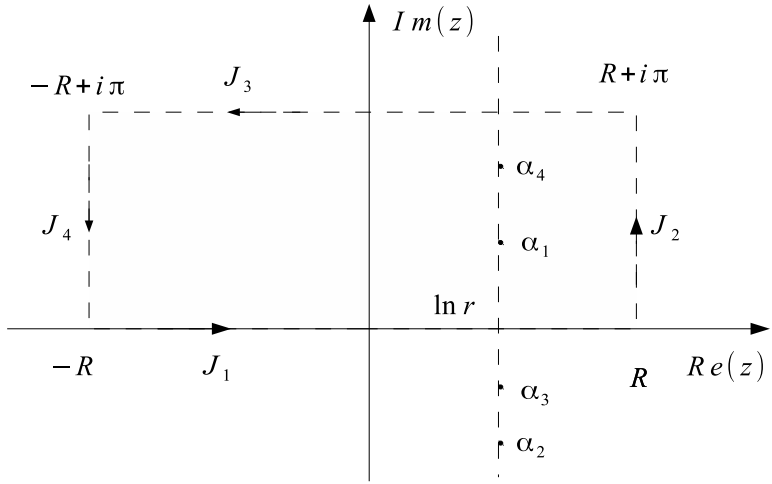}
\caption[Closed path used in the calculation of complex integrals.]
{A rectangle of vertices $-R$, $+R$, $+R+\imath\pi$ and $-R+\imath\pi$, which molds the closed path in a complex plane of $z$. The poles $\alpha_1$ and $\alpha_4$ are encompassed in the upper-half plane. The the direction of integration of the four integrals $J_1$, $J_2$, $J_3$ and $J_4$ are indicated by the arrows. } 
\label{pathn3}
\end{figure}

The angular integral is
\begin{align}
\int_0^\pi  \frac{d \theta \sin \theta}{\left(y^2 + \frac{{\cal A}+1}{2{\cal A}}+ y  \cos \theta \right)^2} 
&=\int_{-1}^{1} \frac{dx}{\left(A+ B x \right)^2} 
 =\frac{-1}{B} \int_{A-B}^{A+B} \frac{dz}{z^2}
 =\frac{-1}{B}\left.\frac{1}{z} \right|_{A-B}^{A+B}, & \nonumber\\
&=\frac{2}{A^2 -B^2}=\frac{2}{\left(y^2+ \frac{{\cal A}+1}{2{\cal A}} \right)^2-y^2} \; , & \nonumber\\
&=\frac{2}{y^4 + \frac{1}{{\cal A}} y^2 + \left(\frac{{\cal A}+1}{2{\cal A}} \right)^2} \; . &  \label{angintn3}
\end{align}
Besides, replacing the spectator functions $\chi_{AA}$ and $\chi_{AB}$ by their asymptotic form, as given in Eq. ~\eqref{chiasymp}, results in
\begin{align}
\chi^{\ast}_{AA}(q_B)&= \frac{c^{\ast}_{AA}}{q_B^2} \sin \left( s \ln \frac{q_B}{q^*} \right) , & \label{xaaasym}\\
\chi_{AB}(q_B \; y)&= \frac{c_{AB}}{y^2\;q_B^2} \sin \left[ s \ln \left( \frac{q_B\;y}{q^*} \right) \right], & \nonumber\\
&= \frac{c_{AB}}{y^2\;q_B^2} \sin \left( s \ln  \frac{q_B}{q^*} + s \ln y  \right), & \nonumber\\
&= \frac{c_{AB}}{y^2\;q_B^2} \left[ \sin \left( s \ln  \frac{q_B}{q^*}  \right) \cos \left(s \ln y  \right) 
								+ \sin \left( s \ln y  \right) \cos \left( s \ln  \frac{q_B}{q^*} \right) \right]. & \label{xabasym}
\end{align}
Inserting Eqs.~\eqref{angintn3} to \eqref{xabasym} into Eq.~\eqref{intne1} gives
\begin{multline}
n_3(q_B)= 8 \pi \frac{c^{\ast}_{AA}\;c_{AB}}{q_B^5}\sin^2(s\;\ln q_B/q^*)
  \int_0^\infty \frac{\cos(s\;\ln y) \ dy }{y^{4}+\frac{1}{{\cal A}}y^{2}+\left(\frac{{\cal
A}+1}{2{\cal A}}\right)^2}  \\* 
+ 8 \pi \frac{c^{\ast}_{AA}\;c_{AB}}{q_B^5}\sin(s\;\ln q_B/q^*)\cos(s\;\ln q_B/q^*)
  \int_0^\infty \frac{\sin(s\;\ln y) \ dy }{y^{4}+\frac{1}{{\cal A}}y^{2}+\left(\frac{{\cal
A}+1}{2{\cal A}}\right)^2}  +c.c. \; .
\label{n3int}
\end{multline}
Averaging out the oscillatory terms, only the first term of Eq.~\eqref{n3int} gives a 
non-vanishing result 
\begin{equation}
\langle n_3(q_B)\rangle= 4 \pi \frac{c^{\ast}_{AA}\;c_{AB}}{q_B^5}
  \int_0^\infty \frac{\cos(s\;\ln y) \ dy }{y^{4}+\frac{1}{{\cal A}}y^{2}+\left(\frac{{\cal
A}+1}{2{\cal A}}\right)^2}+c.c. \; .
\label{n3avg}
\end{equation}
Expressing cosine in the complex exponential form
\begin{align}
\cos(s\;\ln y) = \frac{e^{\imath s\;\ln y}+e^{- \imath s\;\ln y}}{2}=\frac{e^{\ln y^{\imath s}}+e^{\ln y^{-\imath s}}}{2}=
\frac{y^{\imath s}+ y^{-\imath s}}{2} \; ,
\end{align}
the integral in Eq.~\eqref{n3avg} becomes
\begin{align}
I(s)&=\int_0^\infty \frac{\cos(s\;\ln y) \ dy }{y^{4}+\frac{1}{{\cal A}}y^{2}+\left(\frac{{\cal
A}+1}{2{\cal A}}\right)^2} 
=  \frac{1}{2} \int_0^\infty dy \frac{y^{\imath s} + y^{- \imath s}  }{y^{4}+\frac{1}{{\cal A}}y^{2}+\left(\frac{{\cal A}+1}{2{\cal A}}\right)^2} \; . &   
\label{intn3a}
\end{align}
In order to extended the integration to the full real axis, the variables are changed to $y=e^\alpha$, where $dy=e^\alpha d\alpha$ and the integral $I(s)$ is expressed as
\begin{align}
I(s)&=\frac{1}{2}\int_{-\infty}^\infty \frac{d\alpha\;e^{\alpha}\;\left(e^{\imath s \alpha}+e^{-\imath s \alpha}\right)}{e^{4 \alpha}+\frac{1}{{\cal A}}e^{2\alpha}+\left(\frac{{\cal A}+1}{2{\cal A}}\right)^2}=  \operatorname{Re}\;I_\alpha(s) \; , &   
\label{intn3}
\end{align}
where $\operatorname{Re}$ denotes the real part and the integral $I_\alpha(s)$ is explicitly rewritten in terms of its poles as
\begin{align}
I_\alpha(s)&=\int_{-\infty}^\infty{ \frac{e^{\alpha(1+\imath s)} \; d\alpha}{\left(e^\alpha-e^{\alpha_1}\right) \left(e^\alpha-e^{\alpha_2}\right) \left(e^\alpha-e^{\alpha_3}\right) \left(e^\alpha-e^{\alpha_4}\right)}  } \; . &
\end{align}
The integral~\eqref{intn3} is solved using the residues theorem. The next steps are about finding the poles of the integrand $f(\alpha)$ and evaluating the residues of the poles. In order to simplify the calculation of the roots in the denominator of $f(\alpha)$, the variable $x=y^2=e^{2\alpha}$ is introduced. The zeros of the denominator in Eq.~\eqref{intn3} are found from ${ x^2 + \frac{1}{{\cal A}} x + \left(\frac{{\cal A}+1}{2{\cal A}}\right)^2=0}$ and are given by
\begin{equation}
x_{\pm}=-\frac{1}{2 {\cal A}} \pm \frac{1}{2} \sqrt{ \frac{1}{{\cal A}^2} - \left(\frac{{\cal A}+1}{2{\cal A}}\right)^2 } 
= \frac{1}{2 {\cal A}} \left( -1 \pm \imath \sqrt{ {\cal A} ({\cal A}+2} ) \right)\; .
\end{equation}
In the variable $y=\pm \sqrt{x}$ these roots read
\begin{align}
y_1& =+\sqrt{\frac{1}{2 {\cal A}}} \left( -1 + \imath \sqrt{ {\cal A} ({\cal A}+2} ) \right)^{1/2}
     =+\frac{1}{2} + \imath \sqrt{ \frac{{\cal A}+2}{4 {\cal A}} }
     =r e^{\imath \theta_3} \; ,  \label{y1} \\
y_2& =-\sqrt{\frac{1}{2 {\cal A}}} \left( -1 + \imath \sqrt{ {\cal A} ({\cal A}+2} ) \right)^{1/2}
     =-\frac{1}{2} - \imath \sqrt{ \frac{{\cal A}+2}{4 {\cal A}} }
     =r e^{- \imath (\pi-\theta_3)} = - r e^{\imath \theta_3} \; , \label{y2} \\
y_3& =+\sqrt{\frac{1}{2 {\cal A}}} \left( -1 - \imath \sqrt{ {\cal A} ({\cal A}+2} ) \right)^{1/2}
     =+\frac{1}{2} - \imath \sqrt{ \frac{{\cal A}+2}{4 {\cal A}} }
     =r e^{-\imath \theta_3} \; ,  \label{y3} \\
y_4& =-\sqrt{\frac{1}{2 {\cal A}}} \left( -1 - \imath \sqrt{ {\cal A} ({\cal A}+2} ) \right)^{1/2}
     =-\frac{1}{2} + \imath \sqrt{ \frac{{\cal A}+2}{4 {\cal A}} }
     =r e^{ \imath (\pi-\theta_3)} = - r e^{-\imath \theta_3} \; . \label{y4}
\end{align}
Finally, the roots in the variable $\alpha=\ln y$ are
\begin{align}
\alpha_1 &= \ln y_1 = \ln r + \imath \theta_3 \; ,  \label{alpha1} \\
\alpha_2 &= \ln y_2 = \ln r - \imath (\pi-\theta_3) \; , \label{alpha2}\\
\alpha_3 &= \ln y_3 = \ln r - \imath \theta_3 \; ,  \label{alpha3}\\
\alpha_4 &= \ln y_4 = \ln r + \imath (\pi-\theta_3) \; , \label{alpha4} 
\end{align}
showing that all of them
are in the complex plane, out of the real axis. 
The quantities $r$ and $\theta_3$ are defined as
\begin{align}
r&=\sqrt{\frac{{\cal A}+1}{2{\cal A}}} \; , & \label{r3} \\
\tan\theta_3& =\sqrt{\frac{{\cal A}+2}{{\cal A}}} \;\;\; \text{for} \;\;\; 0 \leq \theta_3 \leq \frac{\pi}{2} \; . & \label{theta3}
\end{align}
A rectangle of vertices $-R$, $+R$, $+R+\imath\pi$ and $-R+\imath\pi$ molds a closed path in a complex plane, as shown in Fig.~\ref{pathn3}. Due to the restriction of $\theta_3$, given in Eq.~\eqref{theta3}, this closed path encompasses the poles $\alpha_1$ and $\alpha_4$ in the upper-half plane and then, four integrals have to be worked out, namely $J_1$ which extends along the real axis from $-R$ to $+R$, $J_2$ from $+R$ to $+R+\imath\pi$, $J_3$ from $+R+\imath\pi$ to $-R+\imath\pi$ and $J_4$ from $-R+\imath\pi$ to $-R$. 

The residues theorem is used to calculate $I_\alpha(s)$. The integral in the closed path reads
\begin{align}
\oint \frac{dz\;e^{z}\;e^{\imath s z}}{e^{4 z}+\frac{1}{{\cal A}}e^{2 z}+\left(\frac{{\cal A}+1}{2{\cal A}}\right)^2} = J_1 + J_2 + J_3 + J_4 = 2 \pi \imath  \Bigl[Res(f,\alpha_1)+Res(f,\alpha_4)\Bigr]\; ,
\label{closeint}
\end{align}
where changing variables to the  $z = \alpha + \imath\; 0$, the integral $J_1$ reads
\begin{equation}
J_1= \lim_{R \to \infty} \int_{-R}^{R} \frac{d\alpha\;e^{\alpha}\;e^{\imath s \alpha}}{e^{4 \alpha}+\frac{1}{{\cal A}}e^{2\alpha}+\left(\frac{{\cal A}+1}{2{\cal A}}\right)^2} \to I_\alpha(s) \; . \label{j1n3}
\end{equation}
Analogously, setting $z = R + \imath\; y$ gives $dz =  \imath\; dy$ and the integral $J_2$ becomes
\begin{align}
J_2 &= \lim_{R \to \infty} \int_{R}^{R + \imath \pi} \frac{dz\;e^{z}\;e^{\imath s z}}{e^{4 z}+\frac{1}{{\cal A}}e^{2z}+\left(\frac{{\cal A}+1}{2{\cal A}}\right)^2} 
\; , \nonumber\\ &
=\lim_{R \to \infty} \imath \int_{0}^{\pi} \frac{dy \;  e^{ R + \imath y} \; e^{\imath s (R + \imath y) }}{e^{4 ( R + \imath y) }+\frac{1}{{\cal A}}e^{2 ( R + \imath y) }+\left(\frac{{\cal A}+1}{2{\cal A}}\right)^2}
\; , &\nonumber\\
&
=\lim_{R \to \infty} \frac{e^R}{e^{4R}} \; \imath \int_{0}^{\pi} \frac{dy \; e^{ \imath y} \; e^{\imath s (R + \imath y) }}{e^{4 \imath y }+\frac{}{{\cal A}}e^{2 \imath y } \frac{1}{e^{2 R}}+\left(\frac{{\cal A}+1}{2{\cal A}}\right)^2 \frac{1}{e^{4 R}}}
\to  0 \; . & \label{j2n3}
\end{align}
The integral $J_3$ is similar to $J_1$ and changing variables to $z = \alpha+\imath \pi$ it reads
\begin{align}
J_3 &= \lim_{R \to \infty} \int_{R+\imath \pi}^{-R+\imath \pi} \frac{dz\;e^{z}\;e^{\imath s z}}{e^{4 z}+\frac{1}{{\cal A}}e^{2 z}+\left(\frac{{\cal A}+1}{2{\cal A}}\right)^2} 
\; , &\nonumber\\ &
= \lim_{R \to \infty} \int_{R}^{-R} \frac{d\alpha\;e^{\alpha+\imath \pi}\;e^{\imath s (\alpha+\imath \pi)}}{e^{4 (\alpha+\imath \pi)}+\frac{1}{{\cal A}}e^{2 (\alpha+\imath \pi)}+\left(\frac{{\cal A}+1}{2{\cal A}}\right)^2}  
\; , &\nonumber\\
&
= \lim_{R \to \infty} \int_{-R}^{R} \frac{d\alpha\;e^{\alpha}\;e^{\imath s \alpha} e^{-s \pi} }{e^{4 \alpha}+\frac{1}{{\cal A}}e^{2\alpha}+\left(\frac{{\cal A}+1}{2{\cal A}}\right)^2}  
\to  e^{-s \pi}  J_1 =  e^{-s \pi}  I_\alpha(s) \; , & \label{j3n3}
\end{align}
where $e^{\alpha+\imath \pi}=e^\alpha e^{\imath \pi}=-e^\alpha$.
Then, performing the same transformation as in $J_2$, the last term becomes
\begin{align}
J_4 &= \lim_{R \to \infty} \int_{-R + \imath \pi }^{- R } \frac{d\alpha\;e^{\alpha}\;e^{\imath s \alpha}}{e^{4 \alpha}+\frac{1}{{\cal A}}e^{2\alpha}+\left(\frac{{\cal A}+1}{2{\cal A}}\right)^2} 
\; , \nonumber\\ &
=\lim_{R \to \infty} \imath \int_{\pi}^{0} \frac{dy \;  e^{ R + \imath y} \; e^{\imath s (R + \imath y) }}{e^{4 ( R + \imath y) }+\frac{1}{{\cal A}}e^{2 ( R + \imath y) }+\left(\frac{{\cal A}+1}{2{\cal A}}\right)^2}
\; , &\nonumber\\
&
=- \lim_{R \to \infty} J_2
\to  0 \; , & \label{j4n3}
\end{align}

Summarizing, in the limit $R\to\infty$, the integrals are $J_1=I_\alpha(s)$, $J_3=e^{-s\pi}I_\alpha(s)$ and $J_2,\;J_4\to0$. Then, from Eq.~\eqref{closeint} the integral $I_\alpha(s)$ reads
\begin{align}
I_\alpha(s)&=\frac{2\pi\imath}{1+e^{-\pi s}} \Bigl[Res(f,\alpha_1)+Res(f,\alpha_4)\Bigr] \; . \label{I1b} 
\end{align}
In order to calculate the residues of the poles $\alpha_1$ and $\alpha_4$, it is necessary to expand the $n^{th}$ root, in the limit $\alpha \to \alpha_n$, as
\begin{align}
e^\alpha-e^{\alpha_n} &= e^\alpha\left(1-e^{\alpha_n-\alpha} \right) = e^\alpha\left(1-1-\left(\alpha_n-\alpha \right)-\frac{\left(\alpha_n-\alpha \right)^2}{2 !} - ...   \right) \; , & \nonumber \\
&= e^\alpha \left(\alpha-\alpha_n \right) \left(1+\frac{\left(\alpha-\alpha_n \right)}{2 !} - ...   \right) \; . & 
\label{poleexp1}
\end{align}
Besides, notice that the exponential of each root $\alpha_n$ in Eqs.~\eqref{alpha1} to \eqref{alpha4} is already given in Eqs.~\eqref{y1} to \eqref{y4}. Taking the first order in the expansion, the residues are given by
\begin{align}
Res(f,\alpha_1) &= \lim_{\alpha \to \alpha_1} \left( \alpha- \alpha_1 \right) \frac{e^{\alpha(1+\imath s)} }{ e^\alpha \left(\alpha-{\alpha_1}\right) \left(e^\alpha-e^{\alpha_2}\right) \left(e^\alpha-e^{\alpha_3}\right) \left(e^\alpha-e^{\alpha_4}\right)} \; , &  \nonumber\\
&= \frac{e^{\alpha_1}\;e^{\imath s \alpha_1} }{ e^{\alpha_1} \left(e^{\alpha_1}-e^{\alpha_2}\right) \left(e^{\alpha_1}-e^{\alpha_3}\right) \left(e^{\alpha_1}-e^{\alpha_4}\right)} \; , & \nonumber \\
&=  \frac{e^{\imath s\alpha_1}}{8 \imath r^3 \sin \theta_3 \cos \theta_3 e^{\imath \theta_3}}\; , & \nonumber\\
&=  \frac{1}{2 \imath} \sqrt{\frac{{\cal A}}{{\cal A}+2}} \frac{1}{\cos \theta_3} e^{\imath s \left(\ln r + \imath \theta_3 \right) - \imath \theta_3} \; , & \label{resalpha1}
\end{align}
and
\begin{align}
Res(f,\alpha_4) &= \lim_{\alpha \to \alpha_4} \left( \alpha- \alpha_4 \right) \frac{e^{\alpha(1+\imath s)} }{ e^\alpha \left(\alpha-{\alpha_4}\right) \left(e^\alpha-e^{\alpha_1}\right) \left(e^\alpha-e^{\alpha_2}\right) \left(e^\alpha-e^{\alpha_3}\right) } \; , &  \nonumber\\
&= \frac{e^{\alpha_4}\;e^{\imath s \alpha_4} }{ e^{\alpha_4}  \left(e^{\alpha_4}-e^{\alpha_1}\right) \left(e^{\alpha_4}-e^{\alpha_2}\right) \left(e^{\alpha_4}-e^{\alpha_3}\right) } \; , & \nonumber \\
&=  \frac{e^{\imath s\alpha_4}}{8 \imath r^3 \sin \theta_3 \cos \theta_3 e^{-\imath \theta_3}}\; , & \nonumber\\
&=  \frac{1}{2 \imath} \sqrt{\frac{{\cal A}}{{\cal A}+2}} \frac{1}{\cos \theta_3} e^{\imath s \left[\ln r + \imath (\pi-\theta_3) \right] + \imath \theta_3} \; , & \label{resalpha4}
\end{align}

Inserting the residues of the poles from Eqs.~\eqref{resalpha1} and \eqref{resalpha4} into Eq.~\eqref{I1b} gives
\begin{align}
I_\alpha(s)&=\frac{\pi} {1+e^{-\pi s}}  \sqrt{\frac{{\cal A}}{{\cal A}+2}} \frac{1}{\cos \theta_3}
\left( e^{\imath s \left(\ln r + \imath \theta_3 \right) - \imath \theta_3}
+ e^{\imath s \left[\ln r + \imath (\pi-\theta_3) \right] + \imath \theta_3}\right) \; , & \nonumber \\ 
&=\frac{2 \pi e^{-s \pi/2}} {1+e^{-\pi s}}  \sqrt{\frac{{\cal A}}{{\cal A}+2}} \frac{1}{\cos \theta_3}
e^{\imath s \ln r} \cosh \left[ s \left(\theta_3 - \frac{\pi}{2} \right) + \imath \theta_3 \right] \; , & \nonumber \\ 
&=\frac{ \pi } {\cosh \left( s \frac{\pi}{2} \right) }  \sqrt{\frac{{\cal A}}{{\cal A}+2}} \frac{1}{\cos \theta_3}
\Bigl[\cos \left( s \ln r \right) + \imath \sin \left( s \ln r \right) \Bigr] & \nonumber \\
& \hskip 2cm \times \Bigl\{ \cos \theta_3 \cosh \left[ s \left(\frac{\pi}{2} - \theta_3 \right) \right] - \imath \sin \theta_3 \sinh \left[ s \left(\frac{\pi}{2} - \theta_3 \right) \right] \Bigr\} \; , & \nonumber 
\\ 
&=\frac{ \pi } {\cosh \left( s \frac{\pi}{2} \right) }  \sqrt{\frac{{\cal A}}{{\cal A}+2}} \frac{1}{\cos \theta_3}  \nonumber \\
& \hskip 3.0cm \times \Bigl\{ \cos \theta_3 \cos \left( s \ln r \right)  \cosh \left[ s \left(\frac{\pi}{2} - \theta_3 \right) \right] 
 \nonumber\\* 
&  \hskip 5.0cm 
+ \sin \theta_3 \sin \left( s \ln r \right)  \sinh \left[ s \left(\frac{\pi}{2} - \theta_3 \right) \right] \Bigr\} & \nonumber \\
& \hskip 3.0cm \times - \imath \Bigl\{ \sin \theta_3 \cos \left( s \ln r \right)  \sinh \left[ s \left(\frac{\pi}{2} - \theta_3 \right) \right] \nonumber\\* 
&  \hskip 5.0cm 
+ \cos \theta_3 \sin \left( s \ln r \right)  \cosh \left[ s \left(\frac{\pi}{2} - \theta_3 \right) \right] \Bigr\} \; ,& \nonumber \\
&=\frac{ \pi } {\cosh \left( s \frac{\pi}{2} \right) }  \Bigl\{ \sqrt{\frac{{\cal A}}{{\cal A}+2}} \cos \left( s \ln r \right)  \cosh \left[ s \left(\frac{\pi}{2} - \theta_3 \right) \right] \nonumber\\* 
&  \hskip 4.0cm 
+ \sin \left( s \ln r \right)  \sinh \left[ s \left(\frac{\pi}{2} - \theta_3 \right) \right] \Bigr\} & \nonumber \\
& \hskip 2.0cm \times - \imath \Bigl\{ \cos \left( s \ln r \right)  \sinh \left[ s \left(\frac{\pi}{2} - \theta_3 \right) \right] 
\nonumber\\* 
&  \hskip 3.0cm 
+ \sqrt{\frac{{\cal A}}{{\cal A}+2}} \sin \left( s \ln r \right)  \cosh \left[ s \left(\frac{\pi}{2} - \theta_3 \right) \right] \Bigr\} \; ,& 
\end{align}
where the  the real, $\operatorname{Re}$, and imaginary, $\operatorname{Im}$, parts of $I_\alpha(s)$ are explicitly given by 
\begin{multline}
\operatorname{Re}\;I_\alpha(s)=\frac{\pi}{\cosh\left(\frac{s\pi}{2}\right)}  \left\{  \sqrt{\frac{{\cal A}}{{\cal A}+2}}\cos\left(s\ln r \right)\cosh\left[s\left(\frac{\pi}{2}-\theta_3\right)\right] \right.  \\*
 \left. + \sin\left(s\ln r \right)\sinh\left[s\left(\frac{\pi}{2}-\theta_3\right)\right]\right\} \; ,  \label{n3re}
\end{multline}
\begin{multline}
\operatorname{Im}\;I_\alpha(s)=\frac{-\pi}{\cosh\left(\frac{s\pi}{2}\right)}  \left\{\sqrt{\frac{{\cal A}}{{\cal A}+2}}\sin\left(s\ln r \right)\cosh\left[s\left(\frac{\pi}{2}-\theta_3\right)\right] \right.  \\*
\left. + \cos\left(s\ln r \right)\sinh\left[s\left(\frac{\pi}{2}-\theta_3\right)\right]\right\} \; , \label{n3im}
\end{multline}
with $r$ and $\theta_3$ defined respectively in Eqs.~\eqref{r3} and \eqref{theta3}.
 
Finally, from equations \eqref{n3avg}, \eqref{intn3} and \eqref{n3re}, the non-oscillating part of $n_3(q_B)$ is given by
\begin{multline}
\langle n_3(q_B)\rangle= \frac{4 \pi^2 c_{AA}\;c_{AB}}{q_B^5 \cosh\left(\frac{s\pi}{2}\right)}  \left\{\sqrt{\frac{{\cal A}}{{\cal A}+2}}\cos\left(s\ln \sqrt{\frac{{\cal A}+1}{2{\cal A}}}\right)\cosh\left[s\left(\frac{\pi}{2}-\theta_3\right)\right] \right.  \\
 \left. + \sin\left(s\ln \sqrt{\frac{{\cal A}+1}{2{\cal A}}}\right)\sinh\left[s\left(\frac{\pi}{2}-\theta_3\right)\right]\right\} \; , 
\end{multline}
where $\tan\theta_3=\sqrt{\frac{{\cal A}+2}{{\cal A}}}$ for $0\leq\theta_3\leq\pi/2$. The special case ${\cal A}=1$ yields $\theta_3=\pi/3$ and $\langle n_3(q_B)\rangle= 4 \pi^2 |c_{AA}|^2 \cosh\left(\frac{s\pi}{6}\right) /\left(q_B^5 \sqrt{3} \cosh\left(\frac{s\pi}{2}\right)\right)$.

\subsection{Asymptotic contribution from $n_4(q_B)$} \label{cn4}
As in the 2D case, this is also the most complicated out of the four
additive terms in the one-body momentum density, since the angular
dependence in both spectator arguments of Eq.~\eqref{n4} can not be removed
simultaneously by a change in variables.  Then, although Eqs.~\eqref{n3} and \eqref{n4} are similar, it is not possible to extend the results from the previous case, $n_3(q_B)$, to obtain the non-oscillating part of $n_4(q_B)$. Defining $\mathbf{p}_B=\frac{\mathbf{q}_B}{2}\mathbf{y}$ and dropping the three-body energy, Eq.~\eqref{n4} becomes
\begin{equation}
n_4(q_B)= \frac{4\pi}{q_B}\int_0^\infty{\frac{y^2 dy}{\left(y^2+\frac{{\cal A}+2}{{\cal A}}\right)^2}}\int_{-1}^{+1} dx\;
 \chi^{\ast}_{AB}(q_B x_-)\chi_{AB}(q_B x_+)+c.c. \; ,
 \label{n4varc}
\end{equation}
where $x_\pm=\frac{1}{2}\sqrt{1+y^2\pm2yx}$. The asymptotic spectator function for a shifted argument is given in Eq.~\eqref{xabasym} and  $\chi_{AB}(q_B x_\pm)$ reads
\begin{align}
\chi_{AB}(q_B \; x_\pm)&= \frac{c_{AB}}{x_\pm^2\;q_B^2} \left[ \sin \left( s \ln  \frac{q_B}{q^*}  \right) \cos \left(s \ln x_\pm  \right) 
+ \sin \left( s \ln x_\pm  \right) \cos \left( s \ln  \frac{q_B}{q^*} \right) \right]. & \label{xabasymx}
\end{align}

The spectator function in Eq.~\eqref{n4varc} are replaced by their asymptotic form from Eq.~\eqref{xabasymx}. The integral is then separated in three terms, namely
\begin{align}
n_4(q_B)&= \frac{8\pi |c_{AB}|^2 \sin^2\left(s \ln\frac{q_B}{q^\ast}\right)}{q_B^5} \nonumber \\
& \hskip 2cm \times \int_0^\infty{\frac{y^2 dy}{\left(y^2+\frac{{\cal A}+2}{{\cal A}}\right)^2}}\int_{-1}^{+1} \frac{dx}{x_+^2x_-^2}\cos\left(s \ln x_+\right)\cos\left(s \ln x_-\right)  \nonumber\\
&+\frac{8\pi |c_{AB}|^2 \cos^2\left(s \ln\frac{q_B}{q^\ast}\right)}{q_B^5} \nonumber \\
& \hskip 2cm \times \int_0^\infty{\frac{y^2 dy}{\left(y^2+\frac{{\cal A}+2}{{\cal A}}\right)^2}}\int_{-1}^{+1} \frac{dx}{x_+^2x_-^2}\sin\left(s \ln x_+\right)\sin\left(s \ln x_-\right)  \nonumber \\ 
&+\frac{4\pi |c_{AB}|^2 \sin\left(s \ln\frac{q_B}{q^\ast}\right)\cos\left(s \ln\frac{q_B}{q^\ast}\right)}{q_B^5}  \nonumber \\
& \hskip 2cm \times \int_0^\infty{\frac{y^2 dy}{\left(y^2+\frac{{\cal A}+2}{{\cal A}}\right)^2}}\int_{-1}^{+1} \frac{dx}{x_+^2x_-^2}\sin\left[s\ln\left(x_+ x_-\right)\right] \; . & \label{n4osc}
\end{align} 
As it was done for $n_3(q_B)$, averaging out the oscillatory term, only the two first terms on the right-hand-side of Eq.~\eqref{n4osc} give a non-vanishing contribution. The angular integration is performed using that
\begin{equation}
\int{dx\left(\frac{\beta+x}{\beta-x}\right)^{\pm\imath s/2}\left(\beta^2-x^2\right)^{-1}}=\pm \left(\frac{\beta+x}{\beta-x}\right)^{\pm\imath s/2} \left(\imath \beta s\right)^{-1} \; . \label{angintn4}
\end{equation}
Then, the angular part of Eq.~\eqref{n4osc} has to be written as Eq.~\eqref{angintn4}. The denominator in the integrand of the  first two terms on the right hand side of Eq.~\eqref{n4osc} is
\begin{align}
x_+^2\; x_-^2 & =\frac{1}{4}\left( 1+y^2-2 y x \right) \frac{1}{4}\left( 1+y^2+2 y x \right) \nonumber\\
&=\frac{1}{16} \left(\alpha-v \right) \left(\alpha+v \right)= \frac{1}{16}\left(\alpha^2-v^2 \right) \; , & \label{xpm}
\end{align} 
where $\alpha= 1+y^2$ and $v=2 y x$. In the same notation, the numerator in the integrand of the first two terms on the right hand side of Eq.~\eqref{n4osc} are given by
\begin{align}
D1 &=\cos\left(s \ln x_+\right)\cos\left(s \ln x_-\right) \nonumber \\*
& =\frac{1}{4} \left[ \left( \frac{\alpha + v}{4} \right)^{\imath s/2}+ \left( \frac{\alpha + v}{4} \right)^{-\imath s/2} \right] \left[ \left( \frac{\alpha - v}{4} \right)^{\imath s/2}+ \left( \frac{\alpha - v}{4} \right)^{-\imath s/2} \right] \; , \nonumber\\*
& =\frac{1}{4} \left[ \left( \frac{\alpha^2 - v^2}{16} \right)^{\imath s/2}+ \left( \frac{\alpha + v}{\alpha - v} \right)^{\imath s/2} +  \left( \frac{\alpha + v}{\alpha - v} \right)^{-\imath s/2}+ \left( \frac{\alpha^2 - v^2}{16} \right)^{-\imath s/2} \right] \; , \label{D1}
\end{align} 
and
\begin{align}
D2 &=\sin\left(s \ln x_+\right)\sin\left(s \ln x_-\right) \nonumber \\*
& =\frac{-1}{4} \left[ \left( \frac{\alpha + v}{4} \right)^{\imath s/2} - \left( \frac{\alpha + v}{4} \right)^{-\imath s/2} \right] \left[ \left( \frac{\alpha - v}{4} \right)^{\imath s/2} - \left( \frac{\alpha - v}{4} \right)^{-\imath s/2} \right] \; , \nonumber\\*
& =\frac{-1}{4} \left[ \left( \frac{\alpha^2 - v^2}{16} \right)^{\imath s/2} - \left( \frac{\alpha + v}{\alpha - v} \right)^{\imath s/2} -  \left( \frac{\alpha + v}{\alpha - v} \right)^{-\imath s/2}+ \left( \frac{\alpha^2 - v^2}{16} \right)^{-\imath s/2} \right] \; . \label{D2}
\end{align} 
Collecting Eqs.~\eqref{xpm}, ~\eqref{D1} and ~\eqref{D2} together and inserting them into Eq.~\eqref{n4osc}, the non-oscillating part of $n_4(q_B)$ is given by
\begin{align}
n_4(q_B)&= \frac{ 4 \pi |c_{AB}|^2 }{q_B^5}\int_0^\infty{\frac{y^2 dy}{\left(y^2+\frac{{\cal A}+2}{{\cal A}}\right)^2}} I_x(y)  \; , & \label{n4osc1}
\end{align} 
where the angular integral $I_x(y)$ reads 
\begin{align}
I_x(y)&= \int_{-1}^{+1} \frac{dx}{x_+^2x_-^2} \Bigl[ \cos\left(s \ln x_+\right)\cos\left(s \ln x_-\right) +\sin\left(s \ln x_+\right)\sin\left(s \ln x_-\right) \Bigr] \; , \nonumber \\* 
&= \int_{-1}^{+1} \frac{dx}{2}  
\left[\left( \frac{1+y^2 +2 y x}{1+y^2 - 2 y x} \right)^{\imath s/2} + \left( \frac{1+y^2 +2 y x}{1+y^2 - 2 y x} \right)^{-\imath s/2} \right] \frac{16}{\left(1+y^2\right)^2 - 4 x^2 y^2 } \; ,  \nonumber\\*  
&= \frac{8}{4 y^2} \int_{-1}^{+1} dx  
\left[\left( \frac{\frac{1+y^2}{2y} + x}{\frac{1+y^2}{2y} - x} \right)^{\imath s/2} + \left( \frac{\frac{1+y^2}{2y}- x}{\frac{1+y^2}{2y}- x} \right)^{-\imath s/2} \right] \left[\left(\frac{1+y^2}{2y}\right)^2 -  x^2 \right]^{-1} \; ,  
\end{align} 
which looks like the expression in Eq.~\eqref{angintn4}. Then, the result is
\begin{align}
I_x(y)&= \frac{2}{y^2} \frac{2y}{\imath s (1 + y^2)}
\left\{ \left. \left[ \left( \frac{\frac{1+y^2}{2y} + x}{\frac{1+y^2}{2y} - x} \right)^{\imath s/2}  \right] \right|_{x=-1}^{1} 
-\left. \left[ \left( \frac{\frac{1+y^2}{2y} + x}{\frac{1+y^2}{2y} - x} \right)^{-\imath s/2}  \right] \right|_{x=-1}^{1} \right\}  \; ,  \nonumber \\*
&=\frac{8}{\imath s y (1 + y^2)} 
\left[ \left( \frac{\frac{1+y^2}{2y} + 1}{\frac{1+y^2}{2y} - 1} \right)^{\imath s/2} 
- \left( \frac{\frac{1+y^2}{2y} + 1}{\frac{1+y^2}{2y} - 1} \right)^{-\imath s/2}  \right]  \; ,  \nonumber \\*
&=\frac{8}{\imath s y (1 + y^2)} 
\left[ \left( \sqrt{\frac{(y+1)^2}{(y-1)^2}} \right)^{\imath s} 
- \left( \sqrt{\frac{(y+1)^2}{(y-1)^2}} \right)^{-\imath s}  \right]  \; .  \label{Ix}
\end{align} 

Combining the expressions in Eqs.~\eqref{n4osc1} and \eqref{Ix}, the non-oscillating part of $n_4(q_B)$ reads
\begin{align}
\left\langle n_4(q_B) \right\rangle &=\frac{32\pi |c_{AB}|^2 }{\imath\;s\; q_B^5}\int_0^\infty{\frac{y\; dy}{\left(y^2+\frac{{\cal A}+2}{{\cal A}}\right)^2 (1+y^2)}\left[\left(\frac{y+1}{|y-1|}\right)^{\imath s}-\left(\frac{y+1}{|y-1|}\right)^{-\imath s}\right]} \; , \nonumber\\
&=\frac{64\pi |c_{AB}|^2 }{s\; q_B^5} \int_0^\infty{\frac{y\; dy}{\left(y^2+\frac{{\cal A}+2}{{\cal A}}\right)^2 (1+y^2)} \sin \left( s \ln\left(\frac{y+1}{|y-1|}\right) \right)} \; , \label{n4avg}
\end{align}
however the absolute value complicates the calculation of the integral. Circumventing this problem, the integral is split in two pieces: $y \in [0,1]$ and $y \in \left[\right.1,\infty\left[\right.$ and a new variable is introduced in each piece, i.e., the variable transformation $y=\frac{x-1}{x+1}$ is made in the first piece and $y=\frac{x+1}{x-1}$ in the second one \cite{castinPRA2011}. Notice that in both cases $x \in [1,\infty[$. In detail
\begin{align}
\left\langle n_4(q_B) \right\rangle &=\frac{64\pi |c_{AB}|^2 }{s\; q_B^5} \Bigl[I_<(s) + I_>(s) \Bigr]  \; , 
\label{n4ip}
\end{align}
where
\begin{align}
I_<(s) &= \int_0^1{\frac{y\; dy}{\left(y^2+\frac{{\cal A}+2}{{\cal A}}\right)^2 (1+y^2)} \sin \left( s \ln\left(\frac{y+1}{|y-1|}\right) \right)} \; , \nonumber \\*
&= \int_1^\infty \frac{2 \; dx}{(x+1)^2} \frac{x-1}{x+1} \sin \left(s \ln x \right)  \bigg/ \left\{ \left[\left( \frac{x-1}{x+1} \right)^2  +\frac{{\cal A}+2}{{\cal A}} \right]^2 \left[\left( \frac{x-1}{x+1} \right)^2 + 1 \right] \right\} , \nonumber \\
&= \int_1^\infty \frac{2 (x-1) \; dx}{(x+1)^3}  
\frac{\sin \left(s \ln x \right) (x+1)^6} 
{\left[\left( \frac{{\cal A}+2}{{\cal A}} +1 \right) \left( x^2+1 \right) +  \left( \frac{{\cal A}+2}{{\cal A}} -1 \right) 2 x  \right]^2 2  \left(x^2+1 \right)} \; , \nonumber \\
&= \int_1^\infty dx \frac{(x-1) (x+1)^3 \sin \left(s \ln x \right)}
{\left[ \frac{2}{{\cal A}} \left( {\cal A} +1 \right) \left( x^2+1 \right) + \frac{4}{{\cal A}} x  \right]^2 \left(x^2+1 \right)} \; , \label{I<}
\end{align}
and 
\begin{align}
I_>(s) &= \int_1^\infty{\frac{y\; dy}{\left(y^2+\frac{{\cal A}+2}{{\cal A}}\right)^2 (1+y^2)} \sin \left( s \ln\left(\frac{y+1}{|y-1|}\right) \right)} \; , \nonumber \\*
&= -\int_\infty^1 \frac{2 \; dx}{(x-1)^2} \frac{x+1}{x-1} \sin \left(s \ln x \right) \bigg/ \left\{ \left[\left( \frac{x+1}{x-1} \right)^2  +\frac{{\cal A}+2}{{\cal A}} \right]^2 \left[\left( \frac{x+1}{x-1} \right)^2 + 1 \right] \right\} , \nonumber \\
&= \int_1^\infty \frac{2 (x+1) \; dx}{(x-1)^3}  \frac{\sin \left(s \ln x \right) (x-1)^6} 
{\left[\left( \frac{{\cal A}+2}{{\cal A}} +1 \right) \left( x^2+1 \right) -  \left( \frac{{\cal A}+2}{{\cal A}} -1 \right) 2 x  \right]^2 2  \left(x^2+1 \right)} \; , \nonumber \\
&= \int_1^\infty dx \frac{(x+1)(x-1)^3 \sin \left(s \ln x \right)}
{\left[ \frac{2}{{\cal A}} \left( {\cal A} +1 \right) \left( x^2+1 \right) - \frac{4}{{\cal A}} x  \right]^2 \left(x^2+1 \right)} \; . \label{I>}
\end{align}
The sum of the two integrals in Eqs.~\eqref{I<} and \eqref{I>}, $I_+(s)$, is found to be 
\begin{align}
I_+(s)&= \int_1^\infty dx \frac{(x+1)(x-1) \sin \left(s \ln x \right)}{x^2+1}  \nonumber\\*
& \hskip 2cm \times \left\{ \frac{(x+1)^2}{\left[ \frac{2}{{\cal A}} \left( {\cal A} +1 \right) \left( x^2+1 \right) + \frac{4}{{\cal A}} x  \right]^2 }   + \frac{(x-1)^2}
{\left[ \frac{2}{{\cal A}} \left( {\cal A} +1 \right) \left( x^2+1 \right) - \frac{4}{{\cal A}} x  \right]^2 } \right\} \; , \nonumber\\ 
&= \frac{16}{{\cal A}^2}\int_1^\infty dx \sin \left(s \ln x \right) \frac{ x^2-1 }{x^2+1} \nonumber\\*
& \hskip 0.3cm \times \frac{ 2 \left(x^2+1\right) \left[ \frac{\left({\cal A}+1\right)^2}{4} x^4 + \left( \frac{\left({\cal A}+1\right)^2}{2}+1 \right) x^2 + 1 \right] + 4 x \Bigl[ \left( {\cal A} +1 \right) \left( x^3+x \right) \Bigr] }
{\left[ \frac{4}{{\cal A}^2} \left( {\cal A} +1 \right)^2 \left( x^2+1 \right)^2 - \frac{16}{{\cal A}^2} x^2  \right]^2}  \; , \nonumber\\
&= {\cal A}^2\int_1^\infty dx \sin \left(s \ln x \right) \frac{x^2-1 }{x^2+1} 
\nonumber\\* & \hskip 3cm \times 
\frac{ \frac{\left({\cal A}+1\right)^2}{2} x^6 +\left(\frac{3}{2}{\cal A}^2+ 7 {\cal A} +\frac{15}{2} \right) + \left({\cal A}+3 \right)^2 x^2 + 2} 
{\Bigl\{ \left( {\cal A} +1 \right)^2 x^4 + \bigl[ 2 \left( {\cal A}^2+1 \right)^2-4 \bigr] x^2+ \left( {\cal A}^2 + 1 \right)^2 \Bigr\}^2}  \; , \nonumber\\
&= \frac{{\cal A}^2}{2 \left( {\cal A} +1 \right)^4} \int_1^\infty dx \sin \left(s \ln x \right) \frac{x^2-1 }{x^2+1} 
\nonumber\\* & \hskip 3cm \times 
\frac{ \left({\cal A}+1\right)^2\left(x^6+1 \right) + \left(3 {\cal A}^2 -2 {\cal A} -1 \right) \left( x^4 + x^2 \right)} 
{\left[ x^4 + \left( 2 - \frac{4}{\left( {\cal A} +1 \right)^2} \right) x^2 + 1 \right]^2}  \; . \label{I+}
\end{align}
From Eq.~\eqref{n4ip}, $\left\langle n_4(q_B) \right\rangle$ becomes
\begin{align}
\left\langle n_4(q_B) \right\rangle &=\frac{ 32 \pi |c_{AB}|^2 }{s\; q_B^5}  \frac{{\cal A}^2}{\left( {\cal A} +1 \right)^4} 
\nonumber\\ & \hskip -1cm \times 
\int_1^\infty dx \sin \left(s \ln x \right) \frac{x^2-1 }{x^2+1} 
\frac{ \left({\cal A}+1\right)^2\left(x^6+1 \right) + \left(3 {\cal A}^2 -2 {\cal A} -1 \right) \left( x^4 + x^2 \right)} 
{\left[ x^4 + \left( 2 - \frac{4}{\left( {\cal A} +1 \right)^2} \right) x^2 + 1 \right]^2}  \; .
\label{n4ip1}
\end{align}

The residue theorem is used to calculate the non-oscillating part of $n_4(q_B)$. The poles of Eq.~\eqref{n4ip1} are given by
\begin{align}
x_1 &= \frac{1}{{\cal A}+1}\left(1+ \imath \sqrt{{\cal A} \left( {\cal A}+1 \right)} \right) = \sqrt{\frac{1+{\cal A}^2+ 2 {\cal A}}{\left({\cal A}+1 \right)^2}} e^{\imath \theta_4} = e^{\imath \theta_4} \; , \label{x1n4} \\
x_2 &= \frac{1}{{\cal A}+1}\left(-1+ \imath \sqrt{{\cal A} \left( {\cal A}+1 \right)} \right) =  e^{\imath \left(\pi - \theta_4 \right)} = -e^{- \imath  \theta_4 } \; , \label{x2n4} \\
x_3 &= \frac{1}{{\cal A}+1}\left(1 - \imath \sqrt{{\cal A} \left( {\cal A}+1 \right)} \right) = e^{-\imath \theta_4} \; , \label{x3n4} \\
x_4 &= \frac{1}{{\cal A}+1}\left(-1- \imath \sqrt{{\cal A} \left( {\cal A}+1 \right)} \right) =  e^{-\imath \left(\pi - \theta_4 \right)} = -e^{ \imath  \theta_4 } \; , \label{x4n4} \\
x_5 &= \imath = e^{\imath \frac{\pi}{2}} \; , \label{x5n4} \\
x_6 &= -\imath = e^{-\imath \frac{\pi}{2}} \; , \label{x6n4}
\end{align}
with
\begin{equation}
\tan\theta_4=\sqrt{{\cal A}({\cal A}+2)} \;\;\; \text{for} \;\;\; 0\leq\theta_4\leq\frac{\pi}{2} \; .
\label{theta4}
\end{equation}
Notice that the roots $x_n$ with $1 \leq n \leq 4$ are of order two.
Changing variables to $x=e^\alpha$, the domain of integration is now to the entire real axis, namely
\begin{align}
\left\langle n_4(q_B) \right\rangle &=\frac{ 32 \pi |c_{AB}|^2 }{s\; q_B^5}  \frac{{\cal A}^2}{\left( {\cal A} +1 \right)^4} 
\int_0^\infty d \alpha e^{ \alpha } \sin \left(s \;\alpha \right) \frac{e^{ 2\alpha }-1 }{e^{ 2 \alpha }+1} 
\nonumber\\ & \hskip 3cm \times 
\frac{ \left({\cal A}+1\right)^2\left(e^{ 6 \alpha }+1 \right) + \left(3 {\cal A}^2 -2 {\cal A} -1 \right) \left( e^{ 4 \alpha } + e^{ 2 \alpha } \right)} 
{\left[ e^{ 4 \alpha } + \left( 2 - \frac{4}{\left( {\cal A} +1 \right)^2} \right) e^{ 2 \alpha } + 1 \right]^2}  \; , \nonumber\\
&=\frac{ 16 \pi |c_{AB}|^2 }{s\; q_B^5}  \frac{{\cal A}^2}{\left( {\cal A} +1 \right)^4} 
\operatorname{Im} I_\alpha(s) \; ,
\label{n4ialpha}
\end{align}
where the integral $I_\alpha(s)$ is given by
\begin{align}
I_\alpha(s) & =\int_{-\infty}^\infty d \alpha e^{ \alpha \left(1 + \imath\; s \right) } \frac{e^{ 2\alpha }-1 }{e^{ 2 \alpha }+1} 
\frac{ \left({\cal A}+1\right)^2\left(e^{ 6 \alpha }+1 \right) + \left(3 {\cal A}^2 -2 {\cal A} -1 \right) \left( e^{ 4 \alpha } + e^{ 2 \alpha } \right)} 
{\left[ e^{ 4 \alpha } + \left( 2 - \frac{4}{\left( {\cal A} +1 \right)^2} \right) e^{ 2 \alpha } + 1 \right]^2}  \; . 
\label{Ialphan4}
\end{align}

Writing the integrand $f(\alpha)$ explicitly in terms of its poles gives
\begin{equation}
f(\alpha)=\frac{e^{\alpha(1+\imath s)} \left(e^{2\alpha}-1\right) \Bigl[({\cal A}+1)^2 \left(e^{6\alpha}+1\right)+(3{\cal A}^2-2{\cal A}-1) \left(e^{2\alpha}+e^{4\alpha} \right) \Bigr]} {\left(e^\alpha-e^{\alpha_5}\right) \left(e^\alpha-e^{\alpha_6}\right) 
\Bigl[ \left(e^\alpha-e^{\alpha_1}\right) \left(e^\alpha-e^{\alpha_2}\right) \left(e^\alpha-e^{\alpha_3}\right) \left(e^\alpha-e^{\alpha_4}\right) \Bigr]^2} \; . 
\label{falpha}
\end{equation}
It is possible to see, from Eqs.~\eqref{x1n4} to \eqref{x6n4} that all of the roots on the denominator of the integrand $f(\alpha)$ are  on the imaginary axis. They are given by
\begin{equation}
\alpha_1=\imath \theta_4, \ \alpha_2=\imath (\pi-\theta_4), \ \alpha_3=-\imath \theta_4, \ \alpha_4=-\imath (\pi-\theta_4), \ \alpha_5=\imath \frac{\pi}{2}, \ \alpha_6=-\imath \frac{\pi}{2} \ ,
\label{n4roots}
\end{equation} 
where $\theta_4$ is given in Eq.~\eqref{theta4}. Notice that $\alpha_5$ and $\alpha_6$ are simple poles in Eq.~\eqref{Ialphan4}, while $\alpha_1$, $\alpha_2$, $\alpha_3$, and $\alpha_4$ are poles of second order.

Evaluating the contour integral, the same closed path used in the calculation of $n_3(q_B)$ (see in Fig.~\ref{pathn3}) is chosen, namely a rectangle of vertices $-R$, $+R$, $+R+\imath\pi$ and $-R+\imath\pi$, which now, due to the restriction of $\theta_4$ in Eq.~\eqref{theta4}, encompasses the poles $\alpha_1$, $\alpha_2$ and $\alpha_5$ in the upper-half plane. Once more, four integrals have to worked out, i.e., $J_1$ which extends along the real axis from $-R$ to $+R$, $J_2$ from $+R$ to $+R+\imath\pi$, $J_3$ from $+R+\imath\pi$ to $-R+\imath\pi$ and $J_4$ from $-R+\imath\pi$ to $-R$.  The integral in the closed path reads
\begin{align}
\oint f(z) \; dz = J_1 + J_2 + J_3 + J_4 = 2 \pi \imath  \Bigl[Res(f,\alpha_1)+Res(f,\alpha_2)+Res(f,\alpha_5)\Bigr]\; ,
\label{closeintn4}
\end{align}
with $f(z)$ defined in Eq.~\eqref{falpha}. The four integrals are worked out as in Eqs.~\eqref{j1n3} to \eqref{j4n3} and it turns out that $J_1=I_\alpha(s)$, $J_3=e^{-s\pi}I_\alpha(s)$ and $J_2,\;J_4\to0$.  In this way, the integral $I_\alpha(s)$ from Eq.~\eqref{Ialphan4} reads
\begin{equation}
I_\alpha(s)=\frac{2\pi\imath}{1+e^{-\pi s}}\Bigl[Res(f,\alpha_1)+Res(f,\alpha_2)+Res(f,\alpha_5)\Bigr] \ .
\label{Ialphares}
\end{equation}
Calculating the residues is laborious and the details about the calculation of $I_\alpha(s)$ are given in Appendix~\ref{residues}. The real and imaginary part of $I_\alpha(s)$ are given by
\begin{equation}
\operatorname{Re}\;I_\alpha(s)= \frac{\pi ({\cal A}+1)^3 {\cal A}}{2 \sqrt{{\cal A}({\cal A}+2)} \cosh\left(\frac{s\pi}{2}\right)}  \cosh\left[s\left(\frac{\pi}{2}- \theta_4\right) \right] \; , 
\end{equation}
\begin{multline}
\operatorname{Im}\;I_\alpha(s)=\frac{\pi ({\cal A}+1)^4}{2 \sqrt{{\cal A}({\cal A}+2)} \cosh\left(\frac{s\pi}{2}\right)} \\* \times \left\{\sqrt{{\cal A}({\cal A}+2)}\sinh\left[s\left(\frac{\pi}{2}- \theta_4\right) \right] 
-\frac{s\; {\cal A} }{{\cal A}+1}\cosh\left[s\left(\frac{\pi}{2}- \theta_4\right) \right]\right\} \; . \label{n4im}
\end{multline}

Finally, combining Eqs.~\eqref{n4ialpha} and \eqref{n4im}, the non-oscillating part of $n_4(q_B)$  reads
\begin{multline}
\left\langle n_4(q_B) \right\rangle=\frac{ 8 \pi^2 |c_{AB}|^2 {\cal A}^2 }{s\; q_B^5 \sqrt{{\cal A}({\cal A}+2)} \cosh\left(\frac{s\pi}{2}\right)}  \\*\times \left\{\sqrt{{\cal A}({\cal A}+2)}\sinh\left[s\left(\frac{\pi}{2}- \theta_4\right) \right] -\frac{s\; {\cal A} }{{\cal A}+1}\cosh\left[s\left(\frac{\pi}{2}- \theta_4\right) \right]\right\} \; ,
\end{multline} 
where $\tan\theta_4=\sqrt{{\cal A}({\cal A}+2)}$ for $0\leq\theta_4\leq\pi/2$. The special case ${\cal A}=1$ yields $\theta_4=\pi/3$ and $\langle n_4(q_B)\rangle= 8 \pi^2 |c_{AA}|^2\left[\sinh\left(\frac{s\pi}{6}\right)-s/(2\sqrt{3})\cosh\left(\frac{s\pi}{6}\right) \right]  /\left[s\; q_B^5\;  \cosh\left(\frac{s\pi}{2}\right)\right]$.

\chapter{Detailed calculation of the residue} \label{residues}

It is not expected that the reader try to reproduce the calculation in this Appendix. The purpose here is just to let registered all the steps made in some very inspired days of work. 

The integral $I_\alpha(s)$ in Eq.~\eqref{Ialphares} is given by
\begin{equation}
I_\alpha(s)=\frac{2\pi\imath}{1+e^{-\pi s}}\Bigl[Res(f,\alpha_1)+Res(f,\alpha_2)+Res(f,\alpha_5)\Bigr] \ ,
\label{Ialpharesapp}
\end{equation}
where $f(\alpha)$ is given in Eq.~\eqref{falpha} and $\alpha_1$, $\alpha_2$ and $\alpha_5$ in Eq.~\eqref{n4roots}. Remember that $\alpha_5$ is a simple pole, while $\alpha_1$ and $\alpha_2$ are poles of second order.

Starting with the simplest case and using the pole expansion from Eq.~\eqref{poleexp1}, the last term on the right-hand-side of Eq.~\eqref{Ialpharesapp} reads
\begin{align}
Res(f,\alpha_5) &= \lim_{\alpha \to \alpha_5} \left( \alpha- \alpha_5 \right) 
  \frac{ e^{ \alpha \left(1 + \imath\; s \right) } \left(e^{ 2\alpha }-1\right) }{ e^{ \alpha }\left( \alpha- \alpha_5 \right) \left(e^{ \alpha }-e^{ \alpha_6 }\right)} 
\nonumber\\ & \hskip 2cm \times 
\frac{ \left({\cal A}+1\right)^2\left(e^{ 6 \alpha }+1 \right) + \left(3 {\cal A}^2 -2 {\cal A} -1 \right) \left( e^{ 4 \alpha } + e^{ 2 \alpha } \right)} 
{\Bigl[ \left(e^\alpha-e^{\alpha_1}\right) \left(e^\alpha-e^{\alpha_2}\right) \left(e^\alpha-e^{\alpha_3}\right) \left(e^\alpha-e^{\alpha_4}\right) \Bigr]^2} \; , \nonumber\\
& \hskip -1cm =   \frac{ e^{\alpha_5} e^{\imath s  \alpha_5} \left(e^{ 2 \alpha_5 }-1\right) }{ e^{ \alpha_5 } \left(e^{ \alpha_5 }-e^{ \alpha_6 }\right)} 
\frac{ \left({\cal A}+1\right)^2\left(e^{ 6 \alpha_5 }+1 \right) + \left(3 {\cal A}^2 -2 {\cal A} -1 \right) \left( e^{ 4 \alpha_5 } + e^{ 2 \alpha_{5} } \right)} 
{\Bigl[ \left(e^{\alpha_5}-e^{\alpha_1}\right) \left(e^{\alpha_5}-e^{\alpha_2}\right) \left(e^{\alpha_5}-e^{\alpha_3}\right) \left(e^{\alpha_5}-e^{\alpha_4}\right) \Bigr]^2} \; , \nonumber\\
& \hskip -1cm  =   \frac{ -2 e^{ s  \pi / 2} }{ -2 \imath} 
\frac{ \left({\cal A}+1\right)^2\left(-1+1 \right) + \left(3 {\cal A}^2 -2 {\cal A} -1 \right) \left( 1 -1 \right)} 
{\Bigl[ \left(e^{\alpha_5}-e^{\alpha_1}\right) \left(e^{\alpha_5}-e^{\alpha_2}\right) \left(e^{\alpha_5}-e^{\alpha_3}\right) \left(e^{\alpha_5}-e^{\alpha_4}\right) \Bigr]^2} = 0\; , \nonumber\\
\label{resalpha5n4}
\end{align}
where in the first line it was used that
\begin{align}
e^{ 2\alpha }+1 &= \left(e^{\alpha}-e^{\alpha_5}\right) \left(e^{\alpha}-e^{\alpha_6}\right) 
= e^{2\alpha}-e^{\alpha}e^{\alpha_5}-e^{\alpha}e^{\alpha_6} + e^{\alpha_5+\alpha_6} 
 \; , \nonumber\\ &
 = e^{2\alpha}-\imath e^{\alpha}+\imath e^{\alpha} + 1 
= e^{2\alpha}+ 1 \; .
\end{align}

Since $\alpha_1$ and $\alpha_2$ are poles of second order, the pole expansion of the $n^{th}$ root, in the limit $\alpha \to \alpha_n$, from Eq.~\eqref{poleexp1} is modified to
\begin{align}
\left(e^\alpha-e^{\alpha_n}\right)^2 &= e^{2\alpha}\left(1-e^{\alpha_n-\alpha} \right)^2 = e^{2\alpha}\left(1-1-\left(\alpha_n-\alpha \right)-\frac{\left(\alpha_n-\alpha \right)^2}{2 !} - ...   \right)^2 \; , & \nonumber \\
&= e^{2\alpha} \left(\alpha-\alpha_n \right)^2 \left(1+\frac{\left(\alpha-\alpha_n \right)}{2 !} - ...   \right)^2 \; . & 
\label{poleexp2}
\end{align}

Utilizing the pole expansion in Eq.~\eqref{poleexp2} and a software of algebraic computation, the first term on the right-hand-side of Eq.~\eqref{Ialpharesapp} is given by
\begin{align}
Res(f,\alpha_1) &= \lim_{\alpha \to \alpha_1} \frac{d}{d \alpha} \left\{ \left( \alpha- \alpha_1 \right)^2 
  \frac{ e^{ \alpha \left(1 + \imath\; s \right) } \left(e^{ 2\alpha }-1\right) }{ e^{2 \alpha }\left( \alpha- \alpha_1 \right)^2 \left(e^{ 2\alpha }+1\right)} \right.
\nonumber\\* & \hskip 1cm \times \left.
\frac{({\cal A}+1)^2 \left(e^{6\alpha}+1\right)+(3{\cal A}^2-2{\cal A}-1) \left(e^{2\alpha}+e^{4\alpha} \right)} 
{\Bigl[ \left(e^\alpha-e^{\alpha_2}\right) \left(e^\alpha-e^{\alpha_3}\right) \left(e^\alpha-e^{\alpha_4}\right) \Bigr]^2} \right\} \; ,  \nonumber\\
 &\hskip -1.5cm = \frac{e^{\imath \theta_4- s \theta_4}}{4 \left(-1+ e^{2 \imath \theta_4}\right) \left(1+ e^{2 \imath \theta_4}\right)^3} 
\Bigl[ \imath {\cal A}^2 \left(1+ e^{2 \imath \theta_4}\right)^2 \left(2 \imath + s + e^{2 \imath \theta_4}s\right)+\left(-1+ e^{2 \imath \theta_4}\right) \nonumber \\
&\hskip -1.5cm  \times \left(2 + 6 e^{2 \imath \theta_4}- \imath s+ \imath e^{4 \imath \theta_4} s \right) + 2 {\cal A} \left(-1 + e^{2 \imath \theta_4} \right) \left(2 + 6 e^{2 \imath \theta_4} - \imath s + \imath e^{4 \imath \theta_4} s\right)
\Bigr] \; .
\label{resalpha1n4}
\end{align}
In the same way, the second term on the right-hand-side of Eq.~\eqref{Ialpharesapp} reads
\begin{align}
Res(f,\alpha_2) &= \lim_{\alpha \to \alpha_2} \frac{d}{d \alpha} \left\{ \left( \alpha- \alpha_2 \right)^2 
  \frac{ e^{ \alpha \left(1 + \imath\; s \right) } \left(e^{ 2\alpha }-1\right) }{ e^{2 \alpha }\left( \alpha- \alpha_2 \right)^2 \left(e^{ 2\alpha }+1\right)} \right.
\nonumber\\* & \hskip 1cm \times \left.
\frac{({\cal A}+1)^2 \left(e^{6\alpha}+1\right)+(3{\cal A}^2-2{\cal A}-1) \left(e^{2\alpha}+e^{4\alpha} \right)} 
{\Bigl[ \left(e^\alpha-e^{\alpha_1}\right) \left(e^\alpha-e^{\alpha_3}\right) \left(e^\alpha-e^{\alpha_4}\right) \Bigr]^2} \right\} \; ,  \nonumber\\
& \hskip -1.5cm = \frac{-e^{\imath \theta_4 + s \theta_4 - \pi s}}{4 \left(-1+ e^{2 \imath \theta_4}\right) \left(1+ e^{2 \imath \theta_4}\right)^3} 
\Bigl\{ \left(1+{\cal A}\right)^2 e^{6 \imath \theta_4} \left(2-\imath s \right)+e^{4 \imath \theta_4} \bigl[ 4+{\cal A}\left(8+2 \imath s\right) 
  \nonumber\\* &  +   
{\cal A}^2 \left(4-3 \imath s \right)+\imath s \bigr]-\imath \left(1+{\cal A}\right)^2 s + e^{2 \imath \theta_4} \bigl[-6+{\cal A}^2 \left(2-3 \imath s \right)  + \imath s
  \nonumber\\ &     \hskip 6.5cm +
 2 \imath {\cal A} \left(6 \imath + s \right) \bigr] \Bigr\} \; .
\label{resalpha2n4}
\end{align}

Inserting the residues from Eqs.~\eqref{resalpha5n4}, \eqref{resalpha1n4} and \eqref{resalpha2n4} into Eq.~\eqref{Ialpharesapp}, the integral $I_\alpha(s)$ becomes 
\begin{align}
I_\alpha(s)=& \frac{2\pi\imath}{e^{-\pi s/2} \left(e^{\pi s/2}+e^{-\pi s/2}\right)} \frac{e^{\imath \theta_4 - s \left(\pi + \theta_4 \right)}}{4 \left(-1+ e^{2 \imath \theta_4}\right) \left(1+ e^{2 \imath \theta_4}\right)^3}
\Bigl\{ \imath \left(1+{\cal A}\right)^2 e^{\pi s + 6 \imath \theta_4} s 
\nonumber \\* 
&+ \imath \left(1 + {\cal A} \right)^2 e^{2 s \theta_4} s + \imath \left(1+ {\cal A} \right)^2 e^{\pi s} \left(2 \imath + s \right) 
+ \imath \left(1+{\cal A} \right)^2 e^{2 \left( 3 \imath + s \right) \theta_4} \left(2 \imath + s \right) 
\nonumber \\* 
&+ e^{\pi s + 4 \imath \theta_4} \bigl[6 -2 \left( -6 + {\cal A} \right) {\cal A}+ \imath \left(-1+ {\cal A} \right) \left(1+ 3 {\cal A} \right) s \bigr] + e^{2 \theta_4 \left( \imath + s \right) } 
\nonumber \\* 
& \times \bigl[ 6 -2 \left(-6+ {\cal A} \right) {\cal A} + \imath \left( -1 + {\cal A} \right) \left(1 + 3 {\cal A} \right) s \bigr] +
 \imath \left( e^{\pi s + 2 \imath \theta_4} + e^{2 \left( 2 \imath + s \right) \theta_4} \right) 
\nonumber \\* 
& \times \bigl[4 \imath \left(1+{\cal A}\right)^2+\left(-1+{\cal A}\right)\left(1+3 {\cal A} \right) s \bigr] \Bigr\}   \; , \label{manual1}
\end{align}
which no software of algebraic computation is able to simplify. Manually manipulating Eq.~\eqref{manual1}, $I_\alpha(s)$ is found to be
\begin{align}
I_\alpha(s)=& \frac{2\pi\imath e^{\pi s/2}}{ 2 \cosh \left( \frac{\pi s}{2} \right)} \frac{e^{\imath \theta_4 - s \left(\pi + \theta_4 \right)}}{4 \left(-1+ e^{2 \imath \theta_4}\right) \left(1+ e^{2 \imath \theta_4}\right)^3} \nonumber \\*
&\times \Bigl\{ \imath \left(1+{\cal A}\right)^2  \bigl[ s e^{\pi s + 6 \imath \theta_4}  + s e^{2 s \theta_4} + \left(e^{\pi s} 
+ e^{2 \theta_4 \left( 3 \imath + s \right) } \right)  \left( 2 \imath + s \right) \bigr] \nonumber\\*
&+ \left(e^{\pi s + 4 \imath \theta_4} + e^{2 \theta_4 \left(\imath + s \right) } \right) \bigl[ 6 - 2 \left(-6 + {\cal A} \right) {\cal A} + \imath \left(-1+{\cal A} \right) 
 \left(1+3 {\cal A} \right) s \bigr] \nonumber\\*
 &+ \imath \left(e^{\pi s + 2 \imath \theta_4} + e^{2 \theta_4 \left(2 \imath + s \right) } \right) \bigl[ 4 \imath \left(1+ {\cal A} \right)^2 +\left(-1+{\cal A} \right) \left(1+3 {\cal A} \right) s \bigr] \Bigr\} \; ,
\end{align}
\begin{align}
I_\alpha(s)=& \frac{\pi \imath e^{- \pi s/2} e^{\imath \theta_4} e^{-s \theta_4} }{ 4 \cosh \left( \frac{\pi s}{2} \right) e^{\imath \theta_4} \left(e^{\imath \theta_4}-e^{-\imath \theta_4} \right)  e^{3 \imath \theta_4} \left(e^{\imath \theta_4}+e^{-\imath \theta_4} \right)^3} \nonumber \\*
& \hskip -0.5cm \times \Bigl\{ \imath \left(1+{\cal A}\right)^2  \bigl[ s e^{\pi s}  e^{6 \imath \theta_4}  + s e^{2 s \theta_4} + 2\imath e^{\pi s} +2 \imath e^{6 \imath \theta_4} e^{ 2 \theta_4 s } +s e^{\pi s} +s e^{6 \imath \theta_4} e^{ 2 \theta_4 s } \bigr] \nonumber\\*
& \hskip -0.5cm + \left(e^{\pi s + 2 \imath \theta_4 +  2 \imath \theta_4 } + e^{2 \theta_4 \left(\imath + s \right) } \right) \bigl[ 6 + 12 {\cal A} - 2 {\cal A}^2 + \imath s \left(-1 -2 {\cal A}+ 3 {\cal A}^2  \right) \bigr] \nonumber\\*
& \hskip -0.5cm + \imath \left(e^{\pi s + 2 \imath \theta_4} + e^{ 2 \theta_4 \left( \imath + s + \imath \right) } \right) \bigl[ s \left(-1 -2 {\cal A}+ 3 {\cal A}^2  \right) + \imath \left(4 + 8 {\cal A}+ 4 {\cal A}^2  \right) \bigr] \Bigr\} \; ,
\end{align}
\begin{align}
I_\alpha(s)=& \frac{\pi \imath e^{- \pi s/2} e^{-s \theta_4} e^{-3 \imath \theta_4} }{ 64 \imath  \cosh \left( \frac{\pi s}{2} \right) \sin \theta_4 \cos^3 \theta_4} \nonumber \\*
& \hskip -0.5cm \times \Bigl\{ \imath \left(1+{\cal A}\right)^2  \bigl[ s \left( e^{\pi s}  e^{6 \imath \theta_4}  +  e^{2 s \theta_4} +  e^{\pi s} + e^{6 \imath \theta_4} e^{ 2 \theta_4 s } \right) + 2\imath \left(e^{\pi s} + e^{6 \imath \theta_4} e^{ 2 \theta_4 s } \right) \bigr] \nonumber\\*
&\hskip -0.5cm +e^{\pi s/2} e^{s \theta_4} e^{3 \imath \theta_4} \left(e^{\pi s/2 + \imath \theta_4 -  s \theta_4 } + e^{-\pi s /2 -\imath \theta_4 +s \theta_4 } \right) \nonumber \\*
& \hskip 3.5cm  \times \bigl[ 6 + 12 {\cal A} - 2 {\cal A}^2 + \imath s \left(-1 -2 {\cal A}+ 3 {\cal A}^2  \right) \bigr] \nonumber\\
&\hskip -0.5cm +e^{\pi s/2} e^{s \theta_4} e^{3 \imath \theta_4} \left(e^{\pi s/2 - \imath \theta_4 -  s \theta_4 } + e^{-\pi s /2 +\imath \theta_4 +s \theta_4 } \right) \nonumber \\*
& \hskip 3.5cm  \times  \bigl[ \left(-4 - 8 {\cal A}- 4 {\cal A}^2  \right) + \imath s \left(-1 -2 {\cal A}+ 3 {\cal A}^2  \right)  \bigr] \Bigr\} \; ,
\end{align}
\begin{align}
I_\alpha(s)=& \frac{\pi }{ 64 \cosh \left( \frac{\pi s}{2} \right) \sin \theta_4 \cos^3 \theta_4} \nonumber \\*
& \hskip -1.3cm \times \Bigl\{ \imath \left(1+{\cal A}\right)^2  \bigl[ s \left( e^{ \pi s/2-s \theta_4+3 \imath \theta_4}+ e^{- \pi s/2+s \theta_4-3 \imath \theta_4}+ e^{ \pi s/2-s \theta_4-3 \imath \theta_4} + e^{ -\pi s/2+s \theta_4+3 \imath \theta_4} \right)  \nonumber\\*
&\hskip -1.3cm +2 \imath \left( e^{ \pi s/2-s \theta_4-3 \imath \theta_4}  + e^{- \pi s/2+s \theta_4+3 \imath \theta_4} \right) \bigr] \nonumber\\*
&\hskip -1.3cm + \left(e^{\pi s/2 + \imath \theta_4 -  s \theta_4 } + e^{-\pi s /2 -\imath \theta_4 +s \theta_4 } \right) \bigl[ 6 + 12 {\cal A} - 2 {\cal A}^2 + \imath s \left(-1 -2 {\cal A}+ 3 {\cal A}^2  \right) \bigr] \nonumber\\*
&\hskip -1.3cm +\left(e^{\pi s/2 - \imath \theta_4 -  s \theta_4 } + e^{-\pi s /2 +\imath \theta_4 +s \theta_4 } \right)  \bigl[ -4 - 8 {\cal A}- 4 {\cal A}^2  + \imath s \left(-1 -2 {\cal A}+ 3 {\cal A}^2  \right)  \bigr] \Bigr\} \; ,
\end{align}
\begin{align}
I_\alpha(s)=& \frac{\pi }{ 64 \cosh \left( \frac{\pi s}{2} \right) \sin \theta_4 \cos^3 \theta_4} \nonumber \\*
&\hskip -1.3cm\times \Bigl\{ \imath \left(1+{\cal A}\right)^2  \bigl[ s \left( e^{3 \imath \theta_4} \left(e^{ \pi s/2-s \theta_4}+ e^{- \left( \pi s/2-s \theta_4 \right)} \right) +e^{-3 \imath \theta_4} \left( e^{ \pi s/2-s \theta_4} + e^{ - \left( \pi s/2-s \theta_4\right)} \right)   \right)  \nonumber\\*
&\hskip -1.3cm+2 \imath \left( e^{ \pi s/2-s \theta_4-3 \imath \theta_4}  + e^{- \left( \pi s/2 - s \theta_4 - 3 \imath \theta_4 \right)} \right) \bigr] + \imath s \left(-1 -2 {\cal A}+ 3 {\cal A}^2  \right)  \nonumber\\*
&\hskip -1.3cm \times \bigl[ e^{\imath \theta_4 } \left(e^{ \pi s/2-s \theta_4}+ e^{- \left( \pi s/2-s \theta_4 \right)} \right) + e^{-\imath \theta_4 } \left(e^{ \pi s/2-s \theta_4}+ e^{- \left( \pi s/2-s \theta_4 \right)} \right) \bigr] \nonumber\\*
&\hskip -1.3cm+\left(e^{\pi s/2 + \imath \theta_4 -  s \theta_4 } + e^{-\pi s /2 -\imath \theta_4 +s \theta_4 } \right) \left( 6 + 12 {\cal A} - 2 {\cal A}^2  \right) \nonumber\\*
&\hskip -1.3cm+\left(e^{\pi s/2 - \imath \theta_4 -  s \theta_4 } + e^{-\pi s /2 +\imath \theta_4 +s \theta_4 } \right)  \left( -4 - 8 {\cal A}- 4 {\cal A}^2  \right) \Bigr\} \; ,
\end{align}
\begin{align}
I_\alpha(s)=& \frac{\pi }{ 64 \cosh \left( \frac{\pi s}{2} \right) \sin \theta_4 \cos^3 \theta_4} \nonumber \\*
&\hskip -1.0cm \times \Biggl\{ \imath \left(1+{\cal A}\right)^2  \Bigl\{ 4 s \cosh \left[ s \left( \frac{\pi}{2}-\theta_4 \right)\right] \cos \left( 3 \theta_4 \right) + 4 \imath  \cosh \left[ s \left( \frac{\pi}{2}-\theta_4 \right)-3 \imath \theta_4 \right] \Bigr\} 
\nonumber\\*
&+4 \imath s \left(-1 -2 {\cal A}+ 3 {\cal A}^2  \right) \cosh \left[ s \left( \frac{\pi}{2}-\theta_4 \right)\right] \cos \left(  \theta_4 \right) + 2 \left( 6 + 12 {\cal A} - 2 {\cal A}^2  \right) 
\nonumber\\*
& \times \cosh \left[ s \left( \frac{\pi}{2}-\theta_4 \right) + \imath \theta_4 \right] -8 \left(1 + {\cal A} \right)^2 \cosh \left[ s \left( \frac{\pi}{2}-\theta_4 \right) - \imath \theta_4 \right]  \Biggr\} \; ,
\end{align}
\begin{align}
I_\alpha(s)=& \frac{\pi }{ 64 \cosh \left( \frac{\pi s}{2} \right) \sin \theta_4 \cos^3 \theta_4} \nonumber \\*
&\hskip -1.3cm \times \Biggl\{ \imath \left(1+{\cal A}\right)^2  \biggl\{ 4 s \cosh \left[ s \left( \frac{\pi}{2}-\theta_4 \right)\right] \cos \left( 3 \theta_4 \right) + 4 \imath \Bigl\{ \cosh \left[ s \left( \frac{\pi}{2}-\theta_4 \right)\right] \cos \left( 3 \theta_4 \right) 
\nonumber\\*
&\hskip -1.3cm- \imath \sinh \left[ s \left( \frac{\pi}{2}-\theta_4 \right)\right] \sin \left( 3 \theta_4 \right) \Bigr\} \biggr\} 
+4 \imath s \left(-1 -2 {\cal A}+ 3 {\cal A}^2  \right) \cosh \left[ s \left( \frac{\pi}{2}-\theta_4 \right)\right] 
\nonumber\\*
&\hskip -1.3cm \times \cos \left(  \theta_4 \right) + 2 \left( 6 + 12 {\cal A} - 2 {\cal A}^2  \right) \Bigl\{ \cosh \left[ s \left( \frac{\pi}{2}-\theta_4 \right)\right] \cos \left( \theta_4 \right)+ \imath \sinh \left[ s \left( \frac{\pi}{2}-\theta_4 \right)\right] 
 \nonumber\\* 
& \hskip -1.3cm \times \sin \left(  \theta_4 \right) \Bigr\}
 -8 \left(1 + {\cal A} \right)^2 \Bigl\{ \cosh \left[ s \left( \frac{\pi}{2}-\theta_4 \right)\right] \cos \left( \theta_4 \right) - \imath \sinh \left[ s \left( \frac{\pi}{2}-\theta_4 \right)\right] 
 \nonumber\\* &  \times 
\sin \left(  \theta_4 \right) \Bigr\} \Biggr\} \; ,
\end{align}
where it was used that
\begin{align}
\cosh\left( x \pm \imath y \right) &= \cosh(x) \cosh(\imath y) + \sinh(x) \sinh(\pm\imath y) \; , \nonumber \\
&= \cosh(x)\cos(y) \pm \sinh(x) \sin(y) \; .
\end{align}
Then, continuing to calculate, $I_\alpha(s)$ reads
\begin{align}
I_\alpha(s)=& \frac{\pi }{ 64 \cosh \left( \frac{\pi s}{2} \right) \sin \theta_4 \cos^3 \theta_4} \Bigg\{ 
  4 \imath s \left(1+{\cal A}\right)^2 \cosh \left[ s \left( \frac{\pi}{2}-\theta_4 \right)\right] \cos \left( 3 \theta_4 \right) 
\nonumber\\* &
- 4  \left(1+{\cal A}\right)^2   \cosh \left[ s \left( \frac{\pi}{2}-\theta_4 \right)\right] \cos \left( 3 \theta_4 \right)
+ 4 \imath \sinh \left[ s \left( \frac{\pi}{2}-\theta_4 \right)\right] \sin \left( 3 \theta_4 \right)  
\nonumber\\* &
+ 4 \imath s \left(-1-2 {\cal A}+ 3 {\cal A}^2  \right) \cosh \left[ s \left( \frac{\pi}{2}-\theta_4 \right)\right]\cos\left(\theta_4\right) 
\nonumber\\* &
+ 4 \left( 3 + 6 {\cal A} - {\cal A}^2  \right) \cosh \left[ s \left( \frac{\pi}{2}-\theta_4 \right)\right] \cos \left( \theta_4 \right)
\nonumber\\* &
+ 4 \imath \left( 3+6{\cal A}-{\cal A}^2  \right) \sinh \left[ s \left( \frac{\pi}{2}-\theta_4 \right)\right] \sin\left(\theta_4\right) 
\nonumber\\* &
- 8 \left(1 + {\cal A} \right)^2 \cosh \left[ s \left( \frac{\pi}{2}-\theta_4 \right)\right] \cos \left( \theta_4 \right) 
\nonumber\\* &
+ 8 \imath \left(1 + {\cal A} \right)^2 \sinh \left[ s \left( \frac{\pi}{2}-\theta_4 \right)\right] \sin\left(\theta_4\right) \Biggr\} \; ,
\end{align}
\begin{align}
I_\alpha(s)=& \frac{\pi }{ 64 \cosh \left( \frac{\pi s}{2} \right) \sin \theta_4 \cos^3 \theta_4}  \Bigg\{ \Big\{ 
- 4  \left(1+{\cal A}\right)^2  \cos \left( 3 \theta_4 \right) 
+ \bigl[ 4 \left( 3 + 6 {\cal A} - {\cal A}^2  \right)  
\nonumber\\* & 
\hskip 4cm - 8 \left(1 + {\cal A} \right)^2 \bigr] \cos \left( \theta_4 \right)
\Big\} \cosh \left[ s \left( \frac{\pi}{2}-\theta_4 \right)\right]
\nonumber\\* 
& \hskip -0.5cm + \imath \Bigl\{
  4 s \left(1+{\cal A}\right)^2 \cosh \left[ s \left( \frac{\pi}{2}-\theta_4 \right)\right] \cos \left( 3 \theta_4 \right)   
+ 4 \sinh \left[ s \left( \frac{\pi}{2}-\theta_4 \right)\right] \sin \left( 3 \theta_4 \right)  
\nonumber\\* & 
\hskip -0.5cm+ 4 s \left(-1-2 {\cal A}+ 3 {\cal A}^2  \right) \cosh \left[ s \left( \frac{\pi}{2}-\theta_4 \right)\right]\cos\left(\theta_4\right) 
\nonumber\\* &
\hskip -0.5cm+ 4 \left( 3+6{\cal A}-{\cal A}^2  \right) \sinh \left[ s \left( \frac{\pi}{2}-\theta_4 \right)\right] \sin\left(\theta_4\right) 
\nonumber\\* &
\hskip -0.5cm+ 8 \imath \left(1 + {\cal A} \right)^2 \sinh \left[ s \left( \frac{\pi}{2}-\theta_4 \right)\right] \sin\left(\theta_4\right) 
\Big\} \Biggr\} \; ,
\end{align}
\begin{align}
I_\alpha(s)=& \frac{\pi }{ 64 \cosh \left( \frac{\pi s}{2} \right) \sin \theta_4 \cos^3 \theta_4}  \nonumber \\*
&\hskip -0.8cm \times \Bigg\{ 4 \cosh \left[ s \left( \frac{\pi}{2}-\theta_4 \right)\right] \Big[ 
- \left(1+{\cal A}\right)^2  \cos \left( 3 \theta_4 \right) 
+ \left( -3 {\cal A}^2 + 2 {\cal A} + 1  \right) \cos \left( \theta_4 \right)
\Big] 
\nonumber\\*
&\hskip -0.8cm  + \imath \Bigl\{
  4 s \cosh \left[ s \left( \frac{\pi}{2}-\theta_4 \right)\right] \Big[ 
 \left(1+{\cal A}\right)^2  \cos \left( 3 \theta_4 \right) 
+ \left( 3 {\cal A}^2 - 2 {\cal A} - 1  \right) \cos \left( \theta_4 \right)
\Big] 
\nonumber\\*
&\hskip -0.8cm + 4 \sinh \left[ s \left( \frac{\pi}{2}-\theta_4 \right)\right] \bigl[ \left(1+{\cal A}\right)^2  \sin \left( 3 \theta_4 \right)  +
 \left({\cal A}^2+10{\cal A}+5 \right)  \sin\left(\theta_4\right) \bigr]
\Bigr\} \Biggr\} \; .
\end{align}
Using that
\begin{align}
\sin ( 3 x ) =& 3 \sin(x) - 4 \sin^3(x) \; , \\*
\cos ( 3 x ) =& 4 \cos^3(x) - 3 \cos(x)  \; ,
\end{align}
the integral $I_\alpha(s)$ becomes
\begin{align}
I_\alpha(s)=& \frac{\pi }{ 64 \cosh \left( \frac{\pi s}{2} \right) \sin \theta_4 \cos^3 \theta_4}  \nonumber \\*
& \hskip -0.5cm  \times \Bigg\{ 16 \cosh \left[ s \left( \frac{\pi}{2}-\theta_4 \right)\right] \Big[ 
- \left(1+{\cal A}\right)^2  \cos^3 \left(\theta_4 \right) + \left(1+2 {\cal A}\right) \cos \left(\theta_4 \right)
\Big] 
\nonumber\\* 
& \hskip -0.5cm + \imath \Bigl\{
  16 s \cosh \left[ s \left( \frac{\pi}{2}-\theta_4 \right)\right] \Big[ 
 \left(1+{\cal A}\right)^2  \cos^3 \left( \theta_4 \right) 
- \left(1+  2 {\cal A} \right) \cos \left( \theta_4 \right)
\Big] 
\nonumber\\*
& \hskip -0.5cm + 16 \sinh \left[ s \left( \frac{\pi}{2}-\theta_4 \right)\right] \bigl[ -\left(1+{\cal A}\right)^2  \sin^3 \left( \theta_4 \right)  +
 \left({\cal A}^2+4{\cal A}+2 \right)  \sin\left(\theta_4\right) \bigr]
\Bigr\} \Biggr\} \; .
\end{align}
It is possible to deduce from Eq.~\eqref{theta4} that
\begin{align}
\sin ( \theta_4 ) =& \frac{\sqrt{{\cal A}({\cal A}+2)}}{{\cal A}+1} \;\;\;\; \text{and} \;\;\;\; 
\cos ( \theta_4 ) = \frac{1}{{\cal A}+1}  \; ,
\end{align}
and $I_\alpha(s)$ is
\begin{align}
I_\alpha(s)=& \frac{\pi \left( {\cal A}+1 \right)^4}{ 64 \cosh \left( \frac{\pi s}{2} \right) \sqrt{{\cal A}({\cal A}+2)}}  
 \Bigg\{ 16 \cosh \left[ s \left( \frac{\pi}{2}-\theta_4 \right)\right] \left( 
- \frac{1}{1+{\cal A}}  +\frac{1+2 {\cal A}}{1+{\cal A}}
\right) 
\nonumber\\* 
& + \imath \Bigl\{
  16 s \cosh \left[ s \left( \frac{\pi}{2}-\theta_4 \right)\right] \left( 
 \frac{1}{1+{\cal A}}  - \frac{1+2 {\cal A}}{1+{\cal A}}
\right) 
\nonumber\\*
&+ 16 \sinh \left[ s \left( \frac{\pi}{2}-\theta_4 \right)\right]  \frac{\sqrt{{\cal A}\left({\cal A}+2\right) }}{1+{\cal A}} \bigl[ -{\cal A}\left({\cal A}+2\right)  +  {\cal A}^2+4{\cal A}+2 \bigr] \Bigr\} \Biggr\} \; .
\end{align}
\begin{align}
I_\alpha(s)=& \frac{\pi \left( {\cal A}+1 \right)^4}{ 2 \cosh \left( \frac{\pi s}{2} \right) \sqrt{{\cal A}({\cal A}+2)}}  
 \Bigg\{ \frac{{\cal A}}{1+{\cal A}} \cosh \left[ s \left( \frac{\pi}{2}-\theta_4 \right)\right] 
\nonumber\\* 
& + \imath \Bigl\{
  \sqrt{{\cal A}\left({\cal A}+2\right) } \sinh \left[ s \left( \frac{\pi}{2}-\theta_4 \right)\right]
  - \frac{s {\cal A}}{1+{\cal A}}  \cosh \left[ s \left( \frac{\pi}{2}-\theta_4 \right)\right] 
  \Bigr\} \Biggr\} \; .
\end{align}

Finally, the real and imaginary part of $I_\alpha(s)$ are given by
\begin{equation}
\operatorname{Re}\;I_\alpha(s)= \frac{\pi ({\cal A}+1)^3 {\cal A}}{2 \sqrt{{\cal A}({\cal A}+2)} \cosh\left(\frac{s\pi}{2}\right)}  \cosh\left[s\left(\frac{\pi}{2}- \theta_4\right) \right] \; , 
\end{equation}
\begin{multline}
\operatorname{Im}\;I_\alpha(s)=\frac{\pi ({\cal A}+1)^4}{2 \sqrt{{\cal A}({\cal A}+2)} \cosh\left(\frac{s\pi}{2}\right)} \\* \times \left\{\sqrt{{\cal A}({\cal A}+2)}\sinh\left[s\left(\frac{\pi}{2}- \theta_4\right) \right] 
-\frac{s\; {\cal A} }{{\cal A}+1}\cosh\left[s\left(\frac{\pi}{2}- \theta_4\right) \right]\right\} \; . \label{n4imapp}
\end{multline}



\end{spacing}

\backmatter
\bibliographystyle{ThesisStyle}
\bibliography{referencias_filipe1} \addcontentsline{toc}{chapter}{Bibliography}
\interlinepenalty=10000 

\end{document}